%% file: thesis.tex
\documentclass[a4paper,12pt,times,authoryear,index]{PhDThesisPSnPDF}
\usepackage[grey]{quotchap}
\makeatletter
\renewcommand\chapter{%
  \if@openright\cleardoublepage\else\clearpage\fi
  \thispagestyle{plain}%
  \global\@topnum\z@
  \null\hfill\@printcites\par% NEW
  \@afterindentfalse
  \secdef\@chapter\@schapter
}
\renewcommand{\@makechapterhead}[1]{%
  \chapterheadstartvskip%
  {\size@chapter{\sectfont\raggedright% NEW
    {\chapnumfont
      \ifnum \c@secnumdepth >\m@ne%
      \if@mainmatter\thechapter%
      \fi\fi
      \par\nobreak}%
    {\raggedright\advance\leftmargin10em\interlinepenalty\@M #1\par}}% NEW
  \nobreak\chapterheadendvskip}}
\makeatother
\input{preamble}

\input{bibtex-info}

\newcommand\blfootnote[1]{%
  \begingroup
  \renewcommand\thefootnote{}\footnote{#1}%
  \addtocounter{footnote}{-1}%
  \endgroup
}

\begin{document}

%\frontmatter
%\title{Understanding the Solar Magnetic Field Generation Process, Their Evolution and Irregularities }
\title{Understanding the behavior of the Sun's large scale magnetic field and its relation with the meridional flow }

\submitdate{JULY 2017}
\degree{Doctor of Philosophy}
\dept{Joint Astronomy Programme\\
Department of Physics}
\faculty{Faculty of Science}
\author{Gopal Hazra}

%\begin{titlepage}
  \maketitle
%\end{titlepage}
%--------------------------------------
\null
\thispagestyle{empty} 
\newpage

%--copyright page-----------------------
\vspace*{\fill}
\begin{center}
\large\bf \textcopyright \ Gopal Hazra\\
\large\bf July 2017\\
\large\bf All rights reserved
\end{center}
\vspace*{\fill}
\thispagestyle{empty}

%----------copyright ends---------------------

\setcounter{secnumdepth}{3}
\setcounter{tocdepth}{3}

\pagenumbering{roman}

\include{declaration}
\include{dedication}

\include{acknowledgement}
\include{preface}

\include{publication}

% *********************** Adding TOC and List of Figures ***********************

\tableofcontents
\listoffigures

\listoftables

% \printnomencl[space] space can be set as 2em between symbol and description
%\printnomencl[3em]

%\printnomencl

% ******************************** Main Matter *********************************

\mainmatter
\setcounter{page}{1}
\include{chapter1}

\include{chapter2}

\include{chapter3}

\include{chapter4}
\include{chapter5}

\include{chapter6}

\include{chapter7}

\include{chapter8}

%\begin{spacing}{0.9}
%\bibliographystyle{apalike}
%\bibliographystyle{plainnat} % use this to have URLs listed in References
%\cleardoublepage
%\bibliography{references} % Path to your References.bib file
% ********************************** Appendices ********************************

%\begin{appendices} % Using appendices environment for more functunality

%\include{appendix}

%\end{appendices}
%=======================References============================================
%\bibliographystyle{plainnat}
%\bibliographystyle{LivRevSolar}
\bibliographystyle{mnras}
\bibliography{myref}
\printthesisindex
\end{document}

%% file: bibtex-info.tex
\newcommand{\aap}{    {\it A \& A}}

\newcommand{\apj}{    {\it ApJ}}
\newcommand{\apjl}{   {\it ApJL}}
\newcommand{\apjs}{   {\it ApJ Suppl.}}
\newcommand{\apss}{   {\it Astrophys. Space Sci.}}

\newcommand{\grl}{    {\it Geophys. Res. Lett.}}

\newcommand{\mnras}{  {\it MNRAS}}
\newcommand{\nat}{    {\it Nature}}
\newcommand{\pasp}{   {\it Pub. Astron. Soc. Pac.}}

\newcommand{\solphys}{{\it Sol. Phys.}}

\newcommand{\ssr}{    {\it Space Sci. Rev.}}
\newcommand{\araa}{    {\it Anu Rev. in Astronomy and Astrophysics.}}

%% file: declaration.tex
\begin{declaration}
I hereby declare that the work reported in this doctoral thesis titled ``Understanding the behavior of the Sun's large scale magnetic field and its relation with the meridional flow'' is entirely original and is the result of
investigations carried out by me in the Department of Physics, Indian Institute of Science,
Bangalore, under the supervision of Prof. Arnab Rai Choudhuri and Prof. Dipankar Banerjee.

I further declare that this work has not formed the basis for the award of any degree,
diploma, fellowship, associateship or similar title of any University or Institution.

\end{declaration}

%% file: dedication.tex
% ******************************* Thesis Dedidcation ********************************

\begin{dedication} 
\centering

\LARGE{\it To My Parents}

\end{dedication}

%% file: acknowledgement.tex
\begin{acknowledgements}
The memorable journey to the final stage of Ph.D. and completion of the thesis have only been possible with the support and encouragement of numerous people including my friends, my well-wishers, and colleagues. First and foremost, I thank my supervisors Arnab Rai Choudhuri and Dipankar Banerjee for their continued support, their patience, and their wise advice. They have provided me with opportunities beyond compare. All of the stimulating discussions with them throughout the years have always nurtured my fascination for solar physics. I must mention that the books `The Physics of Fluids and Plasmas' and `Astrophysics for Physicists' written by Arnab Rai Choudhuri have been always with me from the very first day of my Ph.D. and I have learned a lot from these books. I hope to write one day as well as he does regularly.

I wish to thank Mark Miesch for giving me the opportunity to work with him. He has always treated me as one of his own student. His deep insights and positive outlook have contributed greatly to my own improvement. I thank Mausumi Dikpati, Kinfe Teweldebirhan, Hideyuki Hotta, Kyle Augustson, Junfeng Wang and Lokesh Bharti for
those memorable days during my stay in HAO, Boulder. My thanks also go to my senior Bidya Binay Karak,
who walked these steps before me and helped me a lot to find my own way. I also thank the past and present members of Dipu da's group, Tanmoy, Vaibhav, Sudip, Subhamoy, Rakesh, Girju da, Krishna da, Chandru da, Vemareddy and Manjunath in Indian Institute of Astrophysics for various discussions and help.    
  
I would take this opportunity to thank all of the faculty members, postdocs and students of our Astrophysics group for their constant supports and discussions. Special thanks to Prateek Sharma and Chanda J. Jog who taught us Numerical Methods and Galaxy \& ISM respectively during our graduate coursework. Thanks to Prateek again for answering various queries regarding numerical methods during my research.
%And Prateek, you are right: We shouldn't be  afraid of writing small small codes! 
I also thank all of the JAP instructors who taught us in our coursework. 

I also thank my seniors Sujit Kumar Nath, Upasana Das, Indrani Banerjee, Nazma Islam, Arpita Roy, Samyaday Choudhury and Susmitha Anthony for the discussion and help during my Ph.D. days. 

Special thanks to my batchmates- Kartick, Abir, Sreehari, Soumavo, Deovrat, and Mohan. I cherish
many evenings and late nights which we have spent together at Gymkhana and Faculty club. Those days were really
memorable.

Thanks to Soumavo and Prasun for being fantastic friends and supporting me as and when required. Thanks to Debasish da, Siddhartha, Prakriti, Ajay and Tirthankar for their help and discussion in many aspects. 

A big thank you to all the nice people of IISc. IISc is enjoyable and wonderful because of you. I need to finish my thesis, so I would name only a few: Rahool, Krishna Prasad, Sukanya, Kingshuk, Malay, Hemanta, Sumanta, Phani, Gopi, Kaji, Subham, Sudipta, Debasmita, Soumi, Arijit, Rudra, Sourav, Rupak, Pradip, Monojit, Somnath \& whole math group, Debabrata, Soumen, Adhip, Amit, Pushpender, Koushik, Arpan (UG) \& Co., Swarup da, Sudip da, Somnath da, Amiya da, Apurba da, Nafiza di, Bidisha di, Indra da and members of Bandooz group.  

Finally, I would like to thank my parents and my loving sister for their constant patience and support
through the years. They have supported me with love on every step of this adventure. I appreciate the freedom my parents
have given me to pursue my studies in physics. I also thank my elder brother Subrata for encouraging me to pursue study in physics. 

The simulations were performed in SahasraT (SERC, IISc), Yellowstone (NCAR) and Pleiades (NASA). I thank two referees of my thesis, Prof. Paul Charbonneau and Prof. Prasad Subramanian for their helpful suggestions which helped a lot to improve the thesis. Financial supports from Council of Scientific and Industrial Research (India) as well as the J. C. Bose (DST, India) Fellowship awarded to Arnab Rai Choudhuri are also acknowledged.

\end{acknowledgements}

%% file: preface.tex
\begin{abstract}
Our Sun is a variable star. The magnetic fields in the Sun play an important role for the existence of a wide variety of phenomena on the Sun. Among those, sunspots are the slowly evolving features of the Sun but solar flares and coronal mass ejections are highly dynamic phenomena. Hence, the solar magnetic fields could affect the Earth directly or indirectly through the Sun's open magnetic flux, solar wind, solar flare, coronal mass ejections and total solar irradiance variations. These large scale magnetic fields originate due to Magnetohydrodynamic dynamo process inside the solar convection zone converting the kinetic energy of the plasma motions into the magnetic energy. Currently, the most promising model to understand the large scale magnetic fields of the Sun is the Flux Transport Dynamo model. In this thesis, various studies leading to better understanding of the large scale magnetic fields of the Sun are performed using the Flux Transport Dynamo (FTD) models.  

FTD models are mostly axisymmetric models, though the non-axisymmetric 3D FTD models are also started to develop recently. Just like other physical models, FTD models have various assumptions and approximations for different processes which are responsible for the generation of large scale magnetic fields. Some of the assumptions are observationally verified and some of them are not till date. Considering the availability of resources, many approximations have been made in these models on the theoretical basis. Magnetic buoyancy is one of the important processes in these models. We discuss in details about how magnetic buoyancy has been treated in axisymmetric FTD models and the advantages and disadvantages of the different treatments. We finally realize that a proper treatment of the magnetic buoyancy needs a 3D treatment which motivates us to build a more realistic 3D Dynamo model. 

The irregular solar cycles show some interesting properties in their ascending and descending phases. The properties of the solar cycle in the rising phase (e.g., Waldmeier effect) of the cycle have been already well recognized and well explained, but the properties in the decaying phase remain less noticed. We touch upon these properties in the decaying phase in great details. We find some interesting correlations in the decaying phase from Greenwich data and as well as Kodaikanal observatory data. We provide an explanation of this observed correlation from FTD models.  

Recent developments of the helioseismology challenge one of the very important assumptions of the FTD models. FTD models assume a single cell meridional circulation encompassing the whole convection zone of the Sun having a poleward flow on the surface and an equatorward return flow near the bottom of the convection zone. After helioseismology discovers that the meridional circulation might have a different spatial structure in the convection zone, we investigate this issue and justify that FTD models work perfectly fine as long as there is an equatorward propagation of the meridional flow at the bottom of the convection zone. Although the observations about the spatial structure of the meridional flow are found recently, temporal variation of meridional flow with the solar cycle has been found almost a decade ago. Since FTD models operate in the kinematic regime, we are not able to take into account the Lorentz Force feedback. But if we take this Lorentz force feedback into our model, we believe that the observed variation of the meridional circulation would be reproduced. We find that this is indeed the case! We just consider the effect of the Lorentz force feedback on the velocity equation as a perturbation over the steady flow and solve the perturbed equation coupled with the dynamo equations. By doing so, We are able to reproduce the observed variation of the meridional flow with the solar cycle. 

Finally, we present some results with a 3D FTD model. Development of 3D dynamo model is necessary for plenty of reasons. Various processes in the solar dynamo are inherently 3D. The magnetic buoyancy, because of which the magnetic flux tube rises through the solar convection zone and creates sunspots is a 3D process. The decay of sunspots, and corresponding dispersion and migration of the sunspots fields by different convective flows and build up of polar field i.e. altogether the Babcock-Leighton process is also a 3D process. Modeling these 3D processes with an axisymmetric 2D FTD model is difficult. In 2D model, they are treated in a very clever but simplistic way which sometimes is not enough to capture whole physics of the problem. We explain these issues and their modeling procedures in 3D FTD models. We get results which are distinguishable from the results obtained by axisymmetric FTD models. The 3D FTD models open up plenty of windows to understand large scale solar magnetic fields and solar cycle much more realistically. For the first time, we incorporate observed surface convective flows data in 3D FTD model with the highest fidelity. We describe an initial implementation of the data feeding in these models which is very important for realistic treatment of Babcock-Leighton process.            
\end{abstract}

%% file: publication.tex
\begin{publication}
%\large{
\begin{enumerate}
\item {\bf Hazra, G.}, Karak, B. B., \& Choudhuri, A. R., 2014, {\it Is a Deep One-cell Meridional Circulation Essential for the Flux Transport Solar Dynamo?} {\bf ApJ, 782}, 93\\
\item {\bf Hazra, G.}, Karak, B. B., Banerjee, D. \& Choudhuri, A. R. 2015, {\it Correlation Between Decay Rate and Amplitude of Solar Cycles as Revealed from Observations and Dynamo Theory,} {\bf Sol. Phys.
290}, 6\\
\item Choudhuri, A. R. \& {\bf Hazra, G.}, 2016, {\it The treatment of magnetic buoyancy in flux transport dynamo models,} {\bf Advances in Space Research, 58}, 8\\
\item {\bf Hazra, G.}, \& Choudhuri, A. R., \& Miesch, M. S., 2017, {\it A Theoretical Study of the Build-up of the Sun’s Polar Magnetic Field by using a 3D Kinematic Dynamo Model,} {\bf ApJ, 835}, 39\\
\item Mandal, S., Hegde, M., Samanta, T., {\bf Hazra, G.}, Banerjee, D. \& Ravindra, B. 2017, {\it Kodaikanal digitized white-light data archive (1921-2011): Analysis of various solar cycle features,} {\bf Astronomy \& Astrophysics, 601}, 106\\
\item {\bf Hazra, G.} \& Choudhuri, A. R., 2017, {\it A theoretical model of the variation of the meridional
circulation with the solar cycle,} {\bf MNRAS, 472}, 2728\\
\item {\bf Hazra, G.} \& Miesch, M. S., 2017, {\it Incorporating Surface Convection into a 3D Babcock-Leighton Solar Dynamo Model,} {\bf ApJ}, Under review, arXiv:1804.03100\\

\end{enumerate}
%}
\end{publication}

%% file: chapter1.tex
\begin{savequote}[100mm]
``It is not too much to hope that in
a not too distant future we shall be
competent to understand so simple
a thing as a star."

\qauthor{--Sir Arthur Stanley Eddington, 1926}
\end{savequote}
\chapter{Introduction}
\label{C1}
%\begin{quote} \small
% A brief description of your chapter.
%\end{quote}

The importance of the Sun in human life and human civilization is well recognized
from the ancient time. The Indians, the Greeks, the Romans, the Egyptians, and other ancient civilizations always worshiped the Sun, because they believed that the Sun is
the source of our lives on earth. The Sun has always been an important wonder for
mankind from the early days of civilization. Even in the 21st century, where it
would not have been possible to step ahead without the use of technology, the study
of the Sun is much more relevant and important. With the advancement of technology, the realization
about importance of studying the Sun has also increased.   

A great step towards understanding the effect of the Sun on earthly phenomena like Aurora, the geomagnetic storm was
started after the observations of magnetic fluctuations by Anders Celsius and his
assistant Olof Hiorter from Uppsala, Sweden, in 1741, who found that the magnetic
fluctuations occurred at the same local time, as aurorae were sighted. Celsius also 
found that these magnetic disturbances associated with the aurora in Sweden were 
simultaneously observed in England by George Graham, who was a pioneering observer
in geomagnetic variations. With this discovery, it became very clear that the magnetic
disturbances associated with Aurora were global rather than regional character. It was 
around 1806, Alexander Von Humboldt also spent many tedious hours to record the variability 
of the magnetic fields, and in the first-half of the 1800s, it was well appreciated that 
magnetic disturbances associated with aurorae were global in scale and nearly simultaneous 
everywhere. By 1837, Dennison Olmstead argued that the cause for the aurorae must exist
outside of the Earth due to the global scope of the auroral-magnetic phenomenon \citep{heliobook2}. Historically, 
in 1843, Heinrich Schwabe discovered the solar cycle, and in 1852 Sabine showed a 
detailed correlation between the sunspot cycle and the frequency of auroral displays \citep{Sabine1852}. 
This was a remarkable discovery for the Sun and Earth connection, and a space 
influence outside of the Earth was established as the cause of Aurora. Around the same time, 
on 1st September 1859 the amateur English astronomer Richard C. Carrington while 
monitoring the sunspots noticed two rapidly brightening patches of light near the middle
of a sunspot group. This is the first record of solar flare which was observed by human 
and presently known as the two-ribbon flare. The double Great Aurora of August 28 to September2 
1859 was believed to occur because of a pair of Coronal Mass Ejections (CMEs) was ejected 
from the Sun on or about August 27 and September 1. The first CME impacted Earth one day 
later on 28th August and the second faster CME is the Carrington's flare, 
observed on September 1. The observed Carrington's flare and corresponding occurrence of the Aurora 
made it clear that the Sun is influencing the Earth in many ways, and the sunspots which are 
the main drivers for violent flare and CME are mostly responsible for it. 

\begin{figure}[!t]
\centerline{\includegraphics[width=0.75\textwidth,clip=]{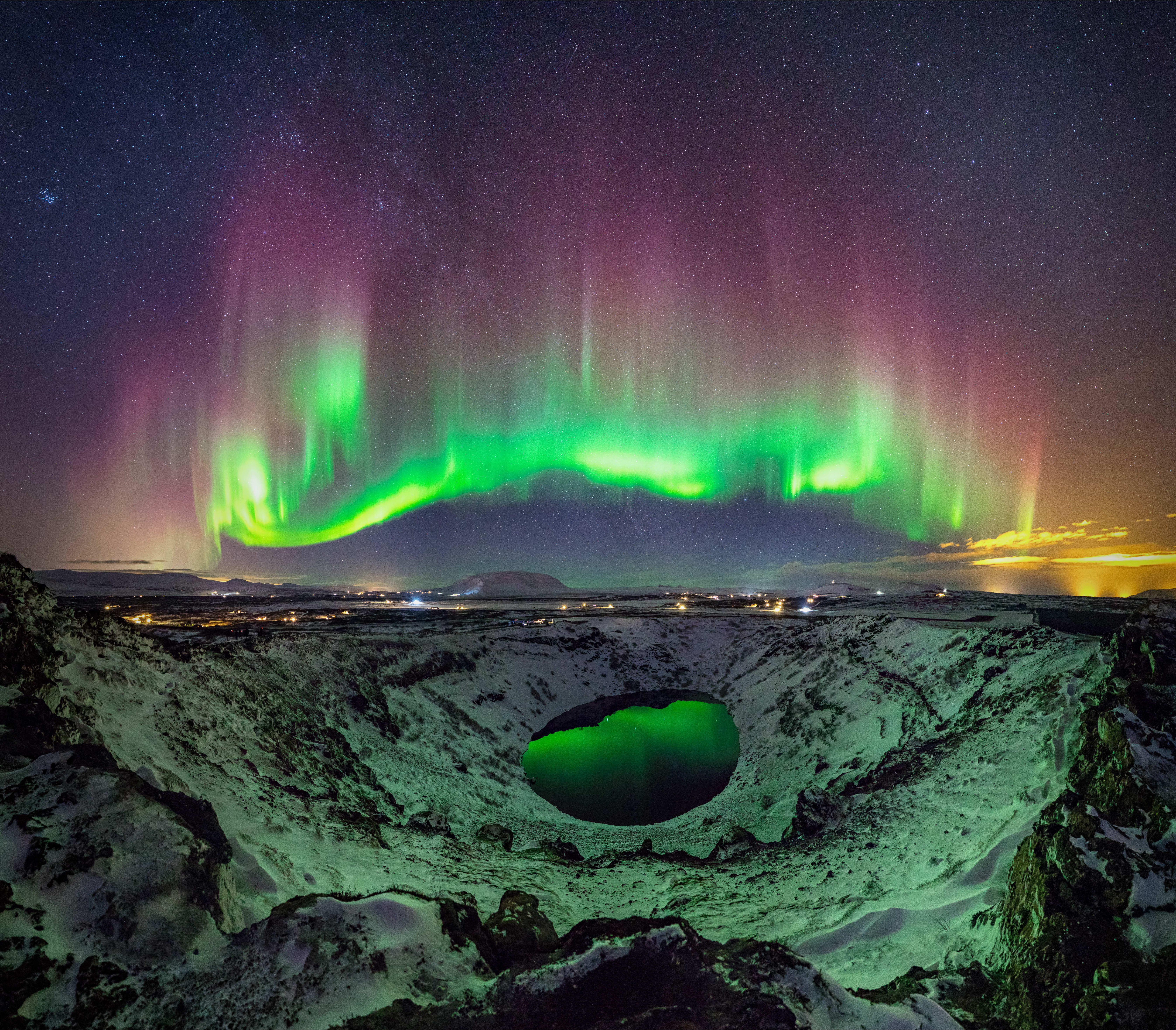}}
\caption[Aurora visible on 6th March, 2017 from Iceland]{Aurora visible on 6th March, 
2017 from Iceland. Picture Courtesy: Sigurdur William Brynjarsson }
\label{figc1:aurora}
\end{figure}

While Scientists were busy to study the interrelation between solar activity 
with aurora and magnetic storms, some other things also kept on 
affecting people's daily life. During eighteenth and nineteenth centuries, 
magnetic compasses were the main high-tech frontier which was used for navigation. 
But in many places, disturbances of several degrees from $2^{\circ}$ to $8^{\circ}$ 
in compass measurement was reported for several hours (Lovering, 1857) making 
navigation unworkable. During great auroral display of September 2, 1859, the 
disturbances of the magnetic needle were almost $4^{\circ}$ in half an hour. 
The worst scenario reported for this disturbance in navigation due to magnetic 
storms was on Sept. 24, 1946. The Los Angeles Times reported that `` Brussels, 
Sept 23 (AP), Budget Minister Josept Merlot today said: abnormal weather 
condition and the aurora borealis might have put the instruments out of order 
on the Sabena airlines plane that crashed near Gander, Newfoundland killing 26 persons". 
Not only the navigation systems but also the electric telegraphy systems got affected 
badly due to solar activity at that time. During an Aurora on Nov 17, 1848, Carlo Matteucci, 
the Director of Telegraphs in Pisa observed an anomalous behavior in telegraph 
connecting Pisa and Florence. He also noticed that electromagnets remained powered 
even without the battery attached, and ceased once the aurora dimmed. 
Solar activity has another impact on humans by disturbing radio communications. 
To communicate from one part with the other part of the Earth, a radio wave must be 
reflected from the ionosphere. During the solar flare, radio transmissions are disrupted 
by changes in the ionosphere layers due to x ray emissions from the flare. Other impacts 
on radio communications happen due to high energy protons coming from the Sun. 
It causes ionization of ionosphere layers, and signals ranging from approximately 
3MHz through 40 MHz get attenuated by the absorption process creating blackouts 
of High Frequency (HF) and Very High Frequency (VHF) radio communications. 
Another direct effect of violent behavior of the Sun is on electrical power grids. 
The most famous outage occurred in Quebec on March 13, 1989, where a complete 
electrical blackout happened affecting 3 million people. Satellites are also 
getting affected by the solar X-ray and flare heating. There is a clear correlation 
between the sunspot cycle and the number of de-orbited satellites in Low-Earth orbit 
\citep{Odenwald05}. Satellite anomalies were also reported when violent solar flare occurred 
in the Sun (e.g, Telstar-1, Anik E1 and E2, GOES-7, TDRS-1). Apart from affecting 
the man-made technologies, solar activity affects the Earth's climate as well. 
The total ultraviolet radiation coming from the Sun varies with the 
solar cycle which changes the Earth's temperature \citep{Shindell99} 
and during 1645-1725, there were no sunspots in the Sun, which is assumed to be responsible for the little ice age 
in Europe and North America \citep{Eddy76}. Although, there are studies showing that other factors (e.g., volcanic eruptions, internal climate variability) are also responsible for the little ice age along with the solar activity \citep{Owens17}.

Therefore, the Sun has a profound impact on human life and their activity 
on the Earth. As time goes on, we are becoming more and more dependent on technology. 
Hence more vulnerable to solar activity. So to control technological damages, 
electrical blackouts, satellite anomalies and long term effect on the Earth's climate, 
we need to understand the violent explosions of the Sun. Solar flares, Coronal mass 
ejections, solar wind and high energy particles coming from the Sun 
are mostly responsible for affecting our Earth, and it is found that these phenomena 
are happening above the sunspot regions. When the Sun has lots of sunspots on its 
surface, it becomes more violent and becomes quiet, when there are fewer sunspots. 
That's why, we see a strong correlation between the sunspots number with the aurora 
frequency, magnetic storms, no. of satellites de-orbited and satellites anomalies. 
In order to understand thoroughly, how the Sun is going to affect our Earth, a study 
is necessary to understand when why and how sunspots are forming on the Sun's surface and how do they evolve. 

Sunspots are the regions of highly concentrated magnetic fields of the Sun. 
After Heinrich Schwabe discovered that sunspot numbers wax and wane with 
a period of roughly 10 years, researchers started thinking the reason behind it. 
In 1908, George Ellery Hale first discovered the magnetic field in the sunspots 
\citep{Hale1909} and it became very clear that the sunspots cycles are nothing 
but the magnetic cycle of the Sun. Just like our Sun, other stars also have 
starspots activity cycle \citep{Noyes84a}. After 100 years of discovery of magnetic 
fields in sunspots, we are able to understand many things about the sunspots 
and its cycle, but still many things about the Sun and sunspots remain unknown 
and we keep on wondering about them.                

\section{History of sunspot observations} 
\label{C1.S1}
As we already mentioned in the previous section that sunspots are the regions 
of strong concentrated magnetic fields in the photosphere of the Sun, they 
appear as very tiny dark blemishes in the solar disc. The size of the sunspots 
are also very small, and it is difficult to observe by the naked eye. 
But some of the sunspots are large enough or many sunspots can appear in a 
group, allowing them to be observed in the naked eye with suitable viewing conditions, 
where the Sun can be partially obscured by cloud or fog or thick mist or during 
sunrise and sunset time. The first naked eye observations of sunspots were made by 
Chinese in 800 BC. During that time, astronomers at the court of the Chinese and 
Korean emperors made regular notes of sunspots because of their possible astrological 
significance \citep{Stephenson90}. After that in the pre-telescopic era, sunspots 
observations were recorded by many others starting with Theophrastus (374–287 B.C.). 
But first sunspot drawing was found in the Chronicles of John of Worcester from a 
sighting on Saturday, 8 December 1128. Before the discovery of the telescope, there 
was confusion whether these sunspots are part of the Sun or some planets 
which are blocking the light to reach the Earth. In 1610, Galileo first 
observed the sunspots using his newly discovered telescope and it was found 
that the sunspots were associated with the Sun itself. 230 years after the 
telescopic discovery of the sunspots, it was Heinrich Schwabe, 
an amateur astronomer who recorded the sunspots for the period 1826 to 1843 
and pointed out that sunspots follow a period of roughly 10 years between maxima in their number. 
In between telescopic discovery of sunspots and discovery of its cycle, the Sun has gone through a prolonged 
minimum where no sunspots were seen from 1645 to 1715. In 1887 and 1889, German astronomer 
Gustav Sp\"{o}rer published two papers showing a remarkable 70 years interruption in the ordinary 
course of the solar cycle. While studying the latitudinal distribution of sunspots, Sp\"{o}rer had 
found evidence that the numbers of spots in the northern and southern hemisphere of the Sun 
were not always balanced, and to check this observation he consulted historical records of 
sunspots from 17th and early 18th century, and discovered 70 years of interruption in sunspots number. 
After Sp\"{o}rer died, E. W. Maunder took the responsibility to investigate more about 
Sp\"{o}rer findings. In 1894, he pointed out in his article entitled 
``Prolonged Sunspot Minimum" providing more details and acknowledging Sp\"{o}rer findings  
that the sunspots were really not seen in that period (1645-1715) and it may happen in the Sun again \citep{Eddy76}. 

\begin{figure}[!htbp]
\centerline{\includegraphics[width=1.0\textwidth,clip=]{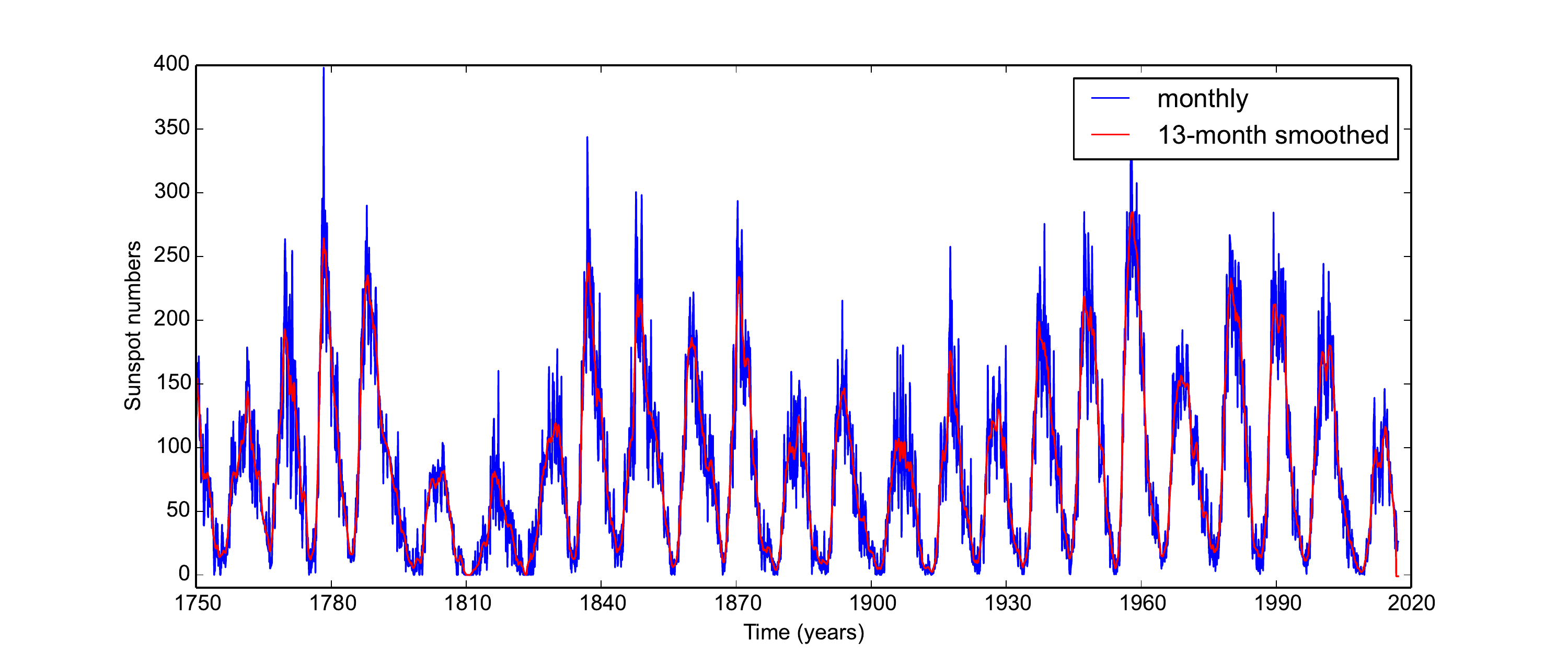}}
\caption[Time series of Sunspot Numbers]{Monthly sunspot numbers are plotted 
in blue lines with time from SIDC database. Yearly smoothed monthly sunspots 
numbers are over plotted with red line}
\label{figc1:ssn}
\end{figure}

After Schwabe's discovery of solar cycle, Rudolf Wolf, director of the Observatory 
at Bern and later at Zurich, organized a number of European observatories to record 
sunspots on regular basis and by a standard scheme. Wolf defined the relative sunspot number as
\begin{equation}
\label{eq:C1.1}
r = k(f + 10g)
\end{equation} 
where f is the number of individual spots including those distinguishable within 
groups on the Sun, g is the number of sunspots group and k is a correction factor 
which varies from one observer to another. Of course for Wolf's own observation k is used as 1. 
This procedure is still used for sunspot number calculations and this relative 
sunspot number is known as Wolf's sunspot number. After defining the Wolf's sunspot number 
and inaugurating an international effort which still continues today, Wolf started 
reconstructing the sunspot number from historical data and successfully reconstructed 
the data of sunspot number as far as the the 1755–1766 cycle, which conventionally 
known as cycle 1. According to this convention, we are in the declining phase of 
cycle 24 today (7th March, 2017). Rudolf Wolf's novel effort and initiative to make 
a universal method of sunspot monitoring helped a lot to better understand the behavior 
of sunspots with time. In Fig~\ref{figc1:ssn}, the entirely revised historical sunspot 
number series is shown from SIDC \citep{Clette14}. Recently a century long sunspot area 
data acquired from persistent observation at the Kodaikanal observatory in India are also available \citep{Sudip17}.
Apart from the sunspot numbers and sunspot area, a number of other indices are now available
to measure solar activity. They include various plage indices and the 10.7 cm solar radio flux \citep{Hathaway10a}. The latter 
have advantage particularly because they can be recorded for any type of weather conditions, but they are 
only available for last 5 cycles from 1946 (see chapter \ref{C3}).

\section{Magnetic properties of Sunspots and Solar Cycle}
\label{C1:S2}
In 1908, George Ellery Hale observed a splitting in sunspot spectral line 
and by measuring Zeeman splitting in the observed spectral line, 
he estimated the value of magnetic field in the sunspots is approximately 
3000 G \citep{Hale1909}. This discovery accelerated the understanding 
of the sunspots and solar cycle. It was understood then that the solar 
cycle is nothing but the magnetic cycle of the Sun and the magnetic field 
is continuously generated inside the Sun by some mechanism to maintain its activity cycle. About a decade after 
this ground breaking discovery, George Ellery Hale and his collaborators made 
another extremely important discovery. They found that sunspots are often seen 
side by side with opposite polarity i.e. the sunspots are bipolar in nature. 
The magnetic polarity of sunspots are also completely opposite in the two 
hemispheres \citep{Hale19} and these polarities in the two hemispheres also 
change with the solar cycle giving magnetic cycle period twice of the period 
of sunspot cycle. This is known as the Hale's polarity law. In the very same paper \citep{Hale19}, 
it was shown that the line joining of the two poles of bipolar sunspots 
always had an inclination with the East-West line of the Sun with the leading 
polarity sunspots are closed to the equator and following polarity sunspots 
are away from the equator. This inclination angle also known as tilt angle is 
found to vary linearly with latitudes. As latitudes increase, 
the tilt angles also increase from $3^{\circ}$ near equator to 
$11^{\circ}$ at $30^{\circ}$ latitude. This latitudinal dependence of 
tilt angle is known as Joy's law, and as we will see in 
latter chapters, it has a huge implication in solar dynamo theory. 
There is also an interesting fact about the latitudinal distribution of sunspots. 
In the beginning of the cycle, they mostly come in higher latitudes 
($30^{\circ}$ to $40^{\circ}$) and as cycle progresses, they are drifted more 
and more towards lower latitudes. This latitudinal drift of sunspots was 
first noticed by Richard Carrington during 1853-1861 which was known 
as Sp\"{o}rer's law and Edward Maunder in 1904, published the time 
latitude plot of sunspots for two cycles starting from 1877 to 1902 
which is now known as the butterfly diagram.

\begin{figure}[!b]
\centerline{\includegraphics[width=0.5\textwidth,clip=]{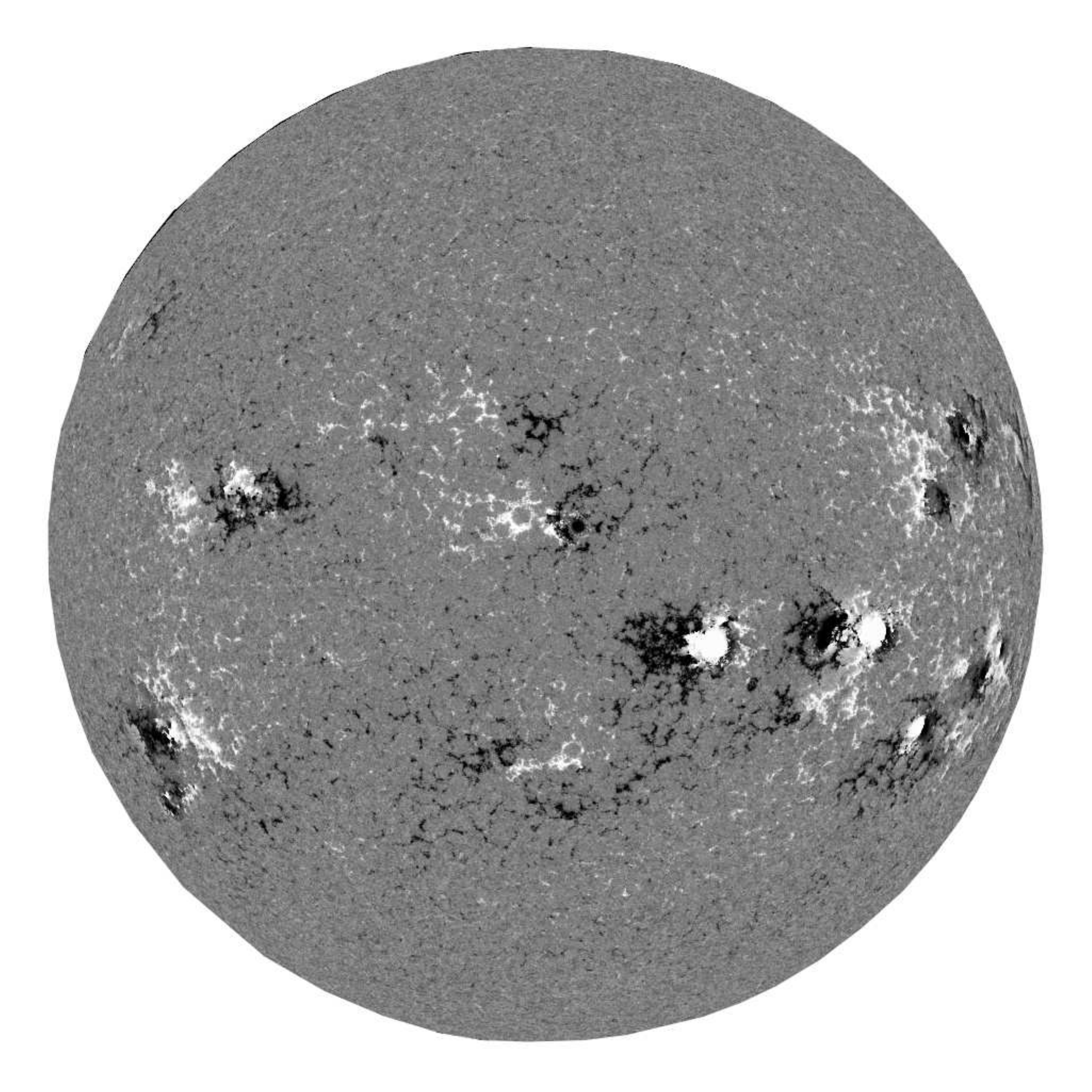}}
\caption[Solar Magnetogram]{Solar Magnetogram was taken on 7th July, 2014 by HMI on board 
SDO during solar maxima. The black color represents the magnetic fields which are directed
towards us and white color represent the magnetic field away from us. Grey color is used 
for diffuse field in the solar photosphere}
\label{figc1:mgrm}
\end{figure}

Hale's pioneering work about the measurement of magnetic field in sunspots 
galvanized interest about the measurement of the weak diffuse general magnetic
fields on the surface of the Sun. The spectroheliograph designed by George Ellery Hale was able to measure
the magnetic field about 2-3 kilo Gauss value i.e. mainly the magnetic field in the 
sunspots. But to measure the very weak diffuse magnetic field, a more-precise 
instrument was necessary which could detect very tiny line shift in the 
Fraunhofer line. For example, a typical sunspot field (2-3 kG) would give 
the line displacement about is $0.1 \text{\AA}$ which is easily detectable but 
for magnetic field of 1 Gauss, line displacement for Fraunhofer line is approximately 
$8\times10^{-5}$ \text{\AA} which was extremely difficult to detect \citep{Babcock53}. 
However, in the early 1950s, H.D Babcock and H. W Babcock, a father
son collaboration successfully designed a magnetogram which could 
readily measure very tiny line shift corresponding the weak
general fields of the Sun having a value of 1-20 Gauss. This was also a remarkable invention.
With this magnetogram, they were able to measure definite poloidal field on the surface of the
Sun. They found a weak positive field in the north polar cap ($\theta > 55^{\circ}$) and a weak negative
field in the southern polar cap \citep{Babcock55} giving an antisymmetric dipolar structure of the 
Sun's magnetic fields. The reversal of polar fields also 
found during the sunspot maxima suggesting a strong correlation between
poloidal field and the sunspots cycle.  
  
\begin{figure}[!t]
\centerline{\includegraphics[width=0.9\textwidth,clip=]{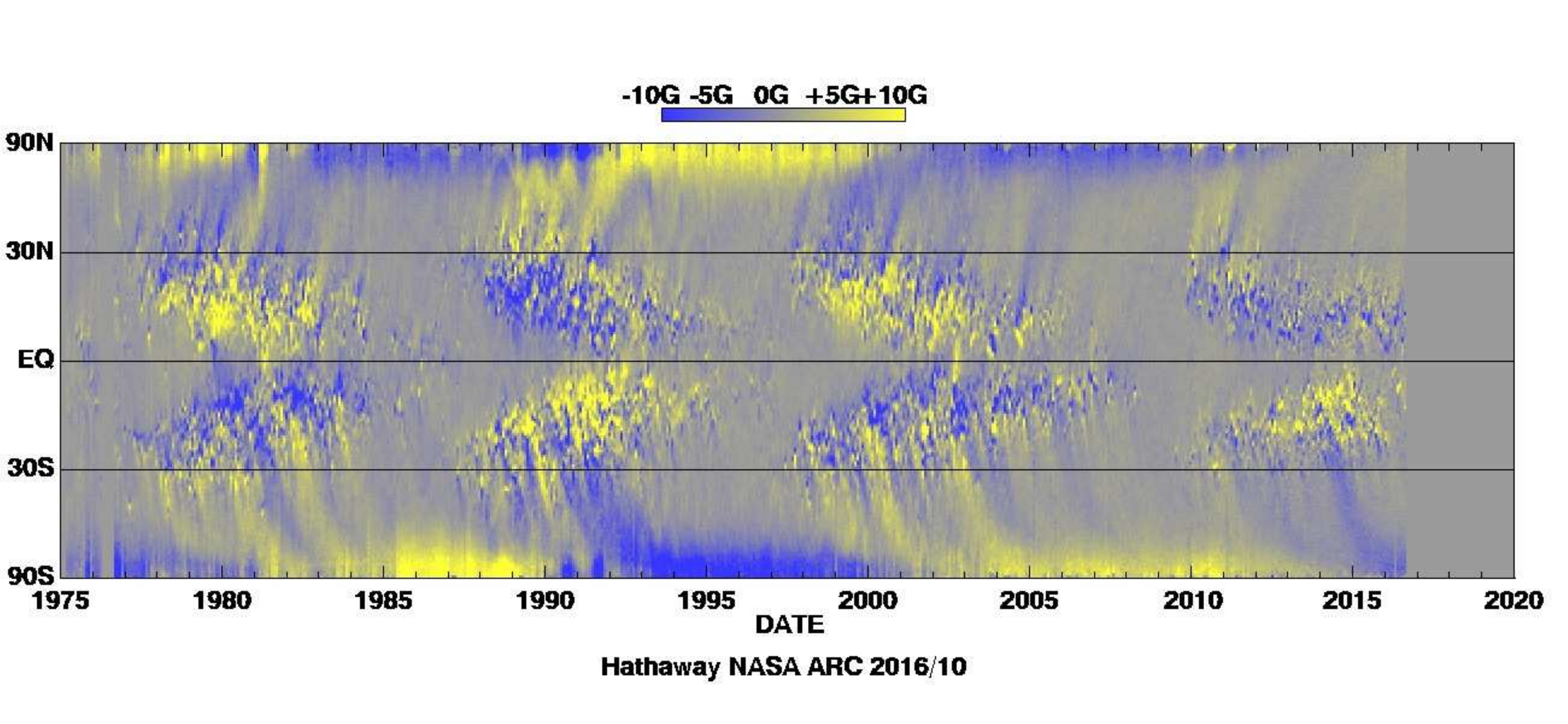}}
\caption[Observed butterfly diagram]{Time latitude plot of radial fields as observed 
on the surface of the Sun. Courtesy: David Hathaway}
\label{figc1:bfly}
\end{figure}

In Fig.~\ref{figc1:mgrm}, a magnetogram is shown from HMI of onboard SDO 
during solar maxima on 7th of July 2014. In both of the hemispheres near low latitude, 
sunspots fields are shown in white and black colors. Black represents the positive
polarity sunspots and white represents the negative polarity sunspots. If we make a 
time-latitude plot by taking an azimuthal average of
magnetic fields as shown in magnetogram, it gives us the butterfly diagram as shown in Fig.~\ref{figc1:bfly}.  
A careful scrutiny of full solar disc magnetogram shows us 
some very small fibril type of structures throughout the solar disc 
apart from the large black and white spots near the equator. These small
structures come due to small scale magnetic fields which mostly lie in the granular and supergranular
boundaries. The large scale structures or the large scale magnetic fields are believed to be generated
by a hydromagnetic dynamo inside the solar convection zone, whereas the small scale magnetic fields are
generated due to the interaction of convection with the turbulent motion in the photosphere \citep{Cattaneo99}.
The equatorward migration of sunspot fields are very clear from the Figure~\ref{figc1:bfly} 
along with the fields from the following polarity migrating towards the pole. These poleward migrating
fields are very important because while migrating towards the pole, they reverse the existing polar field from 
the previous cycle. These photospheric features are the only key to understand what is happening inside 
the solar convection zone. A successful theoretical model should be able to explain all the 
characteristic of the butterfly diagram. 
In the next section (Section~\ref{C1.S3}), we describe theoretical background to understand 
the generation mechanism of magnetic field in the Sun.

Let's discuss some properties of the solar cycle now. The average period of the solar cycle
is 11 years but a solar cycle period can vary from minimum 9 years to maximum 14 years. The amplitude
of the solar cycle is also not same for each cycle (see Fig~\ref{figc1:ssn}). The peak amplitudes of sunspot numbers 
vary from cycle to cycle. The weak cycle has less number of sunspots and the strong cycle has more number of sunspots. This irregular
behavior of the solar cycle makes it very difficult and interesting to predict the next cycle maximum. There are mainly
two widely used methods to predict the amplitude of the next cycle. First, the precursor method, where the information about
some of the cycle characteristics is studied in a cycle to predict the amplitude of the following cycle \citep{Petrovay10}. 
Another method relies on the more physical ground. They are mostly based on the observationally data driven theoretical 
predictions \citep{CCJ07}. Among some of the characteristics of the solar cycle, Waldmeier effect is very famous and
important one \citep{Waldmeier35}. As we see in Fig~\ref{figc1:ssn} that the solar cycles are mostly asymmetric in nature. 
They take 3-4 years to reach its maxima after minimum and take 7-8 years to decay from maxima to its next minimum. In 1935,
Waldmeier first noticed an anticorrelation between rise time of the cycle and its maxima. The stronger cycle takes less
time to reach its maxima after the minimum and vice versa. While this correlation (rise time vs cycle maxima) is not robust in 
all of the solar activity indices (e.g., sunspot area), though see \citet{Sudip17}, the rise rate, and cycle maxima has a very good
strong correlation. The theoretical explanation for these correlations is given in \citet{KarakChou11} based on 
flux transport dynamo theory. Although the rise time of the cycle shows a good correlation with its maxima, decay time
shows exactly no correlation with its maxima or even with the next cycle maxima, rather decay rate of the cycle has a good 
correlation with its maxima and with next cycle maxima \citep{HKBC15}. The detail description of correlations of the decay rate with
the maxima of the same cycle and next cycle and their theoretical explanation are given in chapter~\ref{C3}.   

\section{Theoretical Background}  
\label{C1.S3}
While various facts about sunspots and solar cycle were being dug out by the
observations, some theoretical models were necessary to understand them. Two very
important discipline of physics `Magnetohydrodynamics' and `Magnetoconvection' were
concurrently being developed which explained many of these observational features of 
magnetic fields of the Sun.
 
\subsection{Basic equations of Magnetohydrodynamics}
\label{C1:S3_1}
Magnetohydrodynamics is a discipline of physics which studies the 
motion of plasma in presence of magnetic field.
As we know that the systems where the length scale is larger than the Debye length and
time scale much larger than the inverse of plasma frequency, the charge separation can
be neglected and we can apply MHD approximation there,
provided the plasma velocities are non-relativistic. It is basically one fluid approximation 
of plasma in presence of magnetic fields. The plasma motions in the Sun are non-relativistic 
and slowly varying under the mechanical and magnetic forces making a perfect place 
to apply MHD approximations. The velocity fields and magnetic fields are governed by the following
coupled equations:
\begin{equation}
\label{eqnc1:MHD1}
\frac{\partial {\bf v} }{\partial t} + ({\bf v}.\nabla){\bf v}  = {\bf F} - \frac{1}{\rho}\nabla p + \frac{1}{\rho c}{\bf j}\times{\bf B} + \nu \nabla^2{\bf v}
\end{equation}
\begin{equation}
\label{eqnc1:MHD2}
\frac{\partial {\bf B}}{\partial t} = \nabla \times ({\bf v}\times{\bf B}) + \eta\nabla^2{\bf B}
\end{equation}
The Equation~\ref{eqnc1:MHD1} is the magnetic Naviour-Stokes equation similar to what we get in
hydrodynamic case but only differs by the presence of magnetic force. The equation~\ref{eqnc1:MHD2} 
is a whole new equation known as induction equation which determines the behavior of
the magnetic field in presence of velocity fields and in arriving this equation, the scalar form of Ohm's law is used. All the transport coefficients 
(e.g., viscosity and magnetic diffusivity) in the equations~\ref{eqnc1:MHD1} 
and \ref{eqnc1:MHD2}, are assumed to be scalar and constant in space. 
But in the complex realistic astrophysical scenario those are mostly tensor
with spatial dependence, and easy to include those complex terms in equations~\ref{eqnc1:MHD1} 
and \ref{eqnc1:MHD2}. The RHS of the equation~\ref{eqnc1:MHD1} includes
the body force (F), the pressure force, the magnetic force and the viscous force. 
For non-relativistic plasma, we can write $\nabla \times {\bf B} = \frac{4\pi}{c}{\bf j}$ 
and magnetic force term becomes $\frac{1}{4\pi\rho}{(\nabla\times{\bf B}) \times{\bf B}}$. 
After rearranging the magnetic force term, the new velocity equation can be written as:
\begin{equation}
\label{eqnc1:MHD3}
\frac{\partial {\bf v} }{\partial t} + ({\bf v}.\nabla){\bf v}  = {\bf F} - \frac{1}{\rho}\nabla\left(p + \frac{B^2}{8\pi}\right)+\frac{({\bf B}.\nabla){\bf B}}{4\pi\rho} + \nu \nabla^2{\bf v}
\end{equation}
It is very clear from equation~\ref{eqnc1:MHD3} that the magnetic force gives
rise of magnetic pressure $\frac{B^2}{8\pi}$, and a magnetic tension $({\bf B}.\nabla){\bf B}$ 
along the magnetic field lines. The magnetic pressure has a huge importance in the dynamics of 
flux tube as we see in the next subsection~\ref{C1:S3_2}.

The generation and evolution of magnetic field are carried out by the induction equation
(Equation~\ref{eqnc1:MHD2}) due to electrically conducting fluid motions. The RHS of induction
equation has two terms and both of them have very important significance. First term,
$\nabla\times({\bf v}\times{\bf B})$ is mainly responsible for generation of the magnetic
field (source term) and the second term $\eta\nabla^2{\bf B}$ helps to decay the field (decay term). 
It is similar to an electrical system where varying current gives rise to the magnetic field and magnetic field decays due 
ohmic dissipation of current. But instead of the current, the plasma motions are doing the job here. The ratio of these two terms (1st term to 2nd term)
is known as magnetic Reynolds number $R_m = \frac{LV}{\eta}$, where $L$ and $V$ represents the typical length scale and velocity,
and $\eta$ is magnetic diffusivity of the system. The value of $R_m$ determines the relative importance of these two terms.
For astrophysical systems, the $R_m$ is very high and the first term dominates over the second term. 
Hence, For $R_m >> 1$ the induction equation can be written as:
\begin{equation}
\label{eqnc1:highrm}
\frac{\partial {\bf B}}{\partial t} = \nabla \times ({\bf v}\times{\bf B})
\end{equation}
This equation has very important consequences for the astrophysical dynamo. It was Hannes Olof G\"{o}sta Alfvén, 
a pioneer in the study of the astrophysical magnetic fields who first pointed out that for astrophysical systems
with high $R_m$ value, magnetic field is almost frozen in the plasma and moves with the plasma flows. 
Mathematically, $$\frac{d}{dt}\int_S {\bf B}.d{\bf S} = 0.$$ 
This theorem is known as Alfven's theorem of flux freezing and has a huge implication in the study of the astrophysical 
dynamo as well as in solar dynamo. For this contribution in the MHD, he was awarded Nobel prize in 1970. 
Therefore to study a system, more specifically to study the magnetic field 
generation process self consistently in a system due to plasma motion inside it, 
we need to solve the MHD Equations~\ref{eqnc1:MHD1} and \ref{eqnc1:MHD2}
along with an appropriate equation of state, energy equation and the equation of continuity. 

\begin{figure}[!t]
\centerline{\includegraphics[width=0.97\textwidth,clip=]{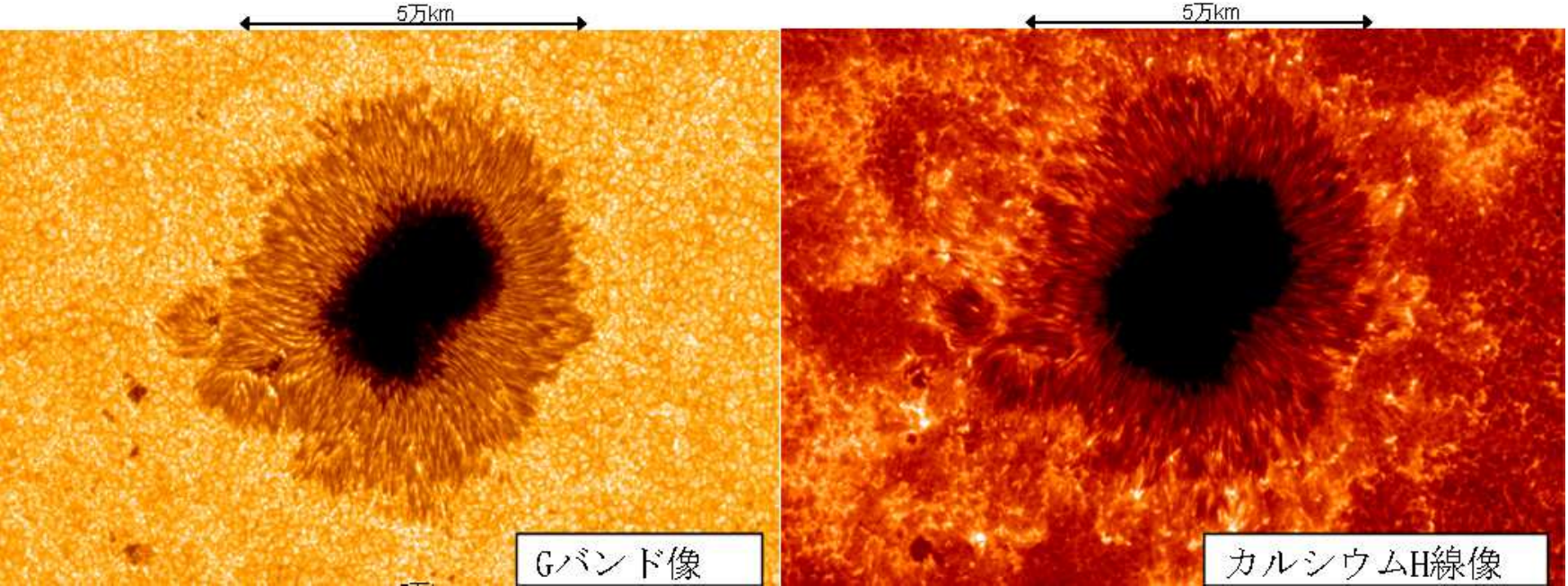}}
\caption[Image of sunspot in G band and in calcium H line]{Sunspot images are shown from Solar Optical Telescope, Hinode.
The left panel is the photospheric image of the sunspot taken in the wavelength of the G band (430 nm). 
The chromospheric images in calcium H (396 nm) line is shown in right panel. 
The scales are shown with a black arrow have a length of 50000 km.
Courtesy: Hinode website}
\label{figc1:sunspots}
\end{figure}

\subsection{Magnetoconvection}
\label{C1:S3_2}
Magnetoconvection studies interaction of magnetic field with the thermal convection. The study of magnetoconvection
was started after the discovery of strong magnetic field in the sunspots and realizing that the sunspots are relatively
cooler than its surroundings. The study of Magnetoconvection is almost on the same footing as the thermal convection. But the MHD 
equations are perturbed here instead of perturbing hydrodynamic equations (as done in Rayleigh-Benard convection) in order 
to find the conditions for which the magnetofluid becomes unstable for convection. No doubt, this stability analysis is much 
more complicated than the case without magnetic fields. The calculations of Rayleigh-Benard convection in presence 
of magnetic field was first carried out by \citet{Thomson51} and \citet{Chandra52}. From their analysis, 
it was confirmed that the presence of magnetic field suppresses the convection or in other words, strong magnetic 
fields make magnetofluid more stable against convection. The physical reason behind this is the magnetic tension. 
As convection starts distorting the magnetic field lines, convective motions are opposed by the magnetic tension. 
Therefore, a steeper temperature gradient is necessary to drive the convection in presence of strong magnetic field.

The perturbative analysis can give us the condition for convection to start. But detail understanding about the nature of 
magnetoconvection needs rigorous simulation using all nonlinear equations with the parameters above the critical 
parameters for the onset of convection. The sunspots structure and surroundings granular structure as shown in Figure~\ref{figc1:sunspots}
provide a very good environment to study magnetoconvection. Also, it is found from the solar magnetogram study \citep{Solanki02c} that the magnetic field is vertical at the center of the spot and becomes increasingly inclined towards the periphery of it. The convective flows
are inhibited by the magnetic tension and make it cooler than its surroundings. The small scale structures i.e., the cellular patterns surrounding 
the sunspots are known as the supergranular and granular structure (left panel of Fig~\ref{figc1:sunspots}). These small scale structures are confined
in the photosphere and not visible in the chromospheric image (right panel of Fig~\ref{figc1:sunspots}). 
As magnetic tension opposes the convection, in supergranules and granules also, there is a partitioning of space in between magnetic 
field and convection. The dark lines in around the cells are the small inter-granular size magnetic field elements, 
where the convection is suppressed. For detail simulation of sunspots using magnetoconvection, see \citet{Rempel2011}.

\begin{figure}[!htbp]
\centerline{\includegraphics[width=0.65\textwidth,clip=]{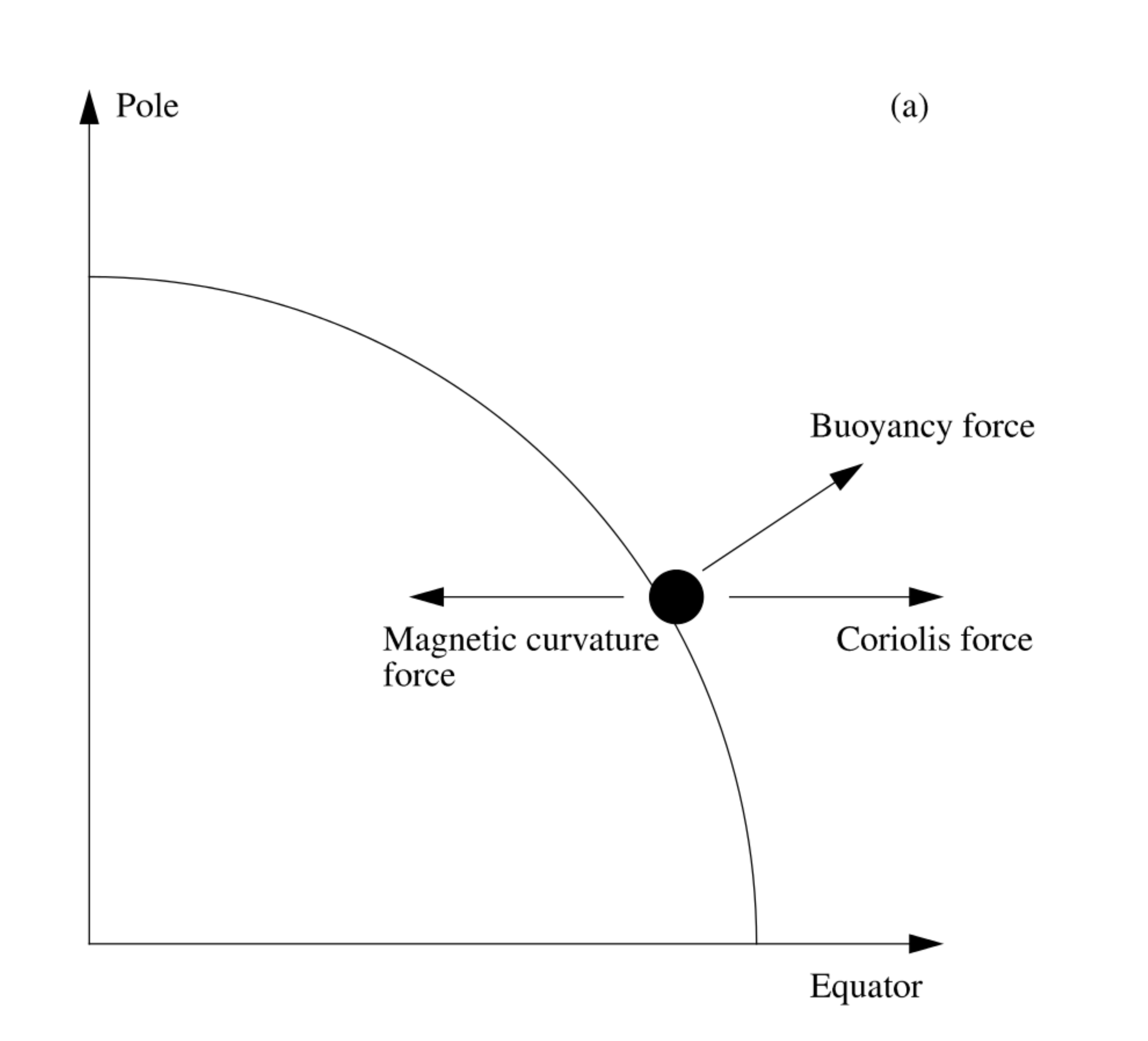}}
\caption[Isolated Flux tube]{Schematic diagram based on \citet{SM02} from \citet{Fan09} showing various
forces acting on an isolated toroidal flux ring at the base of solar convection zone}
\label{figc1:flux1}
\end{figure}

\subsection{Flux tube dynamics and field amplification}
\label{C1:S3_3}
\begin{figure}[!t]
%\centerline{\includegraphics[width=0.75\textwidth,clip=]{Figures/chap1/flux_buoyant.pdf}}
\centerline{\includegraphics[width=0.85\textwidth,clip=]{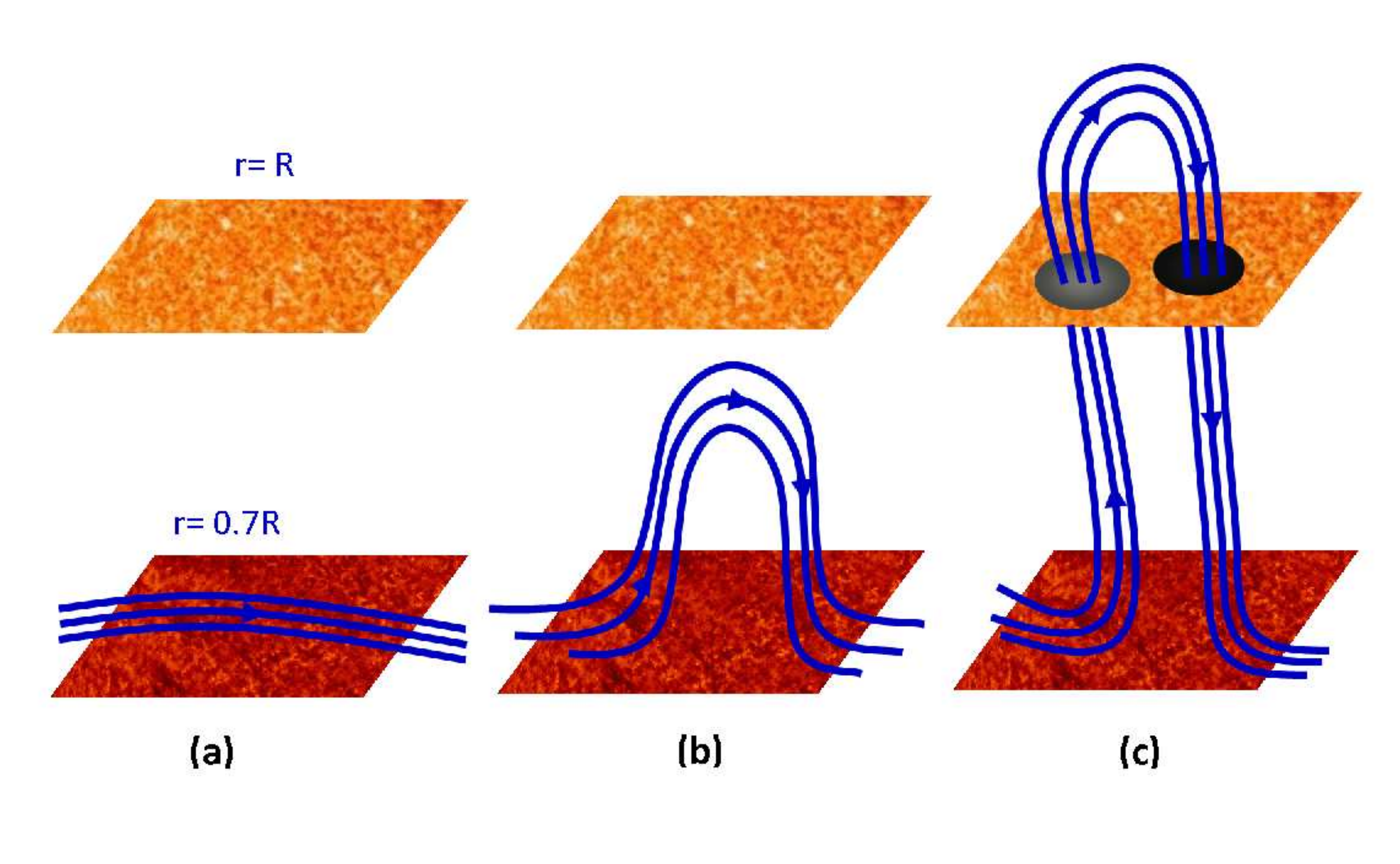}}
\caption[Buoyant rising of flux tube]{Buoyant rise of a flux tube from the bottom of the convection zone to 
the surface. (a) Nearly horizontal flux tube under the solar surface near base of the convection zone. 
(b) Flux tube after its upper part rises through the convection zone and (c) The flux tube emerges 
above the solar surface and creates Bipolar sunspots.}
\label{figc1:flux2}
\end{figure}

Probing the magnetic field inside the solar convection zone is an extremely difficult task 
and still, it is not possible with available modern resources. The observed east west 
orientation of the sunspots and Hale's polarity law of sunspots (Section~\ref{C1:S2}), 
make it very clear that the subsurface fields are well organized, coherent and in the azimuthal 
direction. This well-organized nature of solar active regions also confirms that the magnetic 
fields are not disrupted by the abrupt convective motions inside the solar convection zone, 
so they must be very strong at least same order of magnitude or more than the equipartition 
value ($10^4$ G) of the convective motions. Another very important behavior of magnetic flux 
is they appear as intense intermittent elements in the flux-free, or nearly flux-free plasma. 
As the plasma inside solar convection zone is very highly conducting, magnetic fields are 
almost frozen in plasma which helps to develop this flux tube like structure \citep{Cattaneo06}. 
Let us consider an isolated flux tube in the solar convection zone as shown in Figure~\ref{figc1:flux1}.
Different forces acting on it have also been shown. As we see from the Figure~\ref{figc1:flux1} that the magnetic 
curvature force is balancing the Coriolis force, and the buoyancy force is the only 
deciding factor whether the flux tube is in equilibrium or going to rise. If the pressure inside the 
flux tube is $p_i$ and the pressure in the outside ambient medium is $p_e$, then we can write,
\begin{equation}
\label{eqnc1:buoy}
p_e = p_i + \frac{B^2}{8\pi}
\end{equation} 
where $\frac{B^2}{8\pi}$ is the magnetic pressure exerted by the magnetic fields inside the flux tube. 
It is evident from Equation~\ref{eqnc1:buoy} that $p_i \leq P_e$, which often but not always implies that the density $\rho_i$ inside the flux tube is less than the density $\rho_e$ outside the flux tube. When this happens to a part of a flux tube, it becomes buoyant and rises up against gravitational field. A buoyant flux tube near the bottom of the solar convection
zone is shown in the figure~\ref{figc1:flux2}(a).
As we see in figure~\ref{figc1:flux2}(a), the whole flux tube may not be magnetically buoyant but some part of 
it becomes very light compared to its surroundings (For details see \citet{Chou98,Fan09}) and rises up to the surface against the gravity
(figure~\ref{figc1:flux2}(b)) and creates bipolar sunspots as shown in Figure~\ref{figc1:flux2}(c).

\begin{figure}[!t]
\centerline{\includegraphics[width=0.75\textwidth,clip=]{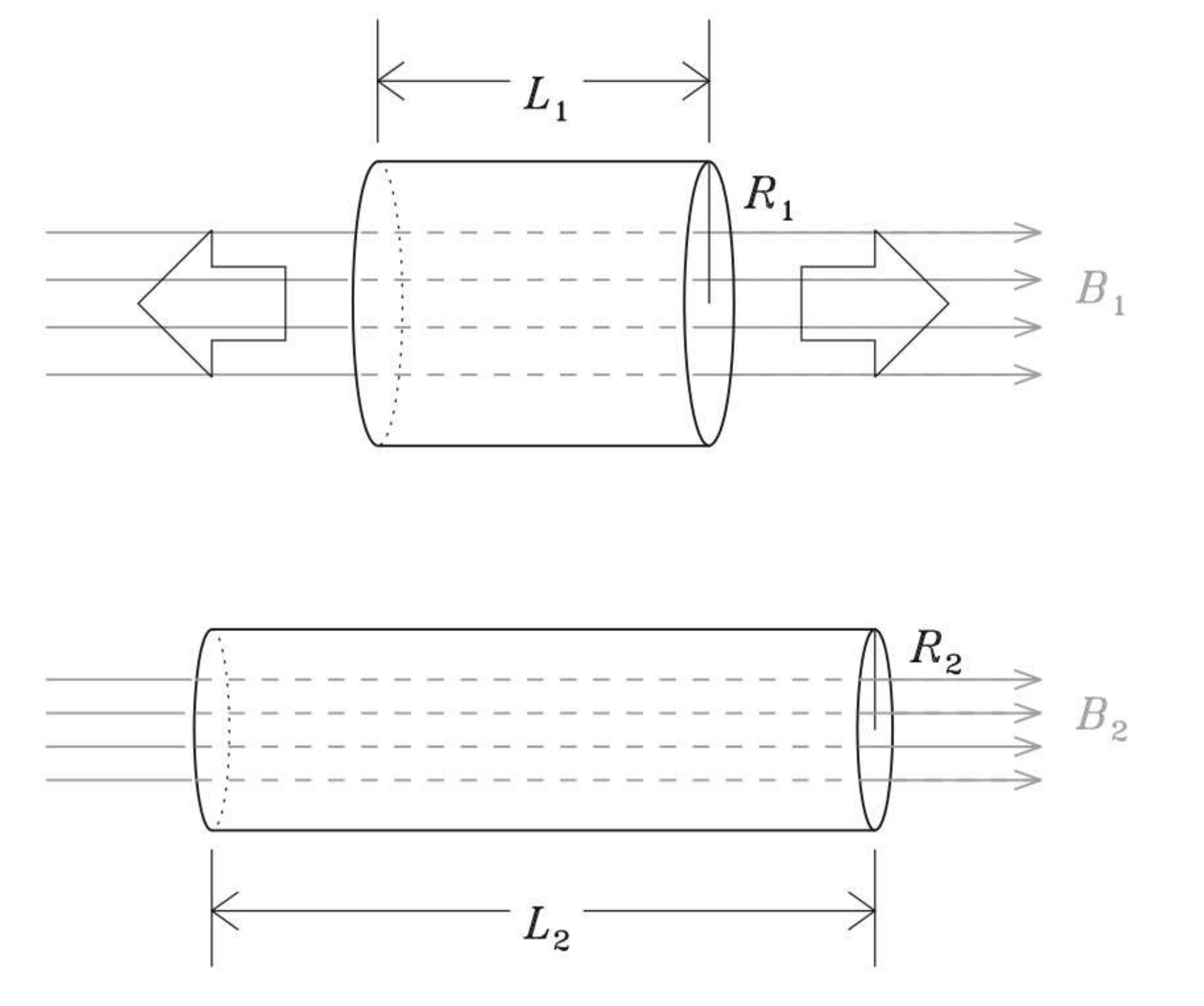}}
\caption[Schematic diagram for amplification of magnetic field]{Schematic diagram for amplification 
of magnetic field. Upper panel shows the magnetic field $B_1$ is frozen inside a cylindrical 
plasma motion of radius $R_1$ and length $L_1$. The lower panel shows the configuration 
after the plasma motion are compressed in the perpendicular direction of the plasma motions. 
$L_2$, $R_2$ and $B_2$ are the length, radius, and magnetic field strength after the stretching.}
\label{figc1:field_amplify}
\end{figure}  

Direct implication of Alfven's flux freezing theorem is found in flux amplification i.e. the growth of the magnetic
field because of inductive effect of plasma flow ${\bf v}$. The induction equation (Eq.~\ref{eqnc1:highrm})
with high $R_m$ for incompressible flow $(\nabla.{\bf v}=0)$ can be rewritten as
\begin{equation}
\label{eqnc1:induct}
\left(\frac{\partial}{\partial t} + \bf v.\nabla \right){\bf B} = {\bf B.\nabla\bf v}
\end{equation}
The LHS is the Lagrangian time variation of magnetic field where the field is moving with the flow and RHS represents the
field amplification due to local shear of the flow (${\bf B.\nabla\bf v}$). 
%In case the 
%diffusivity ($\eta$) is important for a system, the field amplification by the local shear must be faster than 
%the ohmic dissipation of the field, otherwise the fields can not be sustained. 

Let us consider a very simplistic scenario to understand the field amplification as shown in Fig.~\ref{figc1:field_amplify} \citep{Charbonneau13}.
An uniform magnetic field $(B_1)$ is frozen in cylindrical
incompressible plasma flow of length $L_1$. Now if plasma flow undergoes some stretching motion along its central axis
such that length of the cylindrical plasma flow increases to $L_2$, then from mass conservation we have,
\begin{equation}
\label{}
\frac{R_2}{R_1} = \sqrt{\frac{L_1}{L_2}}
\end{equation} 
Now as the flux $(\pi R^2B)$ is also conserved inside the system, so that leads to,
\begin{equation}
\label{eqnc1:stretching}
\frac{B_2}{B_1} = \left({\frac{R_1}{R_2}}\right)^2 = \frac{L_2}{L_1}
\end{equation}
Here $B_2$ is the magnetic field after the cylindrical flow is stretched. As the flow is stretched, 
it also amplifies the magnetic field, and the amplification of field strength is directly proportional to 
the level of stretching (Equation~\ref{eqnc1:stretching}). In this case, two things are very crucial to 
amplify the fields. One, the magnetic fields must be frozen in the plasma (high $R_m$ regime), and second, 
the stretching motion along the tube axis must be accompanied by a squeezing fluid motion perpendicular to 
the axis in order to satisfy the mass conservation. Since the horizontal fluid motions are in same direction as
the magnetic field, so it can not induct any magnetic field in the system according Equation~\ref{eqnc1:induct}. 
It is the perpendicular compressing fluid motion which is actually responsible for amplification of the field. 

As mentioned above, stretching of a flux tube can amplify the magnetic field but it has to fulfill
certain conditions and can be done with proper flows. Also, the magnetic flux remains constant in field amplification by
stretching. But there is another kind of flow motion which can amplify the magnetic field and magnetic flux
both. It is the shearing of the flow. The field amplification by sheared flow is very robust and it does not
require any proper direction of flow motion for the field to amplify like stretching case. This field amplification
by sheared flow is very common in astrophysical objects and 
we will describe it in details, in Section~\ref{C1:S4}.  

\subsection{Helioseismology}
\label{C1:S3_4}
In standard Astrophysics, the structure of the stars can be understood by the models of stellar
structures and evolutions. These models are computed based on the assumed physical conditions 
in stellar interiors, including thermodynamical properties, the interaction between radiation and 
matter and the nuclear reaction that powers the stars. Whereas these models and observations 
relevant to the stellar interiors provide limited constraints on the detailed properties 
of the stars, Helioseismology and Asteroseismology provide a greater window to study the solar 
and stellar interiors by detecting waves on the surface of the Sun and stars. Helioseismology 
is a relatively new field, and it studies the oscillations of acoustic modes for the entire Sun which 
are trapped just below the photosphere. 
 
In the 1960s, \citet{Leighton62} discovered the oscillation on the surface of the Sun by analyzing series of 
dopplergram obtained at the Mount Wilson Observatory. They found two type of oscillation in the velocity
field: large-scale horizontal cellular motion which is known as supergranulation and vertical quasi-periodic
oscillations with a period of 5 min. In that time, these 5-minute oscillations were interpreted as local
phenomena in the solar atmosphere but later, it was realized that these are the modes of acoustic oscillation
in the Sun \citep{Kosovichev11,Dalsgaard02}. In presence of rotation and magnetic field, the frequencies of the 
mode of oscillations can be written as:
\begin{equation}
\label{eqnc1:seismo}
\nu_{n,l,m} = \nu_{n,l} + \sum_jc_j^{n,l}P_j^{n,l}(m)
\end{equation}
Here $\nu_{l,m}$ is the mean frequency, $c_j^{n,l}$ are the splitting coefficients and $P_j^{n,l}(m)$ 
are the orthogonal polynomials set of degree $j$ in $m$. In case, the Sun is spherically symmetric the 
frequency of the mode of oscillation would be equal to mean frequency which is determined by the horizontally 
averaged structure of the Sun. But the presence of rotation and magnetic field put the Sun away from spherical 
symmetry and lift up the degeneracy of frequencies in $m$. Since forces due to rotation and magnetic 
field are much smaller than pressure and gravity forces, they can be treated as small perturbation over 
the spherically symmetric structure and can be determined by the splitting coefficients $c_j^{n,l}$ 
as the departure from the spherical symmetry \citep{Antia05}. From various ground based (GONG, BiSON) and 
space based (MDI, HMI) instruments, the mean frequency has been determined with high accuracy. 
Depending upon the frequency and degree, the acoustic modes are classified. The $n=0$ modes are 
referred as the fundamental or f-mode (for large value of $l$, they are mainly surface gravity waves), 
$n > 0 $ modes are p-modes or pressure modes (pressure gradients are the  main restoring force) and 
$n < 0$ are the gravity modes of g-modes (reliable observation of this mode is still not found). 
The p modes are trapped in the different region of the solar interior depending upon the frequency and 
degree, and they are the very good probe for the internal structure of the Sun. As temperature inside the 
Sun increases, the speed of the sound wave traveling towards the center also increases and its path is successively bent,
such that it is refracted back to the Sun's surface. The acoustic wave encounters a steep drop in density at the surface and reflect back inwards. The point from where they refracted back to the Sun's surface defines the critical layer and above it, the wave mode is trapped. 
Critical layer depends on the angle of inclination, and the angle of inclination of the surface with the radial 
direction is decided by the horizontal wavelength or the degree $l$. The modes with smaller $l$ can 
penetrate deeper layers before they reflected outwards, whereas modes with larger $l$ are trapped mostly 
nearly below the surface layers. $l = 0$ modes can penetrate to the center of the Sun. Hence, the p-mode analysis
is a good tracer for studying the internal structure of the Sun. Though the analysis of global mode frequencies of
the Sun provides an unprecedented opportunity to study the interior of the Sun, this has some limitation. 
Especially it studies the longitudinal average of internal structures. So, to study more localized inferences, 
we need to have various local helioseismic analysis. It includes ring analysis and time-distance 
helioseismology \citep{Thompson04}. 

Helioseismological findings of solar rotation and meridional flow helped a lot to understand 
and model the solar magnetic field generation process and the solar cycle. Measuring the splitting 
of acoustic eigen modes of same $n$ \& $l$ but $m$ values of opposite sign as mentioned in Equation~\ref{eqnc1:seismo}, the differential rotation profile of the Sun is estimated 
\citep{Thompson96,Schou98}. Furthermore, beside the existence of differential rotation throughout the solar convection zone,
helioseismology finds that the rotation rate does not depend on the latitudes below the convection zone and hence 
it rotates almost like a solid body there. This transition region where differential rotation changes 
to solid body rotation is known as the tachocline \citep{Spiegel92}. This tachocline has huge 
importance in the solar dynamo and the toroidal field is mainly generated in this transition region due to differential rotation \citep{CSD95}. 
Since we have helioseismic data for almost two decades, the 
temporal variation of solar rotation rate is also studied by the helioseismology. 
It is found that the frequency splitting of acoustic eigenmodes (same $n, l$ but $m$ values of opposite sign) varies with time, and interestingly 
correlated with solar activity cycle. This is known as the solar torsional oscillation which
has significant importance for the study of solar dynamo theory \citep{AB00,Zhao04,CCC09}.

\begin{figure}[!t]
\centerline{\includegraphics[width=0.97\textwidth,clip=]{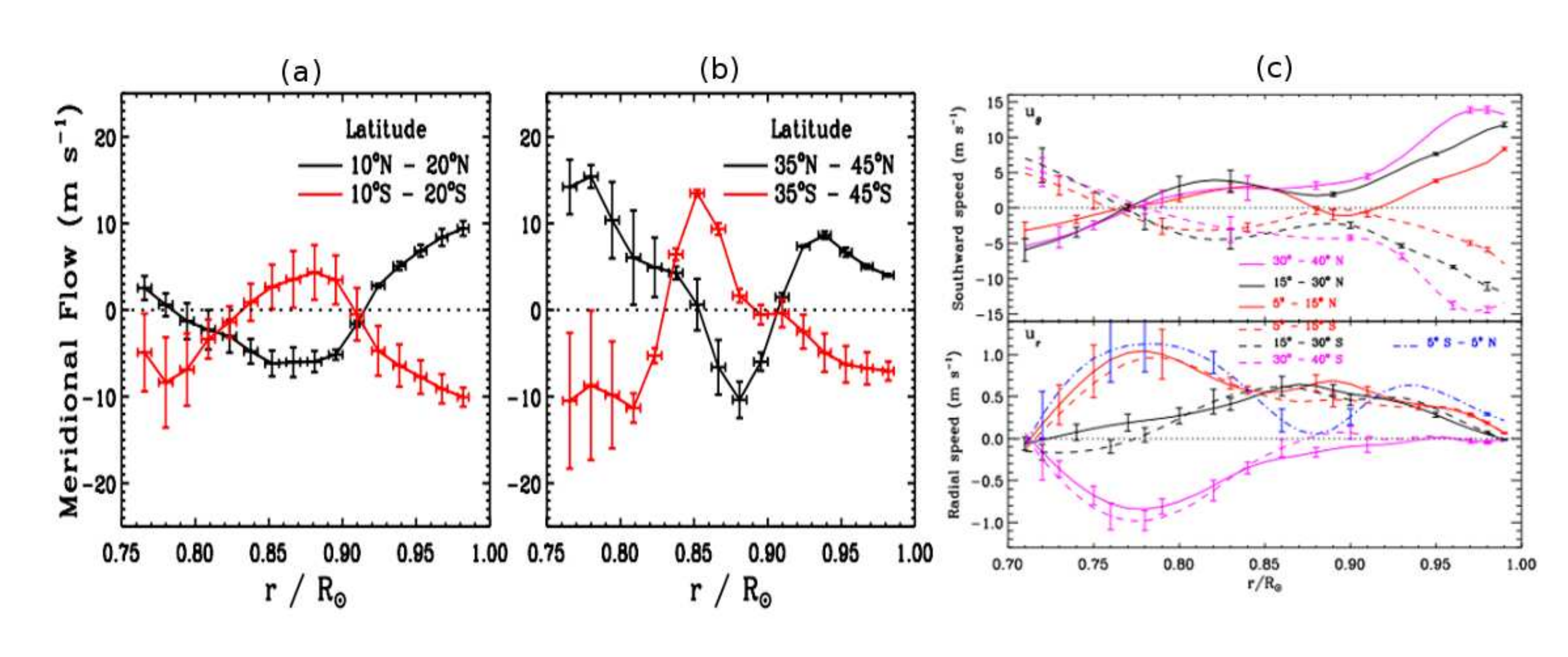}}
\caption[Meridional flow deep interior of sun using HMI data]{Helioseismological measurement of meridional flow 
in deep interior of the Sun. (a) and (b) are taken from \citet{Zhao13} and show the meridional flow variation with radius of
the sun at different latitudes. (c) Meridional flow are shown inside the solar convection zone, as found by \citet{RA15}. 
\citet{RA15} uses same techniques and same HMI data as \citet{Zhao13} but constraints by mass conservations.}
\label{figc1:mc_helio}
\end{figure}

The helioseismic measurements for the meridional flow are also available. The properties of the meridional flow near 
the surface are well established which is poleward and has speed around 20.0 m s$^{-1}$ \citep{Zhao04, BA00}. 
Going deeper inside solar convection zone introduces lots of systematic noise in helioseismic data and it is extremely difficult to 
measure amplitude and direction of meridional flow there. Efforts have been made to measure the meridional flow 
in deeper interior by several groups and they have reported different results \citep{Schad13, Zhao13, Jackiewicz15, RA15}.
Please note that in most of the dynamo models, a single cell profile of meridional flow is assumed which has a poleward flow as observed
and an equatorward return flow near the bottom of the convection zone. This equatorward propagation of meridional flow lacks observational support. 
Figure~\ref{figc1:mc_helio} is taken from \citet{Zhao13} and \citet{RA15}. Figure~\ref{figc1:mc_helio}(a) and (b) show the
meridional flow as inferred from  analysis of HMI data by \citet{Zhao13}. It shows that meridional flow has an equatorward return flow
at the depth $0.82R_{\odot}-0.91R_{\odot}$ and then followed by a poleward flow up to depth $0.77R_{\odot}$. With the same data 
set and using time-distance helioseismology, \citet{RA15} showed that an equatorward flow must exist below $0.77R_{\odot}$. Though latter
have considered the mass conservations. Interestingly \citet{Schad13,Jackiewicz15} have reported different results about meridional
flow than \citet{Zhao13, RA15}. \citet{Schad13} from their global helioseismic inversion got multi cell meridional circulation in solar convection zone whereas, \citet{Jackiewicz15} found a shallow meridional circulation. Numerical simulation of solar convection, whether hydrodynamical or magnetohydrodynamical, all predict multicell meridional flow profiles at Rossby numbers comparable to the Sun \citep{Miesch09}. In respect to the above scenario, it is extremely difficult to say what
is the exact profile of meridional circulation our Sun has in its convection zone. Since meridional flow is very important for solar dynamo 
to operate, we have considered different profile of meridional circulation in order to see its effect on dynamo (see Chapter~\ref{C4})
and we find that the equatorward propagation is very important for dynamo to work (Though see Section~\ref{C4:S6} in Chapter~\ref{C4}). The time 
variation of meridional flow is also found from the Helioseismology \citep{CD01}. It is found that the meridional flow slows down during the 
solar maxima and becomes faster during the solar minima. A theoretical explanation for this variation of meridional circulation with the solar cycle 
is given in Chapter~\ref{C5}.

\section{Dynamo Theory}
\label{C1:S4}
The goal of dynamo theory is to understand, how do astronomical objects sustain their magnetic field 
against resistive decay or ohmic dissipation. 
%If a system can sustain its magnetic field against
%the ohmic dissipation for a time scale larger than the advective and diffusion time scale of the system,
%we generally say that a dynamo mechanism is going on inside the system. 
For most of the astrophysical objects, the timescale for ohmic decay is very large, and once a field is generated inside the system that remains there for a longer time
even if there is no mechanism to sustain it. For example, typical time scale for decay of magnetic field in 
Earth, Sun and a typical galaxy in presence of no plasma motions in the interior of them are 
$3\times10^5$, $10^{11}$ and $3\times10^{23}$ years respectively \citep{Chou98}. Certainly, the Earth's magnetic
field still exists longer than its decay time and for galaxies, in spite of being larger decay than age of the universe, various effects like
differential rotation can wind up and displace the magnetic field, unless some mechanism is there to prevent it. But
for the Sun the story is different. The Sun has a magnetic cycle with 11 years periodicity representing oscillatory nature of magnetic field (Section~\ref{C1:S2}).  
Hence to sustain oscillatory nature of solar magnetic field, some mechanism is needed irrespective of its large decay time scale. Therefore, all astrophysical magnetic fields have to
be continuously generated by the dynamo process, and in this thesis, we are interested in understanding the ongoing dynamo 
mechanism inside the Sun much more realistically, which regulates its 11 years periodic solar cycle.

As pointed out in the Section~\ref{C1:S3_1} that our Sun follows the MHD approximations of plasma, and how the
plasma flow and magnetic field would behave in the Sun can be understood by solving the MHD Equations~(\ref{eqnc1:MHD1}
and \ref{eqnc1:MHD2}) along with proper transport coefficients and equations of state. There are two approaches
to understand the solar dynamo problem. First, the kinematic approach and Second, fully dynamical calculations.
In kinematic approach, the velocity field is considered to be given and can be incorporated without any understanding about
underlying dynamics. This approach has significant practical advantages. The dynamo equations becomes linear in {\bf B}
and easy to solve. In fully dynamical calculations, the velocity fields are calculated by solving full set of 
MHD equations (Eqs.~\ref{eqnc1:MHD1} and \ref{eqnc1:MHD2}) with energy equation and a proper equation of state which includes back reaction
of magnetic fields on the flows via Lorentz force. But these nonlinear 
MHD equations (\ref{eqnc1:MHD1} and \ref{eqnc1:MHD2}) are extremely difficult to solve self-consistently
in the solar context, and need huge computation power. Unlike direct numerical 
simulations where large scale flows like differential rotation and meridional 
circulation have to come from the basic MHD equations self consistently, in kinematic approach we can 
invoke these large scale flows in the induction equation directly from helioseismology.
That is why it is relatively easy to make a realistic
solar dynamo model based on kinematic approach.
  
In a seminal paper, \citet{Parker55b} gave a qualitative idea of turbulent dynamo theory. 
The solar magnetic field can be decomposed into two components, the toroidal component 
or the azimuthal component of the magnetic field, $B_\phi$ and the poloidal component of the field ($B_p$). 
Hence total magnetic field can be written as
\begin{equation}
\label{B_decom}
{\bf B} = B_\phi(r,\theta,t) \hat{{\bf e}_\phi} + {\bf {B_p}}(r,\theta,t) = B_\phi(r,\theta,t) \hat{{\bf e}_\phi} +\nabla \times [A(r,\theta,t)\hat{{\bf e}_\phi}]
%B_\phi \hat{{\bf e}_\phi} + B_r\hat{{\bf e}_r} + B_\theta\hat{{\bf e}_\theta} = 
\end{equation}
Where, A is the magnetic vector potential. Parker suggested that there is continuous exchange 
of energy between toroidal component  and poloidal component of magnetic field which makes 
solar magnetic field oscillatory. This exchange of energy between toroidal part 
and poloidal part is possible by the following mechanisms:

\begin{figure}[!t]
\centerline{\includegraphics[width=0.75\textwidth,clip=]{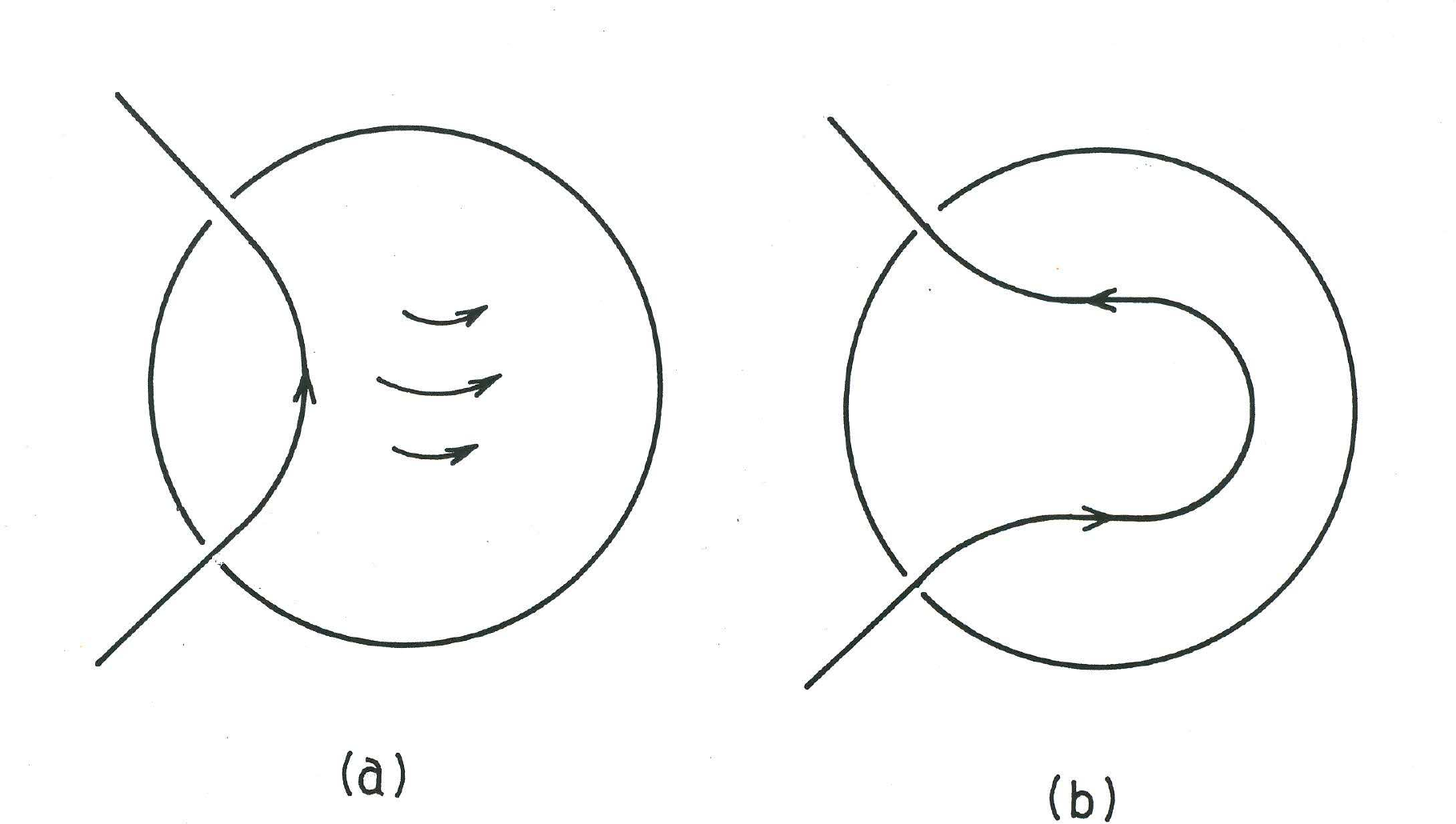}}
\caption[Toroidal field generation by differential rotation]{Generation of toroidal field from the poloidal field by 
differential rotation, Image credit: \citet{Chou98}}
\label{figc1:pol_tor}
\end{figure}

\subsection*{$\Omega$ effect:}
The Sun rotates differentially. Its equator rotates faster than its pole. 
If we have a field line in the meridional ($r-\theta$) plane of the Sun, i.e. the poloidal field line, 
then because of shearing by differential rotation, the poloidal field line will be twisted in the equator 
region as shown in Figure~\ref{figc1:pol_tor}.
Assuming the poloidal field line and differential rotation follow the same rotation axis, 
the flow profile can be taken as ${\bf v}(r,\theta) = r\sin\theta\Omega{\hat{\bf e}}_\phi$.
Invoking the magnetic field {\bf B}, the divergenceless flow {\bf v} and neglecting 
the magnetic diffusion, the induction Equation~\ref{eqnc1:induct} can be written as:
\begin{equation}
\frac{\partial A}{\partial t} = 0
\end{equation}
\begin{equation}
\label{eqnc1:shearing}
\frac{\partial B}{\partial t} = r\sin\theta[\nabla \times A\hat{{\bf e}_\phi}].\nabla \Omega  
\end{equation}
Now as A is constant over time (in reality this is not the case because of magnetic diffusion), integrating Equation~\ref{eqnc1:shearing} gives us
\begin{equation}
B(r,\theta,t) = B (r,\theta,0)+ \left(r\sin\theta[\nabla \times A\hat{{\bf e}_\phi}].\nabla \Omega\right) t 
\end{equation}
So, the toroidal component of magnetic field grows linearly in time and is proportional to the net local shear and the strength of the poloidal field.
In the case of magnetic diffusion plays a role, the poloidal field will decay and toroidal field generation eventually stops unless there is a supply
of poloidal field by some other means. Here, another important thing to note is, if the rotation profile is constant across the poloidal surface, the 
toroidal field generation stops (Ferraro's theorem). In the case of the Sun, the profile of the differential rotation is well mapped from helioseismology and 
the fact that toroidal field is generated by the $\Omega$-effect is well established. Thus, if we have dipolar magnetic field like our Sun with a 
poloidal component as shown in Fig~\ref{figc1:pol_tor}(a), then toroidal field generated via $\Omega$-effect would have opposite sense of polarity in the two
hemispheres which eventually leads to the opposite bipolar sunspots in the two hemispheres.

\subsection*{Turbulent $\alpha$ effect:}
The generation of the toroidal field by differential rotation depends on the strength of
the poloidal field, but due to finite magnetic diffusivity of the Sun, the poloidal field decays and 
eventually, the generation of toroidal field would stop. To sustain the toroidal magnetic field, 
we need to have some mechanism which will generate the poloidal magnetic field in the Sun. 
\citet{Parker55b} gave the crucial idea of how poloidal magnetic field can be generated in 
the Sun. Solar convection zone is highly turbulent. While toroidal field rises through the solar 
convection zone due to magnetic buoyancy, Coriolis force introduces a helical twist to the 
rising plasma blobs. As magnetic field is frozen in the plasma, toroidal field lines are 
twisted because of helical turbulence. This twisting of plasma blobs are happening in the rotating
frame of reference, so the vorticity in the two hemispheres will be in opposite sense and the helical
twist will be opposite sense. But the toroidal field in the two hemispheres is in opposite sense
which makes the magnetic loops have the same sense in two hemispheres. Since there is turbulent diffusion
in the solar convection zone, it smooths out the small loops and generated a large scale poloidal 
field. In this way, poloidal field can be generated from the toroidal field via helical turbulence. This 
procedure is known as classical $\alpha$ effect of helical turbulence. So, the stretching of the poloidal 
field by differential rotation generates the toroidal field and helical turbulence twists the toroidal 
field to generate poloidal field. Hence the poloidal and toroidal field can sustain each other 
through a cyclic feedback process and sustain the solar dynamo.   

After a decade later, \citet{SKR66} first formulated a rigorous and systematic mathematical approach
of Parker's turbulent dynamo theory. Since turbulence plays an important role in the turbulent dynamo, they
have developed a scheme to handle turbulence which is known as mean field magnetohydrodynamics. They
have used the ensemble approach to deal with the statistical properties of turbulence of the system.
The velocity and magnetic field in a turbulent system can be written as summation of mean component and 
fluctuating component.
\begin{equation}
\label{eqnc1:fluc}
{\bf B} = \bar{\bf B} + {\bf B}^\prime, {\bf v} = \bar{\bf v} + {\bf v}^\prime
\end{equation}
Where the bar and the prime represent mean and fluctuating quantities respectively. Invoking Equation~\ref{eqnc1:fluc}
in the induction equation~\ref{eqnc1:MHD2} and taking an ensemble average, we can write:
\begin{equation}
\label{eqnc1:mean_avg}
\frac{\partial \bar{\bf B}}{\partial t} = \nabla \times (\bar{\bf v}\times\bar{\bf B}) + \nabla\times\epsilon + \eta\nabla^2\bar{\bf B}
\end{equation}   
Here, $\eta$ is the molecular diffusivity constant in space and $\epsilon$ is the mean field emf which is defined as
\begin{equation}
\epsilon = \overline{{\bf v}^\prime\times{\bf B}^\prime}
\end{equation}
The beauty of the mean field theory is the build up of the mean emf $(\epsilon)$ from the fluctuating 
part of the turbulent flows and magnetic fields which sustains the the dynamo against the Cowling's Anti-dynamo theorem \citep{Cowling33}. 
Cowling Anti-dynamo theorem says that the axisymmetric magnetic field can not be sustained by axisymmetric flows. 
But Cowling's theorem does not hold for the mean field theory because of this extra mean emf in ensemble average
equation~\ref{eqnc1:mean_avg}. Hence we are able to get axisymmetric dynamo solution by solving mean field 
equation. For isotropic and homogeneous turbulence, the mean emf can be calculated in terms of mean
quantity using the first order smoothing approximation (see \citet{Chou98} for details) as given below
\begin{equation}
\label{eqnc1:emf}
\epsilon = \alpha\bar{\bf B} -\eta_t\nabla \times \bar{\bf B} 
\end{equation}
Where $\alpha = -\frac{1}{3}\overline{{\bf v'}.(\nabla\times {\bf v'})}\tau$, and $\eta_t = \frac{1}{3}\overline{{\bf v'}.{\bf v'}}\tau$
and $\tau$ being the correlation time of the turbulence. Putting \ref{eqnc1:emf} in Equation~\ref{eqnc1:mean_avg} finally we get,
\begin{equation}
\label{eqnc1:mean2}
\frac{\partial \bar{\bf B}}{\partial t} = \nabla \times (\bar{\bf v}\times\bar{\bf B}) + \nabla \times (\alpha\bar{\bf B}) + (\eta +\eta_t)\nabla^2\bar{\bf B}
\end{equation}  
Since the turbulent diffusivity is much larger than molecular diffusivity, we usually neglect the molecular diffusivity in solving mean field equation. Here the turbulent diffusivity is assumed to be constant in space also but for the solar convection zone, it has strong radial dependencies which simply allow some more terms to enter in Eq.~\ref{eqnc1:mean2}.   
It is also found that the $\alpha$ parameter correctly depicts the role of helical turbulence of plasma which generates the poloidal field from
the toroidal field. Please note that, in case of inhomogeneous flow field, there will be other terms in the emf $\epsilon$ (Eq.~\ref{eqnc1:emf}) and \citet{Yokoi16}
showed that they might be important for dynamo in some of the systems like red dwarf stars.

After developing the mathematical formalism  of mean field dynamo theory, \citet{SK69} solved the 
mean field Equation~\ref{eqnc1:mean2} for the Sun. They solved it in spherical geometry with realistic
boundary conditions assuming the angular rotation of the Sun as a
given quantity ${\bf v}(r,\theta) = \Omega r\sin\theta{\hat{\bf e}}_\phi$ and 
were able to get solar like butterfly diagram. This was the first reproduction of
observed butterfly diagram from theoretical and numerical analysis \citep{SK69}.    
While solving the mean field equation, it is found that the propagation 
of dynamo wave depends on the combination of helical turbulence and differential rotation
$\alpha\frac{\partial \Omega}{\partial r}$. In northern hemisphere, if $\alpha\frac{\partial \Omega}{\partial r} < 0$
then dynamo wave propagates towards equator and if $\alpha\frac{\partial \Omega}{\partial r} > 0 $ dynamo wave
propagates towards pole. This is known as the Parker-Yoshimura sign rule \citep{Parker75, Yoshimura75}.

In the 1990s, thin flux tube simulations of flux emergence showed that the strength of magnetic field
in the solar convection zone should be around $\sim$ $10^5$ G in order to match with the observed tilt angle
variation of sunspots on the surface \citep{CG87,Dsilva93, Fan93, Caligari95}. This high strength of magnetic field
in the convection zone is one order of magnitude higher than the equipartition value of convection \citep{Parker79}, 
and it is difficult for helical turbulence to impact a sufficient twist on this strong rising toroidal 
flux tube  and generate the poloidal field. Eventually, another school of thought which was developed in the 1960s
by \citet{Bab61} and \citet{Leighton69} was invoked as a possible candidate for the poloidal field generation from the 
toroidal field which is known as Babcock-Leighton Mechanism (see next section for details). In spite of being a 
potential mechanism for poloidal field generation, invoking this method in dynamo was having some issues. 
Based on the observed tilt angle variation, it is found that the Babcock-Leighton $\alpha$ is positive in 
the northern hemisphere and negative in the southern hemisphere. Also the radial shear $\frac{\partial \Omega}{\partial r}$ 
is positive at low latitude and negative at high latitude in both the hemispheres, as found by helioseismology \citep{Schou98}. 
So, According to Parker-Yoshimura sign rule, we would expect a poleward propagating branch of dynamo wave at low latitudes and 
an equatorward branch at high latitudes which contradict the observations. On the other hand, the toroidal field 
is generated throughout the convection zone by differential rotation, but the magnetic buoyancy destabilizes the storage 
of magnetic field inside convection zone and it is understood that the shear layer between radiative zone and convection 
zone i.e. tachocline is the place where toroidal field can be amplified and stored \citep{Moreno92}. Hence dynamo is confined
in two spatially segregated region of the solar convection zone coupled by diffusion. \citet{WSN91} were the first to include
a meridional circulation having a poleward flow on the surface and an equatorward return flow in the dynamo model
which connects these two spatially separated region of the Sun and explained the equatorward migration of sunspots. 
\citet{CSD95} and \citet{Durney95} have further developed the idea of \citet{WSN91} and made a numerical model with
the single cell meridional circulation. With these calculations, it is found that the equatorward flow of meridional circulation 
near the base of the convection zone indeed helps dynamo wave to overcome poleward propagation (Parker-Yoshimura sign rule) and
migrate sunspots eruptions towards equator. The dynamo model which includes Babcock-Leighton mechanism for poloidal
field generation and where meridional circulation plays a key role 
is known as Flux Transport Dynamo model. In the next section, we discuss this model in detail.
\begin{figure}[!t]
\centering
\begin{tabular}{cc}
\includegraphics*[width=0.5\linewidth]{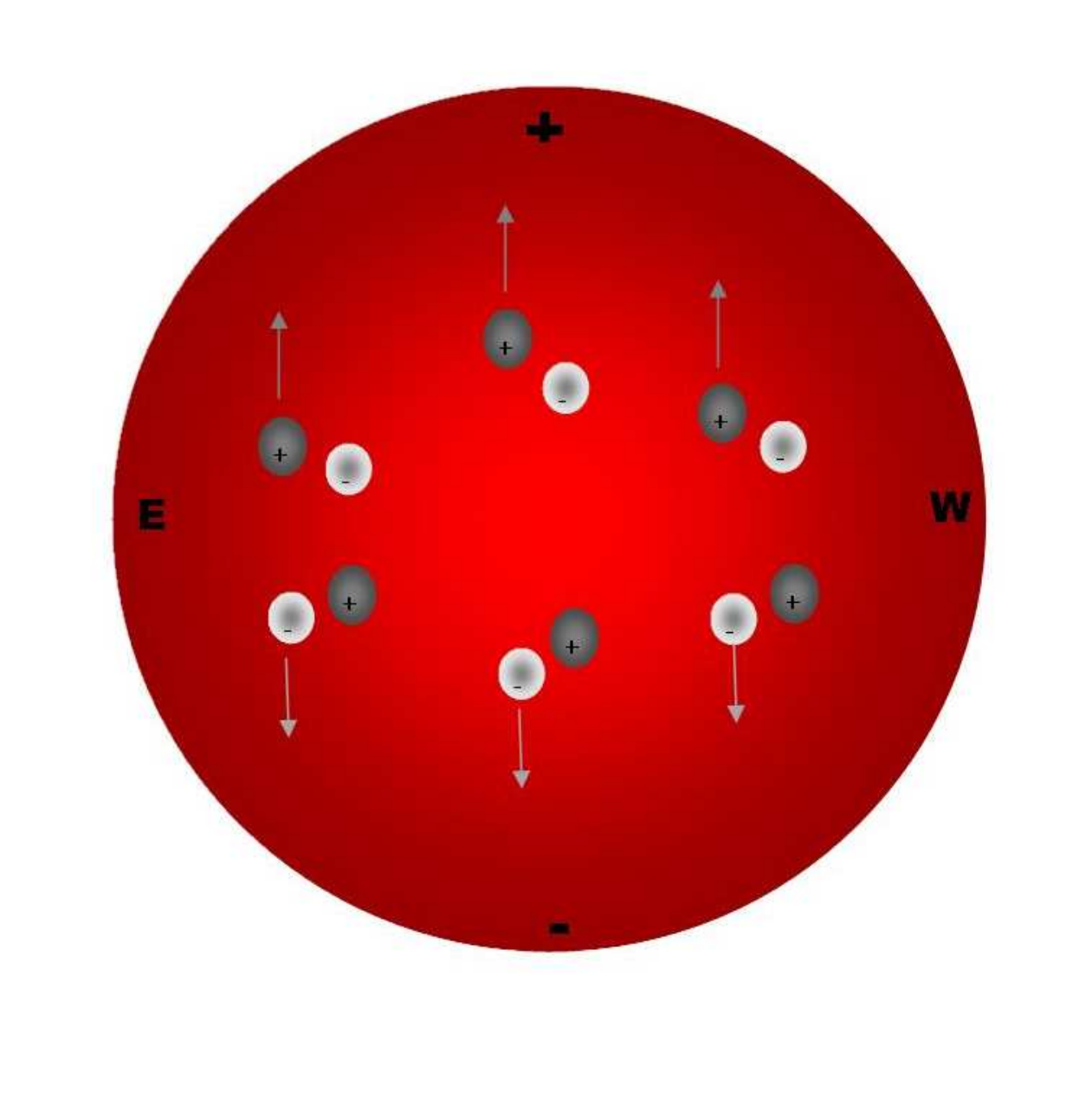} & \includegraphics*[width=0.5\linewidth]{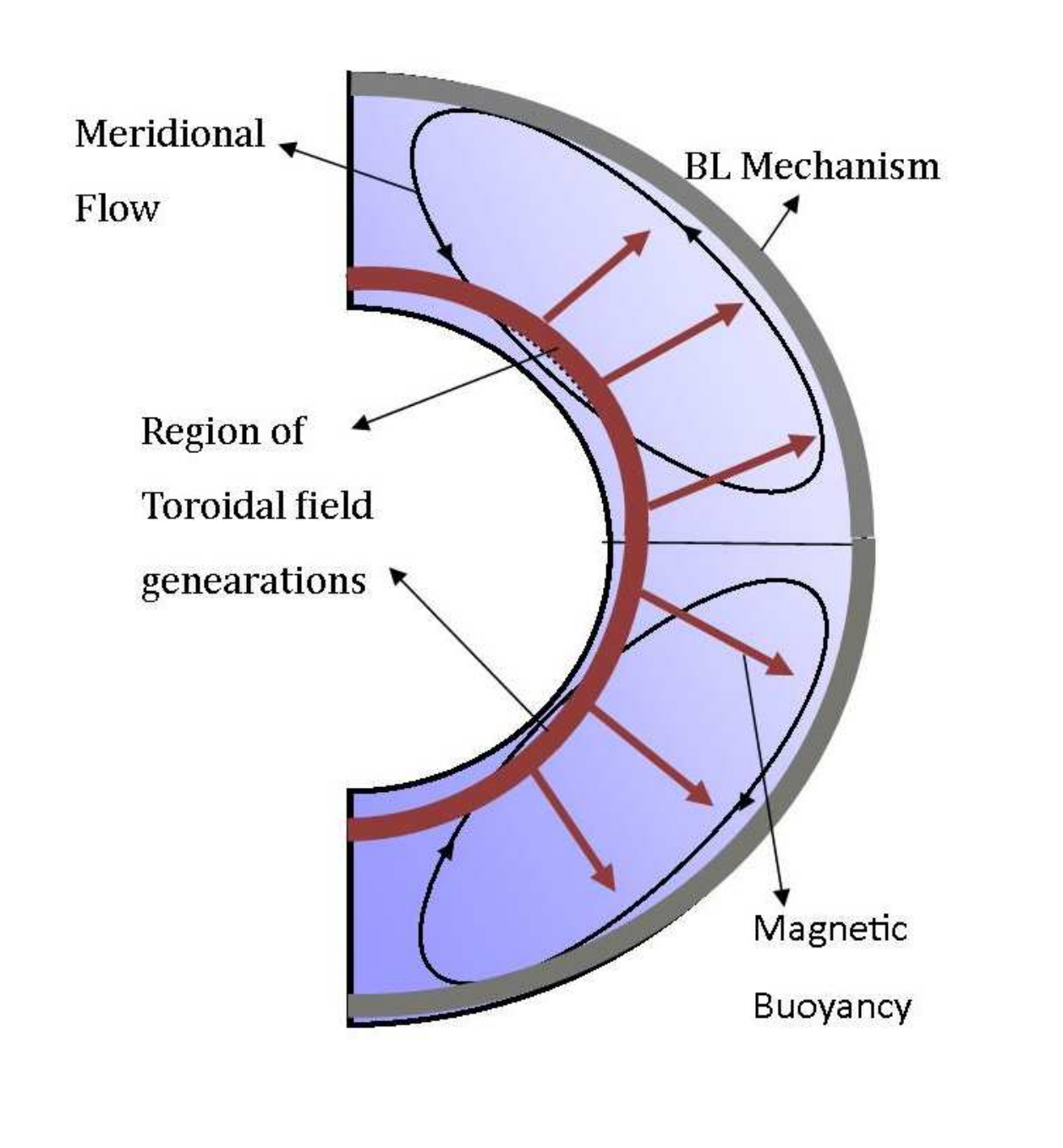}\\
                (a)                                           &                   (b)\\
 \end{tabular}                
\caption[Babcock-Leighton mechanism and Cartoon diagram of Flux Transport Dynamo model]{ (a) A schematic diagram of Babcock-Leighton 
mechanism on the surface of the Sun. The leading polarity sunspots closed to equator are shown in white (black) in the northern 
(southern) hemisphere and trailing polarity sunspots are shown in black (white) in northern (southern) hemisphere. (b) A cartoon diagram
explaining flux transport dynamo model}
\label{figc1:BL}
\end{figure}
\section{Flux Transport Dynamo Model}
In the rest of the chapters of this thesis, we will be dealing with the different aspects of the Flux Transport Dynamo model
to understand solar magnetic field generation process, and properties of the solar cycle more carefully with much more details.
As we already mentioned, this type of models are basically successor of the axisymmetric mean field dynamo models, kinematic, and include 
Babcock-Leighton Mechanism as a poloidal field generation mechanism instead of helical turbulence. Apart from 
the differential rotation of the Sun and turbulent diffusion, a plasma flow in the meridional plane ($r-\theta$ plane) known as meridional
circulation plays an important role in these models. The main ideas of these models are as follows. The poloidal
field is stretched by the differential rotation and generates the toroidal magnetic field. The toroidal
field is then amplified and stored near the base of the convection zone at tachocline (reddish brown region
of the Figure~\ref{figc1:BL}(b)). When the toroidal field becomes
buoyant, it rises through the solar convection zone (as shown by the reddish brown arrow
in the Figure~\ref{figc1:BL}(b)) and creates the bipolar sunspots on the surface. While traveling through
the convection zone, Coriolis force acts on the rising flux tube and produces a tilt between the bipolar sunspots with 
respect to the east west direction of the Sun. As these sunspots are the regions of strong
magnetic field, they diffuse away and leading polarity sunspots (white in northern and 
black in southern hemisphere) move towards the equator from both the hemisphere and cancel 
each other as shown in Figure~\ref{figc1:BL}(a). Whereas the trailing polarity sunspots 
(black in northern and white in southern) are advected towards pole by meridional circulation and generate
a large scale poloidal field. This is known as Babcock-Leighton (BL) process. The poloidal 
field then can be advected to the base of the convection zone by meridional circulation where differential 
rotation can again act on it to generate toroidal field. 
Hence the cycle goes on and solar magnetic field sustains its oscillatory behavior. 
This main idea of flux transport dynamo model is depicted in the cartoon diagram \ref{figc1:BL}(b) where the 
reddish brown colors show the toroidal field generation region and reddish brown arrows show the magnetic buoyancy. The BL mechanism is
shown in grey region on the surface and meridional circulation connecting two spatially segregated region is shown in black 
streamline. The basic equations of these models are
\begin{equation}
\label{eqnc1:Aeq}
\frac{\partial A}{\partial t} + \frac{1}{s}({\bf v}.\nabla)(s A)
= \eta_{p} \left( \nabla^2 - \frac{1}{s^2} \right) A + S(r, \theta, t),
\end{equation}
\begin{equation}
\label{eqnc1:Beq}
\frac{\partial B}{\partial t}
+ \frac{1}{r} \left[ \frac{\partial}{\partial r}
(r v_r B) + \frac{\partial}{\partial \theta}(v_{\theta} B) \right]
= \eta_{t} \left( \nabla^2 - \frac{1}{s^2} \right) B + s({\bf B}_p.{\bf \nabla})\Omega + \frac{1}{r}\frac{d\eta_t}{dr}\frac{\partial{(rB)}}{\partial{r}}
\end{equation}
where ${\bf A}$ is the magnetic vector potential corresponding to the poloidal field, $B$ is the toroidal component of magnetic
field (Equation~\ref{B_decom}) and $s = r \sin \theta$. ${\eta_t}$ and ${\eta_p}$ are the diffusion coefficients corresponding
to poloidal and toroidal magnetic field respectively. $\Omega$ is the differential rotation, and $v_r$ and $v_\theta$ are the
components of meridional circulation. $S(r,\theta)$ is the source term which takes care for the magnetic buoyancy and Babcock-Leighton mechanism. 
We solve Eq.~\ref{eqnc1:Aeq} and ~\ref{eqnc1:Beq} in a meridional plane in the range of 
$0.55R_{\odot} < r < R_{\odot}$ in radius and $0 < \theta < \pi $ in co-latitudes with the following boundary conditions.

At the poles $(\theta =0,\pi)$ we use, $A =0$, and $B =0$. 
In the bottom $(r = R_b)$, we assumed a perfectly conducting solar core and for that we use $A =0$, $B=0$. At the surface $(r = R_\odot)$,
the toroidal field is assumed to be zero $(B=0)$, and poloidal field has to match with potential fields smoothly satisfying the free space
equation.
\begin{equation}
\left(\nabla^2 - \frac{1}{r^2\sin\theta}\right)A =0
\end{equation}
Since we have taken the bottom boundary much below the penetration depth of meridional circulation, 
the solutions do not depend on the bottom boundary conditions too much as shown by \citet{CNC04}. 
From last two decades, these models are successful in explaining various observational properties of 
solar cycle which include equatorward migration of sunspots, 11 years period, irregular properties 
of solar cycle (Maunder minima, Waldmeier effect) and much more. It turns out that these models can also
predict the future solar cycle by assimilating the polar field data during each minima of the cycle \citep{CCJ07}.
In spite of its success, sometime 
doubts have been expressed whether these models are realistic or all the explanation are merely accidental.
To answer this question, we need to check whether all the assumptions and approximations which are made in these models are correct.

\subsection{Mean Flows}
Mean flows i.e. the differential rotation and meridional circulation are the basic building blocks of the 
flux transport dynamo model. Differential rotation is mainly responsible for toroidal field generation and
meridional circulation helps to migrate sunspots towards equator overcoming the Parker-Yoshimura sign rule 
and advects field from the surface to the bottom for low diffusion case. The helioseismology probes the differential rotation inside solar convection zone nicely \citep{Thompson96,Schou98} keeping no space
to doubt on this. The surface meridional circulation is well observed and measured by various techniques
but the equatorward return flow still lacks observational supports. Recently, there are observational
evidences that the return flow can exist in the middle of the convection zone giving double cell meridional
circulation or even multi-cell meridional circulation in the solar convection zone. In Chapter~\ref{C4}, we explain it in details and show that our dynamo model works perfectly fine provided there is an equatorward return 
flow near the bottom of the convection zone. 
\begin{figure}[!t]
\centerline{\includegraphics[width=0.97\textwidth,clip=]{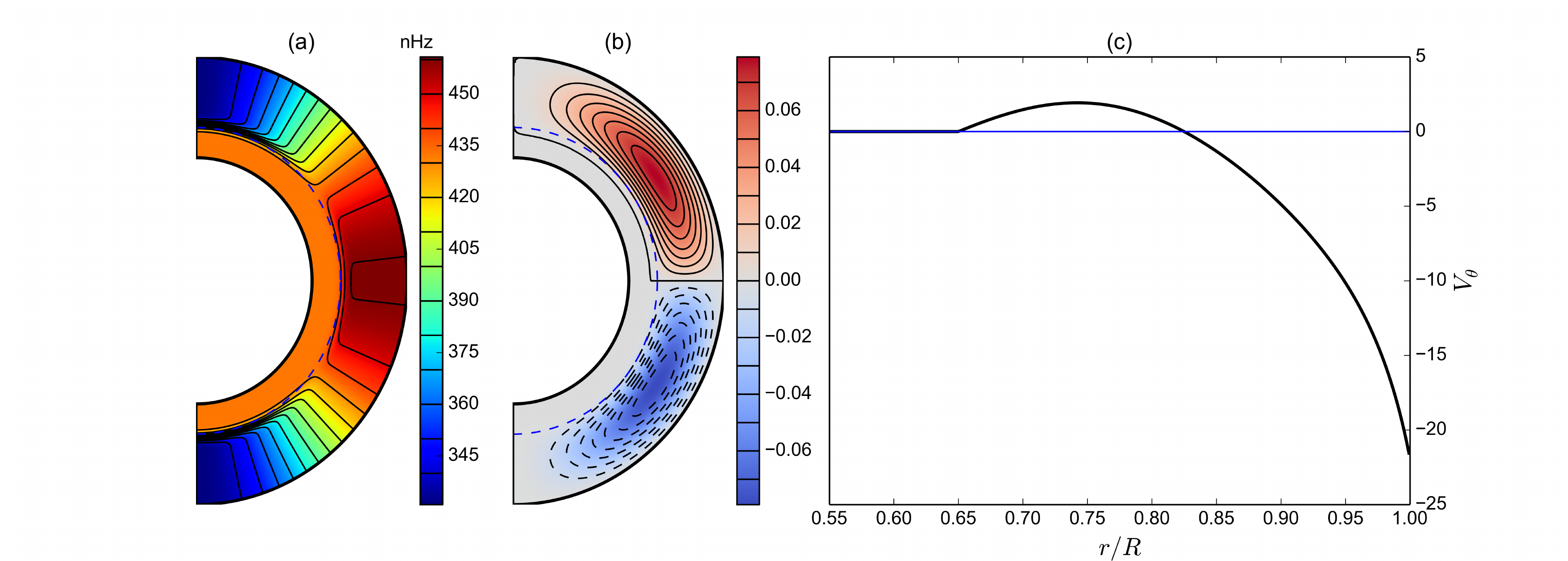}}
\caption[Meanflows used in FTD models]{(a) Differential Rotation, (b) Streamlines of the meridional Circulation. Solid contours with red filled colors show an anti-clockwise flow. Dashed contours with blue colors show clock-wise flow, and (c) $\theta$ component of Meridional flow at $45^{\circ}$ latitude is plotted with radius of the Sun.}
\label{figc1:meanflows}
\end{figure}

An analytical profile has been used in our model to incorporate the differential rotation which fit nicely with the 
helioseismolgy results \citep{DC99}. The near surface shear layer \citep{Corbard02} has not been considered in our model because
as studied by \citet{Dikpati02}, it does not have any effect on the flux transport dynamo. Magnetic
field can be generated near the surface shear layer but can not be stored due to disruption by magnetic buoyancy. In Figure~\ref{figc1:meanflows}(a)
we have shown the differential rotation profile used in our model. The streamlines for single cell meridional circulation is shown in Figure~\ref{figc1:meanflows}(b).
Motivated by observations, we have assumed a poleward flow near the surface and an equatorward flow at the bottom of the convection zone. Though latter
one has no observational evidence, we take it because of mass conservations. 
In Figure~\ref{figc1:meanflows}(c), we have plotted the meridional circulation
with radius at mid-latitude $45^{\circ}N$. The detail parameterization how we can get the meridional profile as shown in Fig~\ref{figc1:meanflows}(b)
is given in Chapter~\ref{C4}. 

Our dynamo models are kinematic in nature and we do not consider the effect of back reaction due to Lorentz force on the large scale
flows. But there are observational evidences that the zonal flows and meridional flows do vary with the solar cycle \citep{BA00,CD01,Komm15,Hathaway10b}.
Whereas there is theoretical explanation considering the back reaction due to Lorentz force for the  variation of zonal flows with the solar cycle known as torsional oscillation \citep{Durney00,CCC09}, the theoretical
explanation why meridional flow varies with the solar cycle is still very primitive. \citet{Rempel06} considered a full mean-field model of both
the dynamo and the large scale flows, and explain both the torsional oscillation and variation of meridional flow with the solar cycle. But they lack details
explanation for the latter. In Chapter~\ref{C5} we have developed a theoretical model to explain the variation of meridional flow with the solar cycle in detail considering 
the back-reaction due to Lorentz force.      

\begin{figure}[!t]
\centering
\begin{tabular}{cc}
\includegraphics*[width=0.5\linewidth]{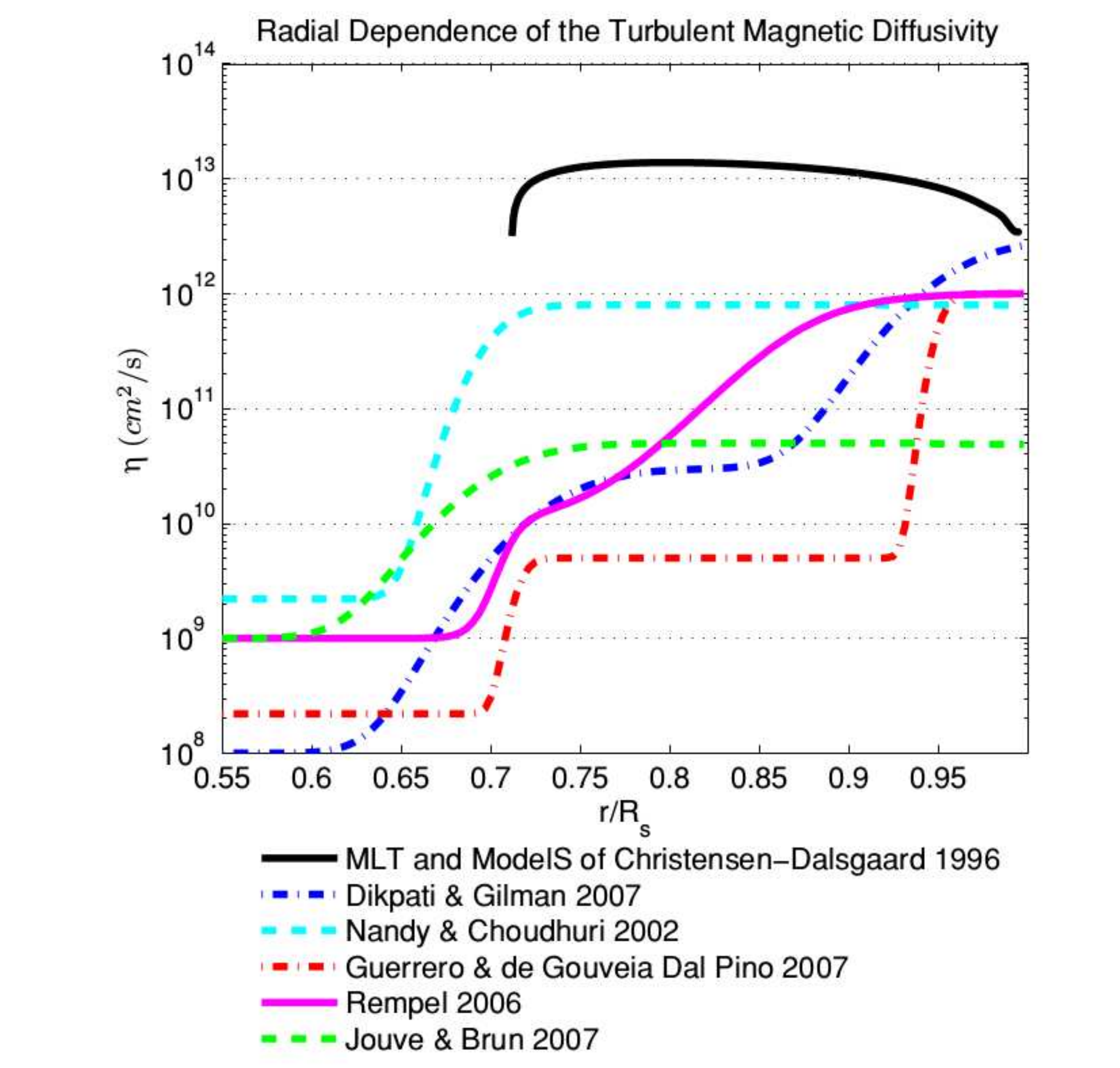} & \includegraphics*[width=0.5\linewidth]{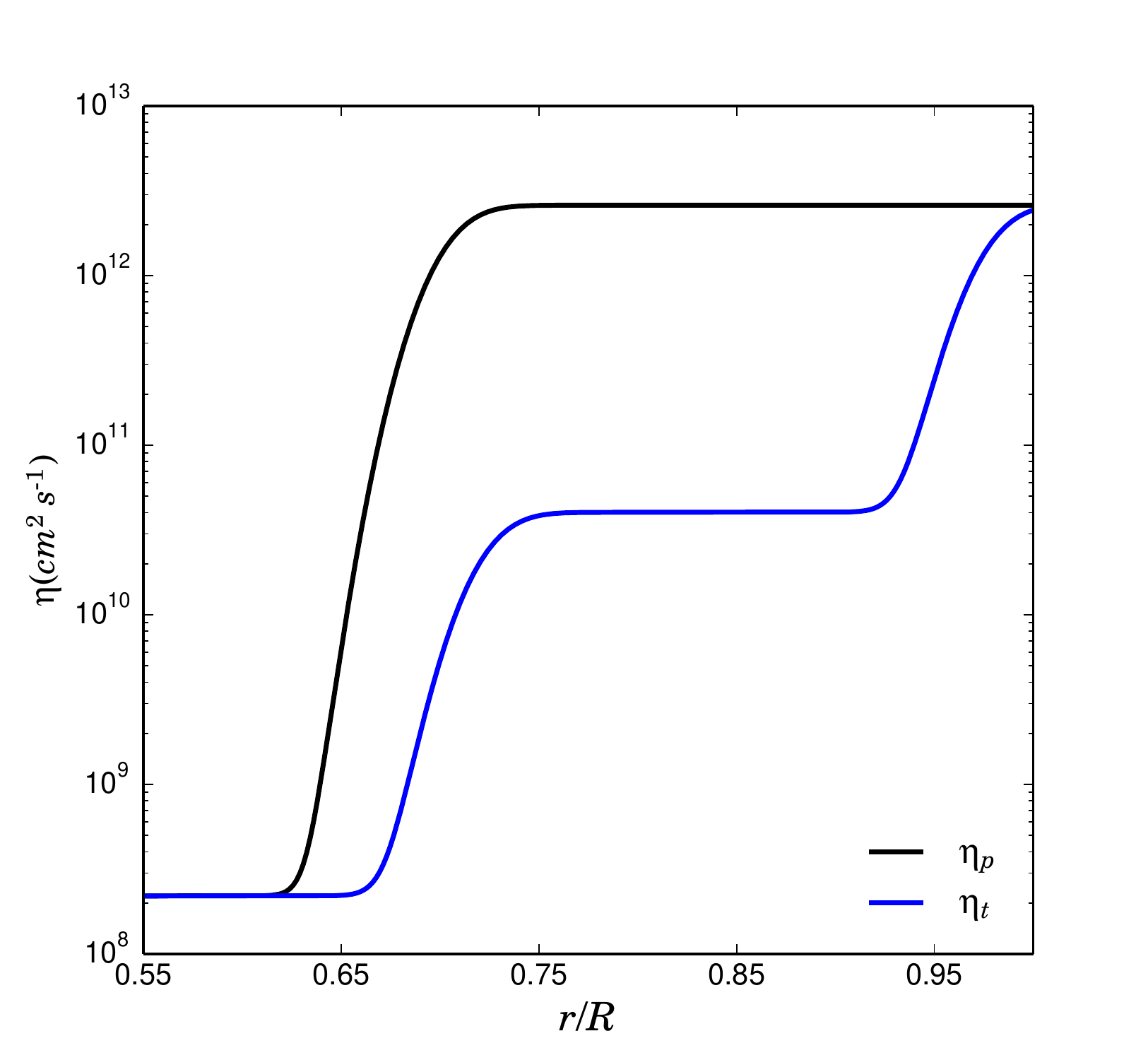}\\
                (a)                                           &                   (b)\\
 \end{tabular}                
\caption[Different diffusivity profiles used in FTD ]{ (a) Different radial dependency of diffusivity profile used in various flux
transport dynamo models (b) The diffusivity profile which we use in our flux transport model}
\label{figc1:diff1}
\end{figure} 
  
\subsection{Turbulent Diffusion}
Turbulent diffusivity is a very important parameter in the flux transport dynamo model, and it is one of
the important ingredients to transport magnetic field in between two spatially segregated regions where dynamo works.
But, it is poorly constrained inside the solar convection zone. From the mean field theory, assuming first order smoothing
approximation, the diffusivity can be written as \citep{Moffatt78}
\begin{equation}
\label{eqnc1:diff1}
\eta = \frac{\tau}{3}<v^2>
\end{equation}
Where $\tau$ is the eddy correlation time and $v$ is the turbulent velocity field. 
Using Mixing Length Theory (MLT) of turbulent convection and the standard SolarModelS 
\citep{Christensen96} we can estimate the order of magnitude of the turbulent 
diffusivity. According to these models, the turbulent diffusivity can be expressed 
in terms of mixing-length parameter $\alpha_p$, the convective velocity $v$ for
different radii, and the pressure scale height $H_p$, and $\eta$ becomes
\begin{equation}
\eta = \frac{1}{3}\alpha_pH_pv
\end{equation}
The estimated turbulent diffusivity from Mixing Length Theory is shown by solid black lines in Fig~\ref{figc1:diff1}(a). The 
diffusivity profile used by the different group of flux transport dynamo models are also shown in Fig~\ref{figc1:diff1}(a) by 
different colors. As it is clear from the figure that, flux transport models use almost two orders of magnitude lesser
value of turbulent diffusivity and almost different radial profiles. This deviation leads to the fact that the dynamo 
model can not operate under the high diffusivity as suggested by the MLT. One possible solution for this is that these dynamo
models are kinematic, and can not take into account the strong back reaction due to Lorentz force on the velocity fields which
results in the suppression of turbulence, and hence quenches the turbulent diffusivity \citep{Munoz11}. Though 
various diffusivity profiles are used in flux transport dynamo models, they qualitatively give solar like solutions. But of course,
the different values of diffusivity in the bulk of the convection zone lead to different memory which leads to the different 
prediction of solar cycle \citep{Yeates08} and different results in explaining irregular properties of the solar cycle 
(e.g, Waldmeier Effect) \citep{KarakChou11}. In order to explain various observational results including dipolar parity 
\citep{CNC04, Hotta10}, lack of hemispheric asymmetry \citep{CC06,GoelChou09}, the correlation between the polar field at cycle minimum with the strength of next cycle \citep{Jiang07}, and the Waldmeier effect \citep{KarakChou11}, a higher value of diffusivity is required, and we use higher value of 
diffusivity in most of our calculations using flux transport dynamo model as shown in Fig~\ref{figc1:diff1}(b). Choosing higher diffusivity profile makes our model so called diffusion dominated model \citep{Yeates08}. Since the toroidal field is much stronger than the poloidal field, we expect the quenching would be different for both of the components and use different diffusivity profile for both of them. Though for our
3D flux transport dynamo model, we use a two-step diffusivity profile (see Chapter~\ref{C6} for details). Hence it is evident that, although
the theory of turbulent diffusion is not well understood but different profiles used in flux transport dynamo do not invalidate the model and give qualitatively good results in agreement with observations. 
 
\subsection{Magnetic Buoyancy}
Magnetic buoyancy is an important physical process in flux transport dynamo which governs the rise of toroidal flux
tube from the bottom of the convection zone to the surface (see section~\ref{C1:S3_2}). It is inherently a 3D process but 
in 2D axisymmetric dynamo model, we can treat them rather using some very crude approximations. Magnetic
buoyancy has been treated using various method in the flux transport dynamo. The various methods about treatment of 
magnetic buoyancy are given in Chapter~\ref{C2} in detail. We have explained that the proper treatment of magnetic buoyancy 
is an important issue in order to understand the irregular features (e.g., Waldmeier effect) of the solar cycle but the 
properties of solar cycle are qualitatively same in most of the magnetic buoyancy treatment. A proper realistic treatment of 
magnetic buoyancy needs a 3D model. \citet{YM13} first developed this kind of
3D model where the buoyant rise of flux tube has been treated by simultaneously applying a radially outward velocity
and a vortical velocity to a localized part of an azimuthal flux
tube at the bottom of the convection zone. We have also developed a 3D dynamo model where the magnetic buoyancy has been treated using 
a SpotMaker algorithm (Chapter~\ref{C6}). Whereas \citet{YM13} have captured the physics of the early rise of the flux tube, Our 3D model
captures the later phase of rising flux tubes more realistically.  
 
\subsection{Babcock-Leighton Mechanism}
Although poloidal field generation from the decay of sunspots is
proposed in the 1960s \citep{Bab61,Leighton69} and invoked as a poloidal field generation 
process to complete dynamo loop in the 1990s (after realizing that turbulent $\alpha$ effect 
is not able to generate poloidal field from the strong toroidal field at the convection 
zone due to helical turbulence), the observational supports is established rather recently 
\citep{DasiEspuig10,Kitchatinov11a}. \citet{DasiEspuig10} use sunspot data and tilt angle 
data from Mound Wilson Observatory and Kodaikanal Observatory for the cycle $15-21$ and 
have found a strong correlation ($r = 0.65$ and $r = 0.70$ for Mount Wilson and Kodaikanal 
data set respectively) between the product of the strength of a cycle with its average tilt 
angle, and the strength of the next cycle. Basically the product of strength of the cycle with 
its average tilt angle represents the measure of the poloidal field due to Babcock-Leighton 
mechanism. If this product is strong, it would give us strong next cycle. 
\citet{Kitchatinov11a} have also calculated the poloidal field from the decay of sunspots 
considering the tilt angle and sunspot area from Pulkovo Astronomical Observatory, and 
found a direct correlation between estimated poloidal field, and the A-index of the
large-scale field \citep{Makarov00} for the solar minima of the following cycles.
These correlations directly support the operation of Babcock-Leighton Mechanism in the Sun.   

The Babcock-Leighton mechanism is an inherently 3D process and real depiction of this process in 2D axisymmetric 
flux transport dynamo model is only possible by very simple and crude approximations. We treat the 
Babcock-Leighton process using a parameter $\alpha$ in our 2D axisymmetric models.
\begin{equation}
\alpha = \frac{\alpha_0\cos\theta}{4} \left[1 + \mathrm{erf}\left(\frac{r-r_1}{d_1}\right)\right] \left[1 - \mathrm{erf}\left(\frac{r-r_2}{d_2}\right)\right]
\end{equation} 
The parameters which we use are $r_1 = 0.95 R_\odot$, $r_2 = R_\odot$, $d1 = d2 =0.025R_\odot$ 
making sure that the Babcock-Leighton process is confined in the surface layer 
$(0.95R_\odot < r < R_\odot)$. Since Babcock-mechanism is very much dependent 
on the tilt angle of the sunspots, and the cause of the tilt angle is Coriolis 
force which has a $\cos\theta$ dependency, we have taken a $\cos\theta$ dependency 
in our $\alpha$ parameter. In literature, some of the groups use $\cos\theta \sin\theta$ 
or even  $\cos\theta \sin^2\theta$ in $\alpha$ parameter to suppress the high latitude 
emergence of the sunspots. Here we should also mention that we use a $\alpha$ quenching 
for dynamo saturation. Since the strong toroidal flux tube is less affected by Coriolis 
force while traveling through the convection zone, it would have a smaller tilt angle than 
the weaker ones. To incorporate these changes in tilt angle, we use a factor 
$\left[1+ \left(\frac{B}{B_0}\right)^2\right]^{-1}$ for $\alpha$ quenching. In 3D model, 
we have used a more sophisticated algorithm to deal with B-L mechanism. To capture B-L
process more realistically in 3D, we first calculate the suitable places at the bottom of the 
convection zone in both longitudes and latitudes where the toroidal flux becomes magnetically buoyant, and put bipolar sunspots 
at those regions but on the surface, and let them evolve under the diffusion and mean flows,
which eventually generate the poloidal field (see chapter~\ref{C6} for details). 
The 3D treatment of BL process makes Flux Transport dynamo model much more realistic and 
in principle, it can capture the build up of polar fields and other observed phenomena in much greater details than before. 

One of the advantages of treating B-L process in 3D is given in Chap~\ref{C7}. The original idea of Babcock-Leighton process involved the dispersal 
and migration of Bipolar sunspots by the supergranular convection motions on the surface which leads to the transport of magnetic
field towards the pole and reverse the existing dipole moment. \citet{Leighton64} suggested that this process
can be considered as a simple random walk on the solar surface and approximated it as a diffusion process. The diffusion
coefficient calculated by \citet{Leighton64} considering the correct time of polar field reversal is $770-1540$ km${^2}$ s$^{-1}$.   
After that various observational measurements have been done to estimate the value of the diffusion coefficient (for details see 
\citet{Schrijver96,Chae08,Jiang_review15}). \citet{Jiang_review15} summarize the current measured values of diffusion coefficient 
on the surface (See Section 4.1 of their paper) which ranges from $100 -600$ km$^2$ s$^{-1}$. Since solar photosphere consists of 
convective motions spanning different length scale (e.g., granular scale to supergranular scale), the diffusion coefficient has different 
value for different length scale in accordance with turbulence cascading \citep{Chae08}. Till now, no Surface Flux Transport model or
Flux Transport model has taken into account this spatial dependence of the diffusion coefficient. One of the possible approach to consider a realistic depiction
of the surface diffusion would be direct incorporation of the random convective cellular flows in the model. \citet{UH1_14} have already implemented this in their
2D Surface Flux Transport model. Since incorporating the convective flows granulation and supergranulation in Flux Transport Dynamo models need extreme higher resolution and a 3D treatment, 
it has not been implemented in 2D axisymmetric Flux Transport dynamo models. 
We are the first to implement the direct observed convective velocity fields from SOHO MDI data in our 3D model to see how does it affect the overall features of solar magnetic fields.
In Chap~\ref{C7}, we show the initial implementation procedure of this observed velocity fields in our model. The results obtained are generally in good agreement with the observed surface flux evolution and with non-convective models that have a turbulent diffusivity on the order of $3 \times 10^{12}$ cm$^2$ s$^{-1}$ (300 km$^2$ s$^{-1}$).  However, we find that the use of a turbulent diffusivity underestimates the dynamo efficiency, producing weaker mean fields than in the convective models.

%% file: chapter2.tex
\begin{savequote}[100mm]
``No one ever accepts criticism so cheerfully. Neither the man who utters it nor the man who invites it really means it."
\qauthor{--R.K. Narayan, Malgudi Days}
\end{savequote}
%\chapter{The treatment of Magnetic buoyancy in flux transport dynamo model\protect\footnote{Based on Choudhuri \& Hazra (2016)}}
\chapter{The Treatment of Magnetic Buoyancy in Flux Transport Dynamo Models}
\label{C2}

\begin{quote}\small
One important ingredient of flux transport dynamo models is
the rise of the toroidal magnetic field through the convection
zone due to magnetic buoyancy to produce bipolar sunspots and
then the generation of the poloidal magnetic field from these
bipolar sunspots due to the Babcock--Leighton mechanism.
Over the years, two methods of treating magnetic buoyancy---a
local method and a non-local method---have been used widely
by different groups in constructing 2D kinematic models of
the flux transport dynamo.  We review both these methods and
conclude that neither of them is fully satisfactory---presumably
because magnetic buoyancy is an inherently 3D process.  We also
point out so far we do not have proper understanding of why
sunspot emergence is restricted to rather low latitudes.
\end{quote}

\section{Introduction}
\label{C2:S1}
From 1990s a type of model has been developed for the solar
dynamo known as the flux transport dynamo model
\citep{WSN91,CSD95,Durney95}. At the present time, 
this model seems to be the most promising and satisfactory
model for explaining different aspects of the solar cycle
\citep{Chou11,Chou15, Charbonneau14}, although still doubts are
sometimes expressed about its validity---especially because
of the uncertainty in our knowledge of the meridional
circulation which plays a crucial role in this model (see, for
example, the Introduction of \citet{HKC14}).
The flux transport dynamo model basically involves
the following three processes.
(i) The strong toroidal field is produced by the
stretching of the poloidal field by differential rotation in
the tachocline. (ii) The toroidal field generated in the
tachocline gives rise to active regions due to magnetic
buoyancy, and the decay of tilted bipolar active regions
produces the poloidal field by the Babcock--Leighton
mechanism. (iii) The
poloidal field produced by the Babcock--Leighton mechanism
is advected by the meridional circulation first to high latitudes 
and then down to the tachocline, while also diffusing down to
the tachocline due to turbulent diffusion.

Most of the calculations of the flux transport dynamo model
have been based on axisymmetric 2D kinematic mean field
equations.  It is completely straightforward to include
the process (i) within such a formalism---especially because
helioseismology gives us the profile of differential rotation.
As far as process (iii) is concerned, there are some uncertainties
in the nature of the meridional circulation \citep{Hathaway12,Zhao13,Schad13,RA15} as well as in the
value of turbulent diffusion \citep{Jiang07,Yeates08}.  Although early models
assumed a simple one-cell form of meridional circulation,
there have been some recent calculations with more
complicated meridional circulation \citep{JouveBrun07,HKC14}. However, once
we specify the form of the meridional circulation and the turbulent
diffusion within the framework of the kinematic model, there
is absolutely no uncertainty in the mathematical forms of
the terms involved in process (iii). Only in the case of
process (ii) involving magnetic buoyancy and the Babcock--Leighton
mechanism, there is considerable
uncertainty at a fundamental level as to how this process should 
be included in a 2D kinematic model.  Magnetic buoyancy involves
the rise of a tilted flux loop through the convection zone and
is an inherently 3D process, which can be included in a 2D dynamo
model only through rather crude approximation procedures.
Over the years, different groups have proposed different
procedures for handling the process (ii). While carrying on
calculations with the flux transport dynamo model, we have
become aware that these different procedures often give significantly
different results and we have tried to understand the physical
reasons behind these differences.  The obvious question is: which
one is the most realistic procedure for treating the process (ii)?
This is not an easy question to settle. Different procedures have
their own strengths and own weaknesses.  We have found that the
subtleties involved in modeling the process (ii) are not
sufficiently appreciated by the scientific community. 
Probably a fully satisfactory treatment of process (ii) is not
possible within the 2D kinematic framework and one has to go
beyond 2D. We discuss 
the different procedures which had been proposed by different
groups and critically examine the aspects of physics which are
covered and which are not covered in these different procedures.
We do not discuss the physics of magnetic buoyancy and, in that
sense, this is not a comprehensive study of the whole topic
of magnetic buoyancy, which is a vast subject.  We restrict
ourselves only to a discussion of how magnetic buoyancy can
be included in kinematic dynamo models. 

\def\vb{{\bf v}}
\def\Bb{{\bf B}}
\def\ol{\overline}
\def\ec{\cal{E}}
\def\pa{\partial}
\def\vf{{\bf v}}
\def\Bf{{\bf B}}

\section{The basic equations}
\label{C2:S2}
We assume both the mean magnetic field and the mean velocity
field to be axisymmetric in 2D kinematic models. The magnetic
field is written as
\begin{equation}
{\bf B} = B (r, \theta, t) {\bf e}_{\phi} + \nabla \times [ A
(r, \theta, t) {\bf e}_{\phi}],
\end{equation}
where $B (r, \theta)$ is the toroidal component and $\Bf_p 
= \nabla \times A (r, \theta, t)$
is the poloidal component. We can write the velocity field
as $\vb + r \sin \theta \, \Omega (r, \theta) {\bf e}_{
\phi}$, where 
$\Omega (r, \theta)$ is the angular velocity in the interior of the
Sun and $\vb$ is the velocity of meridional circulation having components
in $r$ and $\theta$ directions.  Then the main equations telling us
how the poloidal and the toroidal fields evolve with time are
\begin{equation}
\label{eq:C2_A}
\frac{\pa A}{\pa t} + \frac{1}{s}(\vf.\nabla)(s A)
= \lambda_T \left( \nabla^2 - \frac{1}{s^2} \right) A + S(r, \theta, t),
\end{equation}
\begin{equation}
\label{eq:C2_B}
\frac{\pa B}{\pa t} 
+ \frac{1}{r} \left[ \frac{\pa}{\pa r}
(r v_r B) + \frac{\pa}{\pa \theta}(v_{\theta} B) \right]
= \lambda_T \left( \nabla^2 - \frac{1}{s^2} \right) B 
+ s(\Bf_p.\nabla)\Omega + \frac{1}{r}\frac{d\lambda_T}{dr}
\frac{\partial}{\partial{r}}(r B),
\end{equation}
where $s = r \sin \theta$, $\lambda_T$ is the turbulent
diffusivity and $S(r,\theta, t)$ is the dynamo source term.

The term $s(\Bf_p.\nabla)\Omega$ in Equation~\ref{eq:C2_B} corresponds to process
(i) involving the generation of the toroidal field from the
poloidal field involving differential rotation.  On the other
hand, the terms $s^{-1}(\vf.\nabla)(s A)$ and
$\lambda_T \left( \nabla^2 - 1/ s^2 \right) A$ in Equation~\ref{eq:C2_A} correspond
to process (iii) involving the advection of the poloidal field
due to the meridional circulation and the turbulent diffusion
together. It is the source term $S(r,\theta, t)$ which incorporates
the process (ii).  Sometimes we have to do some extra things
to Eq.~\ref{eq:C2_A} as well (as described below) in order to include the
magnetic buoyancy of the toroidal field $B$.

In the early $\alpha \Omega$ dynamo model postulated by
\citet{Parker55a} and \citet{SKR66}, the
source term is $S = \alpha B$ (see, for example, \citet{Chou98}, Chapter 16). Here $\alpha$ is a measure of helical turbulence and is usually referred as the $\alpha$-effect.
Although the Babcock--Leighton mechanism also can be encaptured
by a superficially similar $\alpha$-coefficient, its physical
interpretation is completely different from that of the
$\alpha$-effect, which implies the twisting of the toroidal
field. When it was realized that the toroidal field at the
bottom of the convection zone is much stronger than what was
assumed earlier \citep{CG87,chou89,Dsilva93,Fan93,Caligari95} and the traditional $\alpha$-effect would be suppressed, the Babcock--Leighton mechanism was invoked to
take its place, with a similar-looking $\alpha$-coefficient
having a different interpretation \citep{Durney97}. Since the
Babcock--Leighton mechanism primarily takes place near the surface,
the $\alpha$-coefficient corresponding to it is usually assumed
to be confined near the solar surface.  \citet{CSD95} simply took $S = \alpha B$, with $\alpha$
concentrated at the surface as expected.  Even with such a
source function which did not include magnetic buoyancy
explicitly, they were able to get a periodic solution because
of the term
$$\frac{1}{r} \left[ \frac{\pa}{\pa r}
(r v_r B) + \frac{\pa}{\pa \theta}(v_{\theta} B) \right]$$
in Eq.~\ref{eq:C2_B}, which implied that the toroidal field generated at
the tachocline was advected by the meridional circulation
to the surface where the Babcock--Leighton mechanism operated
on it. \citet{Kuker01} also followed this
approach.

Although it is possible to construct a Babcock--Leighton dynamo
model in this way without explicitly including magnetic buoyancy,
this is certainly not very physical or satisfactory. When the
toroidal field becomes sufficiently strong, its rise time due
to magnetic buoyancy is expected to be much shorter than the
advection time  by the meridional circulation.  So magnetic buoyancy
is expected to dominate over such advection and has to be included
in the model to make it more realistic.  In the next Section,
we discuss two popular procedures for incorporating magnetic
buoyancy---a non-local procedure and a local procedure. We first
discuss how these procedures are used in order to obtain regular dynamo
solutions.  Then we shall point out in Section~\ref{C2:S4} that we get into
added complications when we want to study irregularities of
the solar cycle

\section{Non-local and local treatment of magnetic buoyancy}
\label{C2:S3}
The methods of specifying magnetic buoyancy in 2D kinematic
models of the flux transport dynamo can be broadly classified
into two categories: non-local and local.  Since the non-local
treatment of magnetic buoyancy is somewhat simpler, we discuss
that first.

\subsection{Non-local magnetic buoyancy}
\label{C2:S3_1}
The Babcock--Leighton mechanism essentially involves the
generation of the poloidal field near the surface from the
strong toroidal field at the bottom of the convection zone,
which has risen due to magnetic buoyancy. A simple way of
incorporating it is to take the source term $S(r, \theta,t)$
in Eq.~\ref{eq:C2_A} to have the form
\begin{equation}
\label{eq:nlocal}
S (r,\theta, t) = \alpha (r, \theta) B(r = r_{\rm bot}, \theta,t),
\end{equation}
in which we usually take $\alpha (r, \theta)$ to be significantly
non-zero only near the surface, to ensure that $S(r, \theta,t)$
makes a contribution only near the surface.  Also, to get
$S(r, \theta,t)$ near the surface, we multiply $\alpha (r, \theta)$
not by the toroidal field $B$ there, but by the toroidal field
$B(r= r_{\rm bot}, \theta, t)$ at the bottom of the convection
zone. Since the rise time due to magnetic buoyancy is small
compared to the period of the dynamo, we normally use the same $t$  in $S$ and $B$ without introducing any time delay.

To the best of our knowledge, the method of treating magnetic
buoyancy in this way was first proposed by \citet{CD99} in their study of the evolution of the solar poloidal field. Afterwards, \citet{DC99} adopted it
for their flux transport dynamo model. Some other authors have
followed this procedure in their dynamo calculations since
that time \citep{CD2000,Guerrero04,CSZ05,Hotta10}.  

This method of treating magnetic buoyancy is a simple, robust
and stable method. It is found that dynamo models based on
this method of treating magnetic buoyancy remain stable on
changing the values of basic parameters over wide ranges.
However, in spite of its simplicity and attractiveness, this
method has the following unphysical features.

(1) We expect the toroidal field to be unstable to magnetic
buoyancy only after it has become sufficiently strong.  However,
in the non-local method of treating buoyancy, even a very weak
toroidal field at the bottom of the convection zone starts
contributing to the source term in Eq.~\ref{eq:C2_A}. It is, in principle,
possible to put a threshold on the strength of the magnetic
field even in the non-local treatment, although very few
authors following this approach have done this \citep{CSZ05}.

(2) As a result of buoyant rise from the bottom of the convection
zone, the toroidal field at the bottom of the convection keeps
getting weaker.  The simple method of treating magnetic buoyancy
described above does not incorporate this effect and most of the authors
who treated magnetic buoyancy in this way did not allow
the weakening of the toroidal field due to magnetic buoyancy. 
As we shall see in the next Section, not doing this has serious
consequences when we study irregularities of the solar cycle
such as the Waldmeier effect.

It should be pointed out that the physics of how magnetic flux
depletion takes place in the convection zone is still rather
poorly understood.  The toroidal field of one cycle has to
be removed from the convection zone by the time the toroidal field
of the next cycle starts building up.  If the toroidal field has
a strength of $10^5$ G as suggested by the flux tube rise
simulations, then certainly it cannot be destroyed
by turbulent diffusion, which will be completely suppressed in
the presence of such a strong magnetic field. Then magnetic
buoyancy remains the only viable mechanism for removing the
magnetic field, although how this happens is unclear.  If the
toroidal field exists in the form of a flux ring going around
the rotation axis and a part of it rises to form active regions,
then other portions of the flux ring remain anchored at the
bottom of the convection zone and there is certainly not an
overall reduction of flux there.  In fact, \citet{RS01} have
pointed out that some processes can actually amplify the magnetic
field in the anchored part next to the rising portion of the
flux tube.  A scenario proposed by \citet{Chou03} is that only
some parts of the toroidal field become concentrated to $10^5$
G and rise to form active regions, whereas the magnetic field 
remains more diffuse and much weaker in other regions. If these
weaker fields diffuse by turbulent diffusion, then the buoyant rise of
concentrated magnetic fields may reduce the flux.  Although we
do not understand many aspects of the physics of magnetic fields
at the bottom of the convection zone, the assumption that the
magnetic flux is depleted due to magnetic buoyancy seems 
reasonable and is essential to explain certain aspects of
observational data, as we shall discuss in Section~\ref{C2:S4}.

\subsection{Local magnetic buoyancy}
\label{C2:S3_2}
In this approach, some toroidal magnetic field is transferred
from the bottom of the convection zone to the solar surface
and then this toroidal field at the surface is expected to
produce the poloidal field locally. Magnetic buoyancy is
known to be particularly destabilizing within the convection
zone \citep{Parker75,Moreno83}. So, as the toroidal
field evolves according to Eq.~\ref{eq:C2_B}, we check at periodic intervals
if the toroidal field at any point within the convection zone
becomes stronger than a critical value $B_c$ and, if so, then
some toroidal field from such region is transferred to the
surface.  It may be noted that this procedure introduces a limit
to the growth of the dynamo by not allowing $B$ to grow much
beyond $B_c$.  As a result, the dynamo can exhibit non-growing
oscillatory solutions even without introducing any kind of
quenching.  On the other hand, in the non-local procedure
described in Section~\ref{C2:S3_1}, it is absolutely essential to include
some kind of quenching (a quenching of the $\alpha$-coefficient
being the most common) to stop the runaway growth of the magnetic
field. After the toroidal field is shifted to the top of the
convection zone, we have to prescribe some way of generating
the poloidal field from it.  Two ways of doing this are
described in the next two paragraphs.

{\em Local $\alpha$ parameterization.} One way is to prescribe the 
source term $S(r,\theta,t)$ in Eq.~\ref{eq:C2_A} simply 
as a product of $\alpha(r, \theta)$ confined around the surface and
the toroidal field $B (r, \theta, t)$ there, which has been shifted
there from those regions at the bottom of the convection zone where
 $B$ exceeded $B_c$ in a way that ensured the conservation of
the toroidal flux (i.e.\ the amount of toroidal flux deposited
near the surface has to equal the amount of toroidal flux
removed from the bottom of the convection zone). This method
has been followed in several publications from our group
\citep{Nandy01,Nandy02,CNC04,CCN04,CNC05,CC06,CCJ07,
GoelChou09,CK09,KarakChou11,KarakChou12,KarakChou13,HKBC15}.

\begin{figure}
\begin{center}
\includegraphics[width=10 cm]{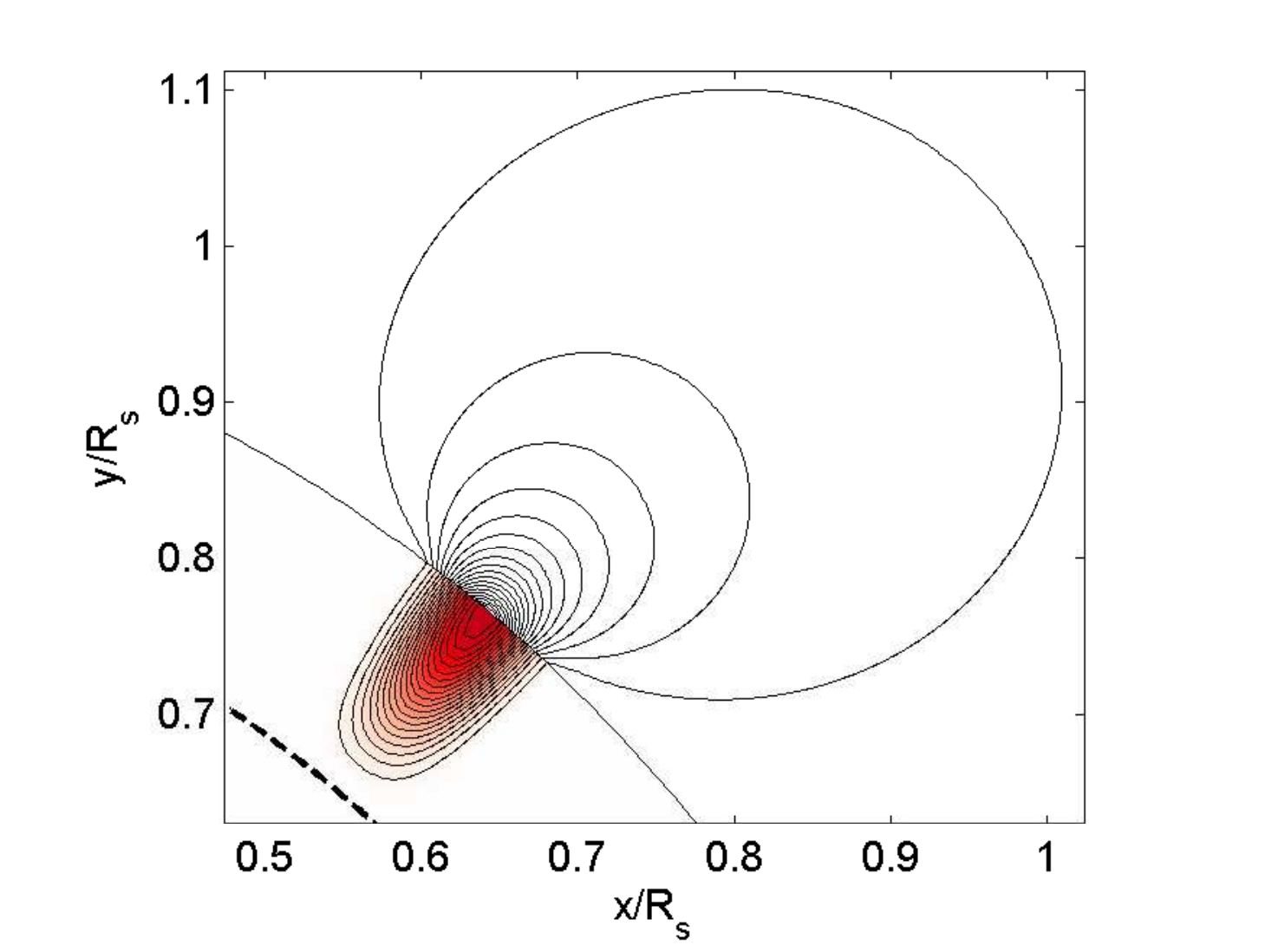}
\end{center} 
\caption[Poloidal field lines in Durney's double ring]{The poloidal field lines of a double ring formed
near the surface due to magnetic buoyancy.
Taken from \citet{Munoz10}.}
\label{figc2:dring}
\end{figure}

{\em {Durney's double ring method.}} The ideas of \citet{Bab61}
and \citet{Leighton69} were followed more closely by \citet{Durney95,Durney97}.
Due to the action of the Coriolis force, the rising
flux tube gets tilted \citep{Dsilva93} and produces
two sunspots of opposite polarity at slightly different
latitudes.  In an axisymmetric 2D formulation, we have to
average over longitude, which gives us two flux rings of
opposite sign at two slightly different latitudes. The generation
of the poloidal field is prescribed in the following way.
Whenever $B$ in some region within the convection exceeds
a critical value $B_c$, we assume that a part of this $B$ gives
rise to the flux ring above the region and we put $A$ appropriate
for this flux ring in Eq.~\ref{eq:C2_A}.  Here we do not discuss the details
of how we find $A$ appropriate for a flux ring, except to
mention that some assumption has to be made about the magnetic
field structure below and above the flux ring.  Presumably
the bipolar sunspot pair eventually gets disconnected from the toroidal
flux ring at the bottom of the convection zone from which it
formed, though it is not clear at the present time how and when
this disconnection takes place \citep{Longcope02}.
Figure~\ref{figc2:dring} from \citet{Munoz10} shows the typical poloidal 
field structure assumed by them in their dynamo model with
Durney's double ring algorithm. 

It is found that the above two methods---local $\alpha$
parameterization and Durney's double ring method---give
qualitatively similar results \citep{Nandy01}.
However, \citet{Munoz10} have argued that the
double ring method is a superior method from the conceptual
point of view.

It may be pointed out that the local treatment of magnetic buoyancy
also has an element of non-locality in it.  Whenever we put
some toroidal flux just below the surface or a double ring
at the surface, non-local considerations (the value of the
toroidal field near the bottom of the convection zone)
determine when and where this should be done.  However, once
the toroidal flux or the double ring has been put near the
surface, its subsequent evolution is governed by local physics.
On the other hand, in the non-local treatment, the source
term explicitly has a dependence on the magnetic field in
a different location, as seen in Eq.~\ref{eq:nlocal}. Whether `local' and
`non-local' are the best terminology to describe these approaches
can be questioned.  However, it is important to distinguish
between these two approaches and we have merely followed the
terminology used by many authors. 

\subsection{Comparison between the non-local and local
methods}
\label{C2:S3_3}
\begin{figure}
\begin{center}
\includegraphics[width=13 cm, height= 16 cm]{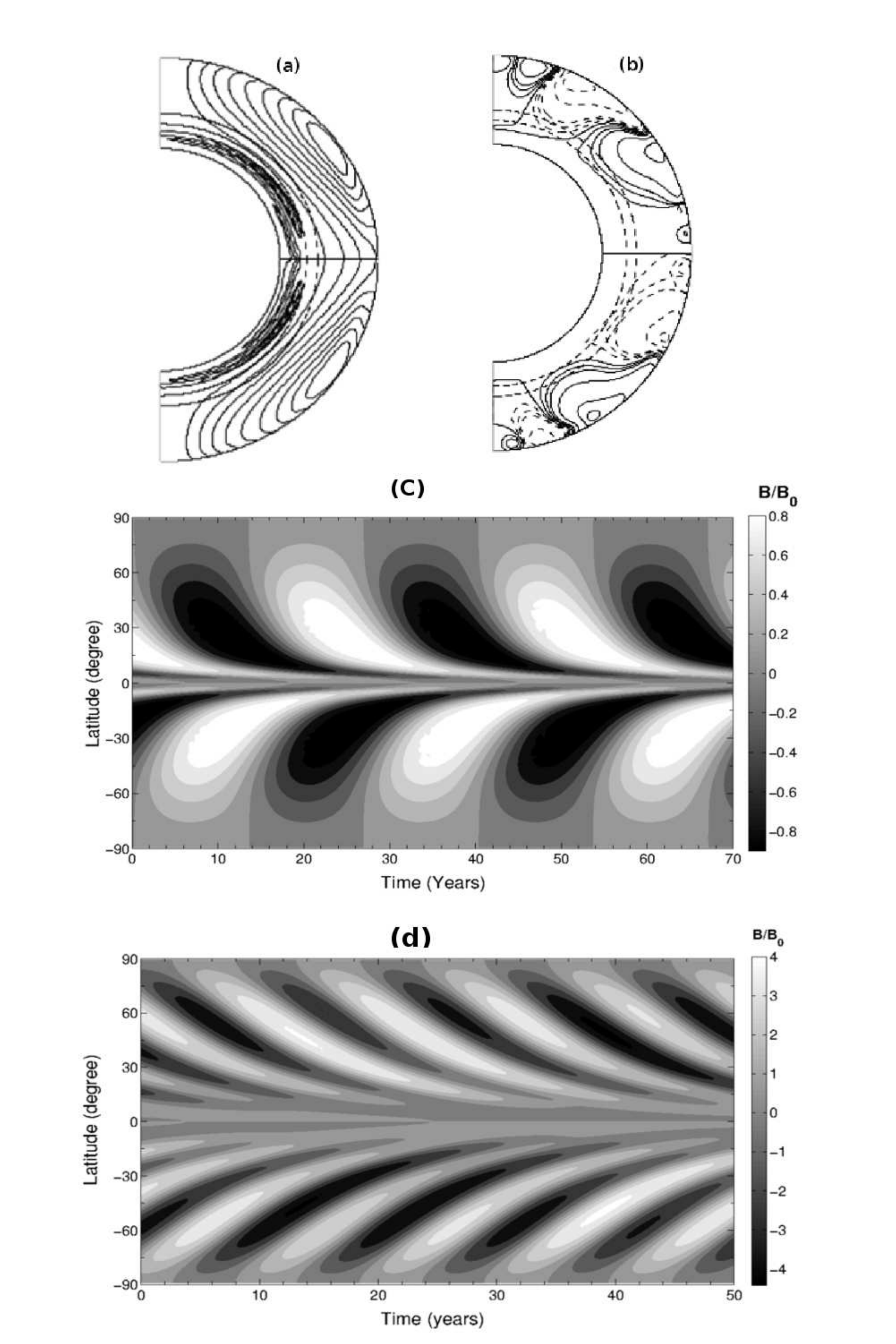} 
\end{center}
\caption[Poloidal and Toroidal magnetic fields for different treatment of magnetic buoyancy]{(a) and (b) Poloidal field lines at a particular instant of
time in two solar dynamo simulations.  Both use the same  combination of parameters which were used to generate the
standard model in \S~4 of \citet{CNC04}.
The only difference between the two runs is that (a) was
generated by using local buoyancy treatment and (b) was generated
by using the non-local buoyancy treatment.  Whereas (a) is
taken from Fig.~14 of \citet{CNC04},
(b) is taken from Fig.~1 of \citet{CNC05}. (c) and (d) are
the time-latitude plots of the toroidal field at the bottom
of the convection zone for these two cases.  It may be noted that,
in the local treatment of magnetic buoyancy adopted by
\citet{CNC04}, the flux eruptions are treated explicitly and
one can generate a theoretical butterfly diagram of sunspots,
as shown in Fig.~13 of \citet{CNC04}. Since this is not possible
in the non-local treatment, we have plotted the values of toroidal
field in both the cases for the sake of easy comparison.}
\label{figc2:comp}
\end{figure}

It would have been wonderful if the non-local and local methods
of treating magnetic buoyancy gave qualitatively similar
results.  Unfortunately, that is not the case! When this was
first discovered by \citet{CNC05}, it
came as a surprise to most of the researchers in the field,
though now from hindsight we feel that this should have been
an expected result.  Figures~\ref{figc2:comp}(a) and \ref{figc2:comp}(b) show the poloidal field configurations
in two dynamo models in which magnetic buoyancy is treated by non-local
and local ($\alpha$ parameterization) methods, but which are 
identical dynamo models in all other respects.  The time-latitude
plots of the toroidal field at the bottom of the convection
zone are shown in Figures~\ref{figc2:comp}(c) and \ref{figc2:comp}(d). When we use
non-local buoyancy, even the weak toroidal fields produced at the
high latitude start giving rise to the poloidal field. This
causes a multi-lobe structure of the poloidal field in this
case, shortening the period of the dynamo.  While the model
with the local $\alpha$ parameterization gives a period of
14 years, we get a period of barely 6.1 years on using the
non-local treatment of magnetic buoyancy. It is possible to
obtain solar-like solutions with the non-local treatment of
magnetic buoyancy with a completely different set of parameters,
as seen in the results of \citet{DC99} and other authors. However,
for the set of parameters which give solar-like solutions
with the local treatment of magnetic buoyancy, we get a totally
different kind of result on changing to a non-local treatment
of magnetic buoyancy (while keeping all the other things same).

Because of these big differences, we cannot avoid the question
as to which of these two methods is better.  We first note that
both the methods are gross over-simplifications of a complicated
3D process.  In this sense, neither of these two methods can be
taken as a realistic depiction of magnetic buoyancy.  We believe
that the local treatment is a more physical and realistic depiction
of what actually happens in the Sun---a conclusion that will
be further reinforced from the discussion of the next Section.
We have used this method of local treatment
in most of the calculations done in our group.  Unfortunately,
this method is less robust and less stable than the non-local method
and seems to give results only within a narrow range of various
important dynamo parameters.  So, when we want to vary our
dynamo parameters over a wide range or when we use unusual
parameter specifications (such as a multi-cell meridional
circulation in Chapter~\ref{C4}), we are often unable to obtain results with the
local $\alpha$ parameterization procedure and are forced to use
the more robust non-local procedure described in Section~\ref{C2:S3_1} 
\citep{KKC14,HKC14}.
\section{Modeling irregularities of the solar cycle}
\label{C2:S4}
The previous section outlined the various ways of treating
magnetic buoyancy and pointed out the differences in the
periodic dynamo solutions we get with them.  Within the last
few years, one of the goals of solar dynamo theory has been to
model the irregularities of the solar cycle. Only very recently
we realized that different formulations of magnetic buoyancy
may give radically different results in this important field
of study---especially when we consider irregularities of the
solar cycle caused by the variations in the meridional 
circulation.

The duration of the solar cycle becomes longer when the meridional
circulation slows down \citep{DC99,Yeates08}.
So it is expected that variations
in the meridional circulation will introduce irregularities
in the cycle.  When this problem is studied with the help
of local $\alpha$ parameterization, the theoretical results
are broadly in agreement with observational data \citep{Karak10,KarakChou11,CK12,KarakChou13,HKBC15}. 
Before explaining how things change on changing the method of treating magnetic
buoyancy, let us say a few words about the basic physics of
the problem.  When a cycle becomes longer due to the slowing
of the meridional circulation, two competing effects take place.
The differential rotation has more time to generate more
toroidal field, trying to make the cycle stronger.  On the
other hand, diffusion has also more time to act on the magnetic
field and tries to make the cycle weaker. Which of these two
effects wins over depends on the assumed value of the turbulent
diffusion coefficient in the convection zone.

Most of the calculations in our group were done with a value
of turbulent diffusion around $10^{12}$ cm$^2$ s$^{-1}$ consistent
with mixing length argument (Parker, 1979, p. 629; \citet{Jiang07}). 
With such a value of turbulent diffusion,
the effect of diffusion trying to make the longer cycle weaker wins
over the effect of differential rotation trying to make longer
cycles stronger.  As a result, longer cycles tend to be weaker.
This naturally leads to an explanation of the Waldmeier effect
that the rise time of the cycle is inversely correlated with
the strength of the cycle \citep{KarakChou11}. 
On the other hand, several papers from Dikpati and her
collaborators \citep{DC99,CD2000,DG06} used a value
of turbulent diffusion about 50 times smaller.  With such a
value of turbulent diffusion, the effect of differential rotation
trying to make longer cycles stronger wins over, giving the
opposite of the Waldmeier effect.

The better agreement with observational data on using the higher
value of turbulent diffusion clearly indicates that this higher
value must be closer to reality.  However, we realized only
recently that an appropriate handling of magnetic buoyancy is
also required to get a match with observational data.  We basically
need diffusion to be dominant so that longer cycles are weaker.
When we use a higher value of diffusion and treat magnetic 
buoyancy through local $\alpha$ parameterization, this happens
and we are able to explain effects like the Waldmeier effect
beautifully.  However, when we use the non-local buoyancy
method, we get into trouble even on taking the higher value of
diffusion.  In our calculations with local $\alpha$ parameterization,
toroidal flux is removed from the bottom of the convection zone
as a result of magnetic buoyancy.  On the other hand, in
the non-local buoyancy formulation used in several papers 
\citep{DC99,CD2000}, the toroidal flux is never depleted as a result of
magnetic buoyancy.  Because of this, the toroidal field keeps
becoming stronger when the cycle is longer and, even with a
high value of diffusion, we find that longer cycles tend to
be stronger, giving the opposite of the Waldmeier.

\begin{figure}
\begin{center}
\includegraphics[width=13cm]{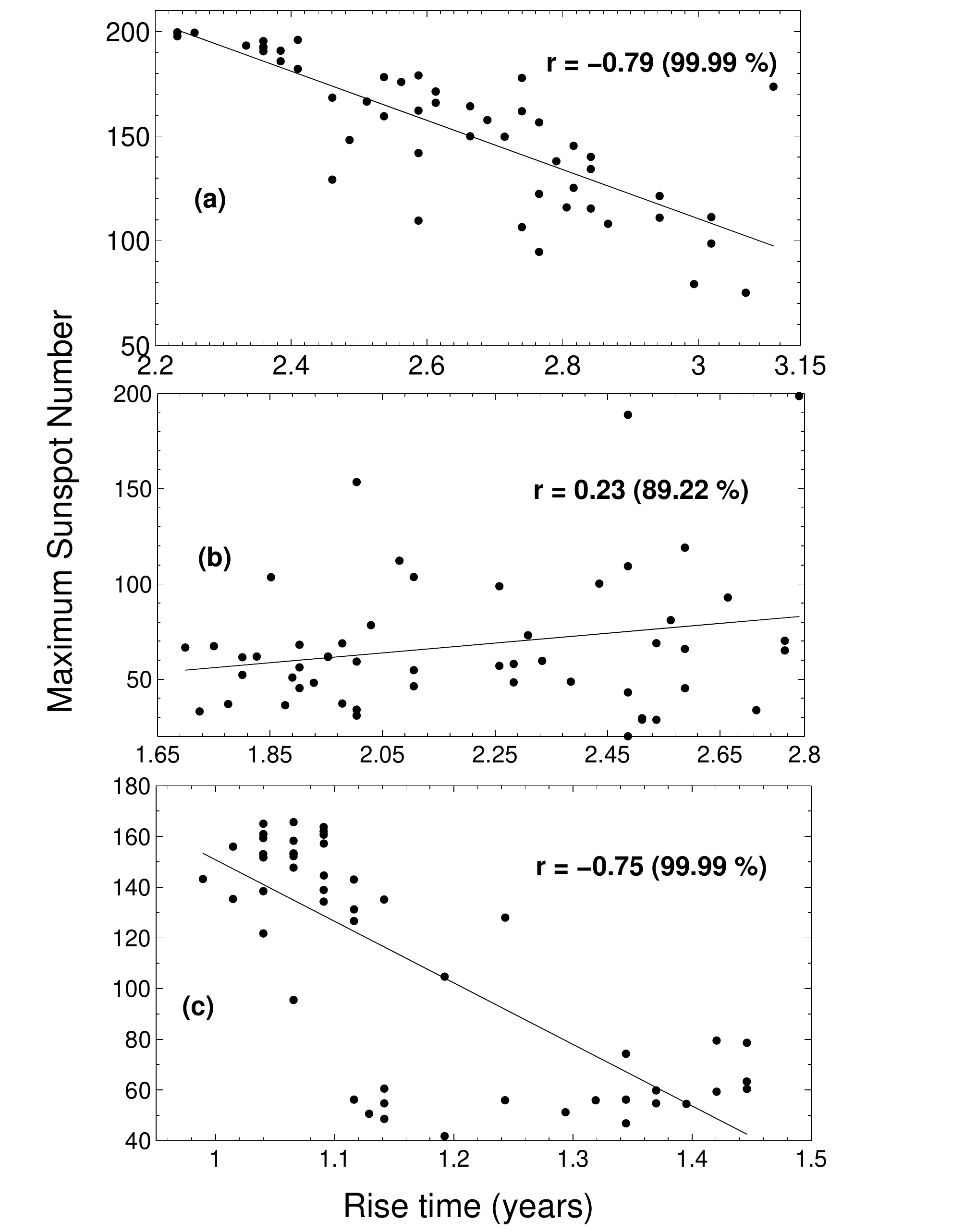} 
\end{center}
\caption[Reproduction of Waldmeier effect using different treatment of magnetic buoyancy]{The correlation between the rise and the strength\
of the solar cycle from a theoretical solar dynamo model using
the same combination of parameters as the high diffusivity model
presented by \citet{KarakChou11}. Only fluctuations
in the meridional circulation are included.  The panels (a)
and (b) are obtained by using respectively the local treatment and
the non-local treatment (without depletion in the toroidal flux)
of magnetic buoyancy. The last panel (c) obtained by using
the non-local treatment of magnetic buoyancy in which the toroidal
field is not allowed to grow larger than a critical value $B_c$.
The panel (a) essentially the same plot
as Fig.~4(a) of \citet{KarakChou11} obtained with a different realization of the randomness.}
% The last panel (c) obtained by using
%the non-local treatment of magnetic buoyancy in which the toroidal
%field is not allowed to grow larger than a critical value $B_c$.}
\label{figc2:corrl}
\end{figure}

Since this aspect of the problem was realized only recently
and has not yet been reported in regular journal publications,
we present some new results here.  Figure~\ref{figc2:corrl}(a) shows the Waldmeier
effect (i.e.\ the anti-correlation between the rise time and
the strength of the cycles) obtained with a dynamo model with
a high value of diffusion in the convection zone on using the
local $\alpha$ parameterization.  If we treat magnetic buoyancy
through the non-local method as given by Eq.~\ref{eq:nlocal}, while keeping
all other things of the dynamo unchanged, then we get the
opposite of the Waldmeier effect, i.e.\ a correlation rather
than an anti-correlation.  This is shown in Figure~\ref{figc2:corrl}(b). To verify
that this is indeed caused by not including the depletion of
the toroidal field by magnetic buoyancy, if we now modify the
non-local method slightly by putting the restriction that the
toroidal field is not allowed to grow beyond a limiting value
$B_c$ (i.e.\ we put a specification in the code that whenever
$B > B_c$ we set $B = B_c$), then we again get back the Waldmeier
effect.  The result is shown in Figure ~\ref{figc2:corrl}(c) which is obtained
by using the non-local buoyancy with the restriction that $B$ is
never allowed to grow beyond $B_c$.
%\begin{figure}
%\begin{center}
%\includegraphics[width=0.75\textwidth]{fig4} 
%\end{center}
%\caption{Same as Figure~3, but obtained by using
%the non-local treatment of magnetic buoyancy in which the toroidal
%field is not allowed to grow larger than a critical value $B_c$}
%\end{figure}
We thus have a rather peculiar situation.  Although the
non-local magnetic buoyancy method is more robust and works
for a wide range of parameters, if this method is used blindly
while studying irregularities caused by the variations in the
meridional circulation, we are likely to get completely wrong
results.  This should make it clear that none of the presently
used methods of treating magnetic buoyancy is completely
satisfactory.  Each method has its limitations and we have
to keep these limitations in mind when interpreting the results
obtained with a particular method. 

\section{The latitudinal distribution problem}
\label{C2:S5}
So far we have discussed problems of a realistic 2D formulation
of magnetic buoyancy, which is an intrinsically 3D process. Now
we mention another problem connected possibly with magnetic
buoyancy, which is still very poorly understood. In a helioseismic
map of differential rotation (see, for example, \citet{Schou98}), 
it is clearly seen that the radial differential rotation is concentrated 
more strongly at high latitudes than at low latitudes---the sign
being different at high and low latitudes.  Because of this, the
generation of the toroidal magnetic field from the poloidal magnetic
field is supposed to be more pronounced at the high latitudes.
At the low latitudes, the differential rotation present there first
has to `unwind' the toroidal field produced by the differential
rotation of the opposite sign at high latitudes and brought to the
low latitudes by the equatorward meridional circulation.  Only after
that, the differential rotation at low latitudes can build up the
toroidal field.  Hence, when we use the differential rotation
discovered by helioseismology, there is a propensity of stronger
toroidal field being produced at the high latitudes. If this
strong toroidal field is allowed to rise to the surface there
due to magnetic buoyancy, then we find sunspots at latitudes
higher than where they are seen.  The first authors to use
helioseismically determined differential rotation already noticed
this problem \citep{DC99, Kuker01}  

\citet{Nandy02} proposed a solution to this problem.
They suggested that the meridional circulation penetrates slightly
below the bottom of the convection zone, where the temperature
gradient is stable against convection and magnetic buoyancy
is suppressed to a large extent.  If this is the case, then
the strong toroidal field created in the high-latitude tachocline
may be pushed below the bottom of the convection zone and may not
be able to rise to the surface due to the suppression of magnetic
buoyancy there, thereby inhibiting formation of sunspots at high
latitudes.  Then the toroidal field would be advected equatorward
by the meridional circulation through layers slightly below the
bottom of the convection.  Then, when the penetrating meridional
circulation again enters the convection zone at low latitudes, the toroidal field
is brought into the convection zone and magnetic buoyancy again
takes over to produce sunspots at low latitudes.  This 
Nandy--Choudhuri hypothesis became a source of 
controversy---\citet{GM04} arguing that it is not
possible for the meridional circulation to penetrate into the
stable layers below the bottom of the convection zone, whereas
\citet{GB08} pointed out that some penetration to a limited extent is
possible. In many dynamo calculations from our group, we have
made the meridional circulation slightly penetrating in order
to confine sunspots to low latitudes. As \citet{CCC09} pointed out, one strong support for the
Nandy--Choudhuri hypothesis comes from the observation that
torsional oscillations begin at higher latitudes before the
start of a sunspot cycle.  Since torsional oscillations are
presumably driven by the Lorentz force of the dynamo-generated
magnetic field, this observation clearly suggests that the
strong toroidal field builds up in the high-latitude tachocline
a few years before this field is advected to low latitudes and
is able to produce sunspots, in accordance with the Nandy--Choudhuri
hypothesis.

In a study of the evolution of the poloidal field (in which
the dynamo equation was not solved), \citet{DC94}
mimicked the behavior of the dynamo by restricting the source
term of the poloidal field to low latitudes.  Within the last
few years, several authors have solved the dynamo equation
by artificially restricting the Babcock--Leighton source term
to low latitudes in this fashion \citep{Hotta10, Munoz10}.
While this procedure produces nice-looking butterfly diagrams with
sunspots restricted to low latitudes, these authors provide
no physical justification for this procedure.  An analysis
by \citet{Sch94} suggested that flux tubes stored at the bottom of
the convection zone are more likely to be unstable at lower
latitudes.  It is possible that such considerations
may provide a justification for restricting the Babcock--Leighton process
to low latitudes.  Curiously, based on their study 
magneto-rotational instability at the bottom of the convection
zone, \citet{PM07} concluded that the lower latitudes are more
stable---which were found to be unstable in the analysis of
\citet{Sch94}.  \citet{PM07} argued that this stability
is important for restricting the dynamo action at lower latitudes.
Certainly this issue needs to be studied in
more detail. In our opinion, we still do not have a
proper understanding of why sunspots are restricted only to
low latitudes.  Although this is an old problem and no significant
progress has been made in the last few years, we decided to
add this Section because the existence of this problem is
nowadays often not appreciated or acknowledged.  We should keep
this problem in mind and should not sweep it under the rug.
 
\section{Conclusion}
\label{C2:S6}
Our discussion has been restricted to kinematic dynamo models
in which the velocity field is assumed to be given.
Although we so far do not have anything that can be called
the standard model of the solar dynamo even within the
framework of kinematic models \citep{Chou08}, 
flux transport dynamo models developed by different groups
have lots of common features. While there is a controversy
on the nature of the meridional circulation at present, it
is still not clear whether a serious revision of existing dynamo
models will be required. Another uncertainty about the value
of turbulent diffusion seems to be resolved now in favor of
higher diffusivity (such that the diffusion time scale is a few
years) because only with such diffusivity it is possible to
model various aspects of cycle irregularities \citep{Chou15}. 
Whether turbulent pumping is important for the solar
dynamo is another question. However, since the inclusion of
turbulent pumping does not change the results of high-diffusion
dynamo models too much \citep{KarakNandy12}, this is presumably
not a large source of uncertainty.  

As we have discussed here, the biggest uncertainty in the flux
transport dynamo at the present time arises from the treatment
of magnetic buoyancy. There have been two widely used procedures
for treating magnetic buoyancy in 2D kinematic models of the flux
transport dynamo: (i) the non-local treatment
in which the toroidal field at the bottom of the convection zone is
assumed to contribute directly to the generation of the poloidal field,
without itself being depleted, and (ii) the local treatment in which 
a part of the toroidal
field from the bottom of the convection zone is shifted to the top
whenever it exceeds a critical value. With magnetic buoyancy treated in either
of these ways, it is possible to arrange the parameters of the model
in such a way that the dynamo solution reproduces various characteristics
of the solar cycle.  However, for the same set of parameters, we get
very different results on using these different treatments of magnetic
buoyancy.  The non-local treatment is more robust and is preferred when
we want to study the behaviour of our system over a wide range of
parameters.  On the other hand, the local treatment with the toroidal
field depletion at the bottom of the convection zone allows the proper
reproduction of results under irregular situations (such as the Waldmeier
effect) and appears the more physically realistic treatment.

This is certainly an unsatisfactory situation and presumably we cannot
model magnetic buoyancy sufficiently well if we restrict ourselves
only to strictly 2D situations.  It may be noted that, in the early
studies of magnetic buoyancy based on thin flux tube equation \citep{Spruit81, Chou90}, 
some physical effects could be studied
through 2D calculations \citep{CG87, CD90, DC91}.
However, a proper study of the tilt of bipolar sunspot regions (which is responsible for
the Babcock--Leighton process) could be carried out only when
the calculations were carried out in 3D \citep{Dsilva93, Fan93, Caligari95}.
Essentially, magnetic buoyancy is inherently a 3D process and
cannot be included very satisfactorily in 2D kinematic dynamo
models.  Further advances in 2D kinematic models are unlikely to
solve the problems we are facing now.

The ultimate goal of solar dynamo models is to construct fully
dynamically consistent 3D models, in which the velocity fields
are also calculated from the fundamental equations \citep{Karakreview14}.  
Beyond 2D kinematic models, however, there is an 
intermediate possibility---3D kinematic models, in which the
mean velocity fields (such as differential rotation and meridional
circulation) are supposed to be axisymmetric and are specified
(i.e.\ not calculated from the basic fluid equations), but the
magnetic field is treated in a full 3D fashion and is allowed to
evolve non-axisymmetrically so that we are able to model the
formation of tilted bipolar sunspots and the Babcock--Leighton
process more realistically.  The relatively few citations (according
to ADS) to the first paper suggesting this approach \citep{YM13} 
indicate that the importance of this approach is not generally
recognized.  But we believe that this is going to be the next
important step in flux transport dynamo theory beyond 2D kinematic
model.  Only a few authors have so far presented results with this approach \citep{MD14}.
This approach, when developed and integrated with a 
realistic dynamo model, should be more satisfactory than the currently widely used two 
treatments of magnetic buoyancy in 2D kinematic models. However,
a 3D kinematic model is not expected to immediately 
solve all the problems connected with magnetic buoyancy 
in the 2D kinematic model. For example, we have
discussed in Section~\ref{C2:S5} the problem of why sunspots appear in lower
latitudes.  This problem will not be solved with the 3D formulation
of magnetic buoyancy.  Rather, we shall need a better understanding
of the physics of the tachocline to address this question.  We
thus expect a combination of more realistic calculation of 
magnetic buoyancy coupled with a better understanding of the
tachocline to put the flux transport dynamo on a more satisfactory
footing.%footnote
\blfootnote{This chapter is based on \citet{CH16}.}

%% file: chapter3.tex
\begin{savequote}[100mm]
``The true adventurer goes forth aimless and uncalculating to meet and greet unknown fate."
\qauthor{--O. Henry}
\end{savequote}
\def\we{Waldmeier effect}
\def\sc{solar cycle}
\def\amp{amplitude}
\def\per{period}
\def\mc{meridional circulation}
\def\pf{poloidal field}
\def\td{turbulent diffusivity}
\def\ftdm{flux transport dynamo model}
\def\Rs{R_{\odot}}
\def\er{\mbox{erf}}
\def\blue{\textcolor{black}}
\newcommand{\Fig}[1]{Figure~\ref{#1}}
\chapter{Explaining various Irregular Properties of the Solar Cycle}
\label{C3}
\begin{quote}\small
Using different proxies of solar activity, we have studied the following features of
solar cycle. ({i}) A linear correlation between the amplitude of cycle and its decay
rate, ({ii}) a linear correlation between the amplitude of cycle $n$ and the decay 
rate of cycle $(n - 1)$ and ({iii}) an anti-correlation between the amplitude of cycle 
$n$ and the period of cycle $(n - 1)$. Features ({ii}) and ({iii}) are very useful because 
they provide precursors for future cycles. From 90 years of persistent Kodaikanal 
sunspot area data, the decay rate correlations (i), (ii) and Waldmeier effects are also reproduced. 
We have explained these features using a flux transport dynamo model with stochastic fluctuations in the
Babcock-Leighton $\alpha$ effect and in the meridional circulation.  
Only when we introduce fluctuations in meridional circulation, we are able to 
reproduce different observed features of solar cycle. We discuss the possible reasons 
for these correlations
\end{quote}
\section{Introduction}
\label{C3:S1}
Solar cycles are  asymmetric with respect to their maxima, the rise time being
shorter than the decay time. While the cycle amplitude (peak value) and the duration have cycle-to-cycle variations, we find some correlations among different
quantities connected with the solar cycle. Since 1935, it has been realized 
that the stronger cycles take less time to rise than the weaker ones 
(Waldmeier, 1935). This anti-correlation between
rise times and peak values of the solar cycle is popularly known as the \we.
\citet{KarakChou11} have defined this aspect of the \we\ as WE1,
whereas the correlation between the rise rates and the peak values is called WE2
(see also \citet{CS08}).
Although WE2 is a more robust feature of the solar cycle,
\citet{KarakChou11} have shown
that  both WE1 and WE2 exist in many proxies of the solar cycle. We have also shown 
that the Waldmeier effects (both WE1 and WE2) exist in the 90 years persistent sunspots area data
from Kodaikanal observatory.
WE2 provides a valuable precursor for predicting solar cycles because one can predict the strength of
a cycle once it has just started (see \citet{Lantos00,Kane08} ).

The declining phase of the cycle also provides important clues for understanding long-term variations.
We find that stronger cycles not only rise rapidly but also fall rapidly (shorter decay time). 
This results in a good correlation between the decay rate and amplitude of the same cycle.
However, defining the decay rate differently, \citet{CS08}
did not find a significant correlation between the decay rate and amplitude.
Furthermore, we find a strong correlation between the decay rate of the current cycle and 
the amplitude of the next cycle, 
which was also found by Yoshida and Yamagishi (2010). 
The decay time, however, is found to 
have no correlation with the amplitude of the same cycle. 
Another important feature observed is that the amplitude of the cycle
is inversely correlated with the period of the previous cycle \citep{Hathaway02, Solanki02b, Ogurtsov11}. 
These two correlations again provide promising precursors 
to predict the strength of the future
cycle \citep{Solanki02b, Watari08}.

Apart from showing these correlations from observational data, we also attempt
to provide theoretical explanations for them. A dynamo mechanism operating in the solar convection zone is believed to be 
responsible for producing the solar cycle. 
It is generally accepted that the strong toroidal field (responsible for the
formation of bipolar sunspots) is produced from the poloidal field by differential rotation
in the solar convection zone \citep{Parker55a}. This is the first part of solar dynamo theory. Due
to magnetic buoyancy \citep{Parker55b} the flux tubes of toroidal field erupt out
through the surface to form bipolar sunspot regions. These bipolar sunspots
acquire tilts due to the action of the Coriolis force during their journey through
the convection zone, giving rise to Joy's law \citep{Dsilva93}. To
complete dynamo action, the toroidal field has to be converted back into the
poloidal field. One possible mechanism for generating the poloidal
field is the Babcock--Leighton (B-L) process \citep{Bab61,Leighton69}, for which we now have strong observational support \citep{DasiEspuig10, Kitchatinov11a, Munoz13}. 
In this process, the fluxes of tilted bipolar active regions spread on the solar surface through
different processes (diffusion, meridional circulation, differential rotation)
to produce the poloidal field.
A model of the solar dynamo that includes a coherent meridional circulation and this B-L mechanism for
the generation of the poloidal field is called the flux transport dynamo model. This model  was proposed in the 1990s
\citep{WSN91,Durney95,CSD95}
and has been successful in reproducing many observed regular
as well as irregular features of the solar cycle \citep{CD2000,Kuker01, Nandy02, CNC04, Guerrero04, CK09, Hotta10, KarakChou13}. Recently \citet{Charbonneau10}
, \citet{Chou11} and \citet{Karakreview14} have reviewed this dynamo model. 
%The dynamo models which include turbulent inductive effects 
%originally proposed by Parker (1955b) for the 
%generation of the poloidal field are also able to produce few observed features of the solar cycle 
%(K\"apyl\"a et al.,\ 2006; Pipin and Kosovichev, 2011).

An important ingredient in flux transport dynamo is the \mc, which is not completely constrained 
either from observations or from theoretical studies.
Until recently not much was known about the detailed structure of the meridional
circulation in the convection zone \citep{Zhao13, Schad13}. Therefore, most of the
dynamo models use a single-cell \mc\ in each hemisphere.
However, we have shown in Chapter~\ref{C4} that a complicated multi-cellular \mc\
also retains many of the attractive features of the flux transport dynamo model if there is an equator-ward
propagating meridional circulation near the bottom of the convection zone or if
there is an equator-ward turbulent pumping \citep{Guerrero08,HN16}.
While most of the calculations in this study are done for a single-cell
meridional circulation, we show that the results remain qualitatively similar
for more complicated meridional circulations.

Since we want to do a theoretical study of the irregularities in the solar cycle, 
let us consider the sources of irregularities in the \ftdm\  that make different solar cycles unequal.
At present we know two major sources: (i) variations in the poloidal field generation due
to fluctuations in the B-L process \citep{CCJ07,GoelChou09} and (ii) variations in the meridional circulation \citep{Karak10, KarakChou11}. 
Direct observations of the polar field during last three cycles 
\citep{SCK05}, as well as its proxies such as the polar faculae and the active network index available for 
about last 100 years \citep{Munoz13,Priyal14}, indicate  large cycle-to-cycle variations of the polar field. 
The poloidal field generation mechanism mainly depends 
on the tilts of active regions, their magnetic fluxes and the meridional circulation,
all of which have temporal variations. Particularly the scatter of tilt angles
around the mean, caused by the effect of convective turbulence on rising flux
tubes \citep{Longcope02}, has
been studied by many authors \citep{WS89,DasiEspuig10}. 
Recently \citet{Jiang14} 
%observed a large scatter of the tilt angledistribution and 
found that the tilt angle scatter led to a variation in the polar field  by
about $30\%$ for cycle 17. In fact, even a single big sunspot group with large tilt angle 
and large area appearing 
near the equator can change the polar field significantly \citep{Cameron13}.
On the other hand, for the meridional circulation, we have some surface measurements for 
about last 20 years, showing significant temporal variations \citep{CD01,Hathaway10b}. 
Although our theoretical understanding of the \mc\ is very limited, all existing spherical global convection simulations do show significant variations
in the \mc\ \citep{PCB12,Karak15}.
Introducing randomness in the poloidal field generation and in the 
\mc, \citet{KarakChou11} have been able to reproduce the Waldmeier effect in their
high diffusivity dynamo model.
When the \mc\ becomes weaker, the cycle period and hence the rise time becomes longer.
The longer cycle period allows the
turbulent diffusion to act for a longer time, making the cycle amplitude weaker
\citep{Yeates08,Karak10} and leading to the Waldmeier effect.
The variation of the meridional circulation is crucial in 
reproducing this effect.

The motivation of the present work is to explore how
the decay rates of cycles are related to their amplitudes in a flux transport dynamo model, with the aim of explaining the observed correlations
%between the falling rates and the peaks of the same cycle and the next cycle, 
%and the anti-correlation between the cycle periods and the peaks of the next cycle
mentioned earlier.
The presentation of the chapter is following. In the next section, we summarize
some of the features of solar cycle that are often considered as precursors
of the solar cycle. In Section~\ref{C3:model},
we present a brief summary of our flux transport dynamo model and
then in Section~\ref{C3:S4} we introduce suitable stochastic fluctuations in the poloidal
field and the meridional circulation, in order to reproduce various
observed features of the solar cycle. 
Finally the last section summarizes our conclusions.

%%---------------------------------------observed part------------------------------------------%%

\section{Observational Studies}
\label{C3:S2}
We have used three different observational data sets: ({i})
Wolf sunspot number{\footnote{http://solarscience.msfc.nasa.gov/greenwch/spot\_num.txt}} 
(cycles 1--23),
({ii}) sunspot area{\footnote{http://solarscience.msfc.nasa.gov/greenwch/sunspot\_area.txt}} 
(cycles 12--23), and ({iii}) $10.7$~cm radio flux{\footnote{http://www.ngdc.noaa.gov/stp/solar/flux.html}} (available only for the last five cycles).
These parameters are very good proxies of magnetic activity and are often used to study the solar cycle (Hathaway {\it {et al.}}\ 2002).
To minimize the noise while keeping the underlying properties unchanged, 
we smooth these monthly data using a Gaussian filter having a full-width at half maximum (FWHM) of 
1~year. We also average the data with FWHM of 2~years to check how the results change with
the filtering. Apart from these three data sets, in Section~\ref{sec_wal}, we present some new irregular properties of solar cycle 
along with some existent irregular properties using recently digitized 90 years sunspot area data from Kodaikanal Observatory, India \citep{Sudip17}.

\begin{figure}[!t]
\centerline{\includegraphics[width=1.0\textwidth,clip=]{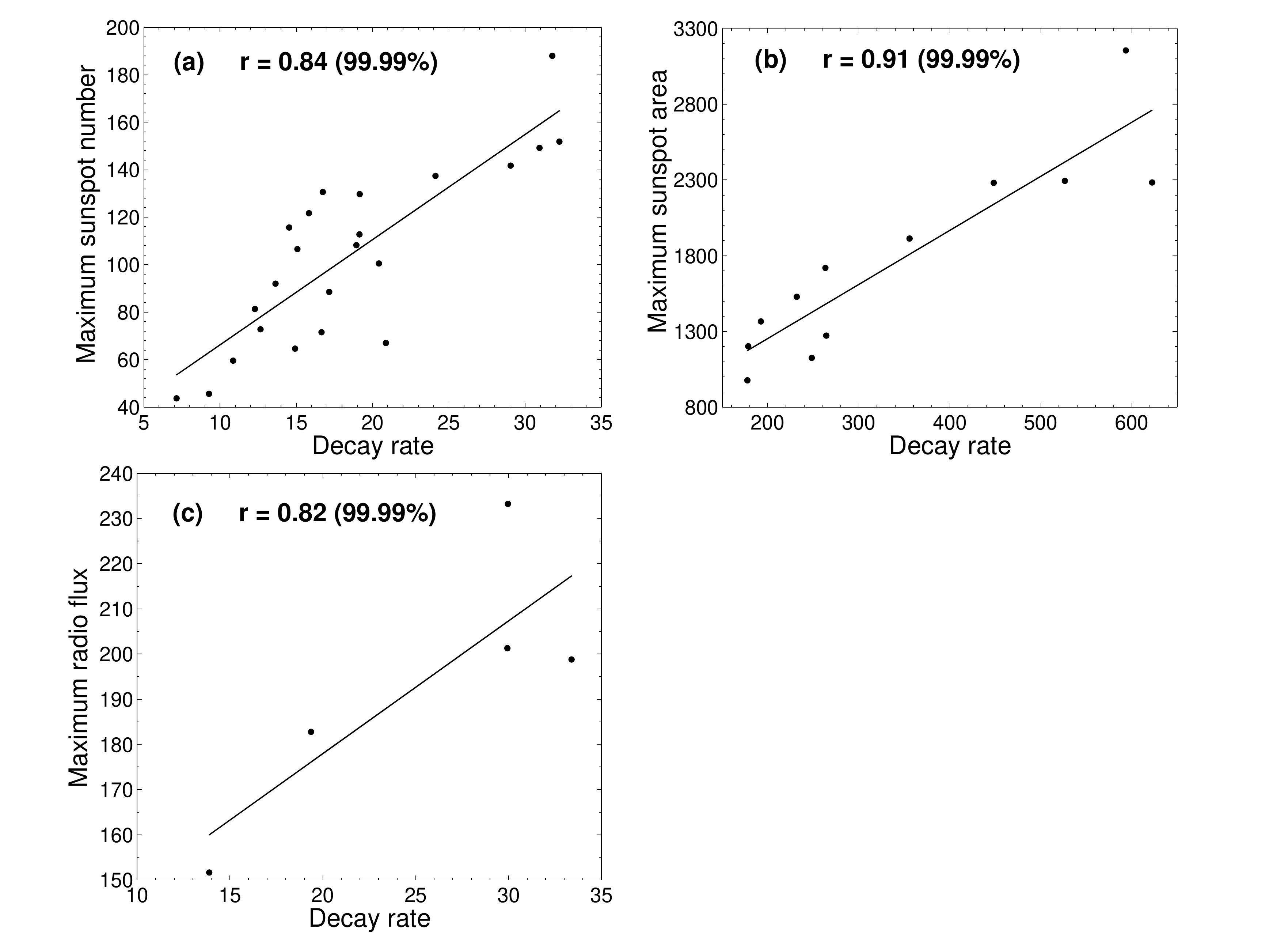} } 
\caption[Observed decay rate correlation with cycle amplitude from different sunspots data]{Scatter plots of the decay rate and the amplitude of the same cycle computed 
from (a) sunspot number, (b) sunspot area, and (c) $10.7$~cm radio flux data. 
In all these cases the original monthly data are smoothed using a Gaussian filter with FWHM of 2~years.The straight line in each plot is best linear fit of data. The correlation
coefficients ($r$) and the significance levels are also given in each plot.}
\label{obs1}
\end{figure}

\subsection{Correlation between the Decay Rate and the Cycle Amplitude}
\label{C3:S2_1}
We have calculated the decay rate at three different phases of the descending phase of the cycle, 
namely, early phase, late phase and entire phase. 
For the early phase, the decay rate is taken as the slope between two points with a separation of 1~year
with the first point one year after the cycle peak, whereas for the late phase the second point 
is taken 1~year before the cycle minimum. 
Here we exclude one year after the maximum when computing decay rate for the early phase
because sometimes the cycle peaks are not so prominent. While computing the decay rate for 
the late phase we also exclude 1~year before the minimum just to avoid the effect of 
overlapping between two cycles during solar minimum.
Finally, the decay rate of the entire decay part ({\it i.e}, entire phase) is taken as the average of the individual
decay rates computed at four different locations with a separation of one year starting from early
phase to the late phase. In \Fig{obs1}~(a), (b) and (c), we show the correlations of the cycle amplitudes with 
the decay rates of the entire phase computed from sunspot number, sunspot area and 
$10.7$~cm radio flux data, respectively.

We would like to point out that \citet{CS08} have computed the decay rate 
from the intervals of two fixed values of solar activity and they did not get 
significant correlation between the decay rate and the amplitude  
(see right column, Figure\ 2 of their paper).
The reason of not finding significant correlation is that they have calculated 
the decay rate in the late phase of the cycle, {\it i.e.} near the tail of the cycle 
where the rate of decay is really very small. 
We find that their values are  comparable with our decay rates computed in the late phase.
In 4th and 5th columns of Table~\ref{tab_obs} we have listed our values and the values computed following 
\citet{CS08} method (hereafter referred as CS08).
It is interesting to note that even for the radio flux data for which we have only five
data points, we get strong correlation; see Table~\ref{tab_obs} for details. 
Therefore we can see that if we determine the 
decay rates from the entire phase of the solar 
cycle or the early phase, we find strong correlation with the amplitude. 
Thus, to determine the decay rate from descending part of the solar cycle, we
need to consider the entire decay phase of the cycle, which provides a better estimate than CS08. 
\begin{table}
\caption{Correlation coefficients between different quantities of the solar cycle.}
\begin{tabular}{ccccccccc}
\hline
&&\multicolumn{6}{c}{Correlation coefficients of the decay rate with}& Correlation \\
&&\multicolumn{6}{c}{the amplitude of}& between the \\
\cline{3-8}
&&\multicolumn{4}{c}{Same cycle}&\multicolumn{2}{c}{Next cycle}& amplitude \\
\cline{3-8}
&&Entire &\multicolumn{2}{c}{Late decay phase}& Early & Entire & Late & and the\\
\cline{4-5}
Data set&FWHM &phase&Our & CS08's &Phase&phase&phase & previous\\
&&& value & value&&& & cycle period \\
\hline
Sunspot&1~yr&0.79&0.21&0.22&0.67 & 0.55 & 0.65 & -0.64\\
number&2~yr&0.86&0.45&--&0.86 & 0.61 & 0.83 & -0.67 \\ 
\hline
Sunspot&1~yr&0.84&0.20&0.11&0.69 & 0.14 & 0.37 & -0.49\\
area&2~yr&0.91&0.53&--&0.92 & 0.39 & 0.66 & -0.60\\ 
\hline
Radio&1~yr&0.86&-0.11&0.14&0.93&-0.42 & 0.64 & 0.11\\
flux &2~yr&0.82&0.24&--&0.95 & -0.43& 0.46 & 0.09 \\ 
\hline
\label{tab_obs}
\end{tabular}
\end{table}

\begin{figure}[!h]   
\centerline{\includegraphics[width=1.0\textwidth,clip=]{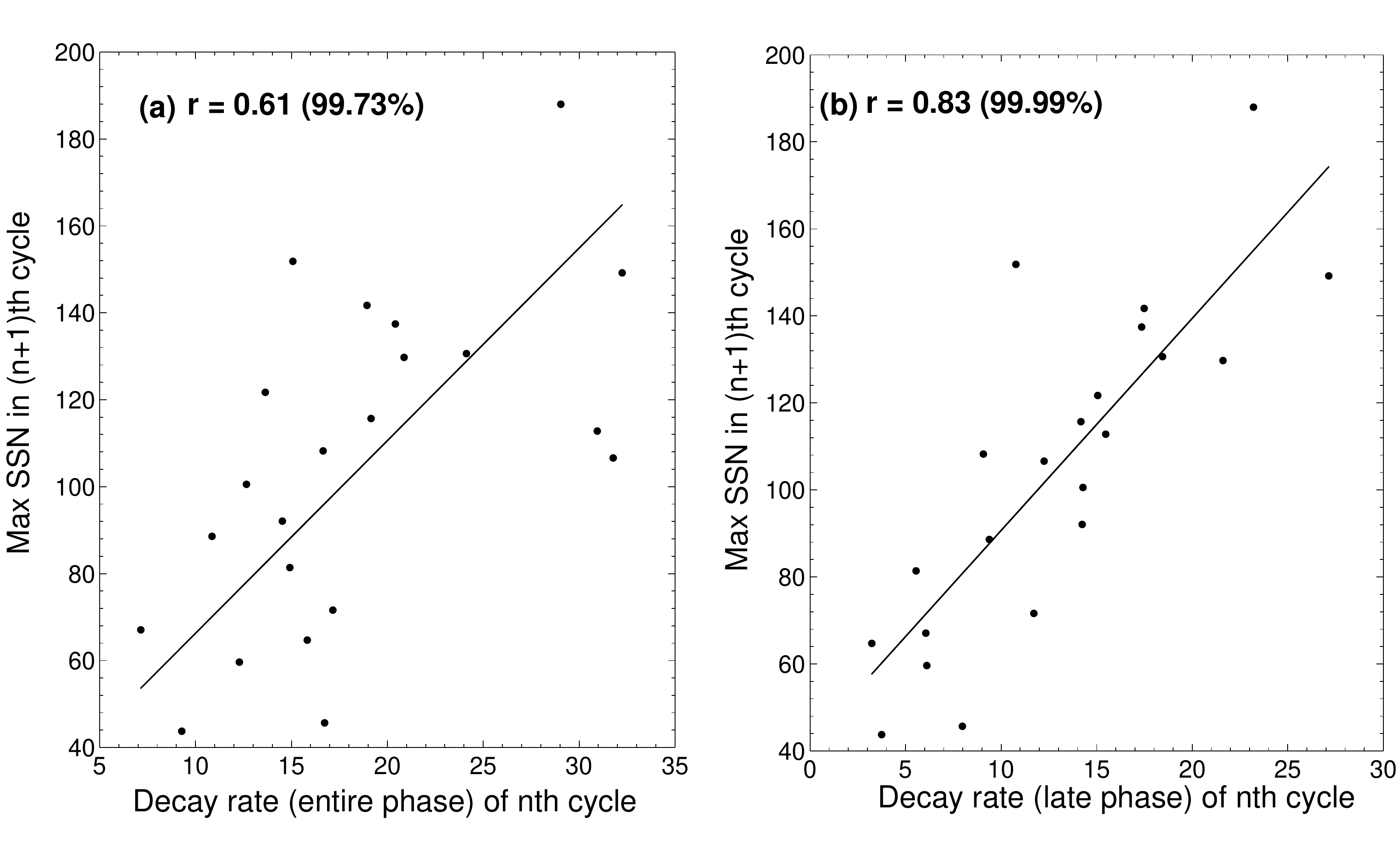} }
\caption[Correlation of decay rate with next cycle amplitude]{ Scatter plots showing the correlation of the amplitude vs. 
the decay rate of the previous cycle computed from sunspot number data (smoothed with FWHM of 2~year).
In (a) the decay rate is computed from the entire decay phase, whereas in (b) it is at 
late decay phase.} 
\label{figc3:obs2}
\end{figure}

\subsection{Correlation between the Decay Rate and the Next Cycle Amplitude}
\label{C3:S2_2}
Next we find that there is a significant correlation between the amplitude of 
cycle and the decay rate of the previous cycle. Again we find this correlation for all the data sets
considered here (see Table~\ref{tab_obs}). However in Figure~\ref{figc3:obs2}(a) we show this correlation only 
for sunspot number. Note that here the decay rates have been calculated from 
the entire decay phase as discussed in Section~\ref{C3:S2_1}.
This correlation suggests that the decay rate of a cycle carries some information 
of the strength of the next cycle.
It is interesting to note that when we look at this correlation with the 
decay rate computed in the late phase, the correlations become even stronger; see Figure~\ref{figc3:obs2}(b).
In 7th and 8th column of Table~\ref{tab_obs}, we show both correlations for all three data sets. 
These results suggest that particularly the late phase of the cycle carries more 
information of the forthcoming cycle.
%response!!!-------------------------------------- 
This correlation of decay rate with the amplitude of succeeding cycle is already reported in {\citet{Yoshida10}}. They have shown
this correlation for only sunspot number data and their methodology for calculation of 
decay rate (rate of decrease in sunspot number over some time) is somewhat different from our methodology.
They have studied decay rate in six different cases (see their Figures 1(a)-(f)). They have obtained the decay rate from
the decrease of sunspot number (SSN) over the period of 1, 2, 3, 4, 5 and 6 years before the minima 
of the cycle in the six different cases of study respectively.  Since solar cycles 
sometimes have overlapping regions during minima and it is difficult to ascertain
the actual minima, there are some uncertainties in the methodology of \citet{Yoshida10}. 
The correlation coefficient ($r$ = 0.70) obtained in the second case of their study (see their Figure 1(b)) 
should be the same with what we obtained during late phase correlation ($r$ = 0.83). Since they have not considered 
the overlapping region between the minima and used monthly smoothed SSN, the value of the correlation coefficient is slightly different.
%-------------------------------------------------------

\citet{CS07} (also see \citep{Brown76}) have observed similar feature that the 
activity level during the solar minimum is an indicator for the strength 
of the next solar cycle and argued that this is caused by overlap between 
two cycles during solar minimum.

In all our theoretical calculations (subsequent section), while studying the correlation between the 
amplitude and the decay rate of the same cycle, we shall consider the decay rate of the 
entire phase, but for the correlation with the next cycle we shall consider only 
the late-phase decay rate.

\begin{figure}[!t]
\centerline{\includegraphics[width=0.8\textwidth,clip=]{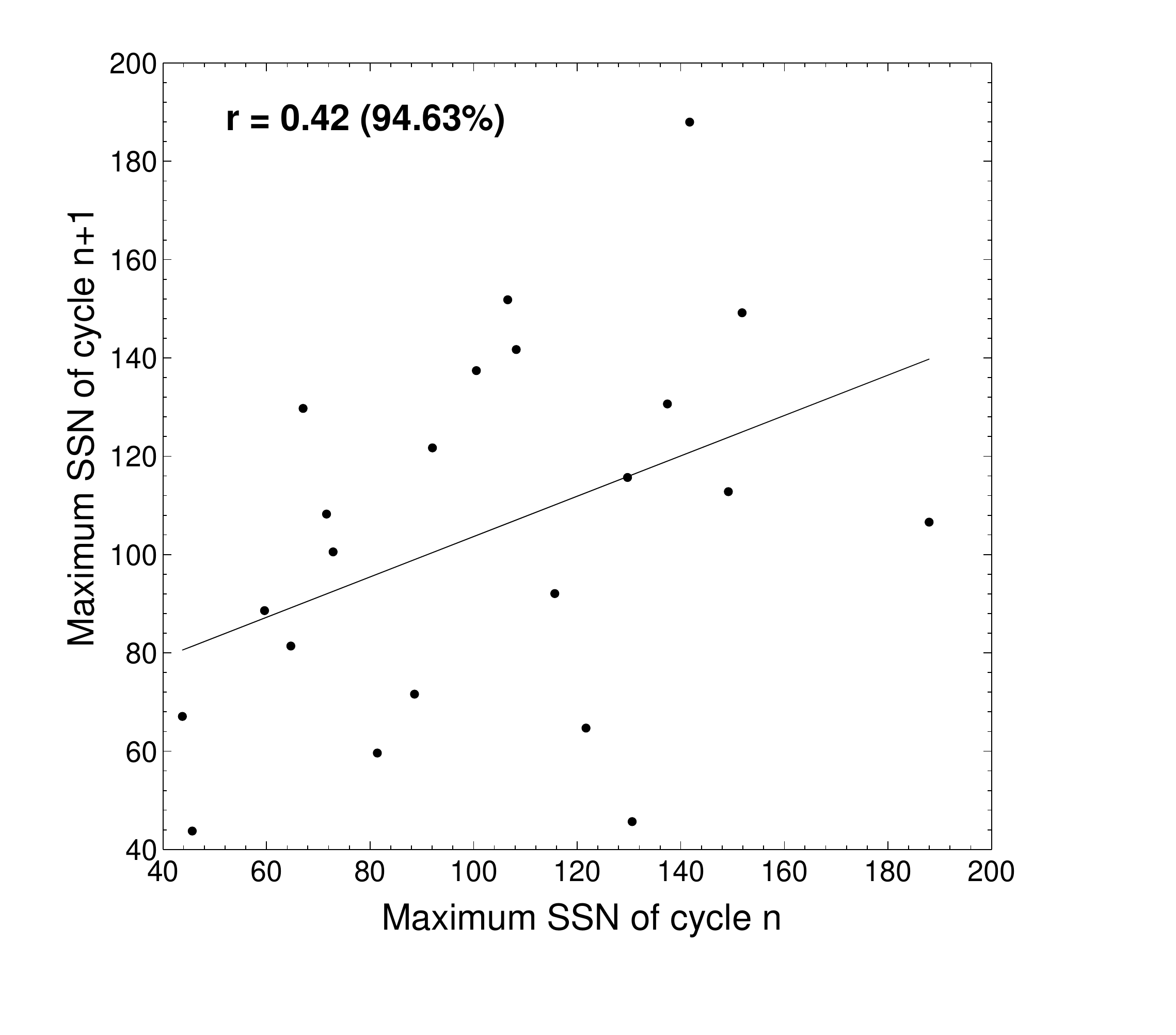}} 
\caption[Correlation plot of $n$th cycle amplitude 
and the amplitude of the next $n+1$ cycle from sunspot number data]{ Scatter plot of $n$th cycle amplitude 
and the amplitude of the next $n+1$ cycle from sunspot number data (smoothed with FWHM of 2~years).}
\label{amplcorl}
\end{figure}

Since the decay rate of the cycle $n$ is correlated both with the amplitude
of cycle $n$ (Figure~\ref{obs1}) and the amplitude of cycle $n+1$ (Figure~\ref{figc3:obs2}), one question that
naturally arises is whether the amplitude of cycle $n$ and the amplitude 
of cycle $n+1$ are themselves correlated.  We show a correlation plot
between these amplitudes in Figure~\ref{amplcorl}, demonstrating that there is not a
significant correlation. The challenge before a theoretical model is, therefore,
to explain how the decay rate of cycle $n$ is correlated both with the amplitude
of cycle $n$ and the amplitude of cycle $n+1$, while these amplitudes themselves
do not have a strong correlation.

\subsection{Correlation between the Cycle Period and the Next Cycle Amplitude}
\label{C3:S2_2}
\begin{figure}[!h]
\centerline{\includegraphics[width=0.75\textwidth,clip=]{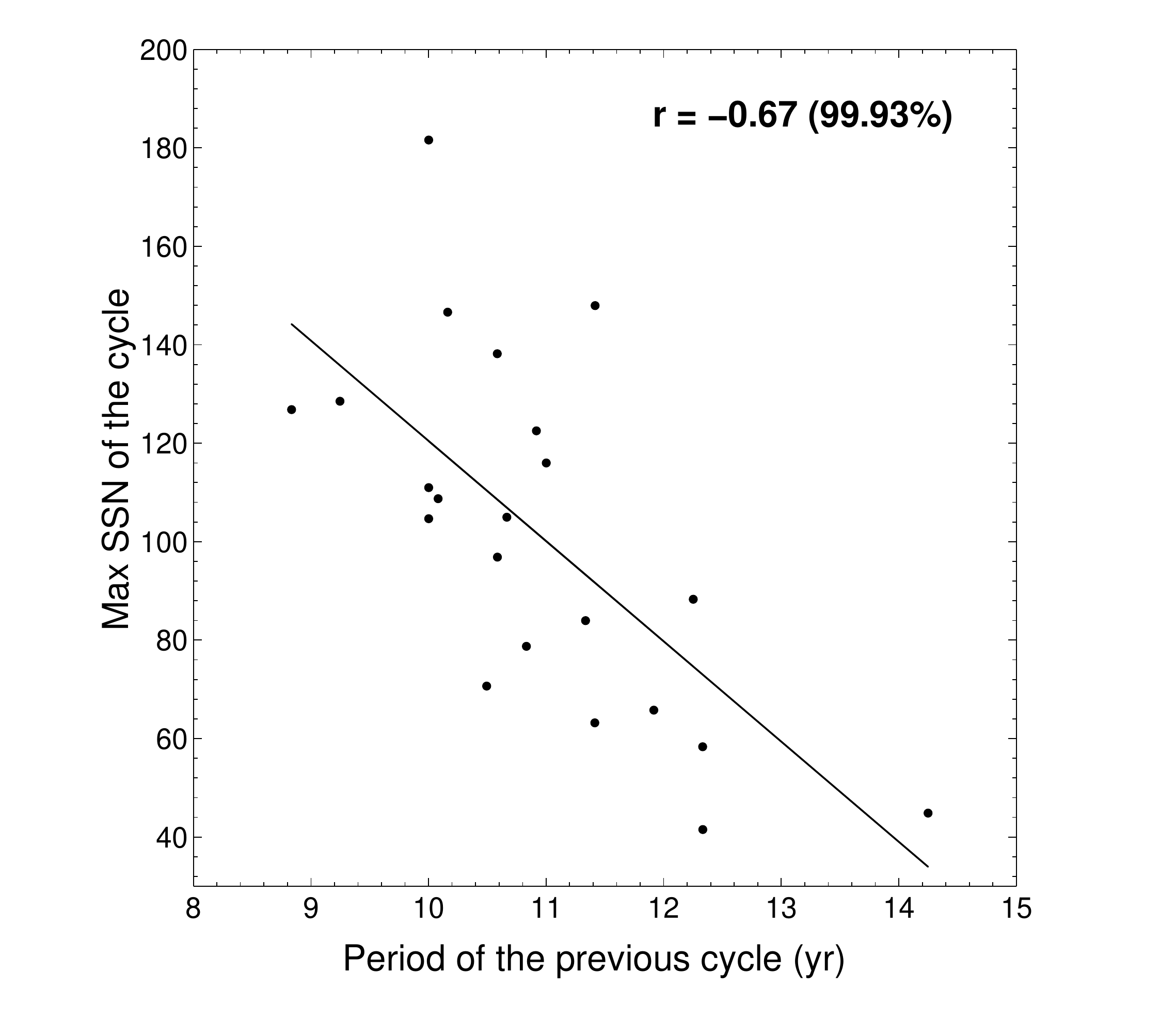}} 
\caption[correlation plot between the cycle amplitude 
and the period of the previous cycle]{Scatter plot showing the anti-correlation between the cycle amplitude 
and the period of the previous cycle from sunspot number data (smoothed with FWHM of 2~years).}
\label{percorl}
\end{figure}

Finally, we also find that the shorter cycles are followed by stronger cycles and 
vice versa. This produces an anti-correlation between the amplitude of a cycle 
and the period of the previous cycle \blue{\citep{Hathaway02, Solanki02b, Ogurtsov11}}. \Fig{percorl} shows this correlation from 
sunspot number data (smoothed using a Gaussian filter with FWHM of 2~years). 
The correlation coefficients  from other data are listed in Table~\ref{tab_obs}. For all data we have taken the period of 
the cycle just as the time difference between two successive minima.

\subsection{Some irregular features of solar cycle from century long sunspot area data of Kodaikanal Observatory}\label{sec_wal}
Kodaikanal Solar Observatory \citep{2010ASSP...19...12H} is observing the Sun since 1904. 
This century long observation has been carried out in three different wavelength bands, 
white-light (since 1904), Ca-K (since 1907) and in H-alpha (since 1912). 
Among these three wavelengths, white-light observations were taken without 
any changes of the telescope optics since 1918. This makes the data set very 
efficient for long term solar studies. Kodaikanal white-light images have been 
used in the past by many authors to study the time evolution of different sunspot parameters. 
\citet{1993SoPh..146...27S} used the white-light data, for a period of 45 years (1940-1985), 
and compared their results with Mount Wilson white-light data. In the subsequent years, 
various aspects of the sunspot and their variation with the solar cycles have been 
studied using this data set. Solar rotation rates \citep{1999SoPh..186...25H,1999SoPh..188..225G}, 
axial tilt of the sunspot groups \citep{1999SoPh..189...69S}, tilt 
angle and size variation of the sunspot groups \citep{2000SoPh..196..333H}, 
rotational rate variation with the age of the sunspot groups \citep{2003SoPh..214...65S} 
are some of the highlighted works produced from this data set. Here, we have calculated 
various correlations during ascending and descending phase of the solar cycles from this 
data set (1921-2011) and reproduced all the irregular features in those phases as given below.
\begin{figure*}[!t]
\centering
\includegraphics[width=.95\textwidth]{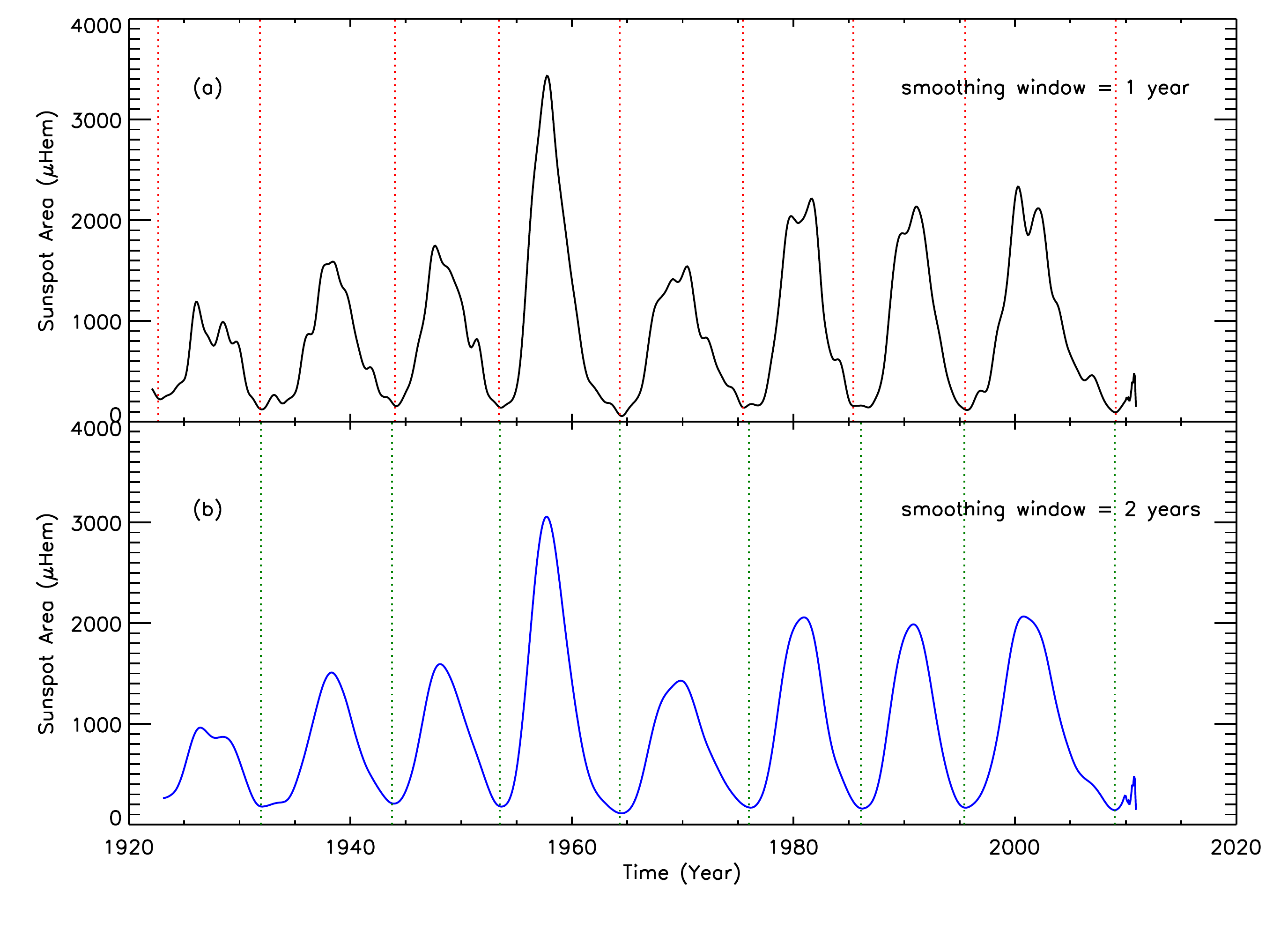} 

\caption{Panel (a) shows the yearly averaged solar cycle from kodaikanal sunspot area data. (b) 2 year averaged solar cycle from kodaikanal sunspot area data}

\label{avg_1_2}
\end{figure*}

%------------------------------------------------
\begin{figure*}[!t]
\centering
\includegraphics[width=1.0\textwidth]{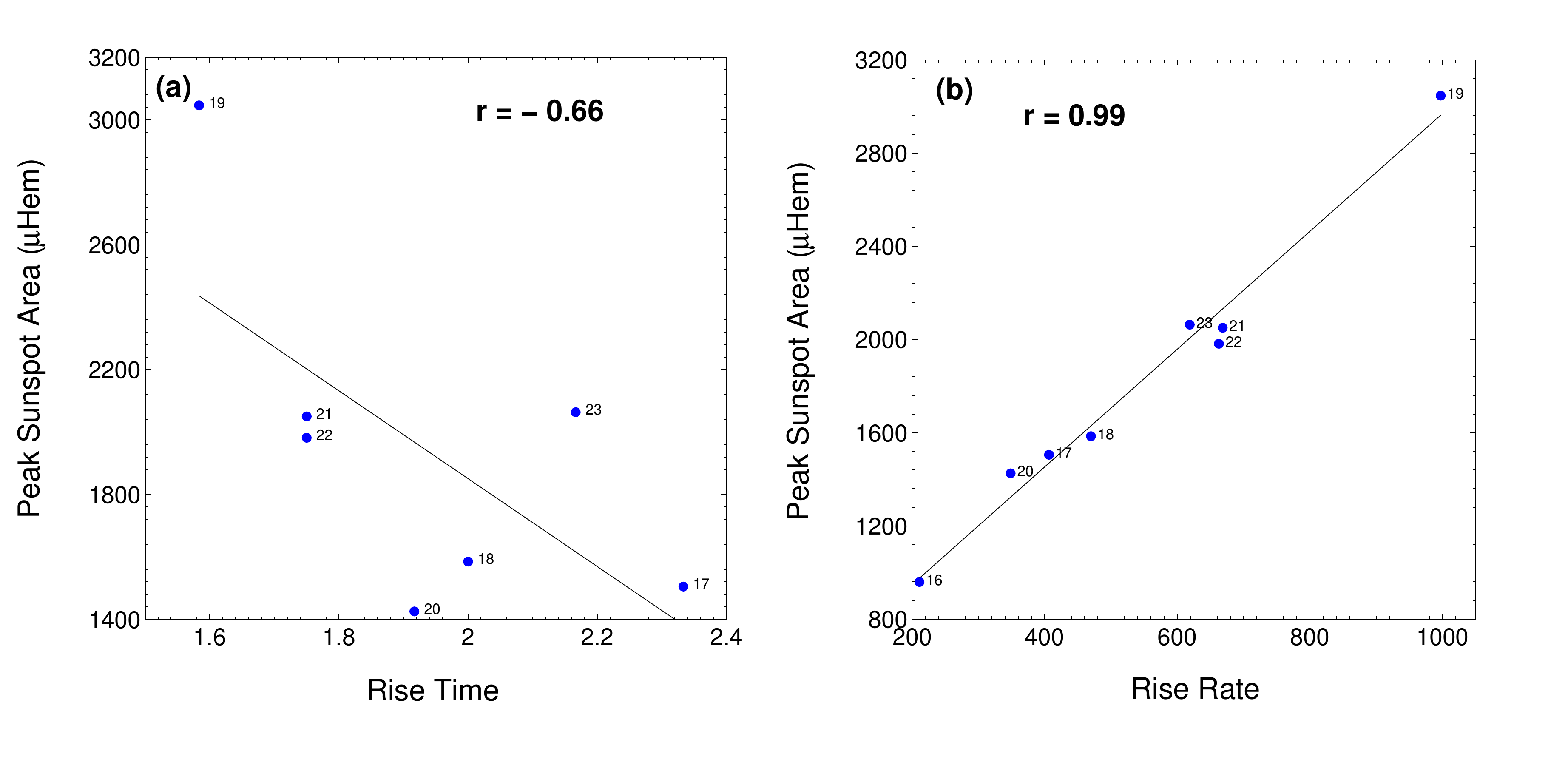} 

\caption[Waldmeier effects from century long Kodaikanal sunspot area data]{Panel (a) shows the scatter plot of rise time (in years) and the peak amplitude of same solar cycle (FWHM=2 year). Solid line represents best linear fit. Scatter plot of rise rate ($\mu$Hem/year) and the peak amplitude of same solar cycle (FWHM=2 year) is shown in panel (b). Solar cycle numbers also marked beside each of the blue circles. }
\label{we11}
\end{figure*}
\subsubsection{Waldmeier Effect: The correlations in ascending phase of the solar cycle}
Although the irregular properties of the solar cycles are well established but 
doubts are always expressed about their existence. Are they really properties of 
solar cycles or just statistical artifacts \citep{Dikpati08}? Time variation of daily 
sunspot area data or even the monthly averaged sunspot area data are so scattered that 
it is very difficult to calculate the required quantities like rise time and rise rate 
to validate their specific properties. So we need do the data smoothing. Here we have also 
smoothed our data by using a Gaussian filter having FWHM =1 year and FWHM = 2years. 
In figure~\ref{avg_1_2}(a) and figure~\ref{avg_1_2}(b) we have shown 1 year averaged and two years
averaged solar cycles respectively. Red vertical dotted lines represents the minima which is 
calculated only for 1 year averaged solar cycle and green vertical dotted line shows the minima 
for 2 years averaged dataset. If we see figure ~\ref{avg_1_2} carefully then we can understand that 
determining solar minima for each cycle is always very sensitive to the filtering window and minima 
calculated using 1 year averaged data is not exactly same with the minima of the 2 years averaged data. 
This is true for determination of solar maxima also. It is also found that there is always an overlap
between two successive cycles and the position of cycle minima depends on this overlapping \citep{CS07}. 
Thus, defining the solar cycle minima and maxima becomes very difficult. So the procedure 
which is followed in literature to determine Waldmeier effect is somewhat arbitrary. 
\citep{Dikpati08} had taken a straight forward approach to calculate rise time and rise rate. 
They have calculated the rise time as the time difference from minima to maxima for each cycle and 
ended up with no significant correlations between rise time and amplitude of the cycle (WE1). 
They concluded that ``{\it the inverse correlation between cycle peak and rise time originally 
found in Wolf sunspot number data by Waldmeier (1935) is not present in another solar cycle index, 
namely sunspot area data}''. \citet{KarakChou11} then slightly changed the procedure of determination 
of rise time and they calculated the rise time by taking the time difference between which the 
solar cycle reaches to $0.8$P from $0.2$P, where P = peak amplitude of the cycle. 
When they defined the rise time in this way they got a good correlation (See fig1(b) of \citet{KarakChou11}) 
which was not found in \citet{Dikpati08} calculations. We want to reconfirm the Waldmeier effect 
from the Kodaikanal sunspot area data. The reconfirmation of Waldmeier effect from 
completely different sunspot area data set is not only important for the prediction of 
the peak amplitude when cycle is in the early stage but also it is important to validate our existing dynamo models. 
Following \citet{KarakChou11}, we have calculated the rise time as the time differences 
between the time where solar cycle reaches to $0.8$P from $0.2$P, where P the peak sunspot area. 
On the other hand rise rate is defined as the slope between two points separated by one year and 
the first point is chosen one year before the cycle maxima. These calculation are done 
for both the 1 year and 2 years average dataset. Panel (a) in Figure~\ref{we11} 
shows the plot  between the rise time and the peak amplitude (WE1) of the cycle for 
2 years average dataset. We see an anti-correlation $(r = -0.66)$ between these two parameters 
and we can confirm that the Waldmeier effect is a true property of the solar cycle. 
Since the minima of the cycle 16 is not captured in 2 years averaged data (see Fig~\ref{avg_1_2}b), 
so we have plotted this correlation (WE1) with 7 data points for 7 cycles only. Similarly in panel (b) 
in Figure~\ref{we11}, the correlation between the rise rate and cycle amplitude is calculated and 
it is evident from the figure that WE2 is also well reproduced with correlation coefficients $r = 0.99$. 
Here we should mention that these correlation coefficient which we have calculated here are 
the normal Pearson correlation coefficient, and normal Pearson correlation coefficients are 
statistically significant when we have a large number of data points but for all our calculations 
we have only data points for 8 cycles. So are these correlation coefficients statistically significant? Yes, they are. 
The normal practice for calculating the correlation coefficient with less data points is to 
calculate the adjusted Pearson correlation coefficient ($r_{adj}$) which is given below
\begin{equation}
r^2_{adj} = 1 - (1 - r^2)\frac{n-1}{n-p-1}
\end{equation}
where r is the normal correlation coefficient, n is the number of data points and p is the independent variables. 
So the calculated values of adjusted coefficient for $r = -0.66$ (WE1) and $r = -0.99$ (WE2) 
become $r_{adj} = -0.57$ and $r_{adj} = 0.99$ respectively. So if you are dealing with a 
statistically high correlation coefficient then adjusted correlation coefficients give almost similar value. 
In Table~\ref{table:kodai} we have summarized the different correlation coefficient values which 
are calculated using Gaussian filter with a FWHM of 1 year. 
The correlations with 2 years averaged data is also enlisted in the Table for comparison. 
It is found that the statistical trends are almost similar for both the cases but the value of correlation coefficients differs slightly. % Please note that since cycle 17 and cycle 18 has almost same strength (Peak sunspot area 1588.36 and 1591.08 ${\mu}$Hem respectively), so its rise time and rise rate is also comparable and they have appeared as a single point in panel (a) and (b) Figure~\ref{we11}.      
%--------------------------------------------
%-------------------------------------------
%----------------------------------------
\begin{figure*}[!t]
\centering
\includegraphics[width=1.0\textwidth]{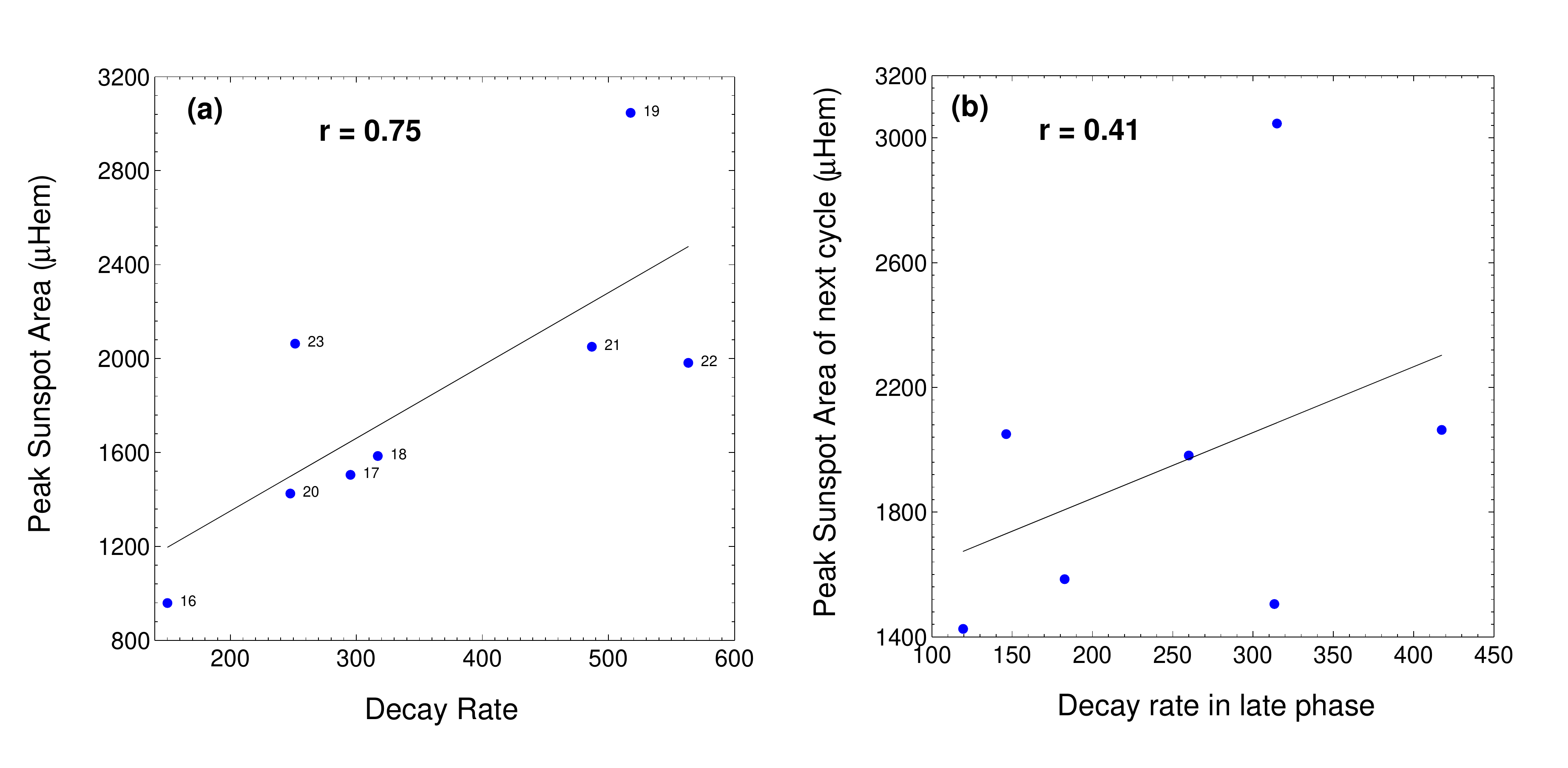}

\caption[The correlation of decay rate with same cycle and next cycle amplitude from Kodaikanal sunspot area data]{ Panel (a) shows the scatter plot of decay rate and the peak amplitude of same solar cycle (FWHM=2 year). Best linear fit is shown with the solid line. Panel (b) shows the scatter plot of decay rate at late phase and the peak amplitude of the next cycle.} 

\label{decay1}
\end{figure*}
%---------------------------------------
\subsubsection{The correlations in descending phase of the solar cycle}

In the previous section we mentioned that the presence of small fluctuations and the double peaks in the solar cycle makes it difficult to identify the actual minima and maxima of a cycle. To avoid this discrepancy, we have excluded the actual cycle minima and the maxima in our decay rate calculation. The final decay rate of a particular cycle has been determined by averaging over the individual decay rates calculated at four different locations as described in Section~\ref{C3:S2_1}. Hence these decay rates are the decay rate for the entire phase of the cycle. For correlations of decay rate with next cycle amplitude, we have calculated decay rate at late phase.
In panel (a) of Figure~\ref{decay1} we have shown the correlations between the final decay rate of a cycle with the cycle amplitude using 2 years average sunspot area data. The correlations of decay rate calculated at the late phase of cycle with the amplitude of the next cycle is shown in panel (b) of Figure~\ref{decay1}.  We obtain a correlation coefficients $r= 0.75$ for the correlation between decay rate and cycle amplitude for same cycle whereas we get $r = 0.41$ for decay rate at late phase with next cycle amplitude. The correlation coefficients obtained from one year averaged data (FWHM = 1 year) is also given in Table~\ref{table:kodai}. From the values given in the Table~\ref{table:kodai}, we notice that the correlation coefficients are of comparable values. During the calculation of the decay rates correlations with less data points one should calculate `adjusted correlation coefficient' as we mentioned in the previous section and when we calculate adjusted correlation coefficient, it does not vary much for high value of correlation coefficient. So we keep the normal Pearson correlation coefficients here. This reconfirmations of various correlations from persistent century long kodaikanal sunspot area data help to understand that they are the true nature of the solar cycle and no statistical artifact.

%Table--------------------------------------------------------------
\begin{table}[!t]
\centering
{\small\renewcommand{\arraystretch}{.8}{
\caption{Correlation coefficients at different phase of solar cycle}
\begin{tabular}{c|c|c|c|c}
\hline
Smoothed & \multicolumn{2}{|c|}{Correlations during rising phase}&\multicolumn{2}{|c|}{Correlations during decaying phase}\\
data with & \multicolumn{2}{|c|}{of the cycle between}&\multicolumn{2}{|c|}{of the cycle between}\\
\cline{2-5}
FWHM&Rise time and cycle & Rise rate and cycle & Decay rate and & Decay rate at late \\
&amplitude (WE1) & amplitude (WE2) &cycle amplitude &  phase and cycle amplitude\\
& & & & of next cycle\\
\hline
1 year & -0.63 & 0.92 & 0.57 & 0.56   \\
&&&&\\
2 years & -0.66 & 0.99 & 0.75 & 0.41 \\ 
\hline
\label{table:kodai} 
\end{tabular}}}
\end{table}

\section{Theoretical Framework of the Dynamo Model}
\label{C3:model}
We carry out our theoretical studies using the flux transport dynamo model
originally presented by \citet{CNC04}. In this model, the evolution of 
the axisymmetric two-dimensional magnetic field is governed by following two equations:
\begin{equation}
\label{eqA}
\frac{\partial A}{\partial t} + \frac{1}{s}({\bf v}.\nabla)(s A)
= \eta_{\rm{p}} \left( \nabla^2 - \frac{1}{s^2} \right) A + S_{\rm{BL}}(r,\theta;B),
\end{equation}
\begin{equation}
\label{eqB}
\frac{\partial B}{\partial t}
+ \frac{1}{r} \left[ \frac{\partial}{\partial r}
(r v_r B) + \frac{\partial}{\partial \theta}(v_{\theta} B) \right]
= \eta_{\rm{t}} \left( \nabla^2 - \frac{1}{s^2} \right) B 
+ s({\bf B}_{\rm{p}}.\nabla)\Omega + \frac{1}{r}\frac{d\eta_{\rm t}}{dr}\frac{\partial{B}}{\partial{r}},
\end{equation}
where $s = r \sin \theta$, $B (r, \theta, t)$ is the toroidal component of the magnetic field , 
$A(r, \theta, t)$ is the vector potential of the poloidal field,
${\bf v}=v_r{\bf \hat r} + v_\theta\hat {\bf \theta}$ is the velocity of the meridional flow, 
$\Omega$ is the internal angular velocity of the Sun and $\eta_{\rm{t}}$, $\eta_{\rm{p}}$ are the
turbulent diffusivities of the toroidal and the poloidal fields. Since the detailed discussion of the 
parameters and boundary conditions are given in \citet{CNC04} and 
\citet{KarakChou11}, here we do not discuss them again. 
We only make a few remarks about magnetic buoyancy and about the term $S_{\rm{BL}}(r,\theta;B)$ appearing in Equation (\ref{eqA}), which captures the longitude averaged B-L
mechanism. 
%(which has its origin in the systematic tilts of the bipolar active regions -- Joy's Law). 

Let us discuss how the magnetic buoyancy is treated in this model.
When the toroidal field above the tachocline ($r= 0.71 \Rs$) at any latitude exceeds a certain
value, a fraction of it is reduced there and the equivalent amount of this field is added on the solar 
surface. Then this local toroidal field near the surface is multiplied by a factor $\alpha$ to 
give the poloidal field. The source term in Equation~(\ref{eqA}), therefore, is
\begin{equation}
S_{\rm{BL}}(r,\theta;B)=\alpha B(r,\theta,t),
\label{alphaH}
\end{equation}
where
\begin{equation}
\alpha =\frac{\alpha_0}{4} \cos \theta 
\left[ 1 + \er \left(\frac{r - 0.95\Rs}{0.03\Rs} \right) \right]
\left[ 1 - \er \left(\frac{r - \Rs}{0.03\Rs} \right) \right],
\end{equation}
with $\alpha_0 = 30$~m~s$^{-1}$.
Now our job is to use this model to study the observed features of solar cycle reported in previous sections.
To study any irregular feature of the solar cycle, we have to make the cycles unequal
by introducing randomness in this regular dynamo model, as we discuss in the following sections.

In most of our calculations, we have followed \citet{CNC04} in assuming the meridional circulation to consist of one cell.  Of late,
this assumption has been questioned, although the exact nature of the meridional
circulation in the deeper layers of the convection zone is still not known. We have
shown in Section~\ref{C3:S4_4} that we can retain the attractive features of our results with
more complicated meridional circulation \citep{HKC14}. We have
also included the near-surface shear layer in the calculations presented in Section~\ref{C3:S4_4}

%-------------------------------------------------------------
\section{Results of Theoretical Modeling}
\label{C3:S4}
\begin{figure}[!h]
\centering{ 
\includegraphics[width=1.0\textwidth,clip=]{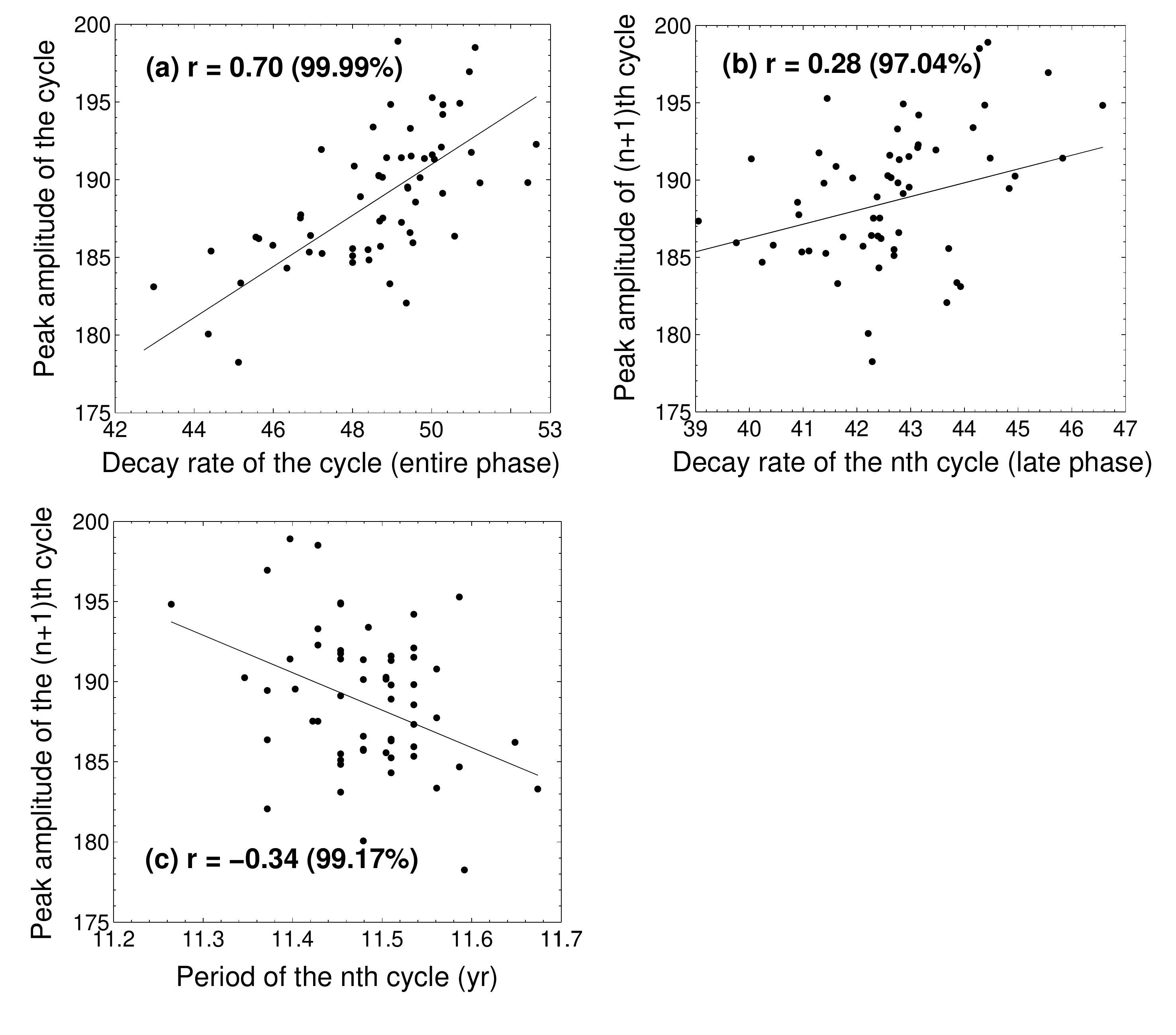}
              }
\caption[Results from stochastically forced dynamo model with B-L $\alpha$ fluctuations]{Results from stochastically forced dynamo model with B-L $\alpha$ fluctuations: 
Scatter plots showing the correlations between (a) the decay rate and 
the amplitude of cycle $n$, (b) the decay rate of cycle $n$ and the amplitude of 
cycle $n+1$, (c) the period of cycle $n$ and the amplitude of cycle $n+1$.}
\label{alflc}
\end{figure}
\subsection{Fluctuations in the Poloidal Field Generation}
\label{C3:S4_1}
We have discussed in the Introduction that the Sun does not produce equal amount of poloidal field 
at the end of every cycle and that the generation of the poloidal field involves randomness. 
Therefore, similar to adding stochastic fluctuations in the traditional
mean-field alpha \citep{Chou92}, adding stochastic fluctuations 
in the B-L $\alpha$ has become a standard practice in the flux transport dynamo community
\citep{CD2000, Jiang07, KarakNandy12}. In the present work, first we introduce stochastic noise in the B-L $\alpha$ in the following way:
\begin{equation}
\alpha_0 \rightarrow \alpha_0 +\sigma(t,\tau) \alpha_0',
\end{equation}\\
where $\tau$ is the coherence time during which the fluctuating component
remains constant and $\sigma$ is a uniformly distributed random number in
the interval [-1, 1]. Considering the typical decay time of the active regions by
surface flux transport process, we fix the coherence time within 0.5 -- 2
months. To see a noticeable effect, we add $75\%$ fluctuations in $\alpha$
({\it i.e.}, $\alpha_0'/\alpha_0=0.75$) with coherence time of $1$ month.
From this stochastically forced model we have to calculate a measure of the 
theoretical sunspot number. We consider the magnetic energy density
($B^2$) of toroidal field at latitude $15^{\circ}$ at the base of
the convection zone ($r = 0.7 \Rs$) as a proxy of sunspot number (this 
was done by \citet{CD2000}). 
Note that absolute value of the theoretical sunspot number does not have any physical meaning.
Therefore, we scale it by an appropriate factor to match it with the observed sunspot number.
From the time series of theoretical sunspot number, we calculate the cycle periods and decay rates 
in the same way as we have done for the observational data.

In Figure~\ref{alflc}(a) we show the correlation between the decay rates  and 
the amplitudes of the same cycles. We see a positive correlation as in the
observed data presented in Figure~\ref{obs1}. It is easy to understand the reason behind getting this positive correlation.
Since we have kept \mc\ fixed, the periods of the solar cycle do not vary much 
but the cycle strengths do vary due to the fluctuations in the poloidal field generation.
Therefore, when the amplitude of a cycle increases while its period remains approximately fixed,
the cycle has to decay rapidly. Hence we find that the stronger cycles decay faster than 
the weaker cycles, producing the positive correlation seen in Figure~\ref{alflc}(a).
However, we see in Figure~\ref{alflc}(b)  that there is not much correlation 
between the decay rate of the cycle $n$ and the amplitude of the next cycle $n+1$
and we are unable to explain the observed correlation seen in Figure~\ref{figc3:obs2}. 
Note that for Figure~\ref{alflc}(a) the decay rates are calculated from the entire 
decaying part of the cycle which is more appropriate definition of the decay rate as we 
argued in Section~\ref{C3:S2}, whereas for Figure~\ref{alflc}(b) it is computed at the late decay phase 
because observationally we find strong correlation when decay rate is computed in late decay phase
only. Finally we see in Figure~\ref{alflc}(c) that in this study the observed anti-correlation between the period of cycle 
$n$ and the amplitude of cycle $n+1$ (shown in \Fig{percorl}) is also not reproduced. Note that the period does not vary too much when the \mc\ is kept constant.

To sum up, when we introduce fluctuations in the poloidal field generation
mechanism, we can explain the observed correlation between the decay rate and the
amplitude of the cycle shown in Figure~\ref{obs1}, but we cannot explain the other observed
correlations presented in Figures~\ref{figc3:obs2} and \ref{percorl}.

\subsection{Fluctuations in the Meridional Circulation}
\label{C3:S4_2}
\begin{figure}[!h]
\centering{ 
\includegraphics[width=1.0\textwidth,clip=]{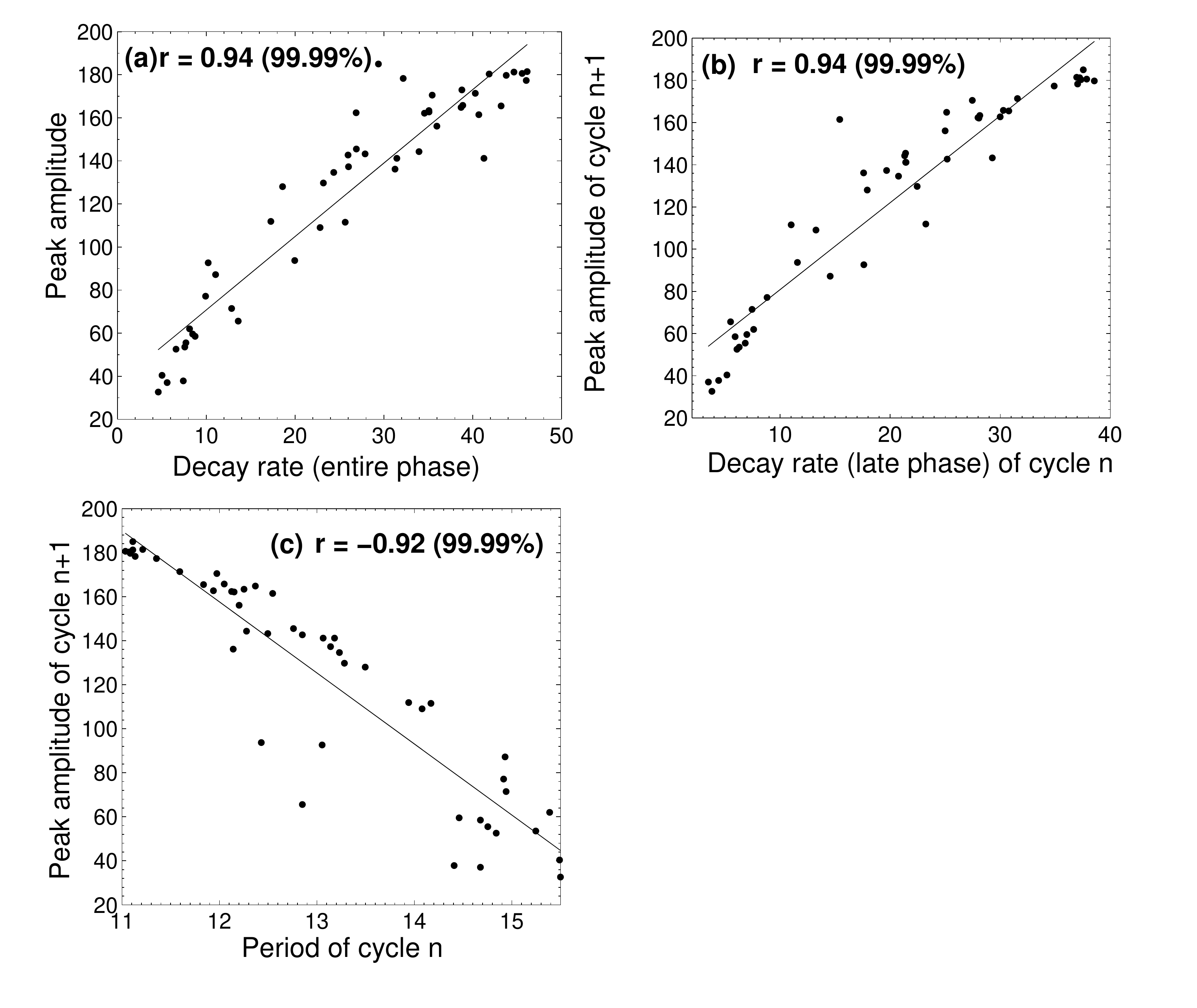}
}
\caption{Same as \Fig{alflc} but with \mc\ fluctuations.}
\label{figmc}
\end{figure}

Next we introduce the other important source of fluctuations in the flux transport dynamo
model, namely, variations of the meridional circulation.
Although we have some observational results of the \mc\ variations near the solar surface for the last 15 -- 20 years,
we do not have long data to conclude the nature of long-term variations
\citep{CD01, Hathaway10b}. However, there are indirect evidences for the
variation of the \mc\ over a long time \citep{LP09, Karak10, PL12}. 
Particularly, \citet{KarakChou11} have used the durations of the past cycles to argue that the \mc\ has long-term variations with the coherence time of probably 20 -- 45~years. There can also be short-term variations
in the \mc\ whose time scale may be related to the convective turnover time of the 
solar convection zone. Such variations with the time scale from a few months to a year are also 
observed in global magnetohydrodynamic simulations \citep{Karak14}.
In this work, we vary the amplitude of the meridional circulation in the same way
as we have done for the $\alpha$ term but with a different coherence
time. We show the results of simulations 
with 30\% fluctuations in the \mc\ with coherence time of 30~years.
\blue{We shall discuss later that various observed correlations can be explained
only if the coherence time is assumed to be not much less than the cycle
period.  While fluctuations of shorter duration (along with spatial variations)
are likely to be present in the meridional circulation, we believe that
they do not play any role in producing the correlations we are studying.} 
With $30\%$ level of fluctuations with a coherence time of 30 years, we get variations
of the \amp\ and of the period in our theoretical model comparable to the observational data. 
As in Section~\ref{C3:S4_1}, we take the time series ($B^2$) 
at latitude $15^{\circ}$ at the base of the convection zone as our proxy
of sunspot activity
and calculate the required correlations from it. 
The relevant correlation plots are shown in Figure~\ref{figmc}. We see in \Fig{figmc}(a) that now the correlation between
the decay rates and the cycle amplitudes has improved. Importantly, the other correlations 
are also correctly reproduced in Figures~\ref{figmc}(b) and \ref{figmc}(c) and can be compared with the
observational plots Figure~\ref{figc3:obs2}(b) and Figure~\ref{percorl}. These correlations did not appear at all when the 
fluctuations in poloidal field generation was introduced ({\it cf.}\ Figures~\ref{alflc}(b) and \ref{alflc}(c)).
To show how the correlations change on changing the correlation time or the level
of fluctuations, we tabulate the values of correlations coefficients under different
situations in Table~\ref{tabmc}. Each correlation coefficient is calculated from a run of
50 cycles.  It should be kept in mind that there is some statistical noise in 
the values of correlation coefficients. If the correlation coefficient for exactly
the same set of parameters is calculated from different independent runs, the values
for different runs will be a little bit different. Keeping this in mind, we note
that there is no clear trend of the correlation coefficients increasing or decreasing
with increasing levels of fluctuations (other things being the same). However,
all the correlation coefficients tend to decrease on decreasing the coherence time.  

It is not difficult to understand how the correlation in \Fig{figmc}(a) arises.
For a stronger cycle, the sunspot number has to decrease by a larger amount
during the decay phase, making the decay rate faster. However, to understand
the physical reason behind the other two correlations seen in Figures~\ref{figmc}(b) and \ref{figmc}(c),
more subtle arguments are needed. \citet{KarakChou11} extended the
arguments of \citet{Yeates08} and pointed out that a weaker \mc, which
makes the cycles longer, will have two effects. Firstly, the differential
rotation has more time to generate more toroidal field and tends to make the cycles stronger.
Secondly, the turbulent diffusivity gets more time to act on the fields
and tends to make the cycles weaker.
When the diffusivity is high (as in our model), the second effect dominates over
the first and the longer cycles are weaker (the opposite is true for
dynamo models with low diffusivity). \citet{KarakChou11} showed that
this led to an explanation of the Waldmeier effect for dynamo models with
high diffusivity.  We now point out that this tendency (longer cycles tending
to be weaker) is also crucial in our understanding of the correlations seen
in Figures~\ref{figmc}(b) and \ref{figmc}(c).

If the \mc\ keeps fluctuating with a coherence time of 30 years, it would
happen very often that the \mc\ would have a certain value during a cycle (say
cycle $n$) and the early rising phase of the next cycle (say cycle $n+1$). This
is less likely to happen when the coherence time is reduced. Suppose the \mc\
is weaker during cycle $n$ and the rising phase of cycle $n+1$.  Then cycle $n$
will tend to be longer and to have a weaker decay rate.  The following cycle
$n+1$ will have a tendency of being weaker.  This will produce the correlations
seen in Figures~\ref{figmc}(b) and \ref{figmc}(c). On decreasing the coherence time, it will happen less
often that the \mc\ will be the same during cycle $n$ and the rising phase of
the next cycle $n+1$.  Hence the correlations degrade on decreasing the coherence
time of the \mc.
\begin{figure}[!t]
\centering{ 
 	\includegraphics[width=1.0\textwidth,clip=]{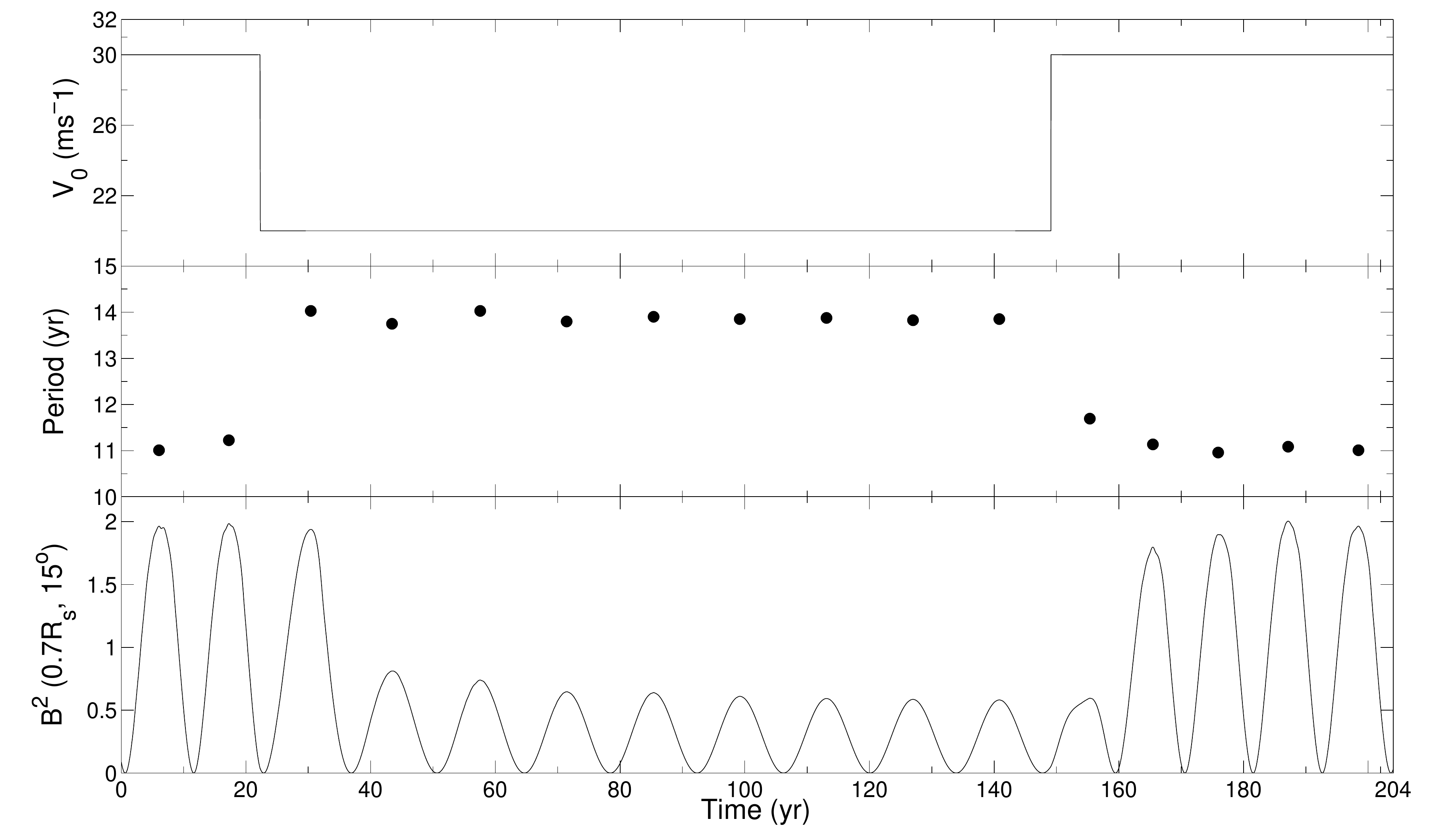}
              }
\caption[Memory of meridional circulation explains the correlation between decay rate and next cycle amplitude]
{Plots showing how the variation of meridional circulation, measured by $v_0$, 
with time (upper panel) changes the period of the cycle (middle panel) and strength of the magnetic field 
(shown by $B^2$ in the lower panel).}
\label{memory}
\end{figure}

We have realized that there is also a memory effect, which enhances the
correlations explained in the previous paragraph. To illustrate this memory
effect, we make a run of our dynamo code in which the \mc\ is decreased
suddenly during a sunspot minimum and then brought back to its original
value during another sunspot minimum a few cycles later.  The \mc\ and
the resulting sunspot activity are plotted in \Fig{memory}.  The periods of
successive cycles are also indicated in the middle panel of \Fig{memory}. On
changing the \mc, it is found that the periods of cycles begin changing
almost immediately.  However, there seems to be a memory effect as far as
the amplitudes of the cycles are concerned.  Even after the \mc\ changes,
the amplitude of the next cycle is very similar to the amplitude corresponding
to the earlier value of the \mc.  This memory effect will certainly enhance
the correlations we are discussing.  Suppose the \mc\ is weaker during the
cycle $n$, making its period longer and decay rate weaker.  Even if the \mc\
becomes stronger by the rising phase of the next cycle $n+1$, the
memory effect will ensure the amplitude of the cycle $n+1$ will still be
weak, thereby producing the correlation.
  
At this point, we would like to mention a misconception behind the correlation between
cycle $n$ period and cycle $n+1$ amplitude. It may be thought that the overlap between 
two cycles during solar minimum is the cause of this correlation. If the next cycle is 
stronger, then it starts early and the overlap with the present cycle is more. This 
makes the present cycle shorter. However we believe that this is not the source of 
 this correlation because if this is so, then we would have seen this correlation 
in Figure~\ref{alflc}(c) also, where cycle strengths were varied by fluctuations in the 
poloidal field generation. So the overlap is not the reason behind this correlation 
and we only get this in high diffusivity dynamo model with fluctuating \mc.
\begin{table}[h!]
\caption{Correlation coefficients at different levels of fluctuations and coherence time of meridional circulation.}
%\begin{tabular}{|c|c|c|c|c|}
\begin{tabular}{ccccc}
\hline
&&\multicolumn{2}{c}{Correlation of decay rate}&Correlation of\\
&&\multicolumn{2}{c}{with cycle amplitude of}&previous cycle \\
\cline{3-4}
Coherence time & Fluctuations& Same cycle & Next cycle& period with \\
(year)&($\%$)&(Entire phase)& (Late phase)& amplitude\\ 
\hline
& 10 & 0.92 & 0.92 & -0.97\\
& 20 & 0.86 & 0.92 & -0.95 \\
30 & 30 & 0.87 & 0.89 & -0.96 \\
& 40 & 0.92 & 0.96 & -0.73 \\
& 50 & 0.87 & 0.91 & -0.94\\
\hline
& 10 & 0.79 & 0.85 & -0.95\\
& 20 & 0.86 & 0.86 & -0.98\\
20 & 30 & 0.93 & 0.96 & -0.97 \\
& 40 & 0.90 & 0.87 & -0.88\\
& 50 & 0.89 & 0.90 & -0.97\\
\hline
& 10 & 0.78 & 0.74 & -0.90\\
& 20 & 0.88 & 0.77 & -0.97\\
11 & 30 & 0.90 & 0.85 & -0.92 \\
& 40 & 0.82 & 0.74 & -0.89\\
& 50 & 0.82 & 0.84 & -0.83 \\
\hline
& 10 & 0.70 & 0.63 & -0.87\\
& 20 & 0.83 & 0.74 & -0.86\\
5.5 & 30 & 0.81 & 0.79 & -0.84 \\
& 40 & 0.81 & 0.57 & -0.85\\
& 50 & 0.80 & 0.81 & -0.78\\
\hline
& 10 & 0.57 & 0.48 & -0.78\\
& 20 & 0.58 & 0.59 & -0.64\\
1 & 30 & 0.61 & 0.67 & -0.80 \\
& 40 & 0.73 & 0.25 & -0.65\\
& 50 & 0.69 & 0.38 & -0.72\\
& 75 & 0.64 & 0.39 & -0.58\\
& 100 & 0.65 & 0.73 & -0.76\\
\hline
& 10 & 0.42 & 0.62 & -0.80\\
& 20 & 0.56 & 0.69 & -0.78\\
0.5 & 30 & 0.68 & 0.47 & -0.74 \\
& 40 & 0.62 & 0.56 & -0.67\\
& 50 & 0.61 & 0.56 & -0.79\\
& 75 & 0.64 & 0.50 & -0.81\\
& 100 & 0.64 & 0.60 & -0.87\\
\hline
\label{tabmc}
\end{tabular}
\end{table}

\subsection{Fluctuations in the Poloidal Field Generation and the Meridional Circulation}
\label{C3:S4_3}
\begin{figure}[!h]
\centering{ 
\includegraphics[width=1.1\textwidth,clip=]{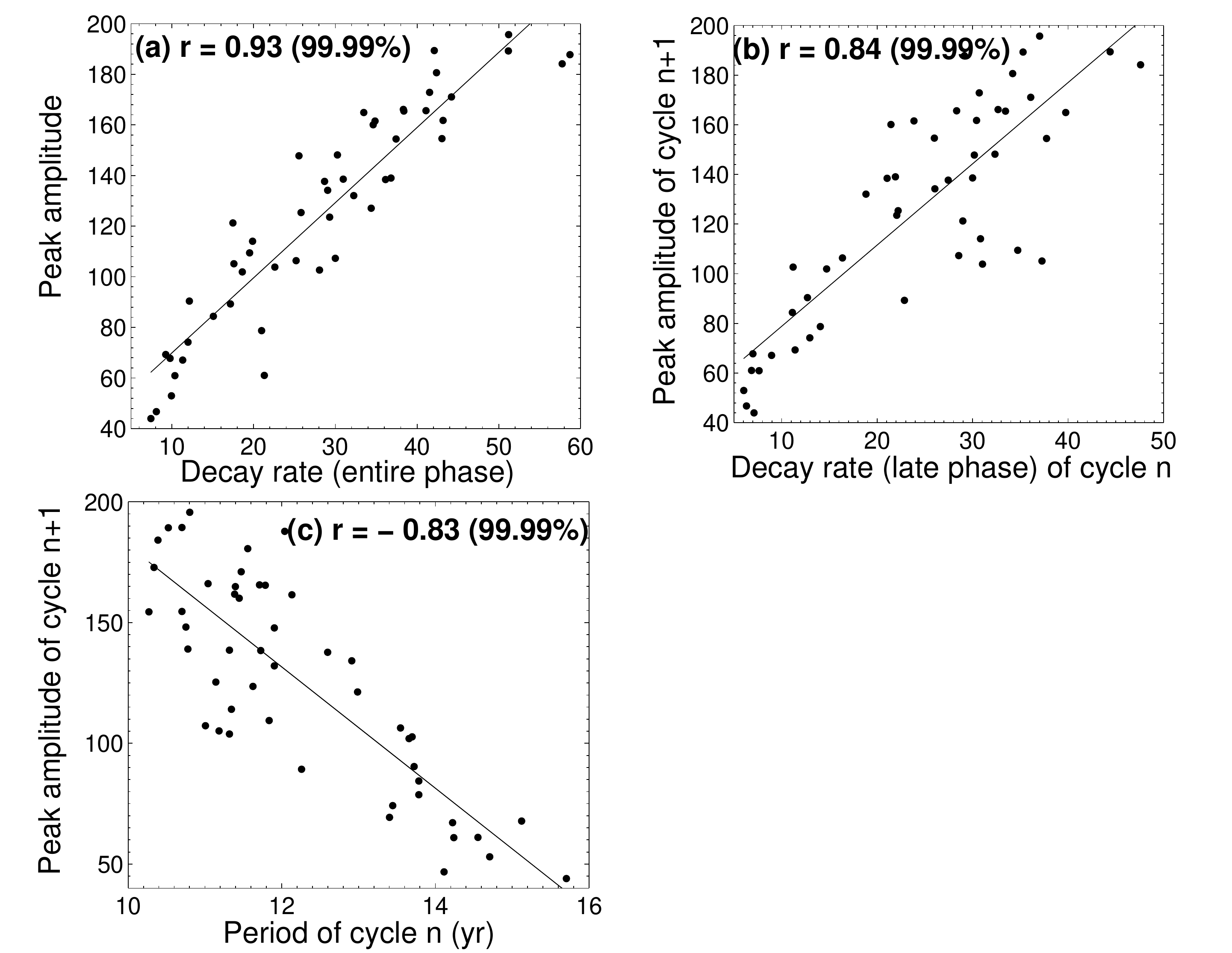}
              }
\caption{Same as \Fig{alflc} but with both B-L $\alpha$ and \mc\ fluctuations.}
\label{figalmc}
\end{figure}

Finally we add fluctuations in both the poloidal field generation process and the meridional 
circulation of the regular model, which is the realistic scenario.
We add the same amount of fluctuations in poloidal field generation
and in meridional circulation that we had added earlier in the individual cases
({\it i.e.}, 75\% fluctuations in the poloidal field generation with
a coherence time of 1 month and 30\% fluctuations in the
meridional circulation with a coherence time of 30~years).
The results are shown in Figures~\ref{figalmc}.
In this figure, we see that the scatters in the correlation plots are very close to 
what we find in actual observations. It is perhaps not a big surprise that all the correlations are reproduced correctly,
because they were already reproduced on introducing fluctuations in \mc\ alone.

\begin{figure}[!h]
\centering{ 
\includegraphics[width=1.0\textwidth,clip=]{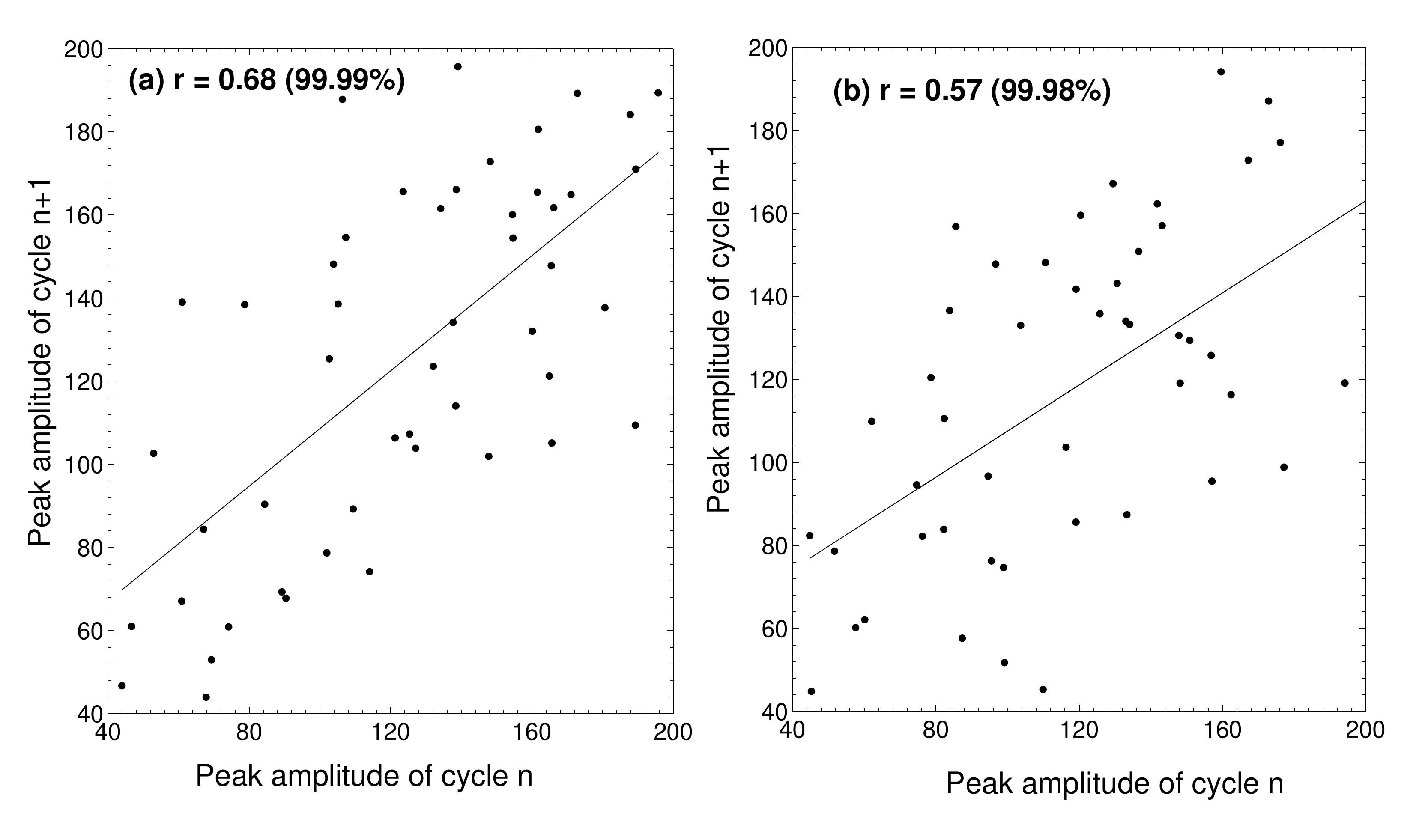}
              }
\caption{(a) Scatter plot of the amplitude of cycle $n$ with the amplitude of cycle $n+1$ with 75\% fluctuation in B-L $\alpha$. (b) Same as (a) but with 100\% fluctuation in B-L $\alpha$.}
\label{theoampl}
\end{figure}

A correct theoretical model also should explain the lack of correlation seen in
Figure~\ref{amplcorl} between peaks of two successive cycles. \Fig{theoampl}(a) shows the 
correlation between the amplitude of cycle $n$ and the amplitude of cycle $n+1$
for the same level of fluctuations which were used to generate \Fig{figalmc},
whereas \Fig{theoampl}(b) gives the same correlation when the fluctuation is B-L
$\alpha$ is raised to 100\% from 75\%. It is found that the correlations between
these amplitudes is weak and becomes weaker still on increasing the fluctuation
in the B-L $\alpha$. A physical interpretation is not difficult to give. A
coherence time of 30 years in \mc\ implies that very often the \mc\ will
be the same during two successive cycles, trying to produce a correlation
between the cycles. On the other hand, a fluctuation in the B-L $\alpha$ 
will definitely try to reduce the correlation. Certainly
this fluctuation would try to reduce the correlations seen in \Fig{figalmc} as well.
However, for our choice of parameters, we are able to theoretically
reproduce the three observed correlations as shown in \Fig{figalmc}, whereas
the correlation between successive cycles is much weaker in conformity
with observations. We may mention that we also get an anti-correlation between the amplitude
of a cycle and its duration. Our theoretical correlation coefficient ($r = -0.65$)
is somewhat stronger than what Charbonneau and Dikpati (2000) obtained from
the observational data ($r = -0.37$).

%\blue{
\subsection{Robustness of the Results on Changing the  Meridional Circulation and Differential Rotation Profiles}
\label{C3:S4_4}
So far, our earlier computations are performed using a single-cell meridional circulation 
in each hemisphere. However, recent observations, helioseismic inversions and convection 
simulations suggest the possibility that
the meridional circulation may have a complicated multi-cellular structure rather
than being single-cellular \citep{Zhao13,Karak15}. In \citet{HKC14}, we have shown that
the flux transport dynamo model can reproduce most of basic features of solar cycle 
using multi-structured meridional circulation as long as there is an equator-ward 
flow near the bottom of the convection zone. Therefore we are curious to know whether the 
correlations studied in this chapter are also reproduced with multi-structured circulation.
To answer this question, we perform a simulation with three radially stacked 
circulation cells, exactly the same as used in Section~3 of \citet{HKC14}. For the differential rotation in all our previous works,
we have used a simplified profile of the observed differential rotation 
that does not capture the near-surface shear layer (see {\it e.g.} Figure~1 of \citet{CNC04}). 
Although it is expected that the near-surface shear layer does not produce significant effect 
on global large-scale fields in the flux transport dynamo \citep{Dikpati02}, 
just for the sake of completeness we use a somewhat improved profile
of differential rotation captured by the following analytical formula
\begin{eqnarray}
\Omega(r,\theta) = \sum_{j=0}^2 \cos\left(2j\left(\frac{\pi}{2}-\theta\right)\right)\,\sum_{i=0}^4 c_{ij} (r/R_\odot)^i.
\end{eqnarray}
For the coefficients $c_{ij}$ see Table 1 of \citet{Belvedere00}, (see also their Figure~1 for the comparison with observed profile).

With these new profiles of the meridional circulation and the differential rotation, we perform 
a dynamo simulation by adding the same amount of stochastic fluctuations 
in B-L $\alpha$ and in meridional circulation as done in the previous section. 
In the results presented earlier, magnetic buoyancy was treated by moving
a part of the toroidal magnetic field to the surface whenever it became larger
than a critical value.  However, as pointed out in \citet{HKC14} and \citet{KKC14},
this way of treating magnetic buoyancy
is not very robust under a large change in parameters and model ingredients.
Therefore, for the computations of this section we use the `non-local' 
magnetic buoyancy as used in \citet{CD2000}, and in many other works.

\begin{figure}[!h]
\centering{ 
\includegraphics[width=1.0\textwidth,clip=]{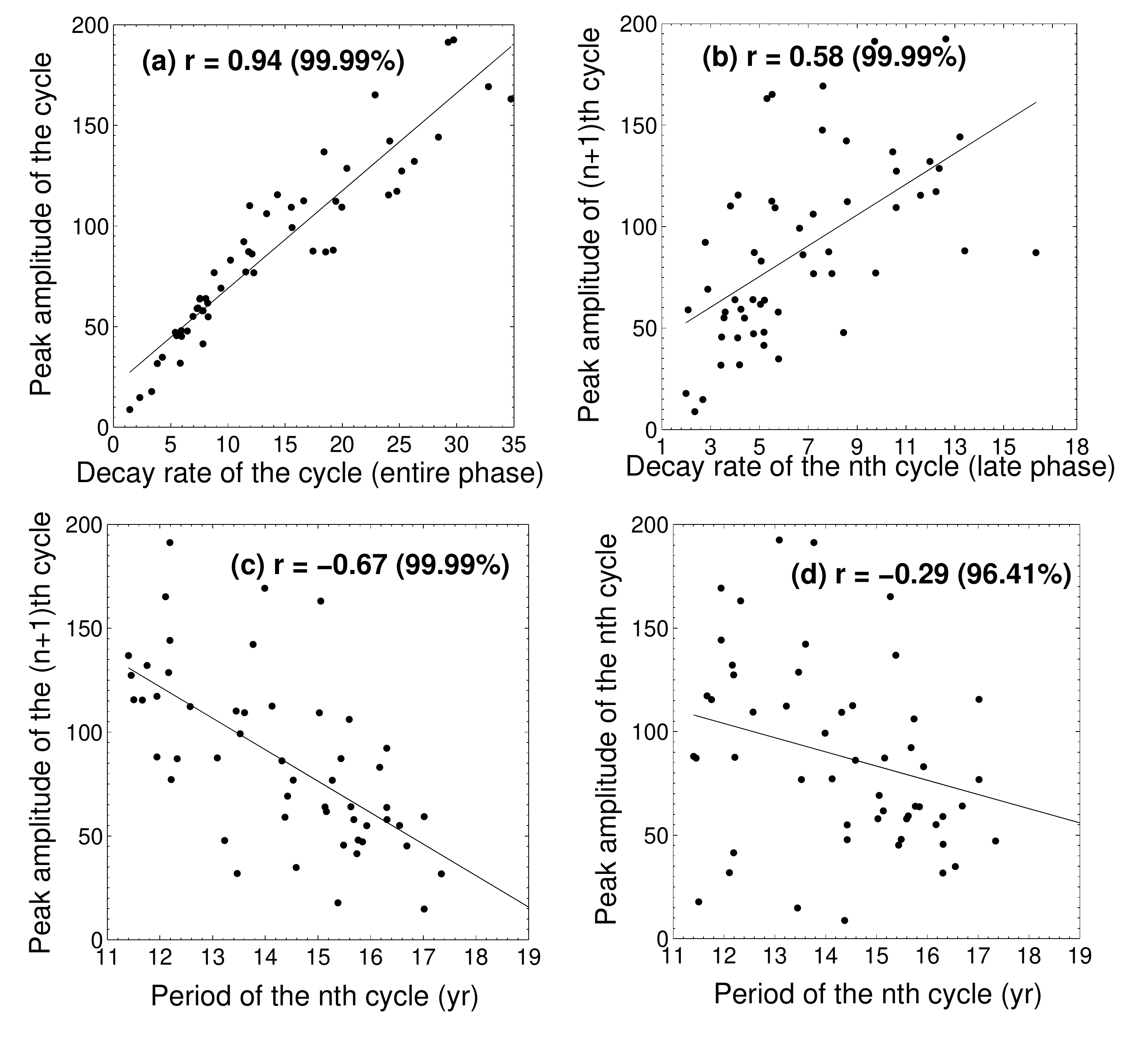}}
\caption{Same as \Fig{figalmc} but in this model, the large-scale flow has three cells radially stacked in the solar 
convection zone and the differential rotation includes near surface shear layer.}
\label{figc3mc}
\end{figure} 

Final results from this computation  are shown in Figure \ref{figc3mc}. We observe that even with
the {\it unconventional} meridional flow profile (three radial cells) and addition of near-surface shear layer, 
the correlations do not disappear. Although the correlations in Figures~\ref{figc3mc}(b) and \ref{figc3mc}(c) 
become a little weaker compared to what we have found for the usual single-cell circulation (Figure \ref{figalmc}), 
they show the correct general features as found in observations. The values of the correlations might be
improved a little bit by tuning the amount of imposed fluctuations; we do not want to do that here,
rather we have used same amount of fluctuations as we used in earlier sections. 

We make a few remarks about the two ways of treating magnetic buoyancy.
The behavior of the dynamo can become substantially different on treating magnetic
buoyancy in these two different ways (see Chapter~\ref{C2} for detail). Since some magnetic
field is removed due to magnetic buoyancy, one would expect the strength of
the toroidal field at the bottom of the convection zone to be depleted due to
the action of magnetic buoyancy.  One unphysical aspect
of the non-local treatment of magnetic buoyancy is that this effect is usually
not taken into account. As we have repeatedly pointed out, one requirement for
obtaining the Waldmeier effect as well as the correlations discussed in this
study is that the effect of diffusivity has to be more important than the effect
of toroidal field generation. Since the first method of treating magnetic buoyancy
(used in the earlier subsections) puts a cap on the strength of the toroidal field
but not the second non-local method, toroidal field generation remains unrealistically
strong in the second method and it is more difficult to get the correlations
properly in this method. We have taken the magnetic energy density ($B^2$)
at latitude $15^{\circ}$ at the bottom of the convection zone as the proxy of the
sunspot number. In the first method of treating magnetic buoyancy (with the single-cell
meridional circulation, as presented in Section~\ref{C3:S4_1}--\ref{C3:S4_3}), we found that
all the correlations come out robustly if we use the magnetic energy density ($B^2$)
in a wide range of latitudes as a proxy of the sunspot number. However, on using
the second method of non-local magnetic buoyancy, we find that the magnetic energy
density ($B^2$) has to be taken in a narrow band of low latitudes, with the
correlations disappearing or even reversing if we use the magnetic energy density at higher
latitudes.  To sum up, the second non-local method of treating magnetic buoyancy
is a more robust method and keeps the dynamo stable over a wide range of
parameters (which is not the case with the first method). However, it is more
difficult to reproduce various observed correlations of the solar cycle with this
non-local buoyancy method because the depletion of magnetic field due to buoyancy
is not included.
%BBK: ended

\section{Conclusion}
\label{C3:S5}
We have discussed three important features of solar cycle -- {i})
a linear correlation between the amplitude of cycle and its decay rate,
{ii}) a linear correlation between the amplitude of cycle $n$ and the decay rate
of cycle $n-1$ and {iii}) an anti-correlation between the amplitude of
cycle $n$ and the period of cycle $n-1$.
We have seen that all these correlations exist in all the data sets considered here.
Last two correlations involve characteristics of one cycle and the amplitude
of the next.  So they provide useful precursors for predicting a future cycle. Just by measuring the
period and the decay rate of a cycle, we can get an idea of the strength of the next cycle. %We will explore later if these can be used for predictions.

We have also explored whether these features can be explained in a B-L type
flux transport dynamo model. We have first introduced stochastic fluctuations in the poloidal
field generation (B-L $\alpha$ term) and we find that only the correlation between the
decay rate and the cycle amplitude is reproduced. However when we added fluctuations
in the \mc, we found that all three correlations are reproduced in qualitative
agreement with observational data.
In our high diffusivity dynamo model, strong \mc\ makes the
period shorter and the decay rate faster, but it also makes the next cycle 
stronger---especially because the cycle strength displays a memory effect,
depending on the \mc\ a few years earlier.
The opposite case happens when \mc\ becomes weaker.
Therefore the fluctuations in the \mc\ are essential to reproduce the observed features.
This study is consistent with earlier studies for modeling the cycle durations and strengths 
of observed cycles \citep{Karak10}, the Waldmeier effect 
\citep{KarakChou11}, grand minima \citep{CK12} and few others
\citep{Passos12} 
which indicate that the variable meridional circulation is crucial in modeling many
aspects of the solar cycle.
We have found that the observed correlations are reproduced even when
the meridional circulation is assumed to be more complicated than the one-cell pattern
used in most flux transport dynamo models. However, the coherence time of the
fluctuations in the meridional circulation has to be not less than the cycle
period in order to produce the correlations.  The correlations disappear on
making the coherence time too short, implying that fluctuations in the meridional
circulation having coherence time of the order of convective turnover time cannot
be the cause of the observed correlations. The theory of meridional circulation
is still very poorly understood and we have no understanding of what may cause
the fluctuations in meridional circulation with long coherence time.  However,
the pattern in the periods of the past cycles indicate the presence of such fluctuations
\citep{KarakChou11} and the fact that only such fluctuations can explain
the various observed correlations of the solar cycle convinces us that such fluctuations
in the meridional circulation with long coherence time must exist.

We have pointed out that the period or the decay rate of a cycle may be used to predict
the next cycle, since these quantities indicate the strength of the \mc\ which
also determines the amplitude of the next cycle a few years later (due to the memory effect).
It seems that the decay rate during the late phase of the cycle is the most
reliable precursor for the next cycle, as found in Figure~\ref{figc3:obs2}(b)---presumably because
the decay rate during this phase is the best indicator of the \mc\ during the
particular interval of time which is most crucial in determining the amplitude of
the next cycle. However, fluctuations in the poloidal field generation process
degrades all the observed correlations.  As a result, even Figure~\ref{figc3:obs2}(b)---displaying
the correlation between the decay rate during the late phase and the amplitude
of the next cycle---has considerable scatter, limiting our ability to predict
the next cycle in this way.

%footnote
\blfootnote{This chapter is based on \citet{HKBC15} and \citet{Sudip17}.}

%% file: chapter4.tex
\begin{savequote}[100mm]
``We should be unwise to trust
scientific inference very far when
it becomes divorced from
opportunity for observational test"
\qauthor{--Sir Arthur Stanley Eddington, The Internal Constitution of the Stars}
\end{savequote}
\def\pf{poloidal field}
\def\bl{Babcock--Leighton}
\def\ftdm{flux transport dynamo model}
\def\Rs{R_{\odot}}
\def\mps{m~s$^{-1}$}
\def\mc{meridional circulation}
\def\ftd{flux transport dynamo}
\def\pa{\partial}
\def\er{\mbox{erf}}
\newcommand{\ov}{\overline}

\chapter{Multi-cell Meridional Circulation in Flux Transport Dynamo Models}
\label{C4}
\begin{quote}\small
The solar activity cycle is successfully modeled by the flux transport
dynamo, in which the \mc\ of the Sun plays an important role.  Most of the
kinematic dynamo simulations assume a one-cell structure of the \mc\ within the
convection zone, with the equatorward return flow at its bottom.
In view of the recent claims that the return flow occurs
at a much shallower depth, we explore whether a \mc\ with such a shallow
return flow can still retain the attractive features of the flux transport
dynamo (such as a proper butterfly diagram, the proper phase relation between
the toroidal and poloidal fields). We consider additional cells of the \mc\
below the shallow return flow---both the case of multiple cells radially
stacked above one another and the case of more complicated cell patterns.
As long as there is an equatorward flow in low latitudes at the bottom of
the convection zone, we find that the solar behavior is approximately
reproduced.  However, if there is either no flow or a poleward flow at the
bottom of the convection zone, then we cannot reproduce solar behavior.
although this conclusion may change if there is suitable turbulent pumping in the equatorward direction.
On making the turbulent diffusivity low, we still find periodic behavior,
although the period of the cycle becomes unrealistically large.  Also, with
a low diffusivity, we do not get the observed correlation between the polar
field at the sunspot minimum and the strength of the next cycle, which is
reproduced when diffusivity is high.  On introducing radially downward
pumping, we get a more reasonable period and more solar-like behavior even
with low diffusivity.
\end{quote}
\section{Introduction}
\label{C4:S1}
The most extensively studied theoretical model of the solar activity cycle
in the last few years is the flux transport dynamo model, originally proposed
in the 1990s \citep{WSN91,CSD95,Durney95}, and recently reviewed by several authors
\citep{Charbonneau10,Chou11}. This model had a remarkable success in explaining various aspects of the solar
cycle and its irregularities.  However, in spite of its success, doubts are often
expressed if this success is merely accidental or if this is the really correct
model. Basically this hinges on the question whether the various assumptions used
in this model are correct.

Let us consider the crucial assumptions of the model.  The toroidal magnetic
field is assumed to be produced from the poloidal field
by the differential rotation that is mapped by
helioseismology, leaving no scope for any doubts.  This toroidal field rises
due to magnetic buoyancy to the solar surface, where the poloidal magnetic field
is produced from it by the Babcock--Leighton mechanism
\citep{Bab61, Leighton64}, for which there is now strong observational support
\citep{DasiEspuig10, Kitchatinov11a, Munoz13}. It is true that magnetic
buoyancy and the Babcock--Leighton mechanism are inherently three-dimensional 
processes and their representation in a two-dimensional kinematic model can
never be fully realistic \citep{Munoz10, YM13}. There are also considerable
uncertainties in the values of some parameters, such as the turbulent diffusivity
inside the convection zone. The Boulder group \citep{DTG04} and the
Bangalore group \citep{CNC04} used values of turbulent
diffusivity which differ by a factor of about 50. The higher diffusivity used
by the Bangalore group has now got strong support due to the success in explaining
various aspects of the observational data \citep{CNC04,CC06, Jiang07,GoelChou09,
Hotta10, Karak10, KarakChou11, KarakChou12, KarakChou13,CK09,CK12,Miesch12, Munoz13}.  Because 
of these uncertainties in the treatment of the Babcock--Leighton
mechanism and in the value of turbulent diffusivity, it is necessary to interpret
the results of the flux transport dynamo model with some degree of caution.
However, these uncertainties do not invalidate the model.  After all, different
treatments of the Babcock--Leighton mechanism and a range of values for the
turbulent diffusivity seem to give qualitatively similar results.  The only
other important ingredient of the flux transport dynamo model is the meridional
circulation.  Because the nature of the \mc\ in the deeper layers of the convection
zone is not yet observationally established, the main source of doubt about the
flux transport dynamo model at the present time concerns the question of whether
the Sun really has the kind of \mc\ which is assumed in the flux transport dynamo
models.

Let us consider the role of the \mc\ in \ftd\ models. In order for sunspots to
form at lower and lower latitudes with the progress of the cycle, a condition known as
the Parker--Yoshimura sign rule was expected to be satisfied (\citet{Parker55a, Yoshimura75}
see also \citet{Chou98}, Section 16.6).  According to this condition, $\alpha \, \pa \Omega/ \pa r$ 
has to be negative in the northern hemisphere.  It
follows from observations of solar surface that $\alpha$ corresponding to the
Babcock--Leighton mechanism is positive in the northern hemisphere.  Helioseismology
shows that $\pa \Omega/ \pa r$ is also positive in the lower latitudes where
sunspots are seen (except in a shear layer just below the solar surface). 
So clearly the Parker--Yoshimura sign rule is not satisfied
and it may be expected that the dynamo wave will propagate in the poleward
direction, contrary to observations. \citet{CSD95}
showed that an equatorward meridional circulation at the bottom of the convection
zone can overcome the Parker--Yoshimura sign rule and make the dynamo wave
propagate in the correct direction.  This is the main role of the \mc\ in the
\ftd\ models.  The second role of the \mc\ is that the poleward \mc\ near the
solar surface advects the poloidal field poleward, as seen in the observations.
In dynamo models with low turbulent diffusivity, the \mc\ has a third important
role.  It brings the poloidal field created near the surface to the bottom of
the convection zone where the strong differential rotation can act on it to
create the toroidal field.  In dynamo models with high turbulent diffusivity,
however, the poloidal field can diffuse from the surface to the bottom of the
convection zone in typically about 5 yr and this third role of the \mc\ is 
redundant \citep{Jiang07}. If there is radial pumping
as suggested by some authors \citep{KarakNandy12}, then that further
eliminates the role of \mc\ for bringing the poloidal field to the bottom
of the convection zone. 
Because we will be using a higher value of turbulent diffusivity in many of our calculations, the twin roles of the \mc\ in our model 
will be the equatorward advection
of the toroidal field at the bottom of the convection zone and the poleward
advection of the poloidal field near the surface.

The simplest kind of \mc\ assumed in most of the 
theoretical models consists of one cell
encompassing a hemisphere of the convection zone, with a poleward flow in the
upper layers and an equatorward flow in the lower layers. This kind of \mc\
successfully plays the twin roles expected of it in a \ftd\ model. 
Observations show a poleward \mc\ near the surface, so there is absolutely no
doubt that this part of the \mc\ advects the poloidal field poleward. The
only remaining question is whether the cell of the \mc\ really penetrates to the 
bottom of the convection zone where the equatorward flow branch has to be
located for the equatorward advection of the toroidal field. Early helioseismic
investigations going to a depth of $0.85 R_{\odot}$ could not find any evidence
of the equatorward return flow until that depth \citep{Giles97, Braun98}. 
However, recently \citet{Hathaway12}, assuming that the supergranules
are advected by the \mc, analyzed the observational data to conclude that the
return flow occurs at depths as shallow as 50--70 Mm. 
\citet{Zhao13} also claim on the basis of their helioseismic inversion
that the equatorward return flow
exists between radii $0.82 R_{\odot}$ and $0.91 R_{\odot}$. On the other 
hand, \citet{Schad13} from the study of global helioseismic analysis 
find the indication of multi-cell meridional circulation in the whole 
convection zone. If these results
are supported by other independent groups and 
really turn out to be true, then the very important question is whether the
attractive aspects of the present \ftd\ models can be retained with such a
\mc. In this chapter, we explore whether additional cells of \mc\ below the
shallow return flow can help us solve the problem.

So far only a few theoretical studies of the \ftd\ with a \mc\ more 
complicated than a single cell have been carried out.  The effects of two
cells in the latitudinal direction have been considered by \citet{Bonanno06}. 
However, we now want to consider a more
complicated structure of the \mc\ in the radial direction, including the
possibility of multiple
cells in the radial direction. Such a study was first carried out by \citet{JouveBrun07}. 
In their calculations, they always had poleward meridional
circulation at the bottom of the convection zone in the lower latitudes
where sunspots are seen.  They were able to get periodic solutions, but
the butterfly diagrams were always in the wrong sense, implying poleward
migration of the toroidal field.  They concluded that ``the resulting butterfly 
diagram and phase relationship between the toroidal and
poloidal fields are affected to a point where it is unlikely that such 
multicellular meridional flows persist for a long period of time
inside the Sun, without having to reconsider the model itself''( \citet{JouveBrun07}, p. 239). If this
conclusion was generally true for any \mc\ more complicated than the
simple single-cell circulation, then the results of \citet{Hathaway12,
Schad13, Zhao13}, if supported by independent investigations by
other groups, would indeed be bad news for
\ftd\ models.  \citet{Guerrero08} considered a single
cell confined to the upper layers of the convection zone.  On introducing
strong radial and latitudinal pumping, they were able to get correct
butterfly diagrams. However, whether such equatorward latitudinal pumping
really exists to give the right kind of butterfly diagram, needs to be studied rigorously.
Another recent attempt of saving the \ftd\ has been made by \citet{Pipin13}, 
who use the near-surface shear layer found in helioseismology
and an equatorward return flow of \mc\ just below it. Since $\pa \Omega/ \pa r$ is 
negative within this shear layer, such a dynamo would have equatorward
propagation according to the Parker--Yoshimura sign rule
even in the absence of an equatorward \mc\ in the right place.
However, we are unable to accept this model of \citet{Pipin13} as a 
satisfactory model of the solar cycle for the following
reasons. It is known for a long time that magnetic buoyancy is particularly
destabilizing in the upper layers of the convection zone and it is impossible
to store magnetic fields generated there for sufficient time for
dynamo amplification \citep{Parker75, Moreno83}. Also, the
scenario that the toroidal field is generated within the tachocline and then
parts of it rise to produce active regions can explain many aspects of active
regions including Joy's law rather elegantly \citep{chou89,Dsilva93,Fan93,Caligari95}. 
We find no compelling reason to discard the scenario that the toroidal magnetic field is produced
in the tachocline where magnetic buoyancy is suppressed in the regions of
sub-adiabatic temperature gradient \citep{Moreno92}. 

The main aim of the present study is to address the question whether a \mc\
with a return flow at a relatively shallow depth would allow us to retain the
attractive features of the \ftd, without introducing such
assumptions as strong equatorward pumping and without abandoning the scenario in which the toroidal
field is generated and stored in the tachocline from where it rises to produce
active regions. If there is a return flow at a shallow depth and there are
no flows underneath it, then we find that the solar cycle cannot be
modeled properly with such a flow.  However, if there are additional
cells of \mc\ below the shallow return flow,
we find that we can retain most of the attractive features
of the \ftd\ model as long as there is a layer of equatorward flow in low latitudes
at the bottom of the convection zone.  The existence of such an equatorward
flow at the bottom of the convection zone is consistent with the findings 
of \citet{Zhao13}, who are unable to extend their inversion below
$0.75 R_{\odot}$ using their limited data set. Also, \citet{Passos15} have shown that there should be an equatorward flow at the base of the convection zone at mid-latitudes, below the current maximum depth helioseismic measures can probe ($0.75 R_{\odot}$) using three-dimensional global simulations of solar convection. Since our knowledge of the
\mc\ either from observational or theoretical considerations is limited,
in this study we take the \mc\ as a free parameter that can be assumed
to have any form involving multiple cells and explore the dynamo problem
with different kinds of \mc.
 
We discuss the mathematical formulation of our dynamo model in Section~\ref{C4:S2}. Then in
Section~\ref{C4:S3} we present our results for several cells of \mc\ in the radial direction,
whereas Section~\ref{C4:S4} presents results for more complicated \mc\ with multiple cells
in both radial and latitudinal directions. Whether the results get modified
for low turbulent diffusivity will be discussed in Section~\ref{C4:S5}. The effect of turbulent
pumping will be discussed in Section~\ref{C4:S6}.  Finally we summarize our conclusions in
Section~\ref{C4:S7}.
%^^^^^^^^^^^^^^^ Mathematical formulation ^^^^^^^^^^^^^^^^^^^^^^^^^^^^^^^^^^^

\section{Mathematical formulation}
\label{C4:S2}
In the two-dimensional kinematic flux transport dynamo model,
we represent the magnetic field as
\begin{equation}
{\bf B} = B \hat{\bf e}_{\phi} + \nabla \times (A \hat{\bf e}_{\phi}),
\end{equation}
where $B (r, \theta, t)$ and $A(r, \theta, t)$ respectively correspond to the 
toroidal and poloidal components which satisfy the following equations:
\begin{equation}
\label{eqc4:Aeq}
\frac{\partial A}{\partial t} + \frac{1}{s}({\bf v}.\nabla)(s A)
= \eta_{p} \left( \nabla^2 - \frac{1}{s^2} \right) A + S(r, \theta, t),
\end{equation}

\begin{eqnarray}
\label{eqc4:Beq}
\frac{\partial B}{\partial t}
+ \frac{1}{r} \left[ \frac{\partial}{\partial r}
(r v_r B) + \frac{\partial}{\partial \theta}(v_{\theta} B) \right]
= \eta_{t} \left( \nabla^2 - \frac{1}{s^2} \right) B
+ s({\bf B}_p.{\bf \nabla})\Omega + \frac{1}{r}\frac{d\eta_t}{dr}\frac{\partial{(rB)}}{\partial{r}}
\end{eqnarray}\\
where $s = r \sin \theta$.  Here ${\bf v}$ is velocity of the meridional flow, $\Omega$
is the internal angular velocity of the Sun, $\eta_p$ and $\eta_t$ are
turbulent diffusivities
and $S(r, \theta, t)$ is the coefficient which describes the generation
of poloidal field at the solar surface from the decay of bipolar
sunspots. These equations have to be solved with the boundary conditions
$A=B=0$ at $\theta = 0, \pi$, whereas at the top boundary $B=0$ and $A$
matches a potential field above \citep{DC95}.
The bottom boundary condition does not affect the solutions as long
as the bottom of the integration region is taken sufficiently below
the bottom of the convection zone. 
Once the parameters $\Omega$, $\eta_p$, $\eta_t$, ${\bf v}$ and
$S(r, \theta, t)$ are specified, Equation \ref{eqc4:Aeq} and \ref{eqc4:Beq} can be solved
with the code {\em Surya} to obtain the behavior of the dynamo
\citep{Nandy02,CNC04}. \citet{CNC04}
present a detailed discussion how the parameters were specified in their
simulations. However, \citet{Karak10} made some small changes in the parameters.
In the calculations in this chapter, we use the $\Omega$, $\eta_p$ and $\eta_t$ as
\citet{Karak10}, except that Section~\ref{C4:S5} and Section~\ref{C4:S6} present some discussion with different diffusivities
which will be explained in Section~\ref{C4:S5}. In this chapter we
carry out dynamo simulations with different kinds of meridional circulation ${\bf v}$.
Before coming to the meridional circulation, let us describe how we specify the
poloidal source term $S(r, \theta, t)$.

The effects of the magnetic buoyancy and the Babcock--Leighton mechanism have
to be incorporated by suitably specifying the poloidal source term $S(r, \theta, t)$.
There are two widely used procedures of specifying magnetic buoyancy.  In the
first procedure, whenever the toroidal field $B$ at the bottom of the convection
zone crosses a critical value, a part of it is brought to the solar surface.
In the second procedure, the Babcock--Leighton coefficient $\alpha$ in the 
source term multiplies the toroidal magnetic field at the bottom of the convection
zone rather than the local toroidal field.  Although the two procedures with all the
other parameters kept the same do not give identical results (see Chapter~\ref{C2} for details),
both the procedures reproduce the qualitative behaviors of the solar cycle.  Since
we believe the first procedure to be more realistic, we had used it in the majority
of calculations from our group \citep{CNC04, CCJ07,
Karak10}. However, it sometimes becomes difficult to introduce this procedure
in a stable way when the meridional circulation is made very complicated. Because
we are studying the behavior of the dynamo with various
complicated meridional circulations, we have opted for the second procedure.  We
specify the poloidal source term in Equation \ref{eqc4:Aeq} in the following way:
\begin{equation}
 S(r, \theta; B) = \frac{\alpha(r,\theta)}{1+(\ov B(r_t,\theta)/B_0)^2} \ov B(r_t,\theta),
\label{source}
\end{equation}
where $\ov B(r_t,\theta)$ is the value of the toroidal field at latitude $\theta$
averaged over the tachocline from $r = 0.685\Rs$ to $r=0.715\Rs$. We take
\begin{eqnarray}
\label{alpha}
\alpha(r,\theta)=\frac{\alpha_0}{4}\left[1+\mathrm{erf}\left(\frac{r-0.95\Rs}{0.05\Rs}\right)\right]\left[1-\mathrm{erf}\left(\frac{r-\Rs}{0.01\Rs}\right)\right]
\sin\theta\cos\theta \\ ~\nonumber
\times\left[\frac{1}{1+e^{-30(\theta-\pi/4)}}\right]
\end{eqnarray}
Note that the last factor in Equation~\ref{alpha} suppresses $\alpha$ in the higher latitudes and constrains
the butterfly diagram from extending to very high latitudes.  We are following
many previous authors who also suppressed $\alpha$ in high
latitudes by such means \citep{Munoz10,Hotta10}. Since this suppression
of $\alpha$ is not based on a clear physical reason, we have not used such a suppression of $\alpha$
in the previous calculations from our group using the first procedure of treating magnetic buoyancy
outlined above.  However, on treating magnetic buoyancy by the second procedure, flux eruptions tend
to occur at higher latitudes \citep{CNC05} and it becomes necessary to suppress eruptions
at high latitudes to get more reasonable butterfly diagrams.
For calculations presented in Section~\ref{C4:S3} and \ref{C4:S4} using high diffusivity, we use $\alpha_0=8.0$ m s$^{-1}$. 
In the low diffusivity case presented in Section~\ref{C4:S5}, we use a lower value $\alpha_0=0.5$ m s$^{-1}$. When 
we include the effect of radial turbulent pumping in Section~\ref{C4:S6} we use $\alpha_0=0.1$ m s$^{-1}$.
Note that the parameter $B_0$ in Equation~\ref{source} introduces the only nonlinearity in the problem and determines the amplitude of
the magnetic field. We will later present magnetic fields in units of $B_0$.

Below we discuss how the meridional circulation is prescribed.
The meridional circulation is always defined in terms of stream function $\psi$ which is given by
\begin{equation}
\rho {\bf v} = \nabla \times [\psi (r, \theta) {\bf e}_{\phi}],
\end{equation}
with the density profile given by
\begin{equation}
\rho = C \left( \frac{R_\odot}{r} - 0.95 \right)^{3/2}, \label{rho}
\end{equation}
We can generate different types of meridional circulation by choosing $\psi$ suitably.
For example, the one-cell meridional circulation used in many of the recent works
from our group \citep{Karak10} is obtained by taking
\begin{eqnarray}
\label{eq:psi}
\psi r \sin \theta = \psi_0 (r - R_p) \sin \left[ \frac{\pi (r - R_p)}{(R_\odot -R_p)} \right]\{ 1 - e^{- \beta_1 \theta^{\epsilon}}\}
\times\{1 - e^{\beta_2 (\theta - \pi/2)} \} \\ ~\nonumber
\times e^{-((r -r_0)/\Gamma)^2} 
\end{eqnarray}\\
with
$\beta_1 = 1.5, \beta_2 = 1.3$, $\epsilon = 2.0000001$, $r_0 = (R_\odot - R_b)/3.5$, $\Gamma =
3.47 \times 10^{8}$ m, $R_p = 0.635 R_\odot$.
The value of $\psi_0/C$ determines the amplitude of the meridional circulation. On
taking $\psi_0/C=0.95{\times}15.0$, the poleward flow near the surface at mid-latitudes peaks around $v_0=15.0$ m s$^{-1}$.
The cell of the meridional circulation is confined between $R_p$ and $R_\odot$. By making $R_p$ larger (but less
than $\Rs$), we can make the meridional circulation confined in the upper layers of the convection zone.

In order to have $N$ cells of meridional circulation, we can take a stream function of the form
\begin{equation}
\psi = \psi_1 + \psi_2 + \ldots + \psi_N,
\end{equation}
where each term in the stream function
gives rise to a cell of meridional circulation. Here we describe how we generate a two-cell
meridional circulation pattern, used in some of our simulations. The details of how we generate three-cell and
more complicated patterns are given in the Appendix. Since \citet{Zhao13} claim that the upper cell of
the meridional circulation is confined above $0.82 R_{\odot}$, we do some calculations with two cells above
and below $R_m = 0.82 R_{\odot}$. To generate such a pattern of meridional circulation, we use the
stream function
\begin{equation}
\label{eq:psi_tot}
{\psi}=\psi_u + \psi_l
\end{equation}\\
The stream function $\psi_u$ which generates the upper cell is given by
\begin{eqnarray}
\label{eq:psiu}
\psi_u = {\psi_{0u}}\left[1-{\rm erf}\left(\frac{r-0.91R_\odot}{1.0}\right)\right](r - R_{m,u})
\sin \left[\frac{\pi (r - R_{m,u})}{(R_\odot -R_{m,u}}\right]\{ 1 - e^{- \beta_1 \theta^{\epsilon}}\}~~\nonumber \\
\times\{1 - e^{\beta_2 (\theta - \pi/2)} \} e^{-((r -r_0)/\Gamma)^2} 
\end{eqnarray}\\
where the parameters have the following values:
$\beta_1 = 3.5, \beta_2 = 3.3$, $r_0 = (R_\odot - R_b)/3.5$, $\Gamma =3.4 \times 10^{8}$ m, 
$R_{m,u} = 0.815 R_\odot$. The stream function $\psi_l$ which generates the lower cell is given by
\begin{eqnarray}
\label{eq:psil}
\psi_l = {\psi_{0l}}\left[1-{\rm erf}\left(\frac{r-0.95R_{m,l}}{1.8}\right)\right](r - R_p)
\sin \left[\frac{\pi (r - R_p)}{(R_{m,l} -R_p)} \right]\{ 1 - e^{- \beta_1 \theta^{\epsilon}}\}\nonumber \\
\times\{1 - e^{\beta_2 (\theta - \pi/2)} \} e^{-((r -r_0)/\Gamma)^2}
\end{eqnarray}\\
where the parameters have the following values:
$\beta_1 = 3.2, \beta_2 = 3.0$, $r_0 = (\Rs - R_b)/3.5$, $\Gamma =3.24 \times 10^{8}$ m, $R_p = 0.65 R_\odot$, $R_{m,l} = 0.825 R_\odot$.  
We choose $\psi_{0u}/C$ and $\psi_{0l}/C$ in such a way that the velocity amplitudes in the upper and lower 
cells are around $20.0$ m s$^{-1}$ and $4.0$ m s$^{-1}$ respectively.

The two-cell meridional circulation given by the above expressions is shown in the upper part of Figure~\ref{fig:radc2}. Figure~\ref{fig:radc2}(a) shows
the streamlines of flow, whereas Figure~\ref{fig:radc2}(b) shows how $v_{\theta}$ varies with $r$ at the mid-latitude. The vertical dashed
lines in Figure~\ref{fig:radc2}(b) indicate bottoms and tops of the two cells.  It may be noted that both the cells have anti-clockwise
flow patterns.  This means that the flows at the bottom of the upper cell and at the top of the lower cell (which are
adjacent to each other) are in opposite directions involving a jump in the value of $v_{\theta}$ from one cell to
the next, as seen in Figure~\ref{fig:radc2}(b). If we replace $\psi_l$ by $-\psi_l$, then we can avoid such a jump
in the value of $v_{\theta}$.  This flow pattern
is shown in the upper part of Figure~\ref{fig:radc2_rev}. However, in this case, the meridional circulation at the tachocline is in the
poleward direction.  We shall see in Section~\ref{C4:S3} that this case will not give solar-like solutions.  If we want the meridional
circulation to be poleward at the surface and equatorward at the tachocline, and additionally we want to avoid a jump in
$v_{\theta}$, then we need at least three cells stacked one above the other in the radial direction. The flows in the top
and bottom cells have to be counter-clockwise, whereas the flow in the middle cell has to be clockwise.  The Appendix
presents the steam function that would give such a flow, which is shown in the upper part of Figure~\ref{fig:radc3}.  The results
with all these flow patterns are presented in the next Section, whereas results with a more complicated flow will be
presented in Section~\ref{C4:S4}. When we discuss the effects of changing the turbulent diffusivity in Section~\ref{C4:S5}, we
shall describe how the diffusivity will be changed. Similarly, in Section~\ref{C4:S6} where we discuss the effects of 
turbulent pumping, we shall explain how pumping is included
in the mathematical theory.

\section{Results with radially stacked cells}
\label{C4:S3}
\begin{figure}[!h]
\centering
\includegraphics[width=0.85\textwidth]{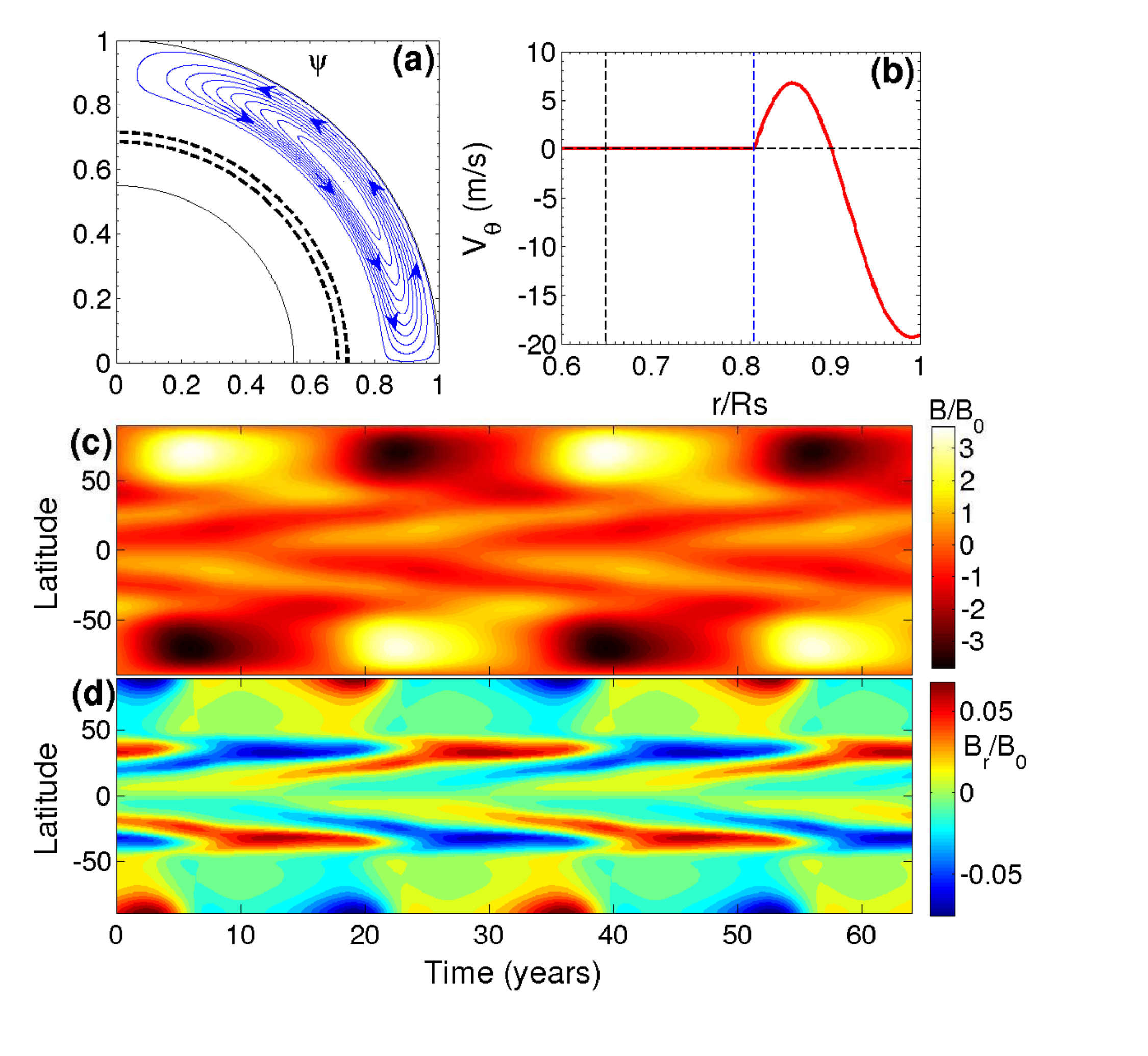}
\caption[Results with shallow meridional circulation]{(a) Streamlines of the shallow meridional circulation with no flow underneath. 
(b) $v_{\theta}$ as a function of ${r}/{\Rs}$ at the mid-latitude $\theta=45^{\circ}$.
(c) Butterfly diagram i.e, the time-latitude plot of the toroidal field at the bottom of the convection zone ($r=0.70\Rs$).
(d) Time-latitude plot of the radial field at the surface of the Sun. All the toroidal and radial fields are in the unit of $B_0$.}
\label{fig:rc2_nlwcll}
\end{figure}

Let us first consider the situation that the \mc\ has a return flow at the middle of the convection zone
and there are no flows underneath it.  We generate such a \mc\ by taking $\psi = \psi_u$ with $\psi_u$ given
by Equation~\ref{eq:psiu}. The upper part of Figure~\ref{fig:rc2_nlwcll} shows the streamlines and the profile of $v_{\theta}$ as a function
of $r$ at the mid-latitude.  The middle part of Figure~\ref{fig:rc2_nlwcll} is the `butterfly diagram', which is essentially
a time-latitude plot of $B$ at the bottom of the convection zone.  The bottom part shows the radial field
at the solar surface as function of time and latitude.  In the butterfly diagram, we find that the belt of
strong $B$ propagates poleward rather than equatorward at the low latitudes, although there is a slight
tendency of equatorward propagation at the high latitudes. This result can be easily understood from the
Parker--Yoshimura sign rule, which holds when there is no flow at the bottom of the convection zone. We
have $\alpha$ positive in the northern hemisphere.  Since $\pa \Omega/ \pa r$ is positive in the low latitudes
and negative in the high latitudes (see Figure~1 of \citet{CNC04} showing the differential rotation
we are using), the Parker--Yoshimura sign rule implies poleward propagation at the low latitudes and equatorward
propagation at the high latitudes. In this case, we are completely unable to reproduce the solar behavior.
It may be noted that \citet{Guerrero08} obtained solar-like behavior with a meridional
circulation similar to what we have used by including equatorward latitudinal pumping at the bottom of the
convection zone. However, observationally it is difficult to find evidence for latitudinal pumping, there are few numerical simulations \citep{Kapyla06, Racine11, Augustson15, Warnecke16} which find latitudinal pumping at low latitudes in the bottom half of the convection zone in highly stratified rotating solar convection.

\begin{figure}[!h]
\centering
\includegraphics[width=0.85\textwidth]{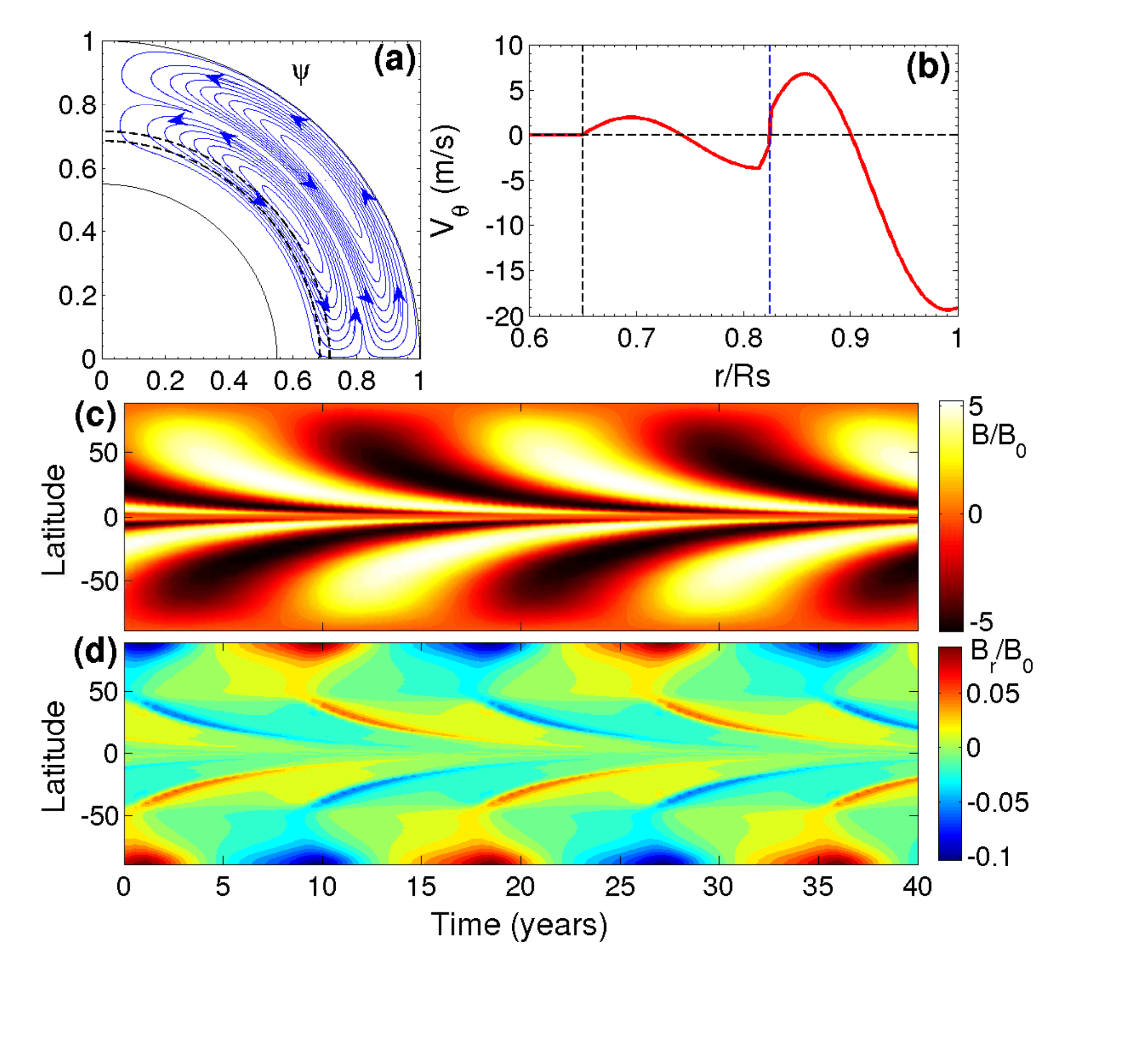}
\caption[Results with radially stacked two cells]{(a) Streamlines for two radially stacked cells of \mc. Arrows show the direction of the flow. 
(b), (c) and (d) are the same plots as in Figure~\ref{fig:rc2_nlwcll}, for this \mc.}
\label{fig:radc2}
\end{figure}

Next we consider the two-cell pattern given by Equations~\ref{eq:psi_tot}--\ref{eq:psil}. For this case, Figure~\ref{fig:radc2} provides plots of the same things 
that are shown in Figure~\ref{fig:rc2_nlwcll} for the earlier case. We see that there is an equatorward flow at the bottom of the
convection  zone, although there is a jump in $v_{\theta}$ between the cells. We find that the equatorward flow at the
bottom forces an equatorward transport of $B$ in accordance with what we see in the Sun. Looking at the lowest
part of Figure~\ref{fig:radc2}, we also see that the polar field changes sign at the time of the sunspot maximum, in 
accordance with the observations.  Thus, on using the two-cell pattern with an equatorward flow at the
bottom of the convection zone, we can reproduce the equatorward migration of the sunspot zone as well as
the correct phase relationship between the toroidal and poloidal fields. It is true that the butterfly
diagram starts at a somewhat high latitude compared to what we see in the Sun.  It is well known that the
butterfly diagram can be confined more to lower latitudes by making the \mc\ a more penetrating
\citep{Nandy02} and playing with other parameters. We had not bothered to fine-tune the 
parameters to achieve this, since our
main aim in this chapter is to study the qualitative behavior of the system under various kinds of \mc.
Note from Figures~\ref{fig:radc2}(c)--(d) that the maximum $B$ at the bottom of the convection zone and
the maximum $B_r$ at the surface bear a ratio of about 100. It should be emphasized that this
ratio corresponds to smoothed mean field values of $B$ and $B_r$, which can have very different
values inside flux concentrations \citep{Chou03}. 
 
\begin{figure}[!h]
\centering
\includegraphics[width=0.85\textwidth]{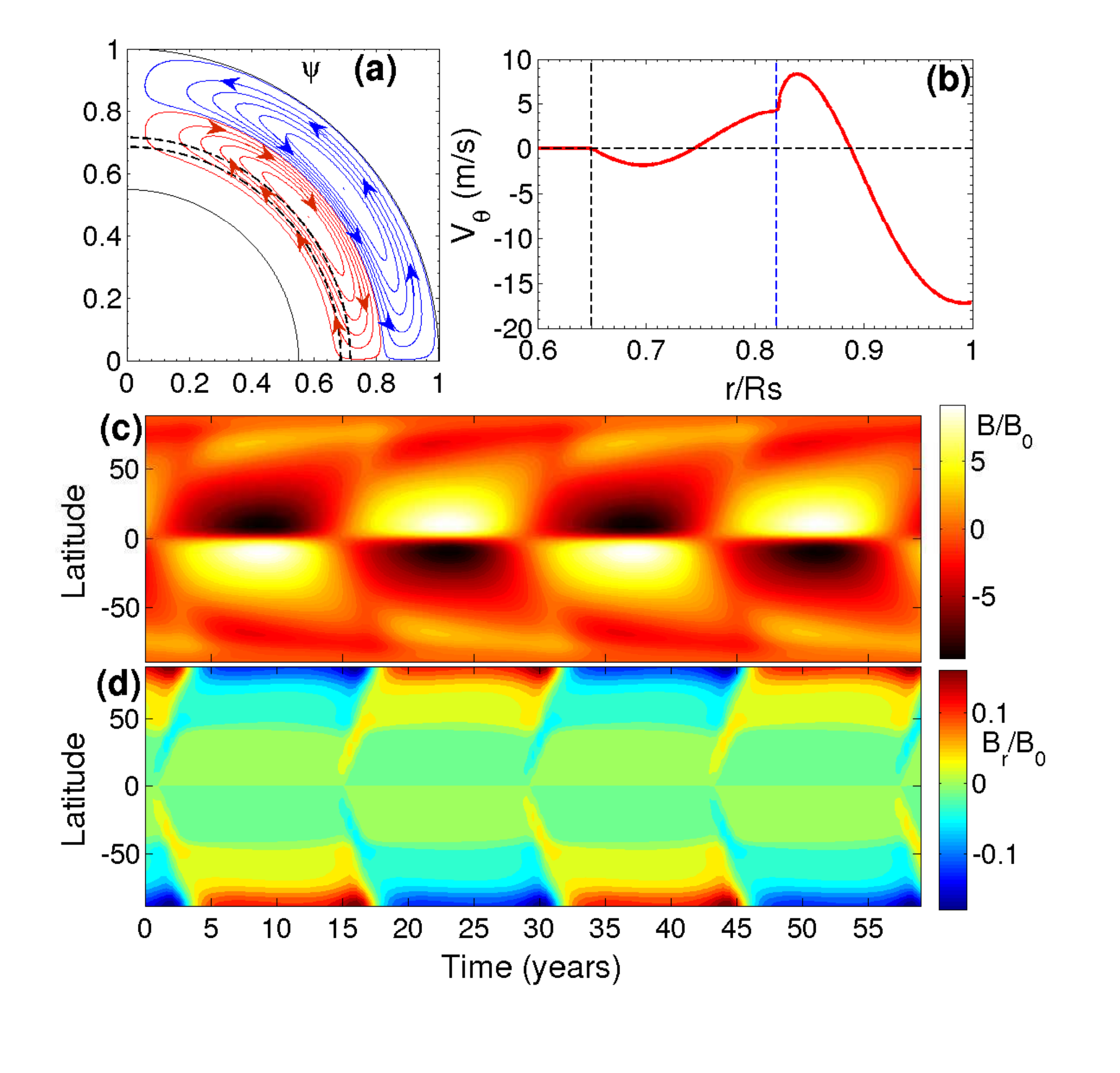}
\caption[Results with two radially stacked cells but opposite sense]{(a) Streamlines for two radially stacked cells of \mc\ with circulations in the opposite
sense. Arrows show the direction of the flow. (b), (c) and (d) are the same plots as in Figure~\ref{fig:rc2_nlwcll}, for this \mc.}
\label{fig:radc2_rev}
\end{figure}

We can avoid the jump in the value of $v_{\theta}$ seen in Figure~\ref{fig:radc2}(b)
by using a two-cell \mc\ in which $\psi_l$ is replaced by $-\psi_l$.
Since the flow at the bottom of the convection zone is poleward in this case, a study of this case
also throws light on the behavior of the dynamo with such a flow.  Our results are shown in Figure~\ref{fig:radc2_rev}.
The butterfly diagram indicates poleward migration and the solar behavior is not reproduced in this
case.  The two-cell meridional circulation we have used is very similar to what was used
by \citet{JouveBrun07} in one of their cases (see their Figure~2 and Figure~3). The butterfly
diagram we have got is quite similar to what they got.

\begin{figure}[!h]
\centering
\includegraphics[width=0.85\textwidth]{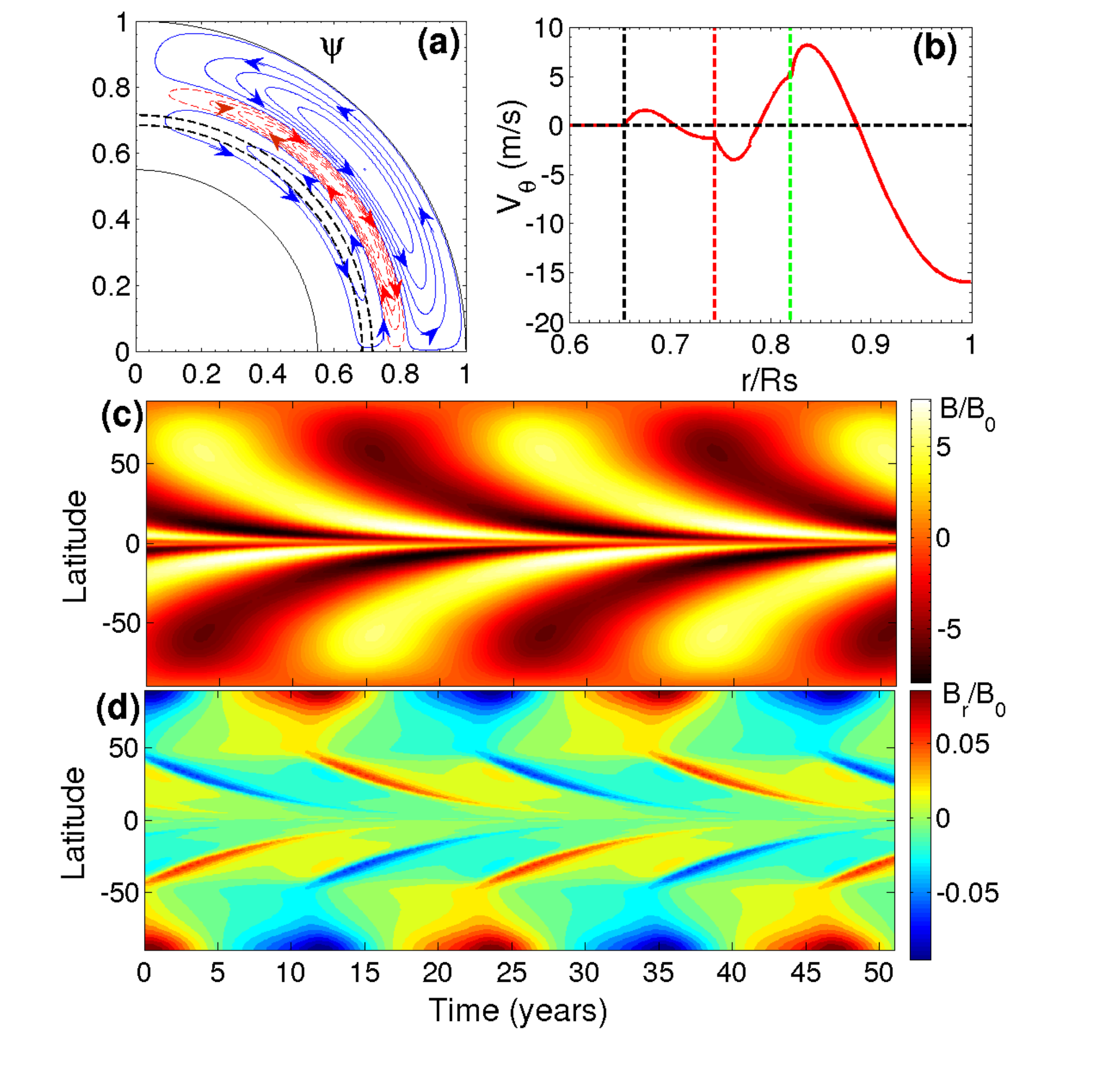}
\caption[Results with radially stacked three cells]{(a) Streamlines for three radially stacked cells of \mc. Directions are shown by arrows. 
(b), (c) and (d) are the same plots as in Figure~\ref{fig:rc2_nlwcll}, for this \mc.}
\label{fig:radc3}
\end{figure}

If we want to avoid a jump in $v_{\theta}$ and also to have a equatorward flow at the bottom
of the convection zone, then we need at least three cells of \mc\ stacked one over the other in the
radial direction.  The Appendix provides the mathematical prescription for generating such a \mc.
Figure~\ref{fig:radc3} presents the results.  Since there is an equatorward flow at the bottom of the convection
zone, we again find that the solar behavior is reproduced, in the sense of having a butterfly diagram showing
equatorward migration as well as
the correct phase relationship between the toroidal and poloidal fields. 

\begin{figure}[!h]
\centering
\includegraphics[width=0.750\textwidth]{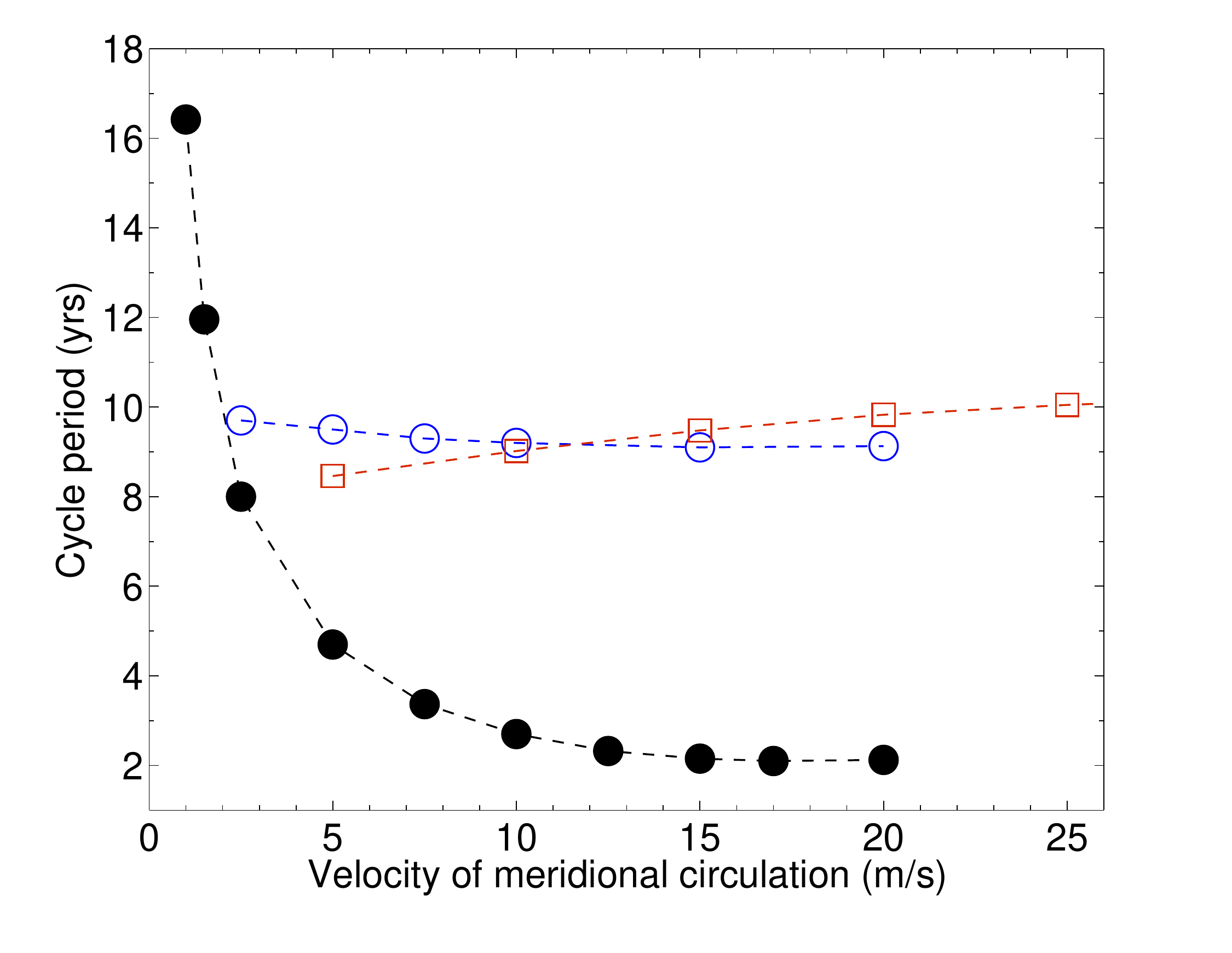}
\caption[Cycle period vs velocity amplitudes of three different cells]{Variation of solar cycle period with the velocity amplitudes of the three different cells shown in 
Figure~\ref{fig:radc3}(a). Black filled circles show the variation of the cycle
period with velocity amplitude of the lower cell while keeping velocities of other cells constant. 
Similarly blue circles show the variation of period with velocity amplitude of middle cell and red boxes for upper cell.  }
\label{fig:perio_vel}
\end{figure}

One of the important results for the flux transport dynamo with a single cell of \mc\ is that
the period of the dynamo decreases when the \mc\ is made faster \citep{DC99,Karak10}.
To explore how the dynamo period depends on the flow velocity in the multi-cell situation, we have
carried out a study for the case of three cells presented in Figure~\ref{fig:radc3}. We have carried out numerical
experiments by varying the flow amplitude of one cell, while keeping the flows in the other two
cells constant.  Figure~\ref{fig:perio_vel} shows how the dynamo period changes with the change of the flow speed in
each of the three cells. It is clearly seen that the flow speeds in the upper two cells have very
minimal effect on the dynamo period.  It is the flow speed in the lowermost cell which determines
the dynamo period and we find the period to decrease with the increase in this flow speed ($T \sim v_{0,l}^{-0.72}$).  This
result can be easily understood from common sense, since the flow in the lowermost cell causes
the equatorward migration of $B$ (giving solar-like butterfly diagram), and it is no wonder that
the period becomes shorter on making this flow faster. 

To sum up, as long as there is an equatorward flow at the bottom of the convection zone (the cases
of Figure~\ref{fig:radc2} and Figure~\ref{fig:radc3}), we are able to get solar-like behavior of the dynamo even if there
is a complicated multi-cell structure of the \mc, the period being determined by the flow
in the cell at the bottom of the convection zone.  Thus, even with a return flow of the \mc\ at a shallow
depth, the flux transport dynamo model can be made to work in this situation.
On the other hand, if there is no flow at
the bottom of the convection zone (the case of Figure~\ref{fig:rc2_nlwcll}) or if there is a poleward flow there
(the case of Figure~\ref{fig:radc2_rev}), then the dynamo model fails to reproduce solar behavior.  This conclusion
was obtained by considering multiple cells only in the radial direction. We consider more
complicated flows in the next Section and show that our main conclusion still holds.

\section{Results with more complicated cells}
\label{C4:S4}
\begin{figure}[!h]
\centering
\includegraphics[width=0.75\textwidth]{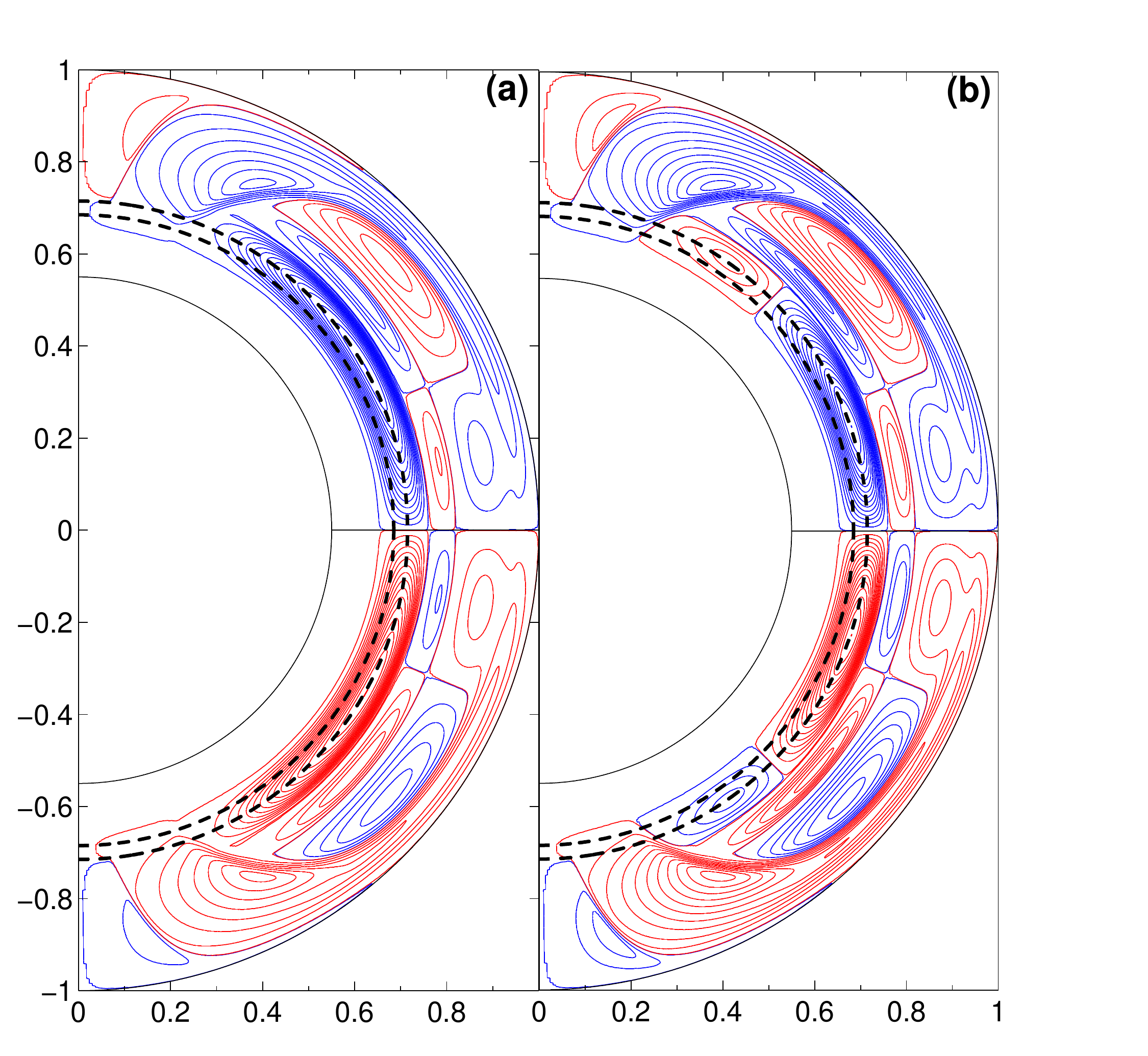}
\caption[Streamlines for two complicated \mc\ patterns]{Streamlines for two complicated \mc\ patterns.  The blue contours imply anti-clockwise
circulation, whereas the red contours imply clockwise circulation. The lowest cell in (a) extends
from the equator to fairly high latitudes, whereas this cell in (b) extends only to mid-latitudes. }
\label{fig:mcomp_cells}
\end{figure}

\begin{figure}[!h]
\centering
\includegraphics[width=0.850\textwidth]{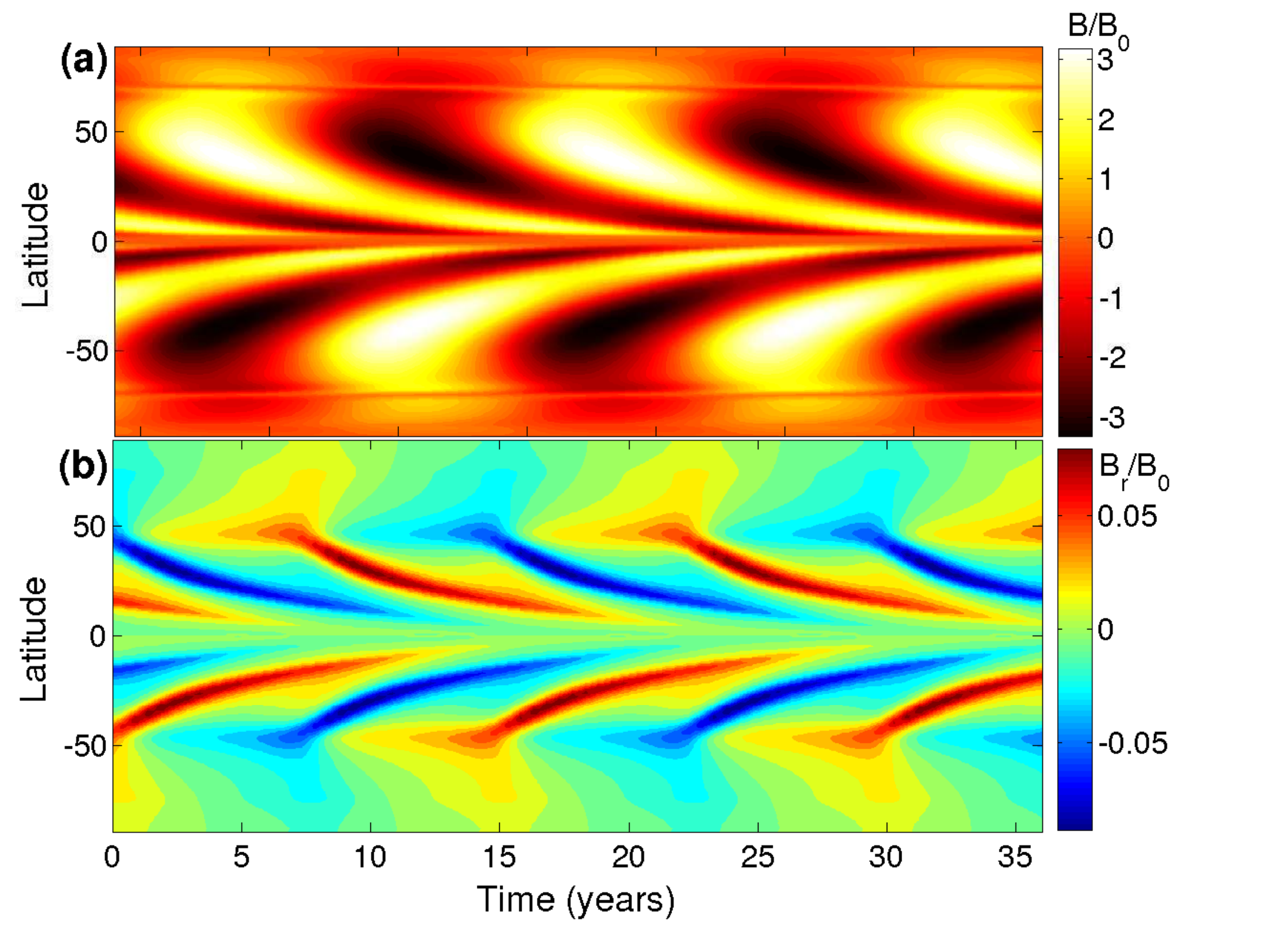}
\caption[Butterfly diagrams with complicated cells (Fig~\ref{fig:mcomp_cells}(a))]{(a) and (b) are the same plots as (c) and (d) in Figure~\ref{fig:rc2_nlwcll}, for the \mc\ given in
Figure~\ref{fig:mcomp_cells}(a).}
\label{fig:most_comp}
\end{figure}

\begin{figure}[!h]
\centering
\includegraphics[width=0.850\textwidth]{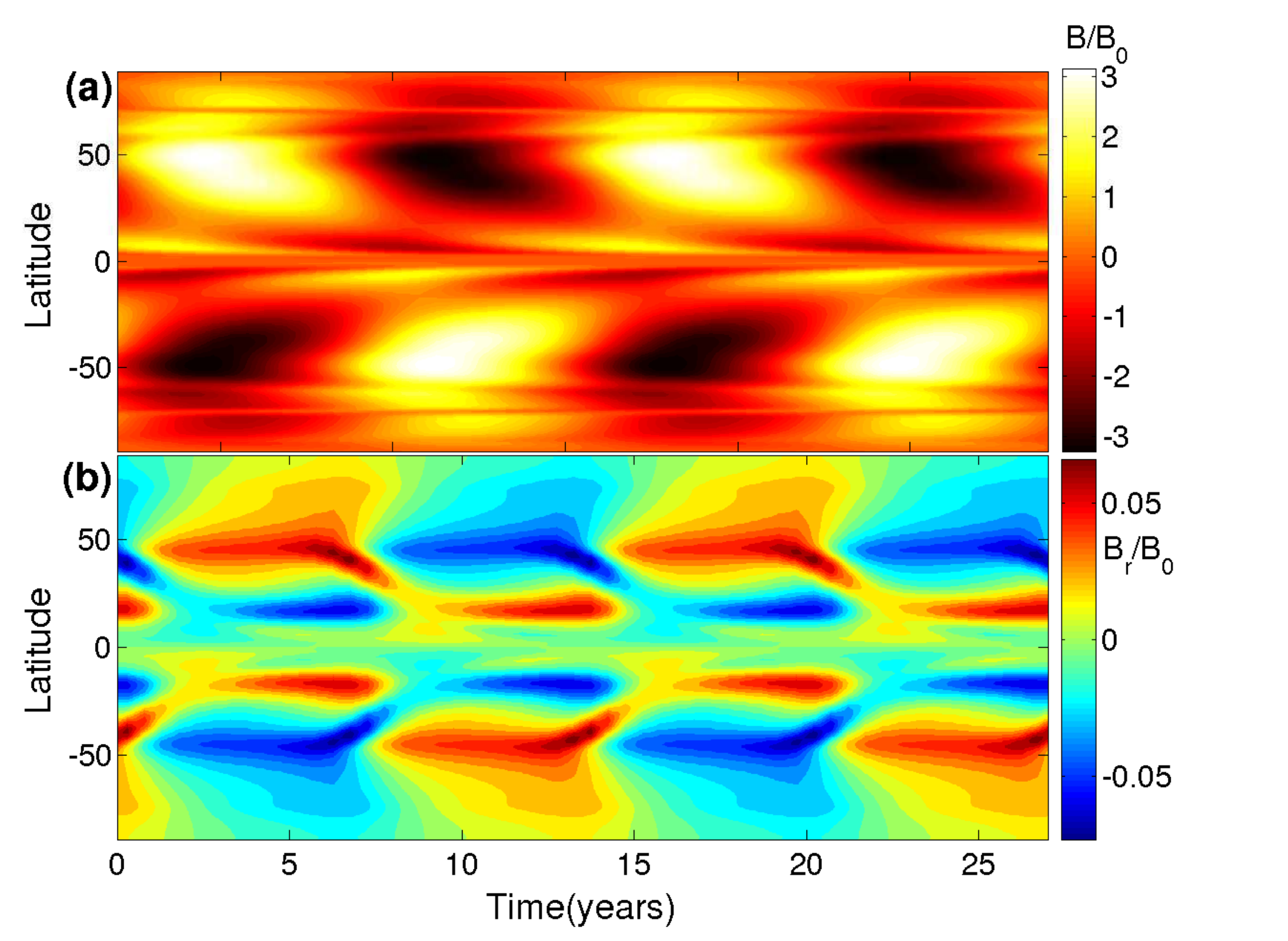}
\caption{Same as Figure~\ref{fig:most_comp}, for the \mc\ given in Figure~\ref{fig:mcomp_cells}(b).}
\label{fig:mcomp_2lowcl}
\end{figure}

We have carried out some simulations with fairly complicated multi-cell \mc, which reinforced
our main conclusion of the previous Section: we can have solar-like dynamo solutions as long
as there is an equatorward flow in low latitudes at the bottom of the convection zone. 
For the very complicated \mc\ pattern shown in Figure~\ref{fig:mcomp_cells}(a), we present the results in Figure~\ref{fig:most_comp}. 
The Appendix indicates how this complicated flow is obtained from a suitable stream function.
Since there is an equatorward flow in low latitudes at the bottom of the convection zone,
we get solar-like butterfly diagrams even with such a flow.  It may be noted that the lowermost
cell in Figure~\ref{fig:mcomp_cells}(a) from the equator does not go all the way to the pole, but the cell ends at
some high latitude.  We find that this cell has to extend sufficiently to reasonably
high latitudes in order to give a solar-like butterfly diagram.  If the cell does not extend beyond
mid-latitudes, then we are unable to get very solar-like butterfly diagrams.
In Figure~\ref{fig:mcomp_cells}(b), we show a \mc\ with the lower cell not extending to high latitudes.  
Results for this case are presented in Figure~\ref{fig:mcomp_2lowcl}. We see that the butterfly diagram
is much less realistic compared to the butterfly diagram presented in Figure~\ref{fig:most_comp}.
It is clear from Figure~\ref{fig:most_comp} and Figure~\ref{fig:mcomp_2lowcl}
that the requirement for a solar-like butterfly diagram is that there has to be an equatorward
flow at the bottom of the convection zone having a sufficient latitudinal extent from the equator
to a reasonably high latitude.

\section{Results for low diffusivity versus high diffusivity}
\label{C4:S5}
\begin{figure}[!h]
\centering
\includegraphics[width=0.750\textwidth]{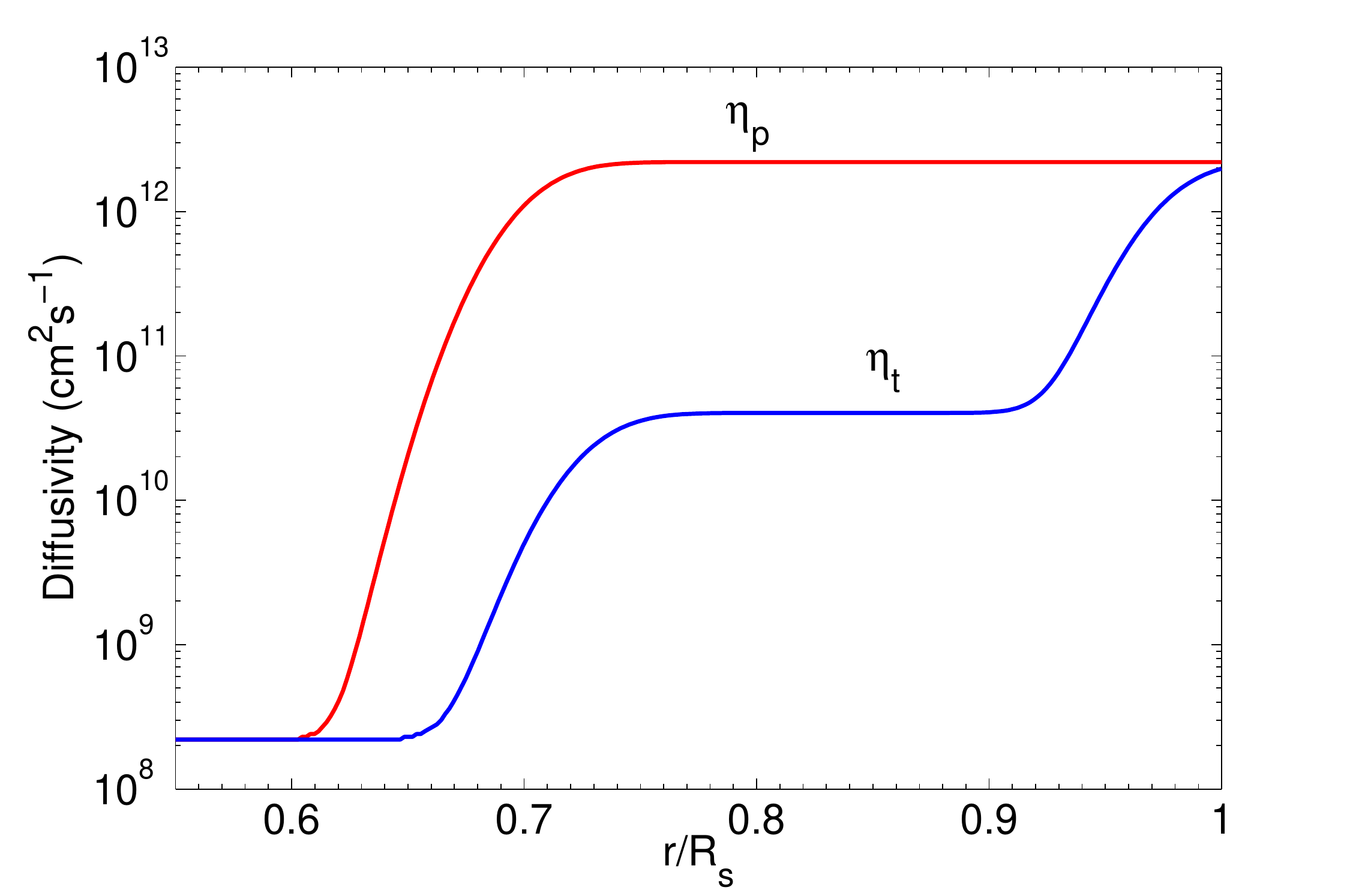}
\caption[Different Turbulent diffusivity used in our model]{Plots of $\eta_p(r)$ and $\eta_t(r)$ as given by (13) and (14). For the low diffusivity case, 
we take $\eta_{p}=\eta_{t}$.}
\label{fig:Highdif}
\end{figure}

We have pointed out that the nature of the dynamo depends quite a bit on whether
the turbulent diffusivity within the convection zone is assumed to be high or low
\citep{Jiang07,Yeates08,Hotta10, Karak10,
KarakChou11}. So far all the calculations in this chapter have been
carried out with a diffusivity on the higher side.  With such diffusivity, the
poloidal field generated near the surface by the Babcock--Leighton mechanism 
reaches the bottom of the convection zone primarily due to diffusion and this
process is not affected by the presence of multiple cells. However, when the
diffusivity is low, it is the \mc\ which has to transport the poloidal field
from the surface to the bottom of the convection zone and such transport becomes
more complicated when there are multiple cells.
Now we come to the question
whether our main conclusion in the previous two sections holds when the diffusivity
is low.  Following \citet{CNC04}, we specify the diffusivity 
for the high diffusivity case in the
following way: 
\begin{equation}
\eta_{p}(r) = \eta_{RZ} + \frac{\eta_{SCZ}}{2}\left[1 + \er \left(\frac{r - 0.7\Rs}
{0.03\Rs}\right) \right]
\label{eq:etap}
\end{equation}
\begin{eqnarray}
\label{eq:etat}
\eta_{t}(r) = \eta_{RZ} + \frac{\eta_{SCZ1}}{2}\left[1 + \er \left(\frac{r - 0.725\Rs}
{0.03\Rs}\right) \right]
+ \frac{\eta_{SCZ}}{2}\left[ 1 + \er \left(\frac{r-0.975\Rs}{0.03\Rs}
\right) \right]~~~
\end{eqnarray} 
Here $\eta_{RZ}$ is the diffusivity below the bottom of the convection zone which is
assumed to be small, whereas $\eta_{SCZ}$ and $\eta_{SCZ1}$ are respectively the
diffusivities of the poloidal and the toroidal components within the body of the
convection zone. Since the toroidal magnetic field is  believed to be much stronger than
the poloidal magnetic field, the diffusivity $\eta_{SCZ1}$ of the toroidal field is
assumed to be less than the diffusivity $\eta_{SCZ}$ of the poloidal field.
For high diffusivity case (all the results presented in Section~\ref{C4:S3} and Section~\ref{C4:S4}), 
the values of the parameters for $\eta_{p}$ are  
$\eta_{RZ}=2.2 \times 10^8$ cm$^2$ s$^{-1}$, $\eta_{SCZ}=2.2\times10^{12}$ cm$^2$ s$^{-1}$,
and for $\eta_{t}$ are $\eta_{SCZ1}=4.0\times10^{10}$ cm$^2$ s$^{-1}$. Figure~\ref{fig:Highdif} shows these diffusivities as
functions of $r$, which have been used in the calculations of Section~\ref{C4:S3} and
Section~\ref{C4:S4}. Now our aim in this Section is to study the case when the diffusivity
of the poloidal field is less.  To achieve this, we now take both $\eta_p$ and $\eta_t$
to be equal to $\eta_t$ in the high diffusivity case, as given by Eq.~\ref{eq:etat}. This means that
the diffusivity of the poloidal field within the main body of the convection zone is
now reduced by a factor of more than 50 (from $2.2\times10^{12}$ cm$^2$ s$^{-1}$ to
$4.0\times10^{10}$ cm$^2$ s$^{-1}$) for the studies presented in this Section.

\begin{figure}[!h]
\centering
\includegraphics[width=0.850\textwidth]{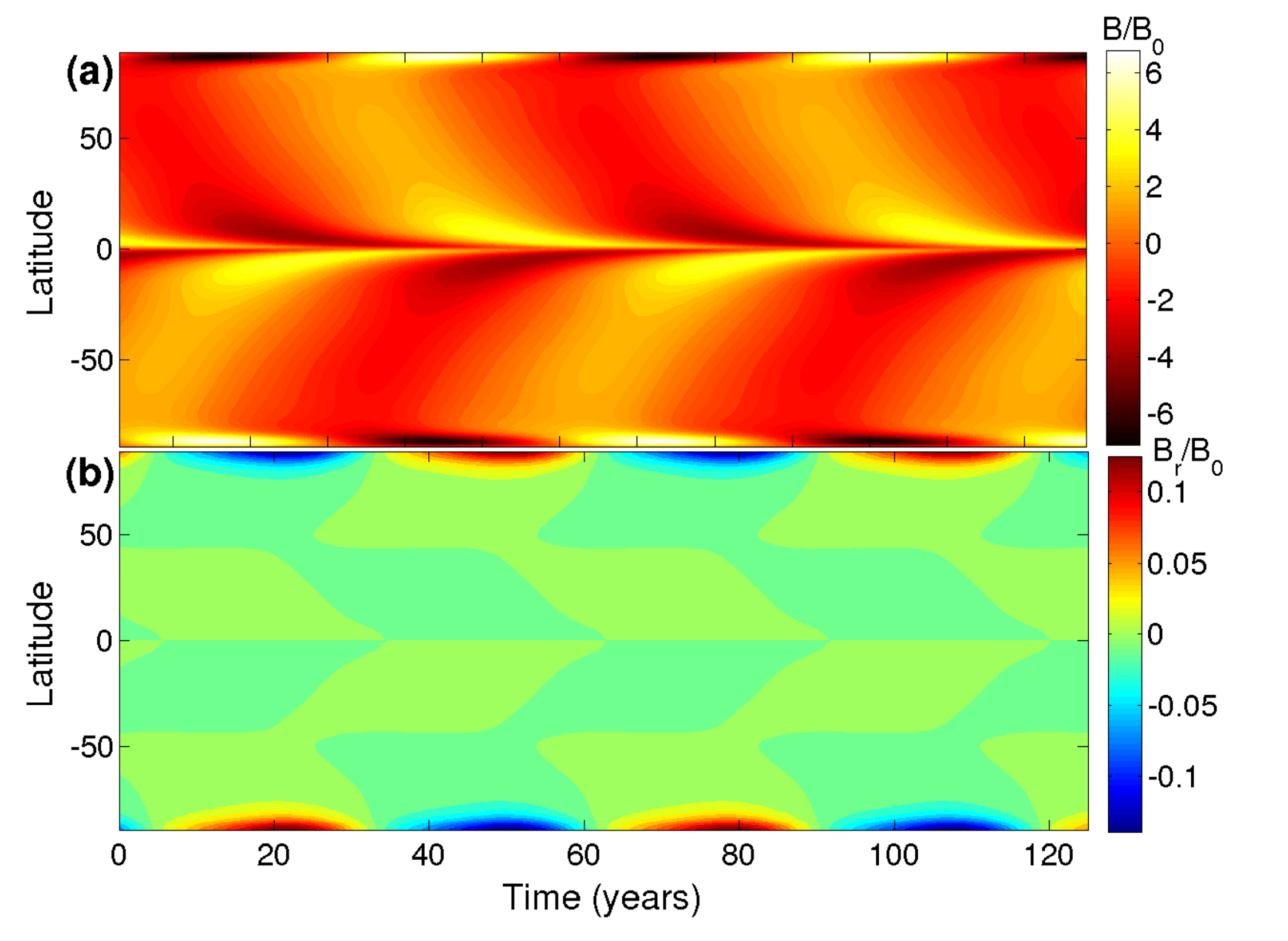}
\caption[Results with low diffusivity for radially three stacked cells]{Same as Figure~\ref{fig:most_comp}, for the case of three radially stacked cells used in Figure~\ref{fig:radc3}
except that the diffusivity of the poloidal field is now lowered by making $\eta_p=\eta_t$.}
\label{fig:rc3lowdif}
\end{figure}

\begin{figure}[!h]
\centering
\includegraphics[width=0.90\textwidth,trim={0 0 7cm 0}]{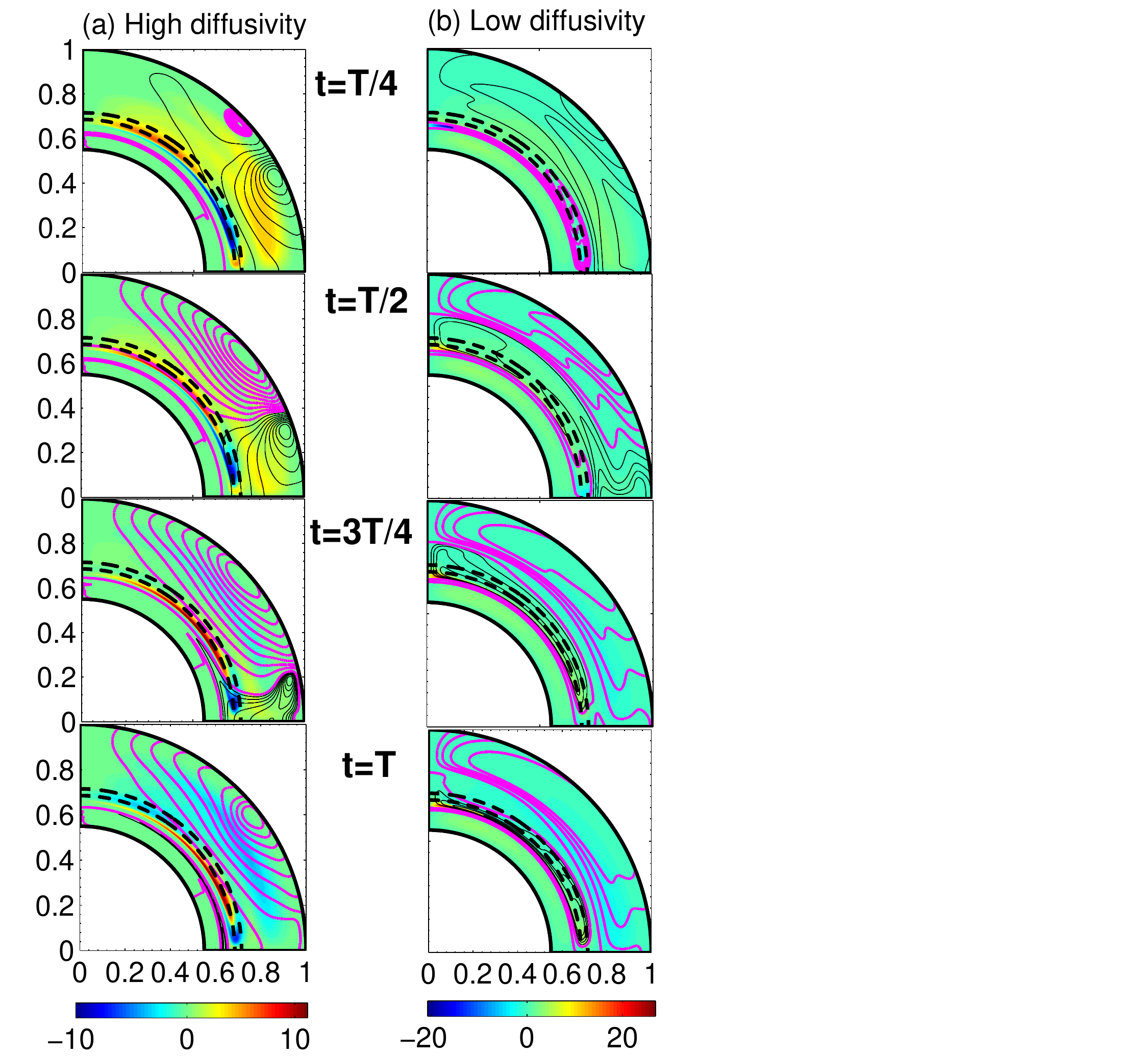}
\caption[Poloidal and Toroidal field lines for different diffusivity]{The poloidal field lines at four different stages of a solar cycle for the cases
of (a) high diffusivity and (b) low diffusivity. The magenta and the black 
colors respectively indicate the clockwise and anti-clockwise sense of field lines.
The background colors indicate the strength of the toroidal field .}
\label{fig:pol_tor}
\end{figure}

To understand the effect of lowering the diffusivity, we carry on calculations
for the case of three radially stacked cells (the case shown in Figure~\ref{fig:radc3}) by
changing the diffusivity from the higher value to the lower value as mentioned
above. While reducing the diffusivity, we also reduce the strength of the
$\alpha$-coefficient as pointed out in Section~\ref{C4:S2}. All the other parameters are kept unchanged. Figure~\ref{fig:rc3lowdif}
presents the results. Although we still find solar-like butterfly diagrams, we find that
the period has become much larger on reducing the diffusivity.  This is not
surprising.  When the diffusivity is low, the poloidal field generated by the
Babcock--Leighton mechanism near the surface is transported to the bottom of
the convection zone (where the toroidal field is generated from it) by the \mc.
If there is only one cell, then this is easily accomplished.  However, when
there are three radially stacked cells as we are considering, the situation
becomes much more complicated.  The uppermost cell brings the poloidal field
from the surface to its bottom. From there, the middle cell has to advect the
poloidal field to its bottom.  Finally, the lowermost cell takes the poloidal
field to the bottom of the convection zone. In this process, the period of the
dynamo gets lengthened.  Figure~\ref{fig:pol_tor} shows how the poloidal field lines evolve
with the cycle for the case of three radially stacked cells---both when the
diffusivity is high (the case of Figure~\ref{fig:radc3}) and when the diffusivity is
low (the case of Figure~\ref{fig:rc3lowdif}). In the high diffusivity case, the poloidal field
generated at the surface is transported downwards to the bottom of the convection
zone by diffusion.  Hence, in this case, we find that the poloidal field lines
still are not very different from what we find in the case of \mc\ with one
cell, as shown in Figure~4 of \citet{Jiang07}.  However, when the 
diffusivity is low, the poloidal field is nearly frozen during a cycle
and is advected by the \mc. In a three-cell \mc, we find that the poloidal
field becomes very complicated, as seen in the right column of Figure~\ref{fig:pol_tor}.

It has been pointed out that, when we introduce fluctuations to model
irregularities of solar cycle, the dynamo models with high and low diffusivities
behave completely differently \citep{Jiang07,KarakChou11}. In the high diffusivity
model, the fluctuations diffuse all over the convection zone in time scale
comparable to the period of the dynamo.  On the other hand, fluctuations
in the low diffusivity model remain frozen during the period of the
dynamo.  \citet{Jiang07} explained how the observed correlation between the
polar field during a sunspot minimum and the strength of the next cycle arises
in the high diffusivity model.
This correlation, which forms the basis of solar cycle prediction in the high
diffusivity model, does not exist in the low diffusivity model.  We now
check if these results hold even when we have multiple cells of the \mc.
\citet{CCJ07} identified the fluctuations in the Babcock--Leighton
process as the main source of irregularity in the sunspot cycles. These fluctuations
arise from the scatter in the tilt angles of sunspots caused by the effect
of convective turbulence on rising flux tubes \citep{Longcope02}. To model
these fluctuations, we introduce stochastic fluctuations in $\alpha_0$ appearing in Equation~\ref{alpha}. 
We set 
\begin{equation}
\alpha_0 \equiv \overline{\alpha_0} [1 \pm 0.75  \sigma (\tau_{\rm{cor}})],
\end{equation} 
where $\sigma$ 
is a uniformly generated random number within 0 to 1 which changes value after a coherence time $\tau_{cor} = 1$ month.
This makes $\alpha_0$ to fluctuate randomly around its mean value $\overline{\alpha_0}$ 
with 75\% amplitude of fluctuations. A simulation with such stochastic fluctuations in $\alpha$
in a traditional $\alpha \Omega$ dynamo model was first presented by \citet{Chou92}.

\begin{figure}[!h]
\centering
\includegraphics[width=0.95\textwidth,trim={0 0 6cm 0}]{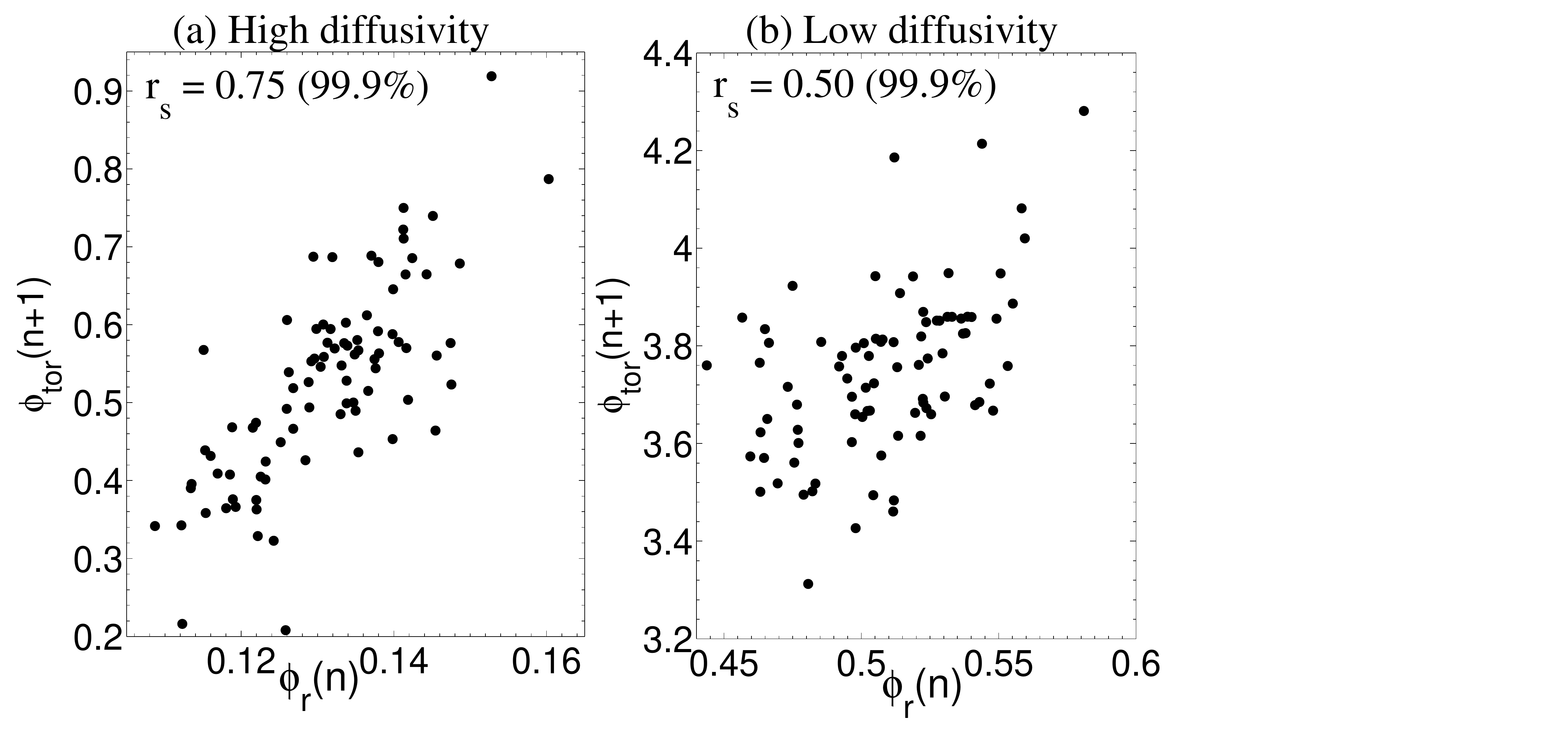}
\caption[Correlaton plot of polar flux with next cycle amplitude for different 
diffusivity]{Correlation between peak polar flux strength at the end of the $n$-th 
cycle and the peak toroidal flux strength of the $(n+1)$-th cycle for (a) high diffusivity and (b) low diffusivity cases.}
\label{fig:corr}
\end{figure}

To study the correlation between the polar field and the strength of the next cycle, we 
consider the procedure of \citet{Yeates08}. We calculate the 
correlation between the peak of the surface radial flux $\phi_{r}$ at high latitudes of a cycle 
with that of the peak value of the deep-seated toroidal flux $\phi_{tor}$ of next cycle. We take 
$\phi_{r}$ as the flux of radial field over the solar surface from latitude $70^{\circ}$ to $89^{\circ}$, 
and $\phi_{tor}$ as the flux of toroidal field over the region $r=0.677\Rs$ -- $0.726\Rs$ and latitude $10^{\circ}$ to $45^{\circ}$.
In the case of a one-cell \mc\ 
(not presented in detail in this study), we get a strong correlation between the high-latitude radial 
flux at the end of a cycle with the toroidal flux of the next cycle, with a correlation coefficient of 0.79, 
which is comparable to the result of \citet{Jiang07} (see their Figure~5) and \citet{Yeates08} (see their Fig.\ 11b).
Interestingly, for radially stacked three cells also, we get a strong correlation 
of 0.75 for the high diffusivity case.
Figure~\ref{fig:corr} shows this result, along with the result for the low 
diffusivity case. For the low diffusivity case, the correlation is substantially poorer.
Thus, a multi-cell \mc\ not only reproduces the regular periodic features of a
simple flux transport dynamo model, it also reproduces some of the irregular
features of the cycle if the diffusivity is high, in accordance with the results presented in Chapter~\ref{C3} with
multi-cell meridional circulation. The methodology for
predicting the next cycle developed by \citet{CCJ07,Jiang07} should work approximately the same way 
in the high diffusivity model even when the \mc\ has
a complicated multi-cell structure.

\section{The effect of turbulent pumping}
\label{C4:S6}
\begin{figure}[!h]
\centering
\includegraphics[width=0.85\textwidth]{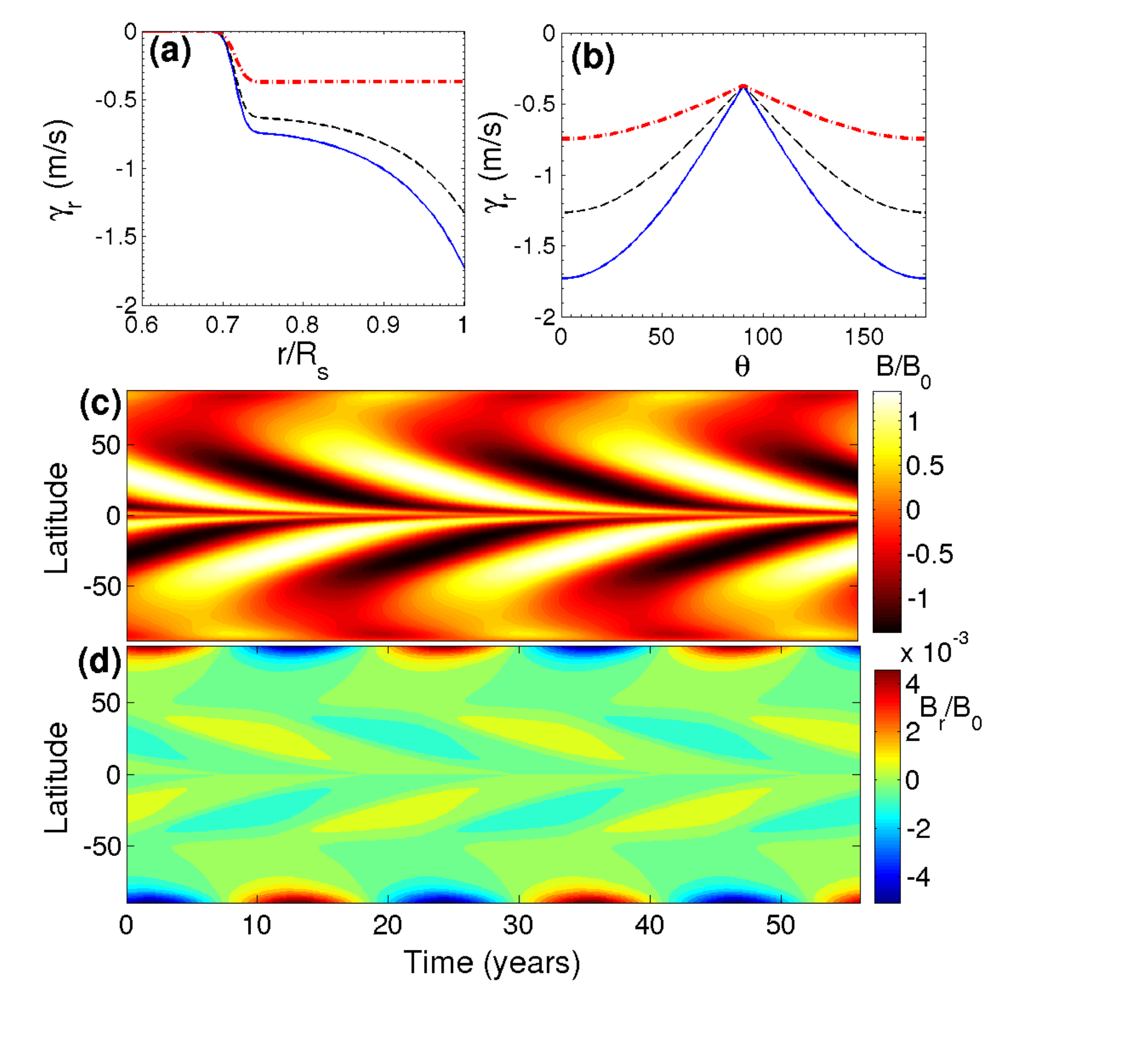}
\caption[Results with radial pumping added in low diffusivity case with radially stacked three cells]
{(a) Radial pumping $\gamma_r$ as a function of radius at different co-latitudes ($\theta$). The solid (blue),
the dashed (black) and the dash-dotted (red) lines correspond to $\theta = 90^{\circ}, 45^{\circ}$ and $0^{\circ}$ respectively.
(b) $\gamma_r$ as a function of $\theta$ at different radii.  The solid (blue),
the dashed (black) and the dash-dotted (red) lines correspond to $r = \Rs, 0.95\Rs$ and $0.75\Rs$ respectively.
(c) and (d) are the same as (a) and (b) in Figure~\ref{fig:rc3lowdif} with the radial pumping added now.}
\label{fig:pumping}
\end{figure}

One possible mechanism for transporting magnetic fields across the solar convection zone which
we have so far not included in our study is turbulent pumping. Many theoretical as well as numerical studies
indicated that, in the strongly stratified solar convection zone, the magnetic fields can 
be pumped preferentially downward towards the base of the convection zone 
\citep{Brandenburg96,Tobias98}. 
Several magnetoconvection simulations have detected a downward pumping speed of a
few meters per second in the solar convection zone \citep{OSBR02,Kapyla06,Racine11}.
Guided by these studies, we now include the effect of turbulent pumping in our dynamo model
by introducing the following downward pumping velocity: 
\begin{eqnarray}
\gamma_r = - 0.1854 \left[ 1 + \rm{erf}\left( \frac{r - 0.715\Rs}{0.015\Rs}\right) \right] 
\left[ \rm{exp}\left( \frac{r-0.715\Rs}{0.25\Rs}\right) ^2 \rm{cos}\theta +1\right],
\label{rpumping}
\end{eqnarray}
the unit being m s$^{-1}$.
The variations of $\gamma_r$ as functions of radius and co-latitude are shown in 
the upper part of Figure~\ref{fig:pumping}.
Turbulent pumping appears as an advective term in the magnetic field equations. Therefore 
in Equations~\ref{eqc4:Aeq} and \ref{eqc4:Beq} we add this extra term $\gamma_r$ in the radial velocity, i.e., we take $v_r \equiv v_r + \gamma_r $.
As in \citet{KarakNandy12, Kitchatinov12, Jiang13}, we 
first present results including only the radial pumping and not the latitudinal pumping.

Since the downward transport of the poloidal field by diffusion is reasonably efficient in
the high diffusivity model, the effect of downward turbulent pumping is not very pronounced
in this model.  However, in the low diffusivity model, the poloidal field is advected by
the \mc\ in the absence of turbulent pumping and the addition of downward pumping can have
quite dramatic effects. \citet{KarakNandy12} found that many of the differences between the
high and the low diffusivity models disappear on inclusion of downward turbulent pumping.
We have seen in Section~\ref{C4:S5} that the low diffusivity model with multi-cell \mc\ gives results which
do not match observations as closely as the results obtained with high diffusivity.
Although the case of \mc\ with three radially stacked cells even with low diffusivity
produces reasonably good equatorward propagation of toroidal field  at low latitude, the solar cycle
period becomes very long (see Figure~\ref{fig:rc3lowdif}).
We now repeat this calculation for the low diffusivity case by including the downward 
pumping.  The butterfly diagram is shown in the middle of Figure~\ref{fig:pumping}.  We find that the
period has become much shorter and the butterfly diagram looks quite similar to the butterfly
diagram of Figure~\ref{fig:radc3} in the high diffusivity case.  Thus, even when multiple cells are
present in the \mc, the inclusion of downward turbulent pumping makes the results of 
the low diffusivity case quite similar to the results of the high diffusivity case.
\begin{figure}[!h]
\centering
\includegraphics[width=0.8\textwidth]{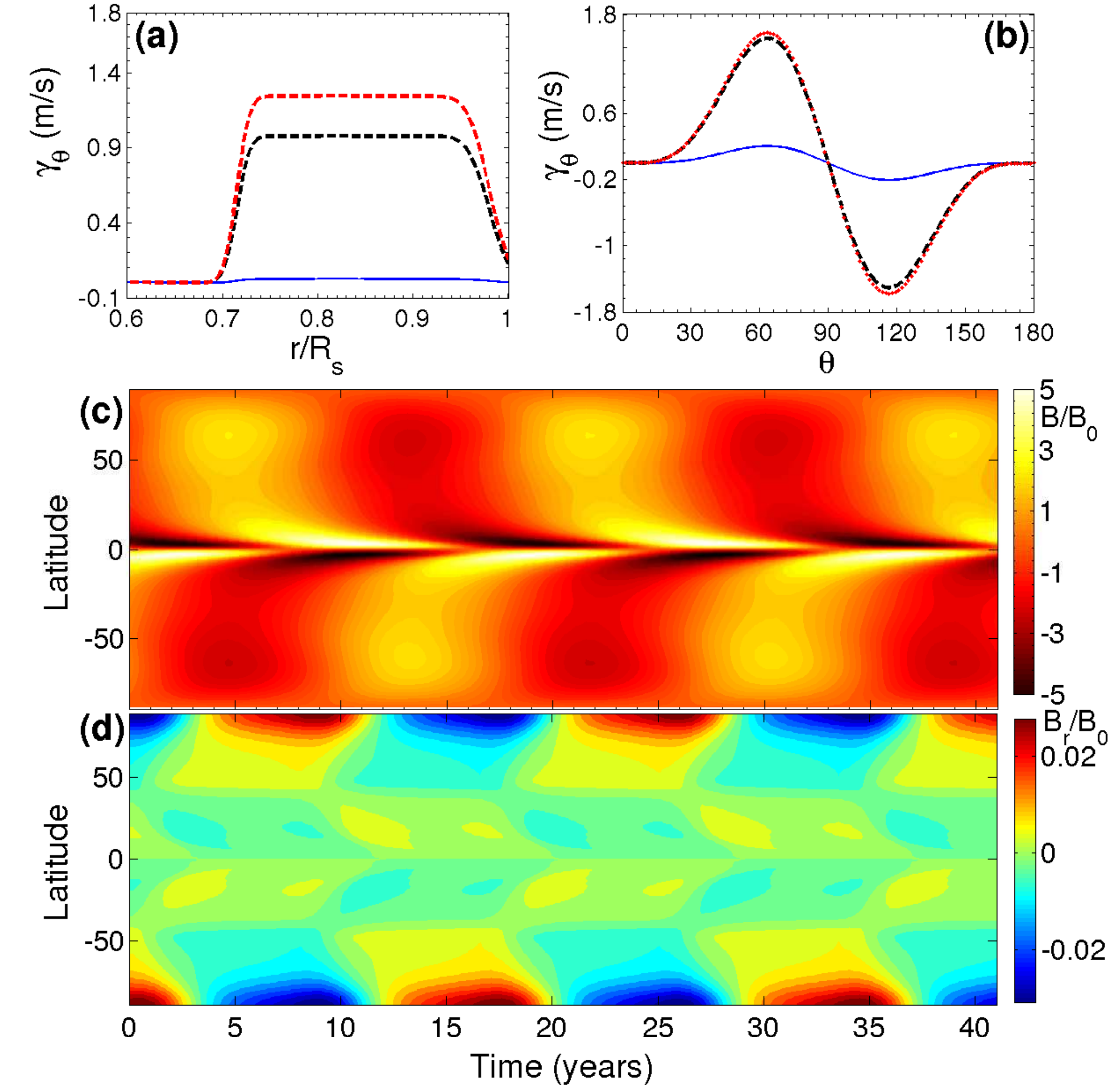}
\caption[Results with latitudinal pumping and shallow meridional circulation]{(a) Latitudinal pumping $\gamma_{\theta}$ as function of radius at different $\theta$. The solid blue, black dashed and red dashed lines correspond to $\theta = 15^{\circ}, 45^{\circ}$ and $75^{\circ}$ respectively. (b) $\gamma_{\theta}$ as a function of $\theta$ at different radii. The solid (blue), the dashed (black) and the dash-dotted (red) lines correspond to $r = \Rs, 0.95\Rs$ and $0.75\Rs$ respectively. (c) and (d) are same as Figure~\ref{fig:rc2_nlwcll}(c) and Figure~\ref{fig:rc2_nlwcll}(d) respectively with radial and latitudinal pumping added. }
\label{fig:lat_pump}
\end{figure}

A few magnetoconvection simulations \citep{OSBR02, Kapyla06, Racine11, Augustson15} have detected a 
latitudinal turbulent pumping when rotation becomes important. Just
to demonstrate that a latitudinal pumping can make many results significantly different, we
show that our model reproduces the results of \citet{Guerrero08}.
We take the equatorward latitudinal pumping profile as given below:
\begin{eqnarray}
\gamma_{\theta}=1.380\left[1+\rm{erf}\left(\frac{r-0.715\Rs}{0.0150\Rs}\right)\right]
\times\left[1-\rm{erf}\left(\frac{r-0.980\Rs}{0.0250\Rs}\right)\right]\cos{\theta}\sin^4{\theta},
\label{eq:lat_pump}
\end{eqnarray}
the unit being m s$^{-1}$.
This is very similar to what \citet{Guerrero08} had taken. We now do
a run for the case of a single cell of shallow meridional circulation (the case of Figure~\ref{fig:rc2_nlwcll}),
now combined with radial pumping given by Eq.~\ref{rpumping} and latitudinal pumping given by Eq.~\ref{eq:lat_pump}. The
profiles of $\gamma_{\theta}$ as well as the results of the run are presented in Figure~\ref{fig:lat_pump}. 
On comparing with Figure~5 of \citet{Guerrero08}, It has been found that we
qualitatively reproduce their results.  In other words, if there is suitable equatorward
pumping at the bottom of the convection zone, that can clearly help in producing solar-like
butterfly diagram. This result is also supported by \citet{HN16}, who use the equatorward latitudinal turbulent pumping profile motivated by the study of magnetoconvection simulation of \citet{Warnecke16}.

\section{Conclusion}
\label{C4:S7}
In the flux transport dynamo model which has been very successful in modeling
different aspects of the solar cycle, the \mc\ of the Sun is a crucial ingredient.
The major uncertainly in the flux transport dynamo model at the present time is
our lack of knowledge about the nature of the \mc\ in the deeper layers of the
convection zone. Although two-dimensional models can never treat magnetic buoyancy
and the Babcock--Leighton mechanism in a fully satisfactory, we believe that
these uncertainties are not so serious because different treatments of magnetic
buoyancy and the Babcock--Leighton mechanism give qualitatively similar
results \citep{Nandy01, CNC05}. Since the \mc\ arises out of a delicate imbalance between the
centrifugal forcing and the thermal wind \citep{Kitchatinov11b}, it is challenging to
model it theoretically. Models of differential rotation based on a mean field
treatment of turbulence give rise to a \mc\ \citep{Kitchatinov95,Rempel05,
Kitchatinov11b}.  MHD simulations of convection with the dynamo
process also produce meridional circulations \citep{Brown10,Racine11,Warnecke13,Kapyla13}.
In such simulations, the \mc\ is often found to have several cells and to vary rapidly
with time.  We are still far from having a definitive theoretical model of the
Sun's \mc.

Most of the flux transport dynamo models are based on the assumption of a
single-cell \mc\ having a return flow at the bottom of the convection zone.
While a support for such a return flow may have been missing, this assumption
of a deeply penetrating single-cell \mc\ was at least consistent with all the
observational data available till about a couple of years ago. The equatorward
propagation of the sunspot belt was indeed regarded as indicative of the \mc\ flow
velocity at the bottom of the convection zone \citep{HathawayNandy03}. Only
recently there are claims that the \mc\ may have a return flow at a much shallower
depth \citep{Hathaway12, Zhao13}. If these claims are corroborated by
independent investigations of other groups, then we shall have to conclude that
the assumption of a deep one-cell \mc\ is not correct.  Since this assumption
was extensively used in most of the kinematic flux transport dynamo models, a
crucial question we face  now is whether this assumption is so essential that
the flux transport dynamo models would not work without this assumption or whether the
flux transport dynamo models can 
still be made to work with a suitable modification of this
assumption. 

On the basis of our studies, we reach the conclusion that, in order to have a
flux transport dynamo giving a solar-like butterfly diagram, we need an equatorward
flow in low latitudes at the bottom of the convection zone. This flow is essential
to overcome the Parker--Yoshimura sign rule and to advect the toroidal field generated
in the tachocline in the equatorward direction.  As long as there is such a flow,
we find that the flux transport dynamo works even if the \mc\ has a much more
complicated structure than what has been assumed in the previous models.  If there
is a return flow at a shallow depth and there are no flows underneath, then the 
flux transport dynamo will not work. If there is a poleward flow at the bottom
of the convection zone, then also we do not get solar-like butterfly diagrams.
However, underneath a shallow return flow if we have multiple cells in such a 
way that there is an equatorward flow in low latitudes at the bottom of the 
convection zone, then the flux transport dynamo works without any serious problem.
The assumption of such a multi-cell \mc\ does not contradict any observational
data available at the present time.  MHD simulations also support the existence of a
complicated multi-cell \mc\ \citep{Brown10,Racine11,Warnecke13,Kapyla13}.
With such a multi-cell \mc, we are able to retain all the attractive features of the
flux transport dynamo model.  The phase relation between the toroidal and the
poloidal fields is correctly reproduced.  The observed correlation between the
polar field during a sunspot minimum and the strength of the next cycle is also
reproduced when the diffusivity is high, although a reduced diffusivity diminishes
this correlation.

One of the important processes in the operation of the flux transport dynamo is
the transport of the poloidal field generated near the surface by the 
Babcock--Leighton mechanism to the bottom of the convection zone where the
differential rotation can act on it.  We have taken the diffusivity on the
higher side in the calculations presented in Section~\ref{C4:S3}--\ref{C4:S4} 
and we find that the poloidal field can diffuse from the surface
to the bottom of the convection zone in a few years.  A complicated multi-cell
\mc\ does not get in the way of this process. However, when the diffusivity
is reduced, this transport has to be done by the \mc.  Interestingly, even
in the case of low diffusivity with a multi-cell \mc, we are still able to
get periodic solutions, although the poloidal field within the convection zone
becomes very complicated and the cycle period is
lengthened. A downward turbulent pumping helps in 
reducing the differences between the high and the low diffusivity models.  
Based on the few magnetoconvection simulations \citep{Kapyla06, Racine11, Augustson15, Warnecke16} 
which found the equatorward latitudinal turbulent pumping near the bottom half of the Convection zone, 
we reproduce the result of \citet{Guerrero08} that an equatorward pumping at the bottom of the
convection zone can make the flux transport dynamo work even in the absence of
a flow there.

To sum up, we do not think that the recent claims of an equatorward return
flow at a shallow depth pose a threat to the flux transport dynamo
model.  Especially, we see no reason to give up the attractive scenario that
the strong toroidal field is produced and stored in the stable regions of the
tachocline, from which parts of this toroidal field break away to rise through
the convection zone and produce sunspots. 
%So we do not agree with the model of the dynamo limited to the upper layers of the convection zone proposed by Pipin \& Kosovichev (2013). 
The crucial assumption needed to make the flux transport
dynamo work is an equatorward flow in low latitudes at the bottom of the convection
zone.  At present, we do not have observational data either supporting or contradicting
it. Since the flux transport dynamo has been so successful in explaining so many
aspects of the solar cycle, we expect this assumption of equatorward flow in low 
latitudes at the bottom of the convection zone to be correct and we hope that
future observations will establish it.  Only if future observations show this
assumption to be incorrect, a drastic revision of our current ideas about the
solar dynamo will be needed at that time.

\bigskip

\section*{Appendix}
\section*{Stream functions for the three-cell and more complicated meridional circulation}

%\subsection*{Meridional circulation with three radially stacked cells:}
To get three radially stacked cells shown in Figure~\ref{fig:radc3}(a), we take the stream function as  
\begin{equation}
\psi = \psi_u + \psi_m + \psi_l
\end{equation}
The stream function which generates the upper cell is given by
\begin{eqnarray}
\psi_u = {\psi_{0u}}\left[1-{\rm erf}\left(\frac{r-0.87R_\odot}{1.5}\right)\right](r - R_{m,u})^{0.3}
\sin \left[ \frac{\pi (r - R_{m,u})}{(R_\odot -R_{m,u})} \right]\{ 1 - e^{- \beta_1 \theta^{\epsilon}}\}\nonumber \\
\times\{1 - e^{\beta_2 (\theta - \pi/2)} \} e^{-((r -r_0)/\Gamma)^2} ~~~~
\end{eqnarray}\\
where the parameters have the following values:
$\beta_1 = 3.5, \beta_2 = 3.3$, $r_0 = (R_\odot - R_b)/3.5$, $\epsilon = 2.0000001$, $\Gamma =3.4 \times 10^{8}$ m, $R_{m,u} = 0.82 R_\odot$.
%***************************************************
The stream function for middle cell is given by
\begin{eqnarray}
\psi_m={\psi_{0m}}\left[1-{\rm erf}\left(\frac{r-0.85R_{m,u}}{1.5}\right)\right](r - R_{m,l})
\sin \left[ \frac{\pi (r - R_{m,l})}{(R_{m,u} -R_{m,l})} \right]\{ 1 - e^{- \beta_1 \theta^{\epsilon}}\}\nonumber \\
\times\{1 - e^{\beta_2 (\theta - \pi/2)} \} e^{-((r -r_0)/\Gamma)^2} ~~~~
\end{eqnarray}\\
where the parameters have the following values:
$\beta_1 = 1.9, \beta_2 = 1.7$, $r_0 = (R_\odot - R_b)/3.5$, $\Gamma =3.4 \times 10^{8}$ m, $R_{m,l} = 0.75 R_\odot$, $R_{m,u}=0.82 R_\odot$.
Finally, the stream function which generates lower cell is
\begin{eqnarray}
\psi_l = {\psi_{0l}}\left[1-{\rm erf}\left(\frac{r-0.75R_{m,l}}{0.8}\right)\right](r - R_p)
\sin \left[ \frac{\pi (r - R_p)}{(R_{m,l} -R_p)} \right]\{ 1 - e^{- \beta_1 \theta^{\epsilon}}\}\nonumber \\
\times\{1 - e^{\beta_2 (\theta - \pi/2)} \} e^{-((r -r_0)/\Gamma)^2} ~~
\end{eqnarray}\\
where the parameters have the following values:
$\beta_1 = 1.5, \beta_2 = 1.3$, $r_0 = (\Rs - R_b)/3.5$, $\Gamma =3.47 \times 10^{8}$ m, $R_p = 0.65 R_\odot$, $R_{m,l} = 0.76 R_\odot$.  
We choose $\psi_{0u}/C$, $\psi_{0m}/C$ and $\psi_{0l}/C$ in such a way that $v_0$ for upper cell, middle cell and lower cell are around $17.0$ m s$^{-1}$, $5.5$ m s$^{-1}$ and $2.0$ m s$^{-1}$ respectively.
%****************************************

In order to get the complicated \mc\ as shown in Figure~\ref{fig:mcomp_cells}, we choose our stream function as given below.
\begin{equation}
\psi = \psi_l + \psi_{lm} + \psi_m +\psi_u + \psi_{uc}
\end{equation}\\
where,$\psi_l, \psi_{lm}, \psi_m, \psi_u$ and $\psi_{uc}$ generate respectively the lower cell, lower middle cells, 
middle cells, the complicated upper cell and the upper corner cell. 
%Expressions of $\psi_l, \psi_{lm}$ and $\psi_m$ are similar as Eq.$\ref{eq:psi}$ multiplied by $\cos(q)$ factor to get latitudinal division of %cells and an $\theta$ dependent error function to include latitudinal dependence. The upper complicated cell {$\psi_u$} is given below
To give an idea about the kind of stream function we use in order to get a complicated cell, we write down the
stream function $\psi_u$ for the most complicated upper cell:
\begin{eqnarray}
\label{eq:com}
\psi_u = {\psi_{0u}}\left[1+{\rm erf}\left(\frac{r-R_c}{0.02\Rs}\right)\right]\sin \left[ \frac{\pi (r - R_p)}{(\Rs -R_p)} \right]
(r - R_p)\{ 1 - e^{- \beta_1 \theta^{\epsilon}}\}\nonumber \\
\times\{1 - e^{\beta_2 (\theta - \pi/2)} \} e^{-((r -r_0)/\Gamma)^2}, ~~~~~
\end{eqnarray}
where 
$$
R_c=\frac{1}{2}\left[1+{\rm erf}\left(\frac{\theta -\pi/24}{\pi/7}\right)\right]{\times}0.95\Rs 
$$
and the parameters  have the following values:

$\beta_1 = 0.45, \beta_2 = 1.3 $, $r_0 = (R_\odot - R_b)/3.5$, $\Gamma =3.1 \times 10^{8}$ m, $R_p = 0.65 R_\odot$. Here do not write down the other stream functions, which are constructed along similar lines.
%***************************************************

%footnote
\blfootnote{This chapter is based on \citet{HKC14}.}

%% file: chapter5.tex
\begin{savequote}[100mm]
``The World is full of obvious things which nobody
by any chance ever observes''
\qauthor{-Arthur Conan Doyle, The Hound of the Baskervilles}
\end{savequote}

\def\MC{meridional circulation}	
\def\vb{{\bf v}}
\def\Fn{{\bf F}_{\nu}}
\def\FL{{\bf F}_L}
\def\pa{\partial}

\chapter{Why does Meridional Circulation Vary with the Solar Cycle?}
\label{C5}
\begin{quote} \small
Observations of the meridional circulation of the Sun,
which plays a key role in the operation of the solar dynamo, indicate that its speed  
varies with the solar cycle, becoming faster during the solar
minima and slower during the solar maxima. To explain this variation of the
meridional circulation with the solar cycle, we construct a theoretical model
by coupling the equation of the meridional circulation (the $\phi$ component of the vorticity
equation within the solar convection zone) with the equations of the flux transport dynamo model. 
We consider the back reaction due to the Lorentz force of the dynamo-generated magnetic fields
and study the perturbations produced in the meridional circulation due to it.
This enables us to model the variations of the meridional circulation without developing a full theory of 
the meridional circulation itself. We obtain results which reproduce the observational data of solar cycle variations 
of the meridional circulation reasonably well. We get the best results on assuming the turbulent
viscosity acting on the velocity field to be comparable to the magnetic diffusivity
(i.e.\ on assuming the magnetic Prandtl number to be close to unity).
We have to assume an appropriate bottom boundary condition to ensure that the Lorentz force cannot drive a flow
in the subadiabatic layers below the bottom of the tachocline. Our results are sensitive to this bottom 
boundary condition.  We also suggest a hypothesis how the observed inward flow towards the active
regions may be produced. 
%We encounter a puzzle that the strength of the Lorentz force has to
%be artificially reduced by about two orders to match observations with theoretical results.
\end{quote}

\section{Introduction}

The meridional circulation is one of the most important large-scale coherent flow
patterns within the solar convection zone, the other such important 
flow pattern being the differential rotation. A strong evidence for the
meridional circulation came when low-resolution magnetograms noted that 
the `diffuse' magnetic field (as it appeared in low resolution) on
the solar surface outside active regions formed unipolar bands which shifted
poleward, implying a poleward flow at the solar surface
\citep{Howard81, WSN89b}. A further confirmation
for such a flow came from the observed poleward migration of filaments
formed over magnetic neutral lines \citep{MFS83,MS89}.
Efforts for direct measurement of the meridional circulation
also began in the 1980s \citep{LH82,Ulrich88}.
The maximum velocity of the meridional circulation at mid-latitudes
is of the order of 20 m s$^{-1}$.

After the development of helioseismology, there were attempts
to determine the nature of the meridional circulation below the
solar surface \citep{Giles97,BF98}.
The determination of the meridional circulation in the lower
half of the convection zone is a particularly challenging problem.
Since the meridional circulation is produced by the turbulent
stresses in the convection zone, most of the flux transport dynamo
models in which the meridional circulation plays a crucial role
assume the return flow of the \MC\ to be at the bottom of the
convection zone.  In the last few years, several authors presented
evidence for a shallower return flow
\citep{Hathaway12, Zhao13, Schad13}. 
However, \citet{RA15} argue that helioseismic
data still cannot rule out the possibility of the return flow
being confined near the bottom of the convection zone.

Helioseismic measurements indicate a variation of the
meridional circulation with the solar cycle.
\citep{CD01,Beck02,BA10,Komm15}.  These results are consistent with
the surface measurements of \citet{Hathaway10b},
who find that the \MC\ at the surface becomes weaker during the sunspot
maximum by an amount of order 5 m s$^{-1}$. One plausible
explanation for this is the back-reaction of the dynamo-generated
magnetic field on the \MC\ due to the Lorentz force.  The aim
of this study is to develop such a model of the variation of
the \MC\ and to show that the results of such a theoretical
model are in broad agreement with observational data.

The theory of the \MC\ is somewhat complicated.  The turbulent
viscosity in the solar convection zone is expected to be anisotropic,
since gravity and rotation introduce two preferred directions at
any point within the convection zone.  A classic paper by 
\citet{Kippenhahn63} showed that an anisotropic viscosity gives rise
to \MC\ along with differential rotation.  This work neglected the
`thermal wind' that would arise if the surfaces of constant density
and constant pressure do not coincide. A more complete model
based on a mean field theory of convective turbulence is due to
\citet{Kitchatinov95} and extended further by \citet{Kitchatinov11b}. 
These authors argue that, in the high Taylor number situation
appropriate for the Sun, the \MC\ should arise from the small imbalance 
between the thermal wind and the centrifugal force due to the variation
of the angular velocity with $z$ (the direction parallel to the rotation
axis).  See the reviews by Kitchatinov \citep{Kitchatinov11review,Kitchatinov13}
for further discussion. If the \MC\ really comes from a delicate
imbalance between two large terms, one would theoretically expect
large fluctuations in the \MC.  Numerical simulations of convection in
a shell presenting the solar convection zone produce \MC\ which is
usually highly variable in space and time \citep{Karak14,FM15,Passos17}.
It is still not understood why the observed \MC\ is much more
coherent and stable than what such theoretical considerations
would suggest. There are also efforts to understand whether mean
field models can produce a meridional circulation more complicated
than a single-cell pattern \citep{BY17}.

The \MC\ plays a very critical role in the flux transport dynamo
model, which has emerged as a popular theoretical model of the solar
cycle. This model started being developed from the 1990s \citep{WSN91,
CSD95,Durney95,DC99, Nandy02, CNC04} and its current
status has been reviewed by several authors \citep{Chou11,Chou14,
Charbonneau14,Karakreview14}.  The poloidal field in this
model is generated near the solar surface from tilted bipolar sunspot
pairs by the Babcock--Leighton
mechanism and then advected by the \MC\ to higher latitudes
\citep{DC94, DC95,CD99}.
This poleward advection of the poloidal field is an important feature
of the flux transport dynamo model and helps in matching various
aspects of the observational data.  The poloidal field eventually
has to be brought to the bottom of the convection zone, where the
differential rotation of the Sun acts on it to produce the toroidal
field.  If the turbulent diffusivity of the convection zone is assumed
low, then the \MC\ is responsible for transporting the poloidal field
to the bottom of the convection zone.  On the other hand, if the
diffusivity is high, then diffusivity can play the dominant role in
the transportation of the poloidal field across the convection zone
\citep{Jiang07,Yeates08}.
We shall discuss this point in more detail later in the chapter. Most
of the authors working on the flux transport dynamo assumed a one-cell
\MC\ with an equatorward flow at the bottom of the convection zone.
This equatorward \MC\ plays a crucial role in the equatorward transport
of the toroidal field generated at the bottom of the convection zone,
leading to solar-like butterfly diagrams.  In the absence of the \MC,
one finds a poleward dynamo wave \citep{CSD95}. When several
authors claimed to find evidence for a shallower \MC, there was a 
concern about the validity of the flux transport dynamo model. \citet{HKC14}
showed that, even if the \MC\ has a much more complicated multi-cell
structure, still we can retain most of the attractive features of the
flux transport dynamo as long as there is a layer of equatorward 
flow at the bottom of the convection zone.  In the present work, we
assume a one-cell \MC.
 
Most of the papers on the flux transport dynamo are of kinematic
nature, in which the only back-reaction of the magnetic field which
is often included is the so-called ``$\alpha$-quenching'', i.e. the
quenching of the generation mechanism of the poloidal field. Usually
kinematic models do not include the back-reaction of the magnetic field
on the large-scale flows. We certainly expect the Lorentz force of the
dynamo-generated magnetic field to react back on the large-scale flows.
From observations, we are aware of two kinds of back-reactions on the
large-scale flows: solar cycle variations of the differential rotation
known as torsional oscillations and solar cycle variations of the \MC\ \citep{Chou11a}.
There have been several calculations showing how torsional oscillations
are produced by the back-reaction of the dynamo-generated magnetic
field \citep{Durney00,Covas00,Bushby06,CCC09}. \citet{Rempel06} presented 
a full mean-field model of both the dynamo and
the large-scale flows. This model produced both torsional oscillations
and the solar cycle variation of the \MC.  However, apart from Figure~4(b)
in the paper showing the time-latitude plot of $v_{\theta}$ near the bottom
of the convection zone, this paper does not study the time variation of the
\MC\ in detail.  To the best of our knowledge, this is the only mean-field
calculation of magnetic back-reaction on the \MC\ before our work. 

The back-reaction on the \MC\ should mainly come from the tension of the
dynamo-generated toroidal magnetic field. Suppose we consider a ring of
toroidal field at the bottom of the
convection zone.  Because of the tension, the ring will try to shrink in
size and this can be achieved most easily by a poleward slip \citep{Van88}. 
It is this tendency of poleward slip that would give
rise to the Lorentz force opposing the equatorward meridional circulation.
We show in our calculations based on a mean field model that this gives rise to a vorticity opposing
the normal vorticity associated with the \MC\ and, as this vorticity diffuses
through the convection zone, there is a reduction in the strength of the \MC\
everywhere including the surface. In order to circumvent the difficulties
in our understanding of the \MC\ itself, we develop a theory of how the Lorentz force
produces a modification of the \MC\ with the solar cycle. The important
difference of our formalism from the formalism of \citet{Rempel06} is the following.
In Rempel's formalism, one requires a proper formulation of turbulent stresses
to calculate the full \MC\ in which one sees a cycle variation due to the 
dynamo-generated Lorentz force.  On the other hand, in our formalism, we have
developed a theory of the modifications of the \MC\ due to the Lorentz force
so that we can study the modifications of the \MC\ without developing
a full theory of the \MC. 

The mathematical formulation of the theory is described in the next Section.
Then Section~\ref{sec:result} presents the results of our simulation.  Our conclusions are
summarized in Section~\ref{sec:conclusion}.

\section{Mathematical Formulation}\label{sec:model}

Our main aim is to study how the Lorentz force due to the dynamo-generated
magnetic field acts on the meridional circulation.  
We need to solve the dynamo equations along with the equation for the \MC.

We write the magnetic field in the form
\begin{equation}
\label{eq:decomp}
{\bf B} = B_\phi(r,\theta,t)\hat{e}_\phi + \nabla\times [A(r,\theta,t) \hat{e}_\phi]
\end{equation}
where $B_\phi$ is the toroidal component of magnetic field and A is the magnetic vector potential 
corresponding to the poloidal component of the field. Then the dynamo equations for
the toroidal and the poloidal components are 
\begin{equation}
\label{eq:Aeq}
\frac{\partial A}{\partial t} + \frac{1}{s}({\bf v}_m.\nabla)(s A)
= \eta_{p} \left( \nabla^2 - \frac{1}{s^2} \right) A + S(r, \theta, t),
\end{equation}

\begin{eqnarray}
\label{eq:Beq}
\frac{\partial B_{\phi}}{\partial t}
+ \frac{1}{r} \left[ \frac{\partial}{\partial r}
(r v_r B_{\phi}) + \frac{\partial}{\partial \theta}(v_{\theta} B_{\phi}) \right]
= \eta_{t} \left( \nabla^2 - \frac{1}{s^2} \right) B_{\phi} %\nonumber \\
+ s({\bf B}_p.{\bf \nabla})\Omega \nonumber \\
+ \frac{1}{r}\frac{d\eta_t}{dr}\frac{\partial{(rB_{\phi})}}{\partial{r}},
\end{eqnarray}\\
where $v_r$ and $v_{\theta}$ are components of the 
meridional flow $\vb_m$, $\Omega$ is the differential rotation and $s=r \sin\theta$, whereas $S(r, \theta, t)$ is the 
source function which incorporates the Babcock-Leighton mechanism and magnetic buoyancy as explained in \citet{CNC04}
and \citet{CH16}.
We have kept most of the parameters the same as in \citet{CNC04}. In the kinematic approach, both $\vb_m = v_r {\bf e}_r + 
v_{\theta} {\bf e}_{\theta}$ and $\Omega$
are assumed to be given.  
%We shall present some calculations going the kinematic approach in which the meridional
%circulation $\vb_m$ altered by the Lorentz-force of the dynamo-generated magnetic field will be used while solving 
%Equations~\ref{eq:Aeq} and \ref{eq:Beq} in future paper.

To understand the effect of the Lorentz force on the meridional circulation, we need to consider
the Navier-Stokes equation with the Lorentz force term, which is
\begin{equation}
\frac{\partial {\bf v}}{\partial t} + ({\bf v}.\nabla){\bf v} = -\frac{1}{\rho}\nabla p + {\bf g} + 
\FL + \Fn (\vb)
\label{eq:basic}
\end{equation} 
where ${\bf v}$ is the total plasma velocity, $\bf g$ is the force due to gravity, $\FL$ is the Lorentz
force term and 
$\Fn (\vb)$ is the turbulent viscosity term corresponding to the velocity
field $\vb$. We would have $\Fn (\vb)= \nu \nabla^2 {\bf v}$ if the turbulent
viscosity $\nu$ is assumed constant in space, but $\Fn (\vb)$ can be more complicated for spatially
varying $\nu$. For simplicity we have considered turbulent viscosity as a scalar 
quantity but, in reality, it is a tensor and can be a quite complicated quantity: see \citet{Kitchatinov11b}.
It may be noted that the tensorial nature of turbulent viscosity is quite crucial in the 
theory of the unperturbed \MC\ and it cannot be treated as a scalar in the complete theory.
To study torsional oscillations, we have to consider the $\phi$ component of Eq.~\ref{eq:basic} as done by
\citet{CCC09}.  To consider variations of the \MC, however, we need to focus our attention on
the $r$ and $\theta$ components of Eq.~\ref{eq:basic}. A particularly convenient approach is to take the curl
of Eq.~\ref{eq:basic} (noting that $({\bf v}.\nabla){\bf v} = - \vb \times (\nabla \times \vb) + \frac{1}{2}\nabla (v^2)$)
and consider the $\phi$ component of the resulting equation.  This gives
\begin{eqnarray}
\label{eq:main}
\frac{\partial \omega_{\phi}}{\partial t} - [\nabla \times \{ \vb \times (\nabla \times \vb)\}]_{\phi} = 
\frac{1}{\rho^2}[\nabla \rho \times \nabla p]_{\phi} %\\ ~~\nonumber
+ [\nabla \times \FL]_{\phi} + [\nabla \times \Fn(\vb_m)]_{\phi}. 
\end{eqnarray} 
where $\omega = \nabla \times \vb$ is the vorticity, of which the $\phi$ component comes from the 
meridional circulation $\vb_m$ only. 
Writing $\vb = \vb_m + r \sin \theta \Omega$, a few steps of straightforward algebra give
\begin{equation}
[\nabla \times \{ \vb \times (\nabla \times \vb)\}]_{\phi} = - r \sin\theta\nabla.\left(\vb_m \frac{\omega_{\phi}}{r \sin\theta}\right)
+ r \sin\theta\frac{\partial \Omega^2}{\partial z}
\end{equation}
Substituting this in Eq.~\ref{eq:main}, we have
\begin{eqnarray}
\label{eq:main1}
\frac{\partial \omega_{\phi}}{\partial t} + s \nabla.\left(\vb_m \frac{\omega_{\phi}}{s}\right) =  
s \frac{\partial \Omega^2}{\partial z}+ \frac{1}{\rho^2}(\nabla \rho \times \nabla p)_{\phi} %\\ ~~\nonumber
+ [\nabla \times \FL]_{\phi} + [\nabla \times \Fn(\vb_m)]_{\phi}.
\end{eqnarray} 

We now break up the the meridional velocity into two parts:
\begin{equation}\label{eq:v_tot}
 \vb_m = \vb_0 + \vb_1,
\end{equation}
where $\vb_0$ is the regular meridional circulation the Sun would have in the absence of
magnetic fields and $\vb_1$ is its modification
due to the Lorentz force of the dynamo-generated magnetic field. The azimuthal vorticity $\omega_{\phi}$
can also be broken into two parts corresponding to these two parts of the \MC:
\begin{equation}
 \omega_{\phi} = \omega_0 + \omega_1.
\end{equation}
It is clear from Eq.~\ref{eq:main1} that the regular \MC\ of the Sun in the absence of magnetic fields, which is assumed
independent of time in a mean field model, should be given by
\begin{eqnarray}
\label{eq:w0}
s \nabla.\left(\vb_0 \frac{\omega_0}{s}\right) =  
s \frac{\partial \Omega^2}{\partial z}+ \frac{1}{\rho^2}(\nabla \rho \times \nabla p)_{\phi} %\\ ~~\nonumber
+ [\nabla \times \Fn(\vb_0)]_{\phi}.
\end{eqnarray} 
This is the basic equation which has to be solved to develop a theory of the regular \MC.
The first term in R.H.S is the centrifugal force term, whereas  
the second term, which can  be written as
\begin{equation}
\label{eq:s}
\frac{1}{\rho^2}(\nabla \rho \times \nabla p)_{\phi} = -\frac{g}{c_pr}\frac{\partial S}{\partial \theta},
\end{equation}
is the thermal wind term. This is the term which couples the theory to the thermodynamics
of the Sun.  So we need to solve the energy transport equation of the Sun in order to
develop a theory of the regular \MC, making the problem particularly difficult.
In the pioneering study of \citet{Kippenhahn63}, the thermal wind term was neglected. However,
we now realize that for the Sun, in which the Taylor number is large, this term has
to be important and of the order of the centrifugal term \citep{Kitchatinov95}. It may
be mentioned that, in the model of the \MC\ due to \citet{Kitchatinov95} and
\citet{Kitchatinov11b}, the advection term on the L.H.S.\ of Eq.~\ref{eq:w0} was neglected. On inclusion
of this term, the results were found not to change much (L.\ Kitchatinov, private
communication).

We now subtract Equation~\ref{eq:w0} from \ref{eq:main1}, which gives
\begin{eqnarray}
\label{eq:main2}
\frac{\partial \omega_1}{\partial t} + s \nabla. \left(\vb_0 \frac{\omega_1}{s}\right) + 
s \nabla. \left( \vb_1 \frac{\omega_0}{s}\right) %\\ ~\nonumber
= [\nabla \times \FL]_{\phi} + [\nabla \times \Fn(\vb_1)]_{\phi}
\end{eqnarray}
on neglecting the quadratic term in perturbed quantities $\vb_1 \omega_1$ and assuming that
$\Fn(\vb)$ is linear in $\vb$, which is the case if we use the standard expressions of the
viscous stress tensor. We have also assumed that the dynamo-generated magnetic fields do not
affect the thermodynamics significantly, making the thermal wind term to drop out of Eq.~\ref{eq:main2}.
This is what makes the theory of the modification of the \MC\ decoupled from the
thermodynamics of the Sun and simpler to handle than the theory of the unperturbed \MC.
We have also not included the centrifugal force term, on the assumption that the temporal variations
in $\Omega$ with the solar cycle do not significantly affect our model of the \MC\ perturbations. The existence
of torsional oscillations indicates that this may not be a fully justifiable assumption.
In a future work, we plan to include torsional oscillations in the theory along
with the \MC\ perturbations.  
%We note one important point here. One can develop a
%theory of torsional oscillations without bothering about modifications of the \MC\ due
%to magnetic stresses: see \citet{CCC09}. However, in a really complete theory of the \MC\
%perturbations, the torsional oscillations have to be taken into account.

\begin{figure*}
\includegraphics[width=0.92\textwidth]{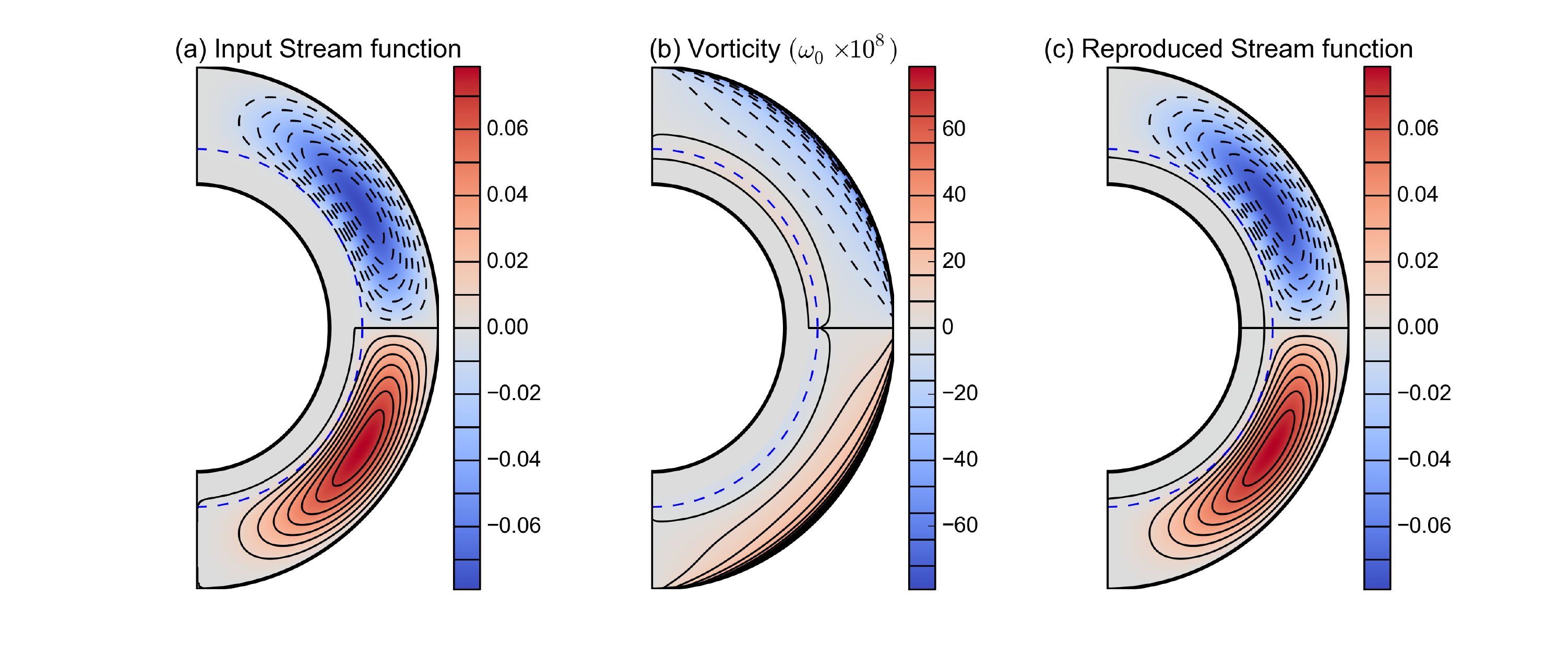}
\caption[Comparison between unperturbed meridional flow and reproduced flow]{(a) The input stream function for the unperturbed meridional circulation which is used in all of our simulations.
The streamlines are indicated by contours---solid contours indicating clockwise flow and dashed contours
indicating anti-clockwise flow.
(b) The unperturbed vorticity calculated from the stream function shown in (a). Solid and dashed
contours indicate positive and negative vorticity. 
(c) The stream function by inverting the vorticity shown in (b) with the help of Eq.~\ref{eq:sv}. This
has to be compared with (a). Blue dashed line represents the tachocline in all plots}
\label{fig:stream_vort}
\end{figure*}

In order to study how the \MC\ evolves with the solar cycle due to the Lorentz force
of the dynamo-generated magnetic fields, we need to solve Equation~\ref{eq:main2} along with
Equations~\ref{eq:Aeq} and \ref{eq:Beq}. A solution of Eq.~\ref{eq:main2} first yields the perturbed vorticity $\omega_1$
at different steps. We can then compute the perturbed velocity $\vb_1$ in the following way.
Since $\nabla\cdot\rho{\bf v}_1 = 0$, we can write $\vb_1$ in terms of a stream
function:
\begin{equation}
 {\bf v}_1 = \frac{1}{\rho} \nabla \times [\psi (r, \theta,t) \hat{e}_\phi].
\end{equation}
Putting this in ${\bf \omega}_1 = \nabla \times {\bf v}_1$, we get
\begin{equation}
\label{eq:sv}
\frac{1}{\rho}[\nabla \rho \times \{\nabla\times ( \psi \hat{e}_\phi)\}]_{\phi} + 
\left(\nabla^2 - \frac{1}{s^2} \right) \psi = -\rho \omega_1.
\end{equation}
This is similar to Poisson's equation and we solve it for a given vorticity to get the stream function. After obtaining
the stream function, it is very trivial to get the velocity fields. There are various methods to solve this
vorticity-stream function equation (Eq.~\ref{eq:sv}), but we choose to solve it using Alternating Direction Implicit (ADI) method. 
The advantage of this method is that it will take less iterations than the other relaxation methods (e.g, Jacobi iteration method) 
to achieve the same accuracy level. To check whether this method is giving us the correct results, 
we take the vorticity of the unperturbed \MC\ that we use in our dynamo calculations. We invert this
vorticity to obtain the velocity field by solving Eq.~\ref{eq:sv} and then cross check if we have succeeded in reproducing
the original input \MC.
The stream function corresponding to the input velocity field is in shown in Figure~\ref{fig:stream_vort}(a). Here black solid 
contours surrounding regions of positive stream function (indicated by red)
imply clockwise flow and black dashed contours 
surrounding regions of negative stream function (indicated by blue) imply anti-clockwise flow.
We calculate the vorticity (Fig~\ref{fig:stream_vort}(b)) from this given velocity field and solve Eq.~\ref{eq:sv} 
to get back the stream function. The stream function obtained in this process is shown in Fig~\ref{fig:stream_vort}(c). As we can see, it is almost
indistinguishable from the original given stream function.
While solving Eq. ~\ref{eq:sv}, we have set the tolerance level to $10^{-5}$, which is found
to be sufficient enough to get accurate results. As meridional circulation cannot penetrate much below the tachocline, we have
used $v_r= 0$ at $r=0.65R_\odot$ as our bottom boundary condition.

It may be noted that a realistic specification of density $\rho$ is more important for studying
the \MC\ than for the dynamo problem.  It is the density stratification which determines how strong
the poleward flow at the surface will be compared to the equatorward flow at the bottom of the
convection zone. 
We use a polytropic, hydrostatic, adiabatic stratification for density that matches the standard
solar model quite well \citep{Jones11}:
\begin{equation}\label{eq:rho}
\overline{\rho} = \rho_i ~ \left(\frac{\zeta(r)}{\zeta(r_1)}\right)^n
\end{equation}
where
\begin{equation}
\zeta(r) = c_0 + c_1 \frac{r_2-r_1}{r}
\end{equation}
\begin{equation}
c_0 = \frac{2 \zeta_0 - \beta - 1}{1-\beta}
\end{equation}
\begin{equation}
c_1 = \frac{(1+\beta)(1-\zeta_0)}{(1-\beta)^2}
\end{equation}
and
\begin{equation}
\zeta_0 = \frac{\beta+1}{\beta \exp(N_\rho/n) + 1}  ~~~.
\end{equation}
Here $\rho_i = \rho(r_1) = 0.1788$ g cm$^{-3}$, $\beta = r_1/r_2 = 0.55$ is the aspect ratio,
$n = 1.5$ is the polytropic index, and $N_\rho = 5$ is the number of density scale heights
across the computational domain, which extends from $r_1 = 0.55R_\odot$ to $r_2 = R_\odot$. 

To solve Eq.~\ref{eq:main2}, we need explicit expressions of $[\nabla \times \FL]_{\phi}$ and $[\nabla \times \Fn(\vb_1)]_{\phi}$.
If the magnetic field is given by Eq.~\ref{eq:decomp}, then we have
\begin{eqnarray}\label{eq:curl_fl}
[\nabla \times \FL]_{\phi} = 
\left[\nabla \times \left( \frac{(\nabla \times {\bf B})\times
{\bf B}}{4\pi\rho}\right)\right]_\phi ~~~~~~~~~~~~~~~~~~~~~~~~\\~\nonumber
=\left[\frac{1}{4\pi\rho}\nabla \times \{(\nabla \times {\bf B})\times {\bf B}\}\right]_\phi
- \left[\{(\nabla \times {\bf B})\times {\bf B}\} \times \nabla \left(\frac{1}{4\pi\rho}
\right)\right]_\phi
\end{eqnarray}
It is straightforward to show that
\begin{eqnarray}\label{eq:tot_lf}
\nabla \times \left[\{(\nabla \times {\bf B})\times {\bf B}\}\right]_\phi ~~~~~~~~~~~~~~~~~~~~~~~~~~~~~~~~~~~~~~~~~~~~~~~~~~~~~~~~~\\ \nonumber
=\left[\frac{1}{r}\frac{\partial}{\partial r}\left(\frac{j_\phi}{s}\right)\frac{\partial (sA)}{\partial \theta}
-\frac{1}{r}\frac{\partial}{\partial \theta}\left(\frac{j_\phi}{s}\right)\frac{\partial (sA)}{\partial r}\right] \\ \nonumber 
~~~~~~~~~~~~~~~~~~~~~~~~~~~~~~~~+\left[\frac{1}{r}\frac{\partial}{\partial \theta}\left(\frac{B_\phi}{s}\right)\frac{\partial(sB_\phi)}{\partial r}
-\frac{1}{r}\frac{\partial}{\partial r}\left(\frac{B_\phi}{s}\right)\frac{\partial (sB_\phi)}{\partial \theta}\right]
\end{eqnarray}
where $j_\phi = (\nabla \times {\bf B})_\phi$ is $\phi$ component of current density, which is
associated only with the poloidal field.  It is clear from Eq.~\ref{eq:tot_lf} that the Lorentz forces due to
the poloidal field and due to the toroidal field clearly separate out.
In typical flux transport dynamo simulations, the toroidal magnetic field turns out to be about
two orders of magnitude stronger than the poloidal magnetic field.  Since the Lorentz force goes as
the square of the magnetic field, the Lorentz force due to the toroidal field should be about
four orders of magnitude stronger than that due to the poloidal field.
So we include the Lorentz force due to the toroidal field alone in our calculations. We also
have to keep in mind that the magnetic field in the convection zone is highly intermittent with
a filling factor, say $f$.  The Lorentz force is significant only inside the flux tubes and
a mean field value of this has to be included in the equations of our mean field theory. This
point was discussed by \citet{CCC09}, who showed that, if we use mean field values of the magnetic
field in our equations, then the mean field value of the quadratic Lorentz force will involve
a division by $f$. Keeping this in mind, the expression of $[\nabla \times \FL]_{\phi}$ to be
used when solving Equation~\ref{eq:main2} is obtained from Equations~\ref{eq:curl_fl} and \ref{eq:tot_lf}:
\begin{eqnarray}\label{eq:tor_lf}
[\nabla \times \FL]_{\phi} = \frac{1}{4\pi\rho f}\left[\frac{1}{r^2}\frac{\partial}{\partial \theta}
(B_{\phi}^2)
- \frac{\cot\theta}{r}\frac{\partial}{\partial r}(B_{\phi}^2) \right] %\\~\nonumber
+ \frac{1}{4\pi\rho^2 f}\frac{d \rho}{d r}\frac{B_\phi}{r \sin\theta}\frac{\partial (\sin\theta B_\phi)}{\partial \theta}
\end{eqnarray} 
The expression for $[\nabla \times \Fn (\vb_1)]_{\phi}$ happens to be rather complicated (even
when the turbulent diffusivity $\nu$ is assumed to be a scalar) and will be discussed
in Appendix~A. We carry on our calculations assuming a simpler form of the viscosity term: 
\begin{equation}
\label{eq:viscous1}
[\nabla \times F_\nu({\bf v_1})]_\phi = \nu\left(\nabla^2-\frac{1}{r^2 \sin^2\theta}\right) \omega_1.
\end{equation}
As we shall show in Appendix~A, this simple expression of  $[\nabla \times \Fn (\vb_1)]_{\phi}$,
which is easy to implement in a numerical code, captures the effect of viscosity reasonably well.
%whether this assumption would be good in the context of our calculations (see Appendix~\ref{sec:appendix}). 

To study the variations of the \MC\ with the solar cycle, we need to solve Equations~\ref{eq:Aeq}, \ref{eq:Beq} and \ref{eq:main2}
simultaneously, with $[\nabla \times \FL]_{\phi}$ and $[\nabla \times \Fn (\vb_1)]_{\phi}$ in 
Eq.~\ref{eq:main2} given by Eq.~\ref{eq:tor_lf} and Eq.~\ref{eq:viscous1} respectively.  It is 
the term  $[\nabla \times \FL]_{\phi}$ in Eq.~\ref{eq:main2} 
which is the source of the perturbed vorticity $\omega_1$ and gives rise the variations of the \MC\
with the solar cycle. The toroidal field
field $B_{\phi}$ obtained at each time step from Eq.~\ref{eq:Beq} is used for calculating  
$[\nabla \times \FL]_{\phi}$ as given by Eq.~\ref{eq:tor_lf}.

The numerical procedure for solving Equations~\ref{eq:Aeq} and \ref{eq:Beq} has been discussed in detail by \citet{CNC04}.  We solve
Eq.~\ref{eq:main2} also in a similar way.  Note that this equation 
consists of two advection terms in the LHS and one diffusion term along with the source term in
the RHS. It is also basically an advection-diffusion equation similar to Eq.~\ref{eq:Aeq} and \ref{eq:Beq}. 
%As we believe that the perturbation in the 
%velocity field will be smaller in comparison to the unperturbed velocity fields, we neglect the advection of unperturbed
%vorticity by the perturbed velocity i.e, the 3rd term in the LHS of Eq.~\ref{eq:main1}). This approximation
%is also justifiable because the unperturbed vorticity plays no role in the dynamics and evolution of the magnetic field 
%untill we incorporates the effect of back reaction in the magnetic equations (Eq. \ref{eq:Aeq}, \ref{eq:Beq}). 
We solve Equation~\ref{eq:main2} to obtain the axisymmetric perturbed vorticity $\omega_1$ 
in the $r-\theta$ plane of the Sun with $256 \times 256$ grid cells in latitudinal and radial directions.
As in the case of Equations~\ref{eq:Aeq} and \ref{eq:Beq}, we solve Eq.~\ref{eq:main2} by using Alternating Direction Implicit (ADI) method of differencing,
treating the diffusion term through the Crank--Nicholson scheme and the advection term
through the Lax--Wendroff scheme (for more detail please see the guide
of {\it Surya} code which is publicly available upon request. Please send e-mail to {\it arnab@physics.iisc.ernet.in}).
We have used $\omega_1=0$ as the boundary condition at the poles $\theta =0$ and $\pi$. 
The radial boundary conditions are also $\omega_1 =0$ on 
the surface $(r = R_\odot)$ and below the tachocline $(r = 0.70R_\odot)$.

We now make one important point. By solving Eq.~\ref{eq:main2}, we obtain the perturbed vorticity $\omega_1$ at each time 
step.  If we need the perturbed velocity $\vb_1$ also at a time step, then we further need to solve Eq.~\ref{eq:sv}
over our $256 \times 256$ grid points.  If this
is done at every time step, then the calculation becomes computationally very expensive.
It is much easier to run the code if we do not need $\vb_1$ at each time step. In a completely
self-consistent theory, the meridional flow appearing in the dynamo Equations~\ref{eq:Aeq} and \ref{eq:Beq}
should be given by Eq.~\ref{eq:v_tot}, with $\vb_1$ inserted at each time step.  As we pointed out, this
is computationally very expensive.  If $\vb_1$ is assumed small compared to $\vb_0$, then
the dynamo Equations \ref{eq:Aeq} and \ref{eq:Beq} can be solved much more easily by using $\vb_m = \vb_0$.
Additionally, $\vb_1$ appears in the third term on the LHS of Eq.~\ref{eq:main2}. 
We discuss this term in Appendix B, where we argue that our results should not change
qualitatively on not including this term.  If we neglect this term along with using $\vb_m = \vb_0$
in the dynamo equations \ref{eq:Aeq} and \ref{eq:Beq}, then we do not require $\vb_1$ at each time step. The
calculations in the present study are done by following this approach, which is computationally
much less intensive.  This has enabled us to explore the parameter space of the problem more
easily and understand the basic physics issues.  We are now involved in the much more
computationally intensive calculations in which $\vb_1$ is calculated at regular time intervals
and the dynamo equations \ref{eq:Aeq} and \ref{eq:Beq} are solved by including $\vb_1$ in the meridional flow.
The results of these calculations will be presented in future work.

Finally, before presenting our results, we want to point out a puzzle
which we are so far unable to resolve.  Our calculations suggest
that the variations of the meridional circulation should be about
two orders of magnitude larger than what they actually are! Since a larger
$f$ would reduce the Lorentz force given by Eq.~\ref{eq:tor_lf}, we can make the
variations of the meridional circulation to have the right magnitude
only if we take the filling factor $f$ larger than unity, which is
clearly unphysical.  We show that this problem becomes apparent even
in a crude order-of-magnitude estimate. We expect $\omega_1$ to be of order
$V/L$, where $V$ is the perturbed velocity amplitude and $L$ is the
length scale.  Then $\partial \omega_1/\partial t$ has to be of order
$V/LT$.  Taking $T$ to be the solar cycle period and $L$ to be the depth
of the convection zone, a value $V \approx 5$ m s$^{-1}$ gives
\begin{equation}\label{eq:dw1dt}
\left| \frac{\partial \omega_1}{\partial t} \right| \approx 10^{-16}
\rm{s}^{-2}.
\end{equation}
The curl of the Lorentz force is the source of this term and should
be of the same amplitude.  Now the curl of the Lorentz force is of
order $|B_\phi^2 /4 \pi f \rho L^2|$.  In a mean field dynamo model, the
magnetic field should be scaled such that the mean polar field comes
out to have amplitude of order 10 G.  Then the mean toroidal field at
the bottom of the toroidal field turns out to be about $10^3$ G. On
using such a value, the curl of the Lorentz force is of order $10^{-15}/f$
s$^{-2}$. This would become equal to $10^{-16}$
s$^{-2}$ given in Eq. \ref{eq:dw1dt} only if $f$ is of order 10---a completely
unphysical result.  If we take $f$ less than 1 as expected, then the 
theoretical value of the perturbed \MC\ would be much larger than
what is observed. One may wonder if the same problem would be there
in the theory of torsional oscillations.  It turns out that the torsional
oscillations have the same amplitude 5 m s$^{-1}$ as variations in the
meridional circulation.  However, torsional oscillations are driven by
the component $\approx |B_r B_\phi/L|$ of the magnetic stress rather than
the component $\approx |B_\phi^2/L|$ relevant for meridional circulation
variations. Since $|B_r|$ turns out to be about 100 times smaller $|B_\phi|$
in mean field dynamo calculations, things come out quite reasonably in
the theory of torsional oscillations \citep{CCC09}. Here is then the puzzle.
Since the variations in the meridional circulation are driven by the
magnetic stress $\approx |B_\phi^2/L|$ which is about 100 times larger
than the magnetic stress $\approx |B_r B_\phi/L|$ driving torsional
oscillations, one would naively expect the meridional circulation variations
to have an amplitude about 100 times larger than the amplitude of torsional
oscillations.  But why are their observational values found comparable?
In the present study, we do not attempt to provide any solution to this
puzzle and merely present this as an issue that requires further study.
We point out that, even in non-magnetic situations, some terms in
the equations tend to produce a much faster meridional circulation
than what is observed \citep{Durney96,Dikpati14}. Presumably, such a
fast meridional circulation would upset the thermal balance and, as a
back reaction, a thermal wind force would arise to ensure that such a large
meridional circulation does not take place. Also, the cyclic alterations of 
meridional flow may occur indirectly (other than the Lorentz force 
associated with the large-scale magnetic field) via magnetically mediated 
changes in other dynamical drivers e.g., Reynolds stresses, Maxwell stresses.
Some MHD simulations (\citet{Beaudoin13} and reference therein) suggest that this is the case for the rotational torsional oscillation and 
since meridional circulation and differential rotation are strongly coupled, one might expect
similar things may happen for variation of the meridional flow.  
  
%We are unable to provide any solution to this puzzle.  We present this as
%something we need to ponder over.

%====================================================
\begin{figure}
\centering{
\includegraphics[width=0.6\textwidth]{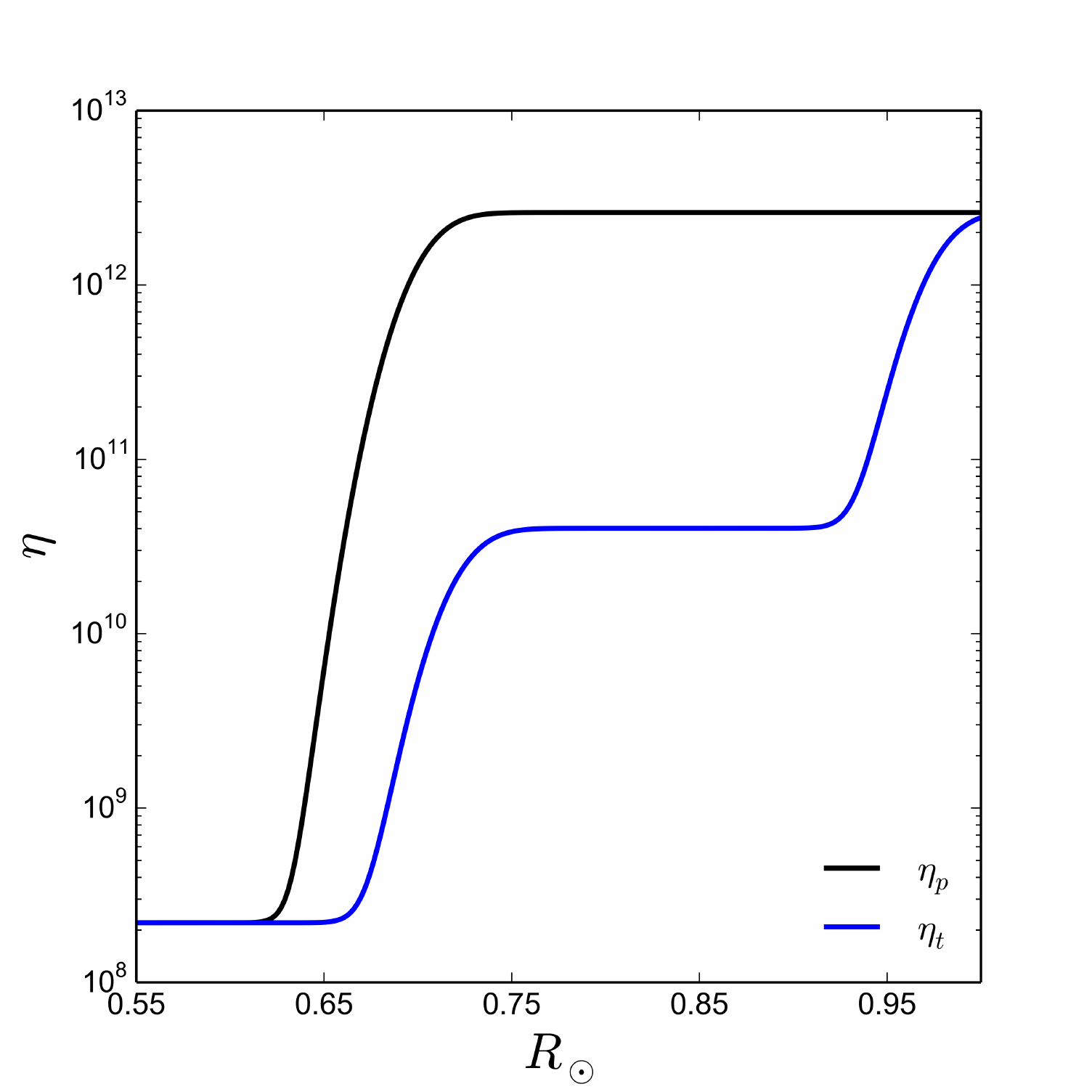}}
\caption[Diffusivity profiles]{Magnetic diffusivities (cm$^2$ s$^{-1}$) used in our flux transport model. The black solid line shows
the turbulent diffusivity for the poloidal magnetic field, and the blue solid line shows the same for
the toroidal magnetic field.}   
\label{figc5:eta}
\end{figure}

\begin{figure}
\centering{
\includegraphics[width=0.55\textwidth]{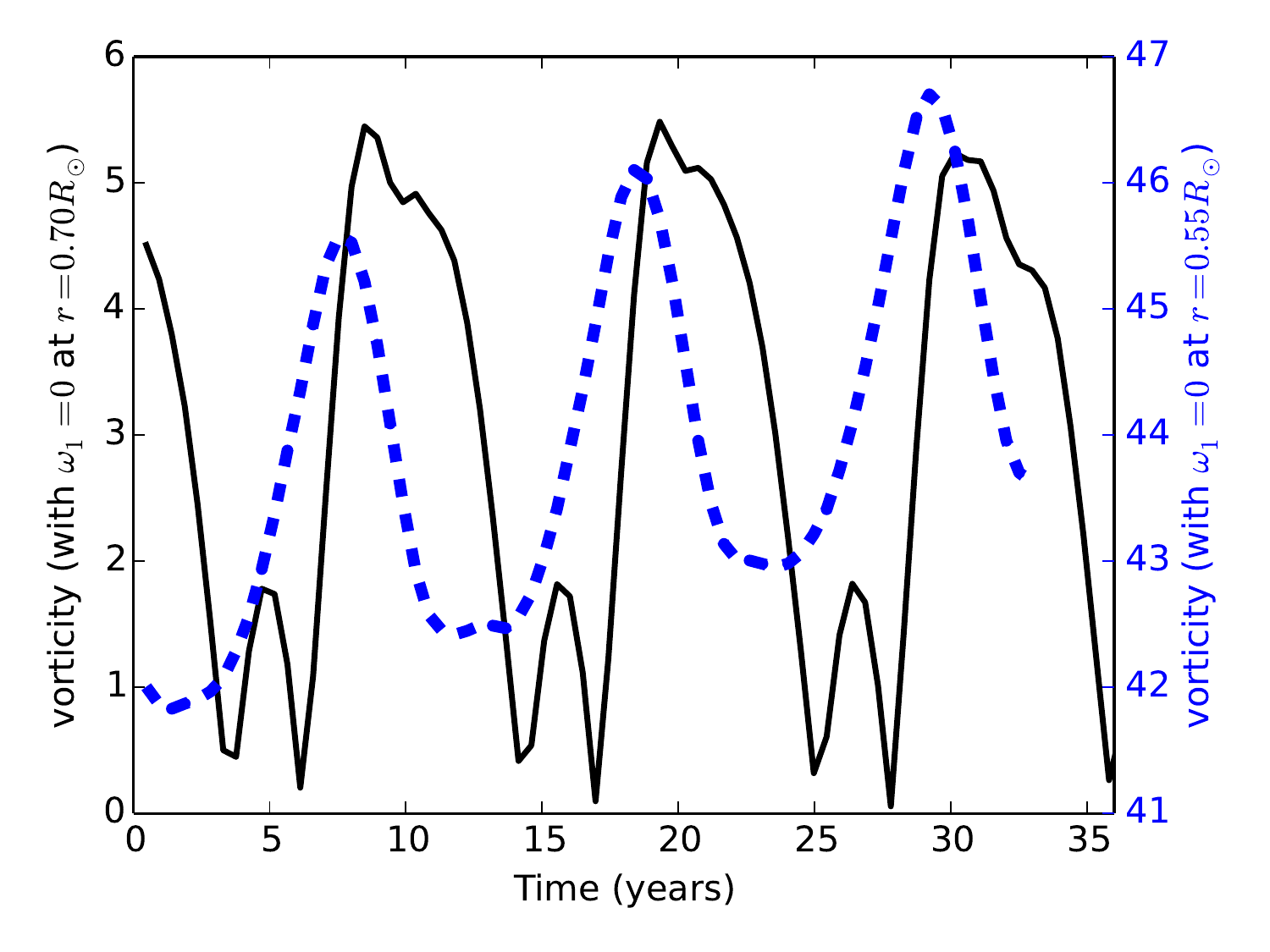}}
\caption[Time evolution of the perturbed vorticity]{The perturbed vorticity at a point within the convection zone is plotted against time for the two cases
with different boundary conditions. The black solid line shows the absolute value of the perturbed vorticity $\omega_1$ at 
15$^{\circ}$ latitude and at depth $0.75R_\odot$, %(where $\omega_1$ has negative values), 
with the  bottom boundary condition $\omega_1 =0$ at $r = 0.7R_\odot$. The blue dashed line shows the absolute value 
of perturbed vorticity $\omega_1$ at $15^{\circ}$ latitude 
and at depth $0.71R_\odot$, with the bottom 
boundary condition $\omega_1 =0$ at $r = 0.55R_\odot$. The values of the vorticity which is shown in blue dashed 
line are given on the right $y$-axis with blue color.}   
\label{fig:w1_time}
\end{figure}
%=====================================================

\section{Results}\label{sec:result}

We now present results obtained by solving the dynamo equations ~\ref{eq:Aeq} and \ref{eq:Beq}
simultaneously with Eq.~\ref{eq:main2} for perturbed vorticity.
The various dynamo parameters---the source function $S(r, \theta, t)$, the differential rotation
$\Omega$, the meridional circulation $\vb_m$ and the turbulent diffusivities $\eta_p$, $\eta_t$---are
specified as in \citet{CNC04}.  As we have pointed out, calculations in this study are based
on using the unperturbed \MC\ $\vb_0$ in the dynamo equations ~\ref{eq:Aeq} and \ref{eq:Beq}.
We also follow \citet{CNC04} in
assuming different diffusivity for the poloidal and the toroidal components of the magnetic field as shown
in Fig.~\ref{figc5:eta}. The justification for the reduced diffusivity of the toroidal component is that
it is much stronger than the poloidal component and the effective turbulent diffusivity experienced by it is expected
to be reduced by the magnetic quenching of turbulence.  The unquenched value of turbulent diffusivity used 
for the poloidal field inside the convection zone is what is expected from simple mixing length arguments
(\citet{Parker79}, p.\ 629). A much smaller value of turbulent diffusivity was assumed in some other flux transport dynamo
models \citep{DC99}. However, a higher value appears much more realistic on the ground that only with
such a higher value it is possible to model many different aspects of observational data: such as the 
dipolar parity \citep{CNC04,Hotta10}, the lack of significant hemispheric asymmetry \citep{CC06,GoelChou09}, the correlation of the cycle strength with the polar field
during the previous minimum \citep{Jiang07} and the Waldmeier effect \citep{KarakChou11}. 
For solving the vorticity Equation~\ref{eq:main2} we need to specify the turbulent viscosity $\nu$.
From simple mixing length arguments, we would expect the turbulent magnetic diffusivity $\eta_p$
and the turbulent viscosity $\nu$ to be comparable, i.e, we would expect the magnetic Prandtl number $P_m=\nu/\eta_p$ to be close
to 1. We first present results obtained with $P_m =1$ in Section~\ref{sec:pm_1}. How our results get modified 
on using a lower value of the magnetic Prandtl number $P_m$ 
will be discussed in the Section~\ref{sec:lowpm}.

We know that the toroidal field $B_{\phi}$ changes its sign from one cycle to the next.  Since
the Lorentz force depends on the square of $B_{\phi}$, as seen in Eq.~\ref{eq:tor_lf}, the Lorentz
force should have the same sign in different cycles.  As a result, one may think that
the perturbed vorticity ${\bf \omega}_1$
driven by the Lorentz force will tend to grow. However, results presented in Sections~\ref{sec:pm_1} and
\ref{sec:lowpm} are obtained with the boundary condition $\omega_1 = 0$ at $r= 0.7 R_{\odot}$, which ensures that
$\omega_1$ cannot grow indefinitely. Asymptotically, we find $\omega_1$ to have oscillations
as the Lorentz force
increases during solar maxima and decreases during solar minima. In Fig.~\ref{fig:w1_time} we have shown
the time variation of perturbed vorticity (at $15^{\circ}$ latitude and $r = 0.75R_\odot$) for this case with the black solid line.
The calculations presented in Sec.~\ref{sec:lowbc} were done by taking the boundary condition $\omega_1 = 0$ at $r= 0.55 R_{\odot}$.
In this situation, we found that $\omega_1$ has oscillations around a mean which keeps growing
with time (see the blue dashed line plot of Fig.~\ref{fig:w1_time} which does not saturate even after
running the code for several cycles). 

\begin{figure*}
\centering{
\includegraphics[width=0.85\textwidth]{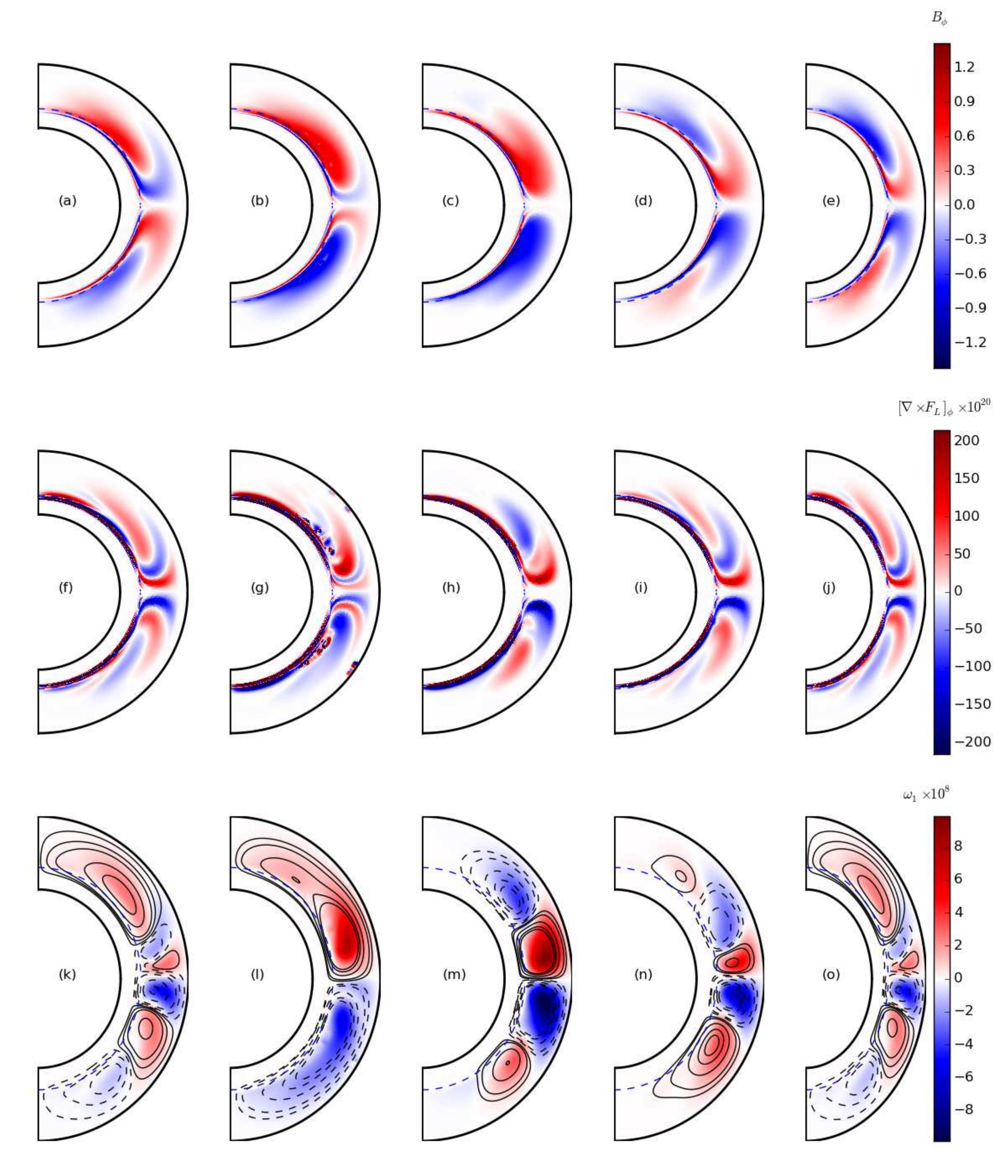}}
\caption[Toroidal field, $\phi$ component of the curl of the Lorentz force, perturbed vorticity and 
perturbed velocity are shown in the meridional plane at different times of a cycle]{Snapshots in the $r-\theta$ plane of different quantities spanning an entire solar cycle: the toroidal field
$B_{\phi}$, the $\phi$ component of the curl of the Lorentz force $[\nabla \times \FL]_{\phi}$
and the perturbed vorticity $\omega_1$ with streamlines are shown in the first row (a)-(e), the second row (f)-(j), and 
the third row (k)-(o) respectively. The five columns represent time instants during
the midst of the rising phase, the solar maximum, the midst of the decay phase and the solar minimum,
followed by the midst of the rising phase again. 
In the third row (k)-(o), the solid black contours represent clockwise flows and the dashed black contours 
represent anti-clockwise flows. The unit of toroidal fields in the first row is given in terms of
$B_c$. The Lorentz forces in the second row also are calculated from toroidal fields in the 
unit of $B_c$. The vorticities are given (third row) in s$^{-1}$. Here $B_c$ is the critical strength 
of magnetic fields above which the fields are magnetically buoyant and create sunspots \citep{CNC04}.
All the plots are for the magnetic Prandtl number $P_m = 1$ with the bottom boundary condition $\omega_1 =0$ at
$r = 0.7 R_{\odot}$.}   
\label{fig:snap_1d12}
\end{figure*}

\begin{figure*}
\includegraphics[width=1.0\textwidth]{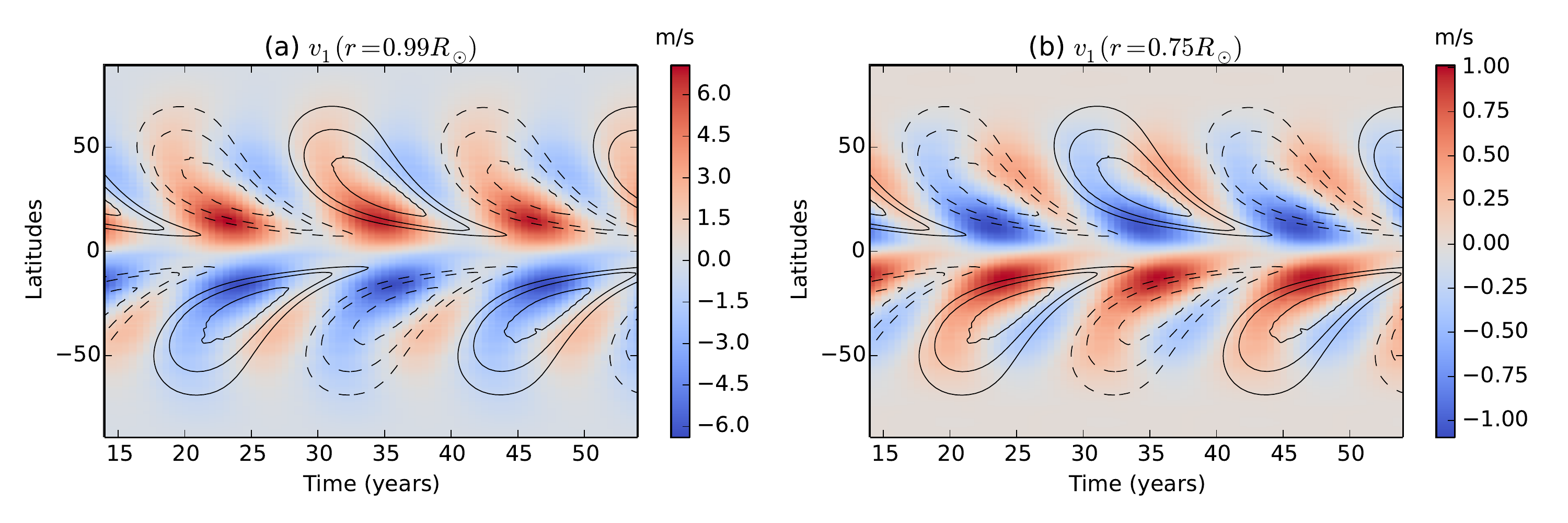}
\caption[Time latitude plot of the perturbed velocities at different radius]{The time--latitude plot of perturbed velocities $v_\theta$ (a) near the surface $(r = 0.99R_\odot)$ and 
(b) at the depth $r = 0.75R_\odot$. 
In the northern hemisphere, positive red color shows perturbed flows towards the equator and the negative blue color 
shows flows towards pole. It is opposite in the
southern hemisphere. Toroidal fields at the bottom of the convection zone are overplotted in both of the plots 
by line contours. Black solid contours indicate positive polarity and black dashed contours are for negative polarity.
All the plots are for the magnetic Prandtl number $P_m = 1$ with the bottom boundary condition $\omega_1 =0$ at
$r = 0.7 R_{\odot}$.}
\label{fig:bfly_1d12}
\end{figure*}

\subsection{Results with $P_m = 1$}\label{sec:pm_1}
The meridional circulation in a poloidal cut of the Sun's northern hemisphere is anti-clockwise, implying
a negative vorticity $\omega_0$. If the meridional circulation is to be weakened by the Lorentz
forces at the time of the sunspot maximum, then we need to generate a positive perturbed vorticity
$\omega_1$ at that time. Fig.~\ref{fig:snap_1d12} explains the basic physics of how this happens.  The three rows
of this figure plot three quantities in the $r-\theta$ plane: the toroidal field $B_\phi$, the 
$\phi$-component of the curl of the Lorentz force $[\nabla \times {\bf F}_L]_{\phi}$ and the perturbed
vorticity $\omega_1$ along with the associated streamlines. The five vertical columns correspond to five
time intervals during a solar cycle.  The second column corresponds to a time close to the sunspot
maximum, whereas the the fourth column corresponds to the sunspot minimum. 

Our discussion of Figure~\ref{fig:snap_1d12} below will refer to the part in the northern hemisphere.
Signs will have to be opposite for some quantities in the southern hemisphere.
We see in the top row that $B_{\phi}$, produced primarily in the tachocline by the strong
differential rotation there, is concentrated in a layer at the bottom of the solar convection
zone. Since $- \partial (B_{\phi}^2)/ \partial r$ will be positive at the top of this layer
and will be negative at the bottom of this layer, we expect from Eq.~\ref{eq:tor_lf} that $[\nabla \times {\bf F}_L]_{\phi}$
will also respectively have positive and negative values in these regions. This expectation in borne out
as seen in the second row of Fig.~\ref{fig:snap_1d12} plotting $[\nabla \times {\bf F}_L]_{\phi}$. Where the layer
of the toroidal field ends at low latitudes, we have negative $\partial (B_{\phi}^2)/ \partial \theta$
and we see from Eq.~\ref{eq:tor_lf} that this will lead to negative $[\nabla \times {\bf F}_L]_{\phi}$.  This is
also seen in the second row of Fig.~\ref{fig:snap_1d12}. We now note from Eq.~\ref{eq:main2} that $[\nabla \times {\bf F}_L]_{\phi}$
is the main source of the perturbed vorticity $\omega_1$. We expect that the positive $[\nabla \times {\bf F}_L]_{\phi}$
above the tachocline (i.e.\
at the top of the layer of concentrated toroidal field) will give rise to positive $\omega_1$,
implying a clockwise flow opposing the regular \MC. We see in the third row (k)-(o) of Fig.~\ref{fig:snap_1d12} that the
distribution of $\omega_1$ inside the convection zone follows the distribution of 
$[\nabla \times {\bf F}_L]_{\phi}$ (as shown in the second row (f)-(j)) fairly closely. We see that
positive $\omega_1$ is particularly dominant within the convection zone near the time of
the sunspot maximum (second column). This corresponds to streamlines with clockwise
flow, which will oppose the regular \MC\ at the time of the 
sunspot maximum and will lead to a decrease in the overall amplitude of the \MC\ at the solar
surface, in accordance with the observational data. It may be interesting to compare Fig.~\ref{fig:snap_1d12} 
with Fig.~3 of \citet{Passos17}, where they present some results on the variation of the meridional circulation
with the solar cycle on the basis of a full numerical simulation.  Our results
obtained from a mean field theory show a spatially smoother perturbed velocity
field (perhaps in better agreement with observational data?) than what is found in the
full simulation.

\begin{figure}
\centering{
\includegraphics[width=0.65\textwidth]{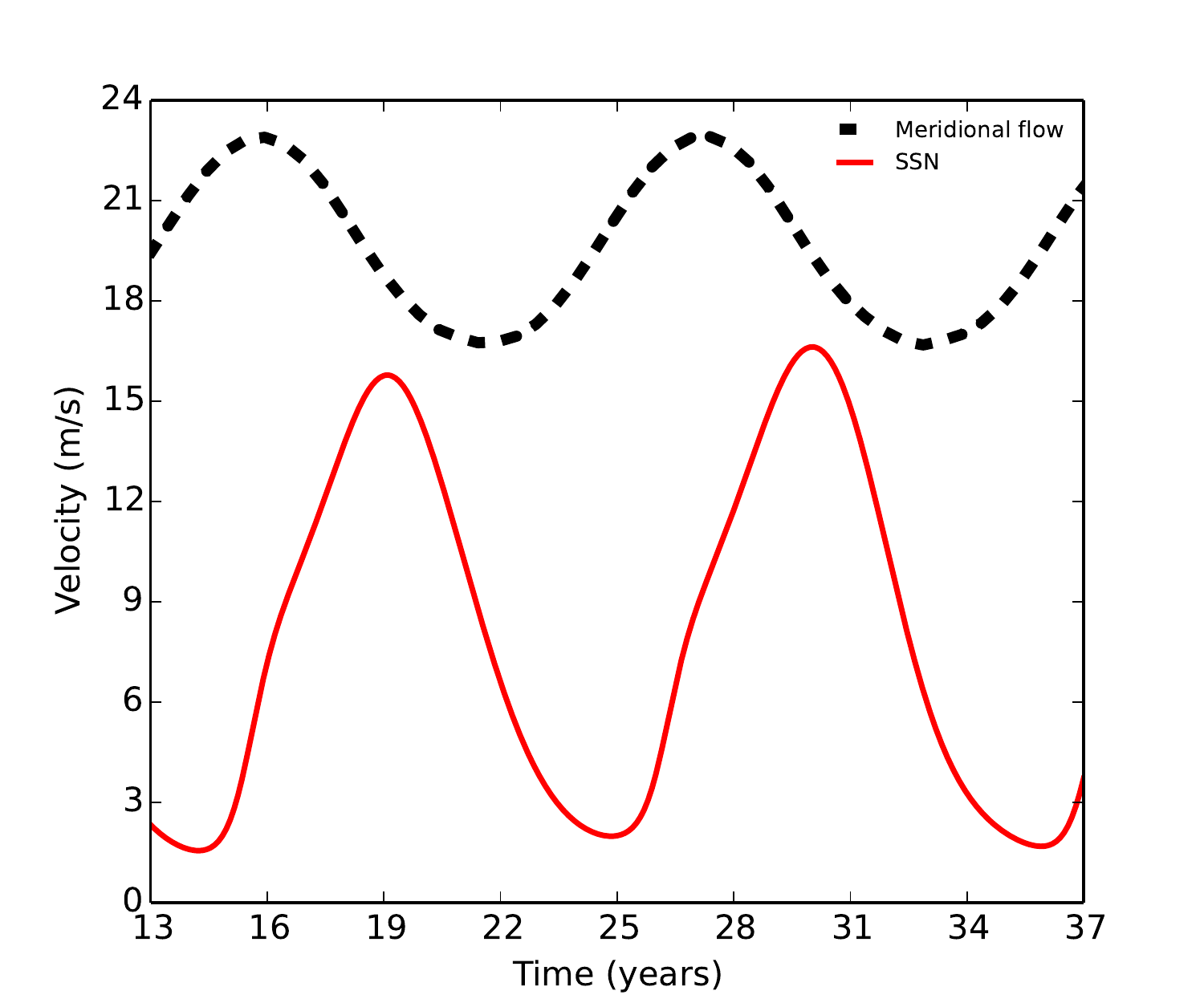}}
\caption[Variation of the meridional circulation with the solar cycle]{The variation of the total meridional circulation with time 
is plotted with solar cycle. The yearly averaged solar cycle is plotted in the red solid line, whereas the black dashed line shows
the total meridional circulation just below the surface at $25^{\circ}$ latitude.}   
\label{fig:tot_mc}
\end{figure}

We wish to make one important point here. A careful look at the second row of Fig.~\ref{fig:snap_1d12} shows
that there is a thin layer of negative $[\nabla \times {\bf F}_L]_{\phi}$ at the bottom of the
tachocline where $- \partial (B_{\phi}^2)/ \partial r$ is negative.  This
layer of negative $[\nabla \times {\bf F}_L]_{\phi}$ will try to create a negative
vorticity driving a counter-clockwise flow below the tachocline.  However, we know that
it is not easy to drive a flow below the tachocline where the temperature gradient is
stable \citep{GM04}. The physics of this is taken into account by using
the bottom boundary condition that $\omega_1 = 0$ at $r = 0.7 R_{\odot}$.  As we shall see
in Section~\ref{sec:lowbc}, results can change dramatically on changing this boundary condition. Our
boundary condition ensures that the negative $[\nabla \times {\bf F}_L]_{\phi}$ at
the bottom of the tachocline cannot produce negative vorticity there (as seen in the 
third row of Fig.~\ref{fig:snap_1d12}) and cannot drive a flow below the tachocline. We thus see that,
while the distribution of $\omega_1$ follows the distribution
of $[\nabla \times {\bf F}_L]_{\phi}$ fairly closely within the convection zone, this
is not the case at the bottom of the tachocline.

The strength of the perturbed \MC\ certainly depends on the strength of $[\nabla \times {\bf F}_L]_{\phi}$,
which is controlled by the filling factor $f$ as seen in Eq.~\ref{eq:tor_lf}. 
We have followed the dynamo model of \citet{CNC04} in which the scale of the magnetic
field is set by the critical value $B_c$ above which the toroidal field becomes buoyant
in the convection zone.  As pointed out by \citet{Jiang07} and followed by \citet{CCC09}, we take
$B_c = 108$ G to ensure that the mean polar field has the right amplitude. With the unit of
the magnetic field fixed in this way, we find that $f = 53$ gives
surface values of the perturbed \MC\ of the order of 5 m s$^{-1}$ comparable to what is seen in the observational
data. Such a value of the filling factor $f$, which should be less than 1, is completely
unphysical.  We anticipated this problem on the basis of the order-of-magnitude estimate
presented in Section~\ref{sec:model}. 
\citet{CCC09} needed $f = 0.067$ to model torsional
oscillations, whereas \citet{Chou03} estimated that it should be about $f\approx 0.02$.
Since $f$ appears in the denominator of the Lorentz force term in Eq.~\ref{eq:tor_lf},
a large value of $f$ means that we have to reduce the Lorentz force in an artificial manner
to match theory with observations. We are not sure about the significance of this. All we
can say now is that, if we reduce the Lorentz force in this manner, then we find theoretical
results to be in good agreement with observational data.

Fig.~\ref{fig:bfly_1d12} shows time-latitude plots of the perturbed $v_\theta$ just below the surface and in
the lower part of the convection zone. Fig.~\ref{fig:bfly_1d12}(a) can be directly compared with Fig.~11 of
\citet{Komm15}. When the perturbed \MC\ near the surface has an amplitude of about 
5 m s$^{-1}$ near the surface (comparable to what we find in observational data),
its amplitude at the bottom of the convection zone turns out to be about 0.75 m s$^{-1}$.
This is a significant fraction of the amplitude of the \MC\ at the bottom of the convection
zone, which is of order 2 m s$^{-1}$. In other words, the ratio of the amplitude at the 
bottom of the convection zone to the amplitude at the top of the convection zone for the 
perturbed \MC\ ($\approx 0.15$) is somewhat larger than this ratio for the regular \MC
($\approx 0.08$). The reason behind this
becomes clear on comparing the distribution of $\omega_0$, as seen in Fig.~\ref{fig:stream_vort}(b), with the 
distribution of $\omega_1$, as seen in Fig.~\ref{fig:snap_1d12}(l), i.e.\
the second figure in the third row of Fig.~\ref{fig:snap_1d12}. 
This ratio is small when the vorticity is concentrated near the upper part of the 
convection zone and streamlines in the lower part of the convection zone are
fairly spread out.  On the other hand, with $\omega_1$ distributed throughout the body
of the convection zone, this ratio for the perturbed \MC\ is not so small.  
The perturbed \MC\ at the bottom of the convection zone (which is in the poleward
direction) will reduce the value of the equatorward \MC\ there considerably at the
time of the sunspot maximum.  This will presumably affect the behavior of the dynamo, in
such ways as lengthening its period (since a slower \MC\ at the bottom of the convection
zone during certain phases of the cycle is expected to lengthen the cycle). We are certainly
not fully justified in solving the dynamo equations~\ref{eq:Aeq}--\ref{eq:Beq} by using the regular
\MC\ $\vb_0$.  However, as we have pointed out in Section~\ref{sec:model}, in order to include the perturbed \MC\
in these equations, we need to evaluate the perturbed velocity field at every time step (i.e.\
not only the perturbed vorticity $\omega_1$), which will require considerably more
computational efforts.  We are currently doing these calculations and will present the results
in a future work.  The aim of the present exploratory study is to elucidate the 
basic physics of the problem and not to construct a very complete model. Figure~\ref{fig:tot_mc} shows
the time evolution of the total $v_{\theta}$ at $25^{\circ}$ degree latitude just below the surface
along with the sunspot number.  We see that the \MC\ reaches its minimum a little after the
sunspot maximum. This figure can be compared with Fig.~4 of \citet{Hathaway10b}.

One important aspect of observational data is an inward flow towards the belt of active
regions at the time of the sunspot maximum \citep{CS10,Komm15}.
We now come to the question whether our model can provide any explanation for this. This
inward flow means that the \MC\ is enhanced in the low latitudes below the sunspot belt
and reduced in the higher latitudes. It is thought that the cooling effect of sunspots
may drive this inward flow \citep{Spruit03,Gizon08}.  Since this idea is not
yet fully established through detailed and rigorous calculations, it is worthwhile to look for 
alternative explanations. We would like to draw the reader's attention to the first figure
in the bottom row of Fig.~\ref{fig:snap_1d12}. In the northern hemisphere, we see positive $\omega_1$ (implying
clockwise flow) extending over latitudes higher than where sunspots are usually seen.
However, we see a region of negative $\omega_1$ (implying counter-clockwise flow) at lower
latitudes.  It is clear that the perturbed \MC\ is of the nature of an inward flow at the 
latitudes between the region of positive $\omega_1$ on the high-latitude side and the region
of negative $\omega_1$ on the low-latitude side.  We are thus able to obtain an inward flow at
the latitudes where sunspots are typically seen, but unfortunately we are getting this at
the wrong time---shortly after the sunspot minimum (when there would be no sunspots in that region)
rather than at the sunspot maximum.  A look at Fig.~\ref{fig:bfly_1d12}(a) also makes it clear that this
inward flow occurs slightly after the sunspot minimum. We are now exploring the question
whether, by changing some parameters of the model, it is possible to get this inward flow
at the right latitudes at the right phase (around the sunspot maximum) of the solar cycle.
If our interpretation is correct, then the inward flow has nothing to do with the actual
physical presence of sunspots.  While the Lorentz force above the tachocline tends to
produce positive vorticity, the low-latitude edge of the toroidal magnetic field belt can
be a source of negative vorticity.  If the sunspot belt merely happens to be a region having
positive $\omega_1$ on the high-latitude side and negative $\omega_1$ on the low-latitude
side, then it will be a region of apparent inward flow.  We suggest this as a tentative
hypothesis which requires further study. This similar pattern for the meridional circulation is also found 
in the MHD simulations of \citet{Passos17} (see Fig~1(F) of their paper) supporting the fact that sunspot may appear in a region having
a positive $\omega_1$ on the high-latitude side and negative $\omega_1$ on the low-latitude
side. We shall argue in Appendix~B that the advection
term $s \nabla. ( \vb_1 \omega_0/s)$ not incorporated in the present study (the third
term on the LHS of Eq.~\ref{eq:main2}) may be quite important for modeling the inward
flow in active regions.

\subsection{Results with low Prandtl number $(P_m < 1)$}\label{sec:lowpm}

In order to study how the nature of the perturbations in the \MC\ will change on
changing the turbulent viscosity $\nu$, we now present results obtained by taking
$\nu = \eta_t$ (shown by the blue solid line in Fig.~\ref{figc5:eta}) rather than
$\nu = \eta_p$ which was the case for the results presented in Section~\ref{sec:pm_1}. This makes
the magnetic Prandtl number defined as $\nu/\eta_p$ (note that $\eta_p$ rather than
$\eta_t$ represents the unquenched turbulent diffusivity) much smaller than
unity within the body of the convection zone.  

\begin{figure*}
\centering{
\includegraphics[width=0.84\textwidth]{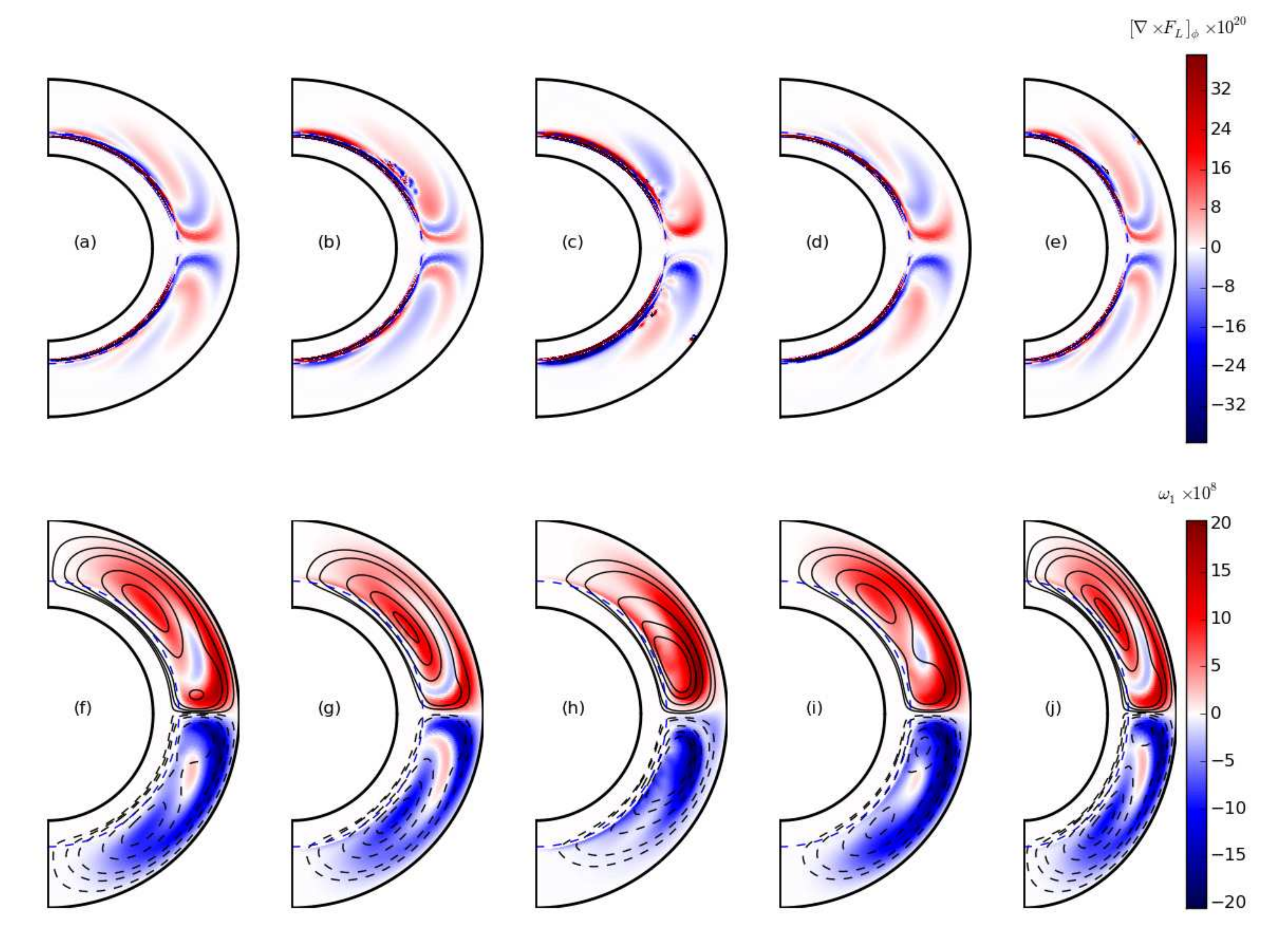}}
\caption[Same as Fig.~\ref{fig:snap_1d12} but with low Prandtl number $P_m$]{Results obtained with low Prandtl number plotted in the same way as Fig.~\ref{fig:snap_1d12}.
The top row plotting $[\nabla \times \FL]_{\phi}$ is the same as the second row of Fig.~\ref{fig:snap_1d12}.
The bottom row shows the vorticity $\omega_1$ with streamlines.}   
\label{fig:snap_5d10}
\end{figure*}

We now want to present our results in the meridional 
plane of the Sun. Since there have been no changes in the dynamo equations ~\ref{eq:Aeq} and \ref{eq:Beq},
the toroidal field $B_\phi$ and the 
$\phi$-component of the curl of the Lorentz force $[\nabla \times {\bf F}_L]_{\phi}$ will
be the same as in the first two rows of Fig.~\ref{fig:snap_1d12} if we take the snapshots
at the same instants of time. To facilitate easy comparison with the perturbed vorticity
$\omega_1$ generated, we again plot $[\nabla \times {\bf F}_L]_{\phi}$ in the top row of
Figure~\ref{fig:snap_5d10}. This is the same as the middle row of Fig.~\ref{fig:snap_1d12}.
Then the bottom row of Figure~\ref{fig:snap_5d10} shows the perturbed vorticity $\omega_1$
along with the streamlines of the perturbed velocity.
This bottom row should be compared to the bottom row of Fig.~\ref{fig:snap_1d12}.
When $\nu$ is higher (the case of Fig.~\ref{fig:snap_1d12}), the perturbed vorticity
$\omega_1$ spreads more evenly within the convection zone.  On the other hand, when
$\nu$ is lower (the case of Figure~\ref{fig:snap_5d10}), the perturbed vorticity $\omega_1$
tends to be frozen where it is created by $[\nabla \times {\bf F}_L]_{\phi}$.  As a result,
$\omega_1$ follows $[\nabla \times {\bf F}_L]_{\phi}$ more closely in Figure~\ref{fig:snap_5d10}
rather than in Fig.~\ref{fig:snap_1d12}. We thus see a more complicated distribution
of $\omega_1$ in Figure~\ref{fig:snap_5d10} implying a more complicated
velocity field.

\begin{figure*}
\centering{
\includegraphics[width=0.9\textwidth]{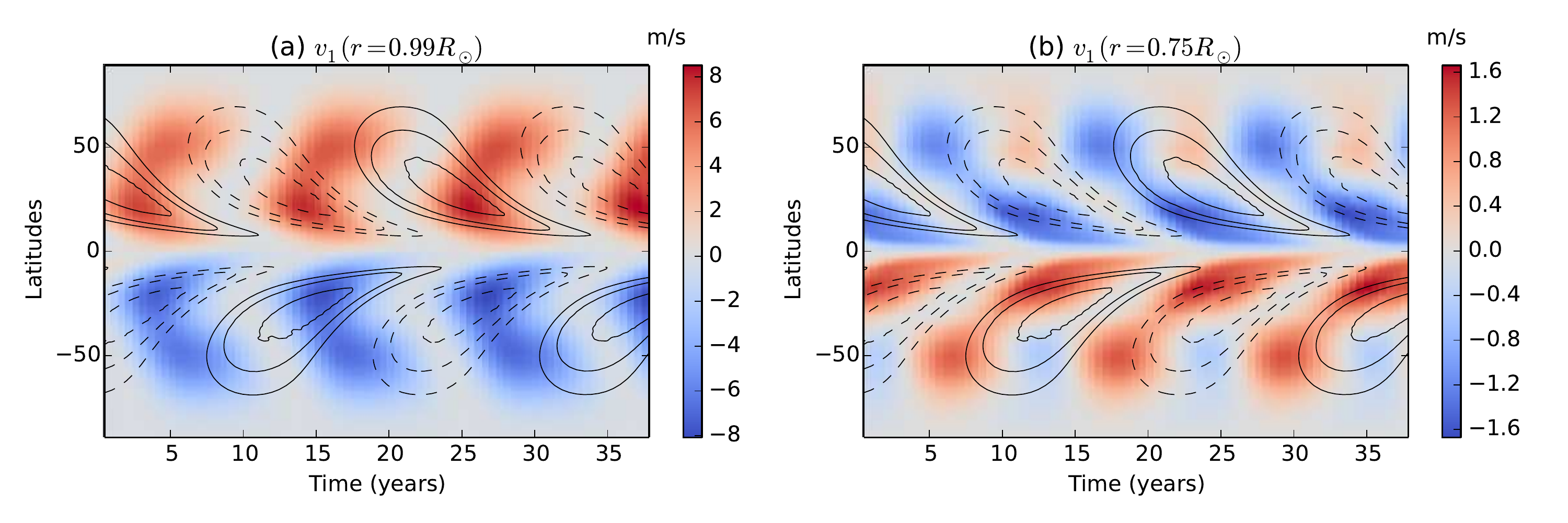}}
\caption{Same as Fig.~\ref{fig:bfly_1d12} but with low Prandtl number $P_m$.}   
\label{fig:bfly_5d10}
\end{figure*}

To make the perturbed meridional circulation near the surface 
comparable to what we find in the observational data, we need to choose $f = 5.3 \times 10^2$
in this case of low $P_m$. This is even more unphysical than what we needed for the case 
$P_m =1$ ($f = 53$) with all the other dynamo parameters same. Basically, on reducing
viscosity, vorticity tends to get piled up and we need to reduce the strength of the
Lorentz force further to match observations.
The perturbed velocities near the surface and 
near the bottom of the convection zone are shown in Fig.~\ref{fig:bfly_5d10}(a) and \ref{fig:bfly_5d10}(b) respectively. 
The requirement that the perturbed velocity at the surface compares with observational
data forces the perturbed velocity at the bottom of the convection zone 
as seen in Fig.~\ref{fig:bfly_5d10}(b) of order 1.5 m s$^{-1}$, which makes it almost comparable
to the unperturbed velocity (though they are in opposite directions). This shows that solving
the dynamo equations ~\ref{eq:Aeq} and \ref{eq:Beq} with the unperturbed meridional circulation
$\vb_0$ is more unjustified in this case than in the case of $P_m = 1$.

We do not know how large the perturbations to the \MC\ at the bottom of the convection
zone actually are. If they were comparable to the unperturbed \MC, then probably we would see some effects
on the dynamo.  Since this problem has not been studied before, we do not know what those
effects may be like.  As, on theoretical grounds also we expect $P_m$ to be of order unity,
we believe that the model with $P_m=1$ presented in Section~\ref{sec:pm_1} is probably a more realistic
depiction of what is happening inside the Sun rather than the model with low $P_m$.

\subsection{Dependence of the results on lower radial boundary condition}\label{sec:lowbc}

While discussing Fig.~\ref{fig:snap_1d12}, we have pointed out that the Lorentz force at the bottom of
the tachocline would have a tendency of producing a perturbed vorticity with the same
sign as the unperturbed vorticity. This will tend to drive a counter-clockwise flow (in the
northern hemisphere) below the tachocline.  Since on physical grounds we do not expect such a
flow to be driven in a region of stable temperature gradient, we have used the boundary
condition $\omega_1 = 0$ at $r = 0.70R_\odot$ in Sections~\ref{sec:pm_1} and \ref{sec:lowpm} to suppress this flow.
We found that our results are quite sensitive to this bottom boundary condition.  We now
discuss what we get on changing the boundary
condition to $\omega_1 = 0$ at $r = 0.55R_\odot$.

\begin{figure*}
\centering{
\includegraphics[width=0.84\textwidth]{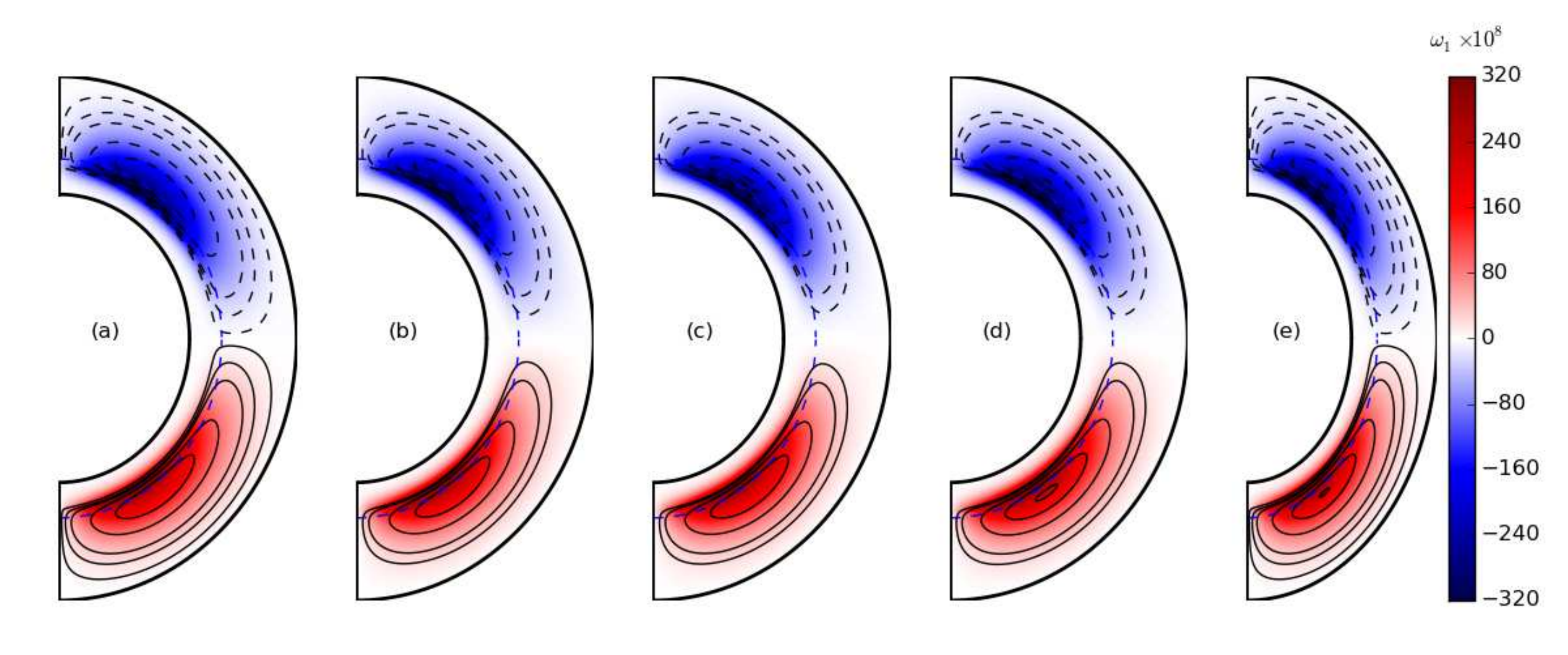}}
\caption{On changing the bottom boundary condition to $\omega_1 =0$ at $r = 0.55R_\odot$, 
the third row of Fig.~\ref{fig:snap_1d12} would get changed to this.}
\label{fig:snap_55}
\end{figure*}

Again the solutions of the dynamo equations ~\ref{eq:Aeq} and \ref{eq:Beq} remain the same
as in Section~\ref{sec:pm_1} and \ref{sec:lowpm}. We present the solutions of the perturbed vorticity 
at the same time instants as the time instants
in Figure~\ref{fig:snap_1d12}. Only the third row of this figure will be 
replaced by what is shown in Figure~\ref{fig:snap_55}. When we had taken the boundary
condition $\omega_1 = 0$ at $r = 0.70R_\odot$ earlier, $\omega_1$ was restrained from
growing indefinitely. Now, on pushing the boundary well below the bottom of the convection
zone, we found that $\omega_1$ kept growing for many cycles for which we ran the code (see Fig.~\ref{fig:w1_time}).  This
is the reason why the values of $\omega_1$ given in the color scale in Figure~\ref{fig:snap_55}
are rather large.  We would expect the growing perturbed vorticity to have the opposite
sign of the original unperturbed vorticity.  But this is not the case now. The growing perturbed vorticity
has the same sign as the original unperturbed vorticity. This means that the Lorentz force
strengthens the original \MC\ (This is also seen in the MHD simulation of \citet{PCB12}) rather than opposing it! Presumably this is because the effect of the Lorentz
force at the bottom of the tachocline (which tries to produce a perturbed vorticity of
the same sign as the original vorticity) overwhelms the effect of the Lorentz force at the top of the
tachocline which opposes the unperturbed \MC.  These results show that the bottom boundary
condition is quite crucial. If we allow the Lorentz force to create a perturbed vorticity
below the bottom of the tachocline, we may get totally unphysical results.

\section{Conclusion}\label{sec:conclusion}     

The \MC\ of the Sun plays a crucial role in the flux transport dynamo
model, the currently favored theoretical model for the solar cycle.
Presumably an equatorward flow at the bottom of the convection zone is
responsible for the equatorward migration of the sunspot belts \citep{CSD95,HKC14}. 
The poleward flow at the surface builds up the polar field of the Sun by bringing
together the poloidal field generated by the Babcock--Leighton mechanism
from the decay of active regions \citep{HCM17}. 
We now have growing evidence that similar flux transport dynamos operate
in other solar-like stars as well \citep{Chou17}. Our lack of understanding
of the \MC\ in such stars is the major stumbling block for building theoretical
models of stellar dynamos \citep{KKC14, Chou17}.
The mathematical theory of the \MC\ involves a centrifugal term and a thermal wind
term.  The centrifugal term couples the theory of the \MC\ to the theory of
differential rotation and the thermal wind term couples the theory to the 
thermodynamics of the star---making it a very complex subject.

We now have evidence that the \MC\ of the Sun varies with the solar cycle, becoming
weaker around the sunspot maximum \citep{CD01,Beck02,Hathaway10b,Komm15}. 
We believe that this is caused by the back-reaction of the Lorentz force of the dynamo-generated
magnetic field. We show that the theory of the perturbations to the \MC\
caused by the Lorentz force can be decoupled from the theory of the 
unperturbed \MC\ itself.  Especially, this theory becomes decoupled from
any thermodynamic considerations if we assume that the magnetic fields do
not affect the thermodynamics of the Sun. This makes the theory of the
perturbations to the \MC\ much simpler than the theory of the \MC\ itself.
Even without having a theory of the unperturbed \MC, we are able to develop
a theory of how the \MC\ gets affected by the Lorentz force of the dynamo-generated
magnetic fields. 

The most convenient way of developing a theory of the \MC\ is by considering
the equation for the $\phi$-component of vorticity $\omega_{\phi}$. We solve
the equation for the perturbed vorticity along with the standard axisymmetric mean field equations of the
flux transport dynamo.  For the unperturbed \MC, $\omega_{\phi}$ is negative
in the northern hemisphere of the Sun.  If the Lorentz force can produce some
positive vorticity there, then we expect the \MC\ to become weaker during the
sunspot maxima. Our calculations show that positive vorticity is indeed produced
near the top of the tachocline.  Our theoretical results are able to explain
many aspects of observational data presented by \citet{Hathaway10b} and \citet{Komm15}
We get best results when the turbulent viscosity $\nu$
is taken to be comparable to the turbulent diffusivity $\eta_p$ of the magnetic
field, i.e.\ when the magnetic Prandtl number $P_m$ is close to unity.  Our
results are sensitive to the boundary condition at the bottom of the convection
zone. At the bottom of the tachocline, the Lorentz force tends to produce vorticity with
the same sign as the unperturbed vorticity and this has to be suppressed with a 
suitable boundary condition to incorporate the important physics that it is
very difficult to drive flows in the stable subadiabatic layers below the
tachocline. 

One intriguing observation is that of an inward flow towards active regions
during the sunspot maxima \citep{CS10, Komm15}. 
While we have not been able to construct a satisfactory model of this inward
flow, we suggest a possibility how this flow may be produced.  We would have
such a flow if the northern hemisphere has a positive perturbed vorticity
at latitudes above the mid-latitudes and a negative perturbed vorticity at lower
latitudes.  We propose a scenario how this can happen.  The Lorentz force at
the top of the tachocline tends to produce positive vorticity, whereas 
the Lorentz force at the edge
of the toroidal field belt at low latitudes tends to produce negative vorticity.
In our model, we found an inward flow towards typical sunspot latitudes
at a wrong phase of the solar cycle. It remains to be seen whether the phase
can be corrected by changing the parameters of the problem.

One puzzle we encounter is that we have to reduce the strength of the Lorentz
force artificially by about two orders of magnitude to make theory fit with
observations. While we do not have a proper explanation for this, we make
one comment.  While the overall strength of the Lorentz force varies with
the solar cycle, it does not change sign between cycles.  As a result, the 
Lorentz force would always tend to drive a very strong clockwise flow in the northern
hemisphere (anti-clockwise in the south), if it is not reduced sufficiently. 
Now, in the theory of the unperturbed \MC, we know that we need some force
trying to drive such a flow---to make the surfaces of constant angular velocity
depart from cylinders and to balance the centrifugal term.  Normally, the thermal
wind caused by anisotropic heat transport in the Sun is believed 
to do this \citep{Kitchatinov95,Kitchatinov11b}. Is it possible that the expected large Lorentz force
also plays an important in the force balance?  To address this question, one
has to carry on a much more complicated analysis than what we have attempted
in this chapter, without separating the equations of unperturbed and perturbed \MC.
While we do not attempt such an analysis, our analysis brings out many basic
physics issues rather clearly and raises intriguing questions like this.

The equation of the perturbed \MC\ gives us the perturbed vorticity at different
times.  In order to get the perturbed velocity from the perturbed vorticity,
we need to solve a Poisson-type equation. If we want to do this at every time
step, then running the code becomes computationally very expensive.  On the hand,
if we are satisfied only with the perturbed vorticity at all steps and do not
require the perturbed velocity, then we need much less computer time.  This is what
we have done in this chapter in which our aim was to carry out a parameter space
study to understand the basic physics.  However, we have to pay a price for not
calculating the perturbed velocity at all time steps.  We keep solving the
dynamo equations by using the unperturbed velocity of the \MC. This is certainly a
questionable approximation. If we want to make the perturbed velocity at the
surface comparable to what is found in observational data ($\approx$ 5 m s$^{-1}$),
then the perturbed velocity at the bottom of the convection zone turns out to
be $\approx$ 0.75 m s$^{-1}$. This is certainly smaller than the unperturbed
velocity there ($\approx$ 2 m s$^{-1}$), but is not completely negligible.
While we feel reassured that the perturbed velocity at the bottom of the
convection zone is smaller than the unperturbed velocity and probably will
not necessitate a major revision of the flux transport dynamo, the effect of this
perturbed velocity on the dynamo needs to be investigated.
We are now in the process of including this perturbed velocity in the
dynamo equations.  This will need finding the perturbed velocity from the
perturbed vorticity at different time intervals---pushing up the computational
requirements significantly. Once we do that, we shall be able to include another
advection term involving ${\bf v}_1$ in the computations which had been left
out in this study because of computational difficulties.  This term may be
important for modeling the inward flow towards active regions.

Not including the back-reactions on the \MC\ in the dynamo equations is not
a limitation of this study alone, but a limitation of nearly all hitherto
published studies on the flux transport dynamo based on 2D kinematic mean
field formulation. To the best of our knowledge, only \citet{KarakChou12}
studied this problem by modeling the back-reaction on the \MC\
due to the Lorentz force through a simple parameterization (though see
\citet{PCB12} who have included back-reaction in a simple low-order dynamo model).  
Although such a back-reaction on the \MC\ can have dramatic effects on the dynamo if the
magnetic diffusivity is low, \citet{KarakChou12} found that the effect
is not much on a high-diffusivity dynamo (like what we present in this study).
However, it is now important to go beyond such simple parameterization and
include the properly computed perturbed velocity in the dynamo equations and
study its effects.  We hope that we shall be able to present our results in the near future.

\begin{figure}[!b]
\centering{
\includegraphics[width = 0.85\textwidth,trim={0 0.75cm 0 0}]{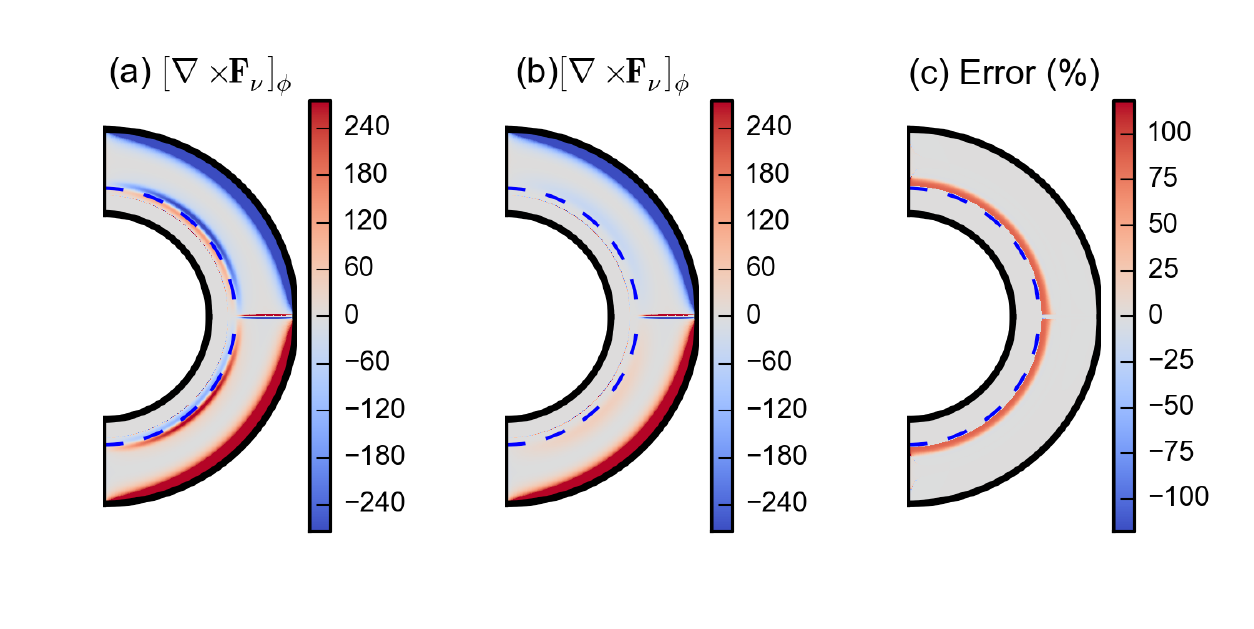}}
\caption[Behavior of various viscous terms]{$[\nabla \times {\bf F}_{\nu}]_{\phi}$ calculated using full expression (Eq.~\ref{eq:A6}) and using approximated expression (Eq.~\ref{eq:viscous1}) are shown
in (a) and (b) respectively. The difference between the values of case (a) and (b) are shown in percentage in (c).}   
\label{fig:v_force}
\end{figure}

\section{Appendix A: Calculation of the Viscosity Term}\label{sec:appendix}
The calculation of the viscous term becomes immensely complicated if
$\nabla. {\vb}$ is not set equal to zero.  Although we are dealing with
a situation in which we rather have $\nabla. (\rho {\vb}) = 0$, the
meridional circulation velocities are everywhere highly subsonic and
the role of compressibility in viscous dissipation is expected to be
negligible.  So we shall assume $\nabla. {\vb} = 0$ to make the calculations
manageable.  For a velocity field 
$$\vb = v_r (r, \theta,t) {\bf e}_r + v_{\theta} (r, \theta,t) {\bf e}_{\theta}$$
the non-zero components of the viscous stress tensor are
\begin{eqnarray}
\sigma_{rr} = 2 \nu \frac{\pa v_r}{\pa r}, \sigma_{\theta \theta}
= 2 \nu \left( \frac{1}{r} \frac{\pa v_{\theta}}{\pa \theta} 
+ \frac{v_r}{r} \right), ~~~~~~~~~~~~~~~~~~~~~~\nonumber\\
\sigma_{\phi \phi}= 2 \nu \left( \frac{v_r}{r}
+\frac{v_{\theta} \cot \theta}{r} \right), 
\sigma_{r \theta} = \nu \left(\frac{1}{r} \frac{\pa v_r}{\pa \theta}
+ \frac{\pa v_{\theta}}{\pa r} - \frac{v_{\theta}}{r} \right)
\label{eq:A1}
\end{eqnarray}
%\eqno(A1)$$
The $r$ and $\theta$ components of the viscosity term in Navier--Stokes
equation are given by
\begin{equation}
F_{\nu,r} = \frac{1}{r^2} \frac{\pa}{\pa r} (r^2 \sigma_{rr}) +
\frac{1}{r \sin \theta} \frac{\pa}{\pa \theta}(\sin \theta \sigma_{r \theta})
-\frac{\sigma_{\theta \theta} + \sigma_{\phi \phi}}{r}, %\eqno(A2)$$
\label{eq:A2}
\end{equation}
\begin{equation}
F_{\nu,\theta} = \frac{1}{r^2} \frac{\pa}{\pa r} (r^2 \sigma_{r \theta}) +
\frac{1}{r \sin \theta} \frac{\pa}{\pa \theta}(\sin \theta \sigma_{\theta \theta})
+ \frac{\sigma_{r \theta}}{r} - \frac{\sigma_{\phi \phi} \cot \theta}{r}, 
\label{eq:A3}
\end{equation}
%\eqno(A3)$$
We now substitute from Eq.~\ref{eq:A1} in Equations~\ref{eq:A2} and \ref{eq:A3}. On making use of 
$\nabla. {\vb} = 0$, we get after a few steps of algebra
\begin{equation}
F_{\nu,r} = \nu \left[ \nabla^2 v_r - \frac{2}{r^2} 
\frac{\pa v_{\theta}}{\pa \theta}
- \frac{2 v_r}{r^2} - \frac {2 \cot \theta v_{\theta}}{r^2} \right]
+ 2 \frac{d \nu}{d r} \frac{\pa v_r}{\pa r},
\label{eq:A4}
\end{equation}
\begin{equation}
F_{\nu,\theta} = \nu \left[ \nabla^2 v_{\theta} + \frac{2}{r^2} 
\frac{\pa v_r}{\pa \theta}
- \frac { v_{\theta}}{r^2 \sin^2 \theta} \right]
+ \frac{d \nu}{d r} \left[\frac{1}{r} \frac{\pa v_r}{\pa \theta}
+ \frac{\pa v_{\theta}}{\pa r} - \frac{v_{\theta}}{r} \right].
\label{eq:A5}
\end{equation}
%\eqno(A5)$$

To solve Equation~\ref{eq:main2}, we need $[\nabla \times {\bf F}_{\nu}]_{\phi}$, which
is given by
\begin{equation}
[\nabla \times {\bf F}_{\nu}]_{\phi} = \frac{1}{r}
\left[ \frac{\pa}{\pa r}(r F_{\nu, \theta}) - \frac{\pa F_{\nu, r}}
{\pa \theta} \right]. 
\label{eq:A6}
\end{equation}
%\eqno(A6)$$
On substituting Equations~\ref{eq:A4} and \ref{eq:A5} into Eq.~\ref{eq:A6}, we get a rather complicated
analytical expression for the viscosity term.  It is extremely challenging
to incorporate this complicated expression in the time evolution code to
solve Eq.~\ref{eq:main2}---especially if we want to handle the diffusion terms through
Crank-Nicholson scheme which is partly implicit, to avoid very small time steps. 
Since there are many uncertainties in this problem---including
the uncertainty in the value of turbulent viscosity $\nu$---we felt that
it is not worthwhile to make an enormous effort to incorporate a completely
accurate expression of the viscous term in our code to solve Eq.~\ref{eq:main2}. In order
to capture the effect of viscosity in the problem, we have used the
simpler approximate expression given in Eq.~\ref{eq:viscous1}. The question is whether
we make a very large error in doing this simplification.  To address this
question, we take the unperturbed meridional circulation shown in Fig.~\ref{fig:stream_vort}(a)
and calculate $[\nabla \times {\bf F}_{\nu}]_{\phi}$ for this velocity
field both by the approximate expression in Eq.~\ref{eq:viscous1} and the full expression
in Eq.~\ref{eq:A6}.  To calculate the value of $[\nabla \times {\bf F}_{\nu}]_{\phi}$ at
different grid points by the full expression in Eq.~\ref{eq:A6}, we first calculate
$F_{\nu,r}$ and $F_{\nu, \theta}$ at different grid points by using
Equations~\ref{eq:A4} and \ref{eq:A5}. Then we use these numerical values of 
$F_{\nu,r}$ and $F_{\nu, \theta}$ to calculate 
$[\nabla \times {\bf F}_{\nu}]_{\phi}$ by using a simple difference
scheme to handle Eq.~\ref{eq:A6}. The values obtained for
$[\nabla \times {\bf F}_{\nu}]_{\phi}$ by using the full expression (Eq.~\ref{eq:A6}) and 
the approximate expression (Eq.~\ref{eq:viscous1}) are shown in 
Figure~\ref{fig:v_force}(a) and Figure~\ref{fig:v_force}(b) respectively. We
see that both the expressions give fairly close values within
the main body of the convection zone.  Only at the bottom of the convection
zone where radial derivatives of $\nu$ included in the full expression
are important, the two expressions give somewhat different values---though
the values are of the same order of magnitude (see Fig~\ref{fig:v_force}(c)). In Figure~\ref{fig:v_force}(c), we have 
shown percentage differences of $[\nabla \times {\bf F}_{\nu}]_{\phi}$ values calculated in 
two cases (Figures~\ref{fig:v_force}(a) and (b)) in our integration region 
above $r > 0.7R_\odot$. Note that things change
much more at the bottom of the convection zone on changing the bottom
boundary condition slightly. 
We thus conclude that,
by using the approximate expression in Eq.~\ref{eq:viscous1} for
$[\nabla \times {\bf F}_{\nu}]_{\phi}$, we do not make a large
error and get everything within the correct order of magnitude.

\begin{figure}[!t]
\centering{
\includegraphics[width = 0.8\textwidth,trim={0 0.75cm 0 0}]{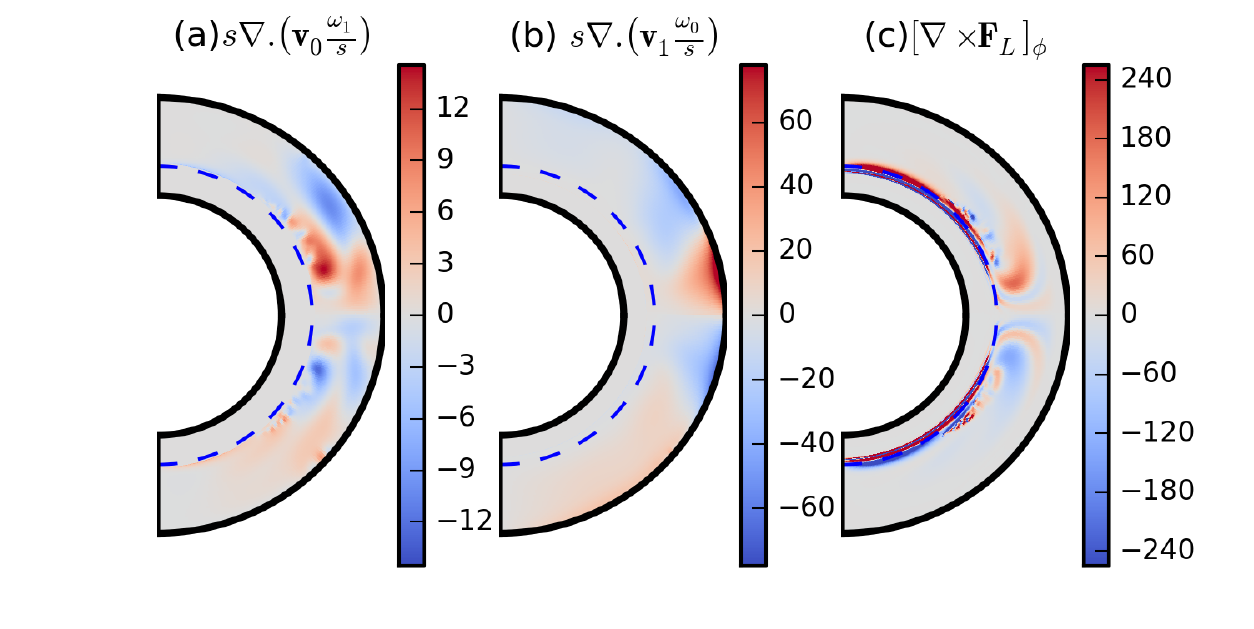}}
\caption[Behavior of various advective terms]{The advective terms in LHS of Eq.~\ref{eq:main2} are shown in (a) and (b), where $s=r\sin\theta$.
The $\phi$ component of the curl of the Lorentz force is shown in (c). All terms are calculated
at a particular time step near a solar maximum.}   
\label{fig:advect}
\end{figure}

\section{Appendix B: Importance of various  advective terms}\label{sec:appendix2}
In the calculations presented in this chapter, 
we have neglected the term $s\nabla.\left({\bf v}_1\frac{\omega_0}{s}\right)$ while solving
Eq.~\ref{eq:main2}, because its evaluation requires the perturbed velocity ${\bf v}_1$ in every time 
step which is computationally
very expensive to obtain. Now we calculate the two advective terms in Eq.~\ref{eq:main2} separately using simple finite 
difference scheme at a particular time step near a solar maximum to see their relative importance
with respect to the Lorentz force 
term $[\nabla \times {\bf F}_L]_{\phi}$. The values of the two advection terms involving 
unperturbed velocity $s\nabla.\left({\bf v}_0\frac{\omega_1}{s}\right)$ 
and involving perturbed velocity $s\nabla.\left({\bf v}_1\frac{\omega_0}{s}\right)$  in the $r - \theta$ plane
are shown in Fig.~\ref{fig:advect}(a) and 
\ref{fig:advect}(b) respectively. The $\phi$ component of the curl of the Lorentz force $[\nabla \times {\bf F}_L]_{\phi}$ 
is shown in Fig.~\ref{fig:advect}(c). The color table for curl of the Lorentz force is clipped to 240 to show its value in 
the convection zone (where it has much smaller values compared to the tachocline) but its maximum value is about 3000. 
Thus, in comparison to the curl of the Lorentz force, the other terms are small but not negligible. 
The advective term $s\nabla.\left({\bf v}_0\frac{\omega_1}{s}\right)$ which we already included
is important within the convection zone, but the other term with perturbed 
velocity $s\nabla.\left({\bf v}_1\frac{\omega_0}{s}\right)$ which we plan to include in our future work
is important near the surface at low latitudes. 
The dynamics of vorticity presented in the Section~\ref{sec:pm_1} clearly reflect the 
fact that the time evolution of vorticity is mostly determined by 
the Lorentz force term $[\nabla \times {\bf F}_L]_{\phi}$ (see Fig.~\ref{fig:snap_1d12}((k)-(o))) even
when the advective term $s\nabla.\left({\bf v}_0\frac{\omega_1}{s}\right)$ is included. However, since the 
$s\nabla.\left({\bf v}_1\frac{\omega_0}{s}\right)$ term becomes appreciable in the regions where we have
inward flow towards active regions during the sunspot maximum, this term may have some importance
in modeling this inward flow. We plan to investigate this further in future.
%footnote
\blfootnote{This chapter is based on \citet{HC17}.}

%% file: chapter6.tex
\begin{savequote}[100mm]
``We must not think of the things we could do with, but only of the things that we can't do without"
\qauthor{--Jerome k Jerome, Three men in a boat}
\end{savequote}

\chapter{3D Flux Transport Dynamo Model and Build up of Solar Polar Field}
\label{C6}
\begin{quote}\small
We develop a three-dimensional kinematic self-sustaining model of the solar
dynamo in which the poloidal field generation is from tilted bipolar sunspot
pairs placed on the solar surface above regions of strong toroidal field by using
the SpotMaker algorithm and then the transport of this poloidal field to the
tachocline is primarily caused by turbulent diffusion. We obtain a dipolar
solution within a certain range of parameters. We use this model to study the
build-up of the polar magnetic field and show that some insights obtained from
surface flux transport (SFT) models have to be revised.  We present results
obtained by putting a single bipolar sunspot pair in a hemisphere and two symmetrical
sunspot pairs in two hemispheres. We find that the polar fields produced by them
disappear due to the upward advection of poloidal flux at low latitudes, which emerges
as oppositely-signed radial flux and which is then advected poleward by the meridional flow.
We also study the effect that a large sunspot pair violating Hale's polarity law would
have on the polar field. We find that there would be some effect---especially if
the anti-Hale pair appears at high latitudes in the mid-phase of the cycle---though
the effect is not very dramatic.
\end{quote}
\section{Introduction}
\label{C6:S1}
The flux transport dynamo model, which started being developed from the 1990s
\citep{WSN91,CSD95,Durney95}, has emerged as
an attractive theoretical model for explaining the solar cycle  and 
has been extensively reviewed by several authors \citep{Chou11,Charbonneau14,Karakreview14}.
In any dynamo model, the toroidal magnetic field
is generated from the poloidal field by the differential rotation, which has
now been mapped by helioseismology \citep{Thompson96}. The distinctive features of
the B-L flux transport dynamo model are that the meridional circulation plays a crucial
role in this model and the poloidal magnetic field is generated by the Babcock--Leighton
(BL) process involving the decay of tilted bipolar sunspots.  Bipolar sunspots
are assumed to form due to the buoyant rise of the toroidal magnetic flux
through the convection zone \citep{Parker55b} and their tilts result from the action of the
Coriolis force on the rising flux tubes \citep{chou89,Dsilva93, Fan93}
leading to Joy's law \citep{Hale19}. When a tilted pair of bipolar sunspots
decays, turbulent diffusion spreads the magnetic flux to produce a poloidal magnetic
component \citep{Bab61,Leighton64}. An over-all poloidal field develops
from the contributions due to many bipolar sunspots and 
is advected to the poles by the meridional circulation,
which is poleward in the upper layers of the convection zone.  The polar magnetic
field of the Sun is built up in this process.

The BL process---which involves the production of tilted bipolar sunspot pairs
and the generation of the poloidal field from their decay---is an inherently 3D
process and can be modeled in 2D only through drastically simplified crude approximations
\citep{CH16}. Still an understanding of how the poloidal field builds
up by the BL process historically came from two distinct classes of 2D theoretical
models: the 2D flux transport dynamo model and the surface flux transport (SFT)
model.  In the 2D flux transport dynamo model, we average over the azimuthal
direction $\phi$ and solve the axisymmetric dynamo equation in the $r$-$\theta$
plane. On the other hand, in the SFT model, we focus our attention
only on the $B_r$ component of the magnetic field
at the solar surface spanned by the $\theta$-$\phi$
coordinates and study its evolution on this surface under the joint action  of diffusion, meridional
circulation and differential rotation. Neither of these approaches provides a fully
satisfactory depiction of the BL process and each approach has its own limitations.

If a tilted bipolar sunspot pair at the solar surface is averaged over the 
azimuthal direction $\phi$, then we get two rings of opposite magnetic polarity
at slightly different latitudes. \citet{Durney95,Durney97} advocated the development
of the flux transport dynamo model by using such double rings as the source of the
poloidal component. However, a more popular approach has been to introduce an $\alpha$-coefficient
reminiscent of the $\alpha$-effect of the mean field dynamo theory (Chapter 16, \citet{Chou98}; \citet{Parker55a,SKR66}), 
although this now has a completely different interpretation.
The source term of the poloidal field is taken as the product of this $\alpha$-coefficient
and the toroidal field that has risen from the tachocline due to magnetic buoyancy.
\citet{CH16} review how different authors achieve this, with references
to the original papers. \citet{Nandy01} showed that the double ring approach
and the treatment through $\alpha$-coefficient give qualitatively similar results,
although \citet{Munoz10} argued that the double  ring approach is more
realistic. In any case, the 2D kinematic dynamo models do not give a detailed picture
of how the poloidal field builds up from the contributions of many individual bipolar
sunspot pairs, since such pairs get smeared over when we average over the azimuthal
direction. Also, as most of these dynamo models rely on a mean field approach,
flux tubes or sunspots are not handled properly in these models \citep{Chou03}.

Starting from the pioneering work of \citet{Wang89a,Wang89b}, the surface flux
transport (SFT) model has been made more sophisticated in several recent studies
\citep{van98,Schrijver02,Baumann04,Baumann06,Cameron10, Jiang14, UH1_14, Jiang15}. 
In this model, recently reviewed by \citet{Jiang_review15}, one can study in detail
how individual sunspot pairs contribute in building up the poloidal field and can address
such questions as to how this process depends on such factors as the latitudinal positions
of the sunspot pairs and the distribution in their tilt angles.  The main limitation
of this model is that several important aspects of physics get left out by ignoring the
vectorial nature of the magnetic field and by not including any subsurface
processes.  By studying the time evolution of an axisymmetric poloidal field,
\citet{DC94,DC95} and \citet{CD99} showed that the subduction of the poloidal field by the meridional circulation sinking
underneath the surface at the polar region plays an important role in the dynamics
of the magnetic field.  Since this process cannot be included in the SFT
models, flux of $B_r$ tends to get piled up in the polar regions
and has to be neutralized by flux of opposite sign advected there. If additional flux
of opposite sign is not brought there, then the polar field may reach an asymptotic
value as seen in Figure~6 of \citet{Jiang14}. When one tries to model several
successive cycles through an SFT model, one may get a `secular
drift' of the polar field if the flux of the succeeding cycle is unable to properly
neutralize the polar flux of the preceding cycle, as seen in Figure~1 of \citet{Baumann06}.
A way of fixing this problem proposed by \citet{Baumann06}
involves adding an ad hoc decay term corresponding to the radial diffusion not
included in the SFT model. 
%Altogether, the surface flux transport
%model runs into serious difficulties when an attempt is made to model several successive
%cycles. 
In spite of the tremendously important historical role the SFT model has
played in elucidating the BL process, this model has the inherent limitation that
it cannot adequately handle the magnetic field dynamics in the Sun's polar region. 

We believe that the next step forward is the 3D kinematic flux transport dynamo model.  
In this model, the fluid motions (differential rotation, meridional circulation) are specified
and the evolution of the magnetic field is calculated in 3D. Such a model has the
promise of incorporating the attractive features of both the 2D flux transport dynamo
model and the SFT model, while being free from the limitations of both
these models. It can handle the BL process much more realistically than the 2D flux
transport dynamo model where we average over the azimuthal direction and cannot include
tilted bipolar sunspots properly.  On the other hand, this model incorporates the vectorial
nature of the magnetic field and the subsurface processes which are left out in the
SFT models.

Efforts of constructing 3D kinematic flux transport dynamo models began only
within the last few years. In a landmark paper,
\citet{YM13} developed a method of treating the buoyant rise
of a flux tube in their 3D dynamo model by simultaneously applying a radially outward
and a vortical velocity to a localized part of an azimuthal flux tube at the bottom of 
the convection zone. Although they did not present a self-excited dynamo solution, 
they simulated a solar cycle by incorporating bipolar sunspot eruptions by this method at the
actual locations where bipolar sunspots were observationally seen.
\citet{MD14} succeeded in producing a self-excited dynamo by identifying the locations (in latitude
and longitude) at the bottom of the convection zone where the toroidal field was
the strongest (in the theoretical model)
and then putting tilted bipolar sunspot pairs above those locations 
with loop-like magnetic structures both above and below the surface (by using an
algorithm which they named SpotMaker). More details of this model have been
given by \citet{MT16}. After a part of the toroidal flux tube
rises to the surface to produce bipolar sunspots, the magnetic field underneath these
sunspots has to be detached at some stage from the bottom of the convection zone before
the magnetic flux of the sunspots is dispersed freely by diffusion and carried poleward
by the meridional circulation \citep{Longcope02}. Our lack of understanding
of this process is the main difficulty in constructing realistic 3D dynamo models
at the present time.  Presumably, the approach of \citet{YM13}
captures the physics of the early phase soon after the bipolar sunspots emerge, whereas
the approach of \citet{MD14} is more appropriate for the later phase
when the magnetic field below the sunspot pairs has become detached from the bottom
of the convection zone.

The aim of this study is to use a modified version
of the model of \citet{MD14} to
study the build-up of the Sun's polar magnetic field by the BL process in more detail.
The model of \citet{MD14} uses values of parameters (such as turbulent
diffusion) which are probably not very realistic for the Sun.  We use the
same dynamo code named as STABLE (i.e.\ Surface flux Transport And Babcock LEighton Model)
to first construct a model of the solar dynamo based
on more realistic values of parameters and then use this model for our study.
Since the BL process has been studied most extensively by the SFT
models, we especially address the question whether the insights gained about various
aspects of this process from the SFT models are borne out by
the 3D model or have to be revised significantly. We shall see that the
accumulation of magnetic flux at the poles seen in the SFT models does
not occur when the low-latitude advection and emergence of oppositely-signed
radial flux is taken into account. Thus, the problem of `secular drift' is 
automatically eliminated. One insight from the SFT models is that the fluxes of
leading sunspots at lower latitudes get canceled across the equator and the fluxes 
from the following sunspots are then advected to the poles, building up a dipole moment of the Sun.
We shall see that this insight also will have to be modified significantly.  SFT models indicate that
even a few large sunspot pairs with anti-Hale or wrong polarity (i.e.\ opposite of what is expected
of sunspot pairs in that cycle) may have significant effect on the polar field \citep{Jiang15}.
We shall be able to study this effect more realistically
in our 3D model.

After discussing the mathematical formulation of the problem in the next section,
the standard model of the solar dynamo which we shall use is presented in Section~\ref{C6:S3}. Then
the build-up of the polar field is studied in Section~\ref{C6:S4}, whereas the effects of large anti-Hale
sunspot pairs are discussed in Section~\ref{C6:S5}. Our conclusions are summarized in Section~\ref{C6:S6}.

\section{Mathematical Formulation}
\label{C6:S2}
In this section, we explain the basic formulation of the STABLE 
dynamo model which is first reported in \citet{MD14} and in more detail in \citet{MT16}. 
This model is a 3D generalization of the pre-existing axisymmetric 2D flux transport dynamo
models and it solves the induction equation in full 3D rotating spherical shell with radius
ranges from $r = 0.69R$ to $r = R$ of the Sun:

\begin{equation}
\label{eqc6:induction1}
\frac{\partial {\bf B}}{\partial t} = \nabla \times ({\bf v}\times{\bf B} -\eta_t \nabla\times {\bf B})
+ S(\theta,\phi,t)
\end{equation}
where ${\bf v}$ is the velocity field, $\eta_t(r)$ is the turbulent diffusion in the solar 
convection zone and $S(\theta,\phi,t)$ is the source function which captures the
effect of the BL mechanism. As we shall discuss in detail later, the source function
$S(\theta,\phi,t)$ is of the nature of an impulsive forcing term which becomes
non-zero only at the instants when we allow a bipolar sunspot pair to be put at
the solar surface. Though this model is fully 3D and no axisymmetric assumption is considered 
but still this model is kinematic and we provide the velocity field motivated from  
helioseismology and observations. We solve equation (\ref{eqc6:induction1}) using Anelastic 
Spherical Harmonic (ASH) code \citep{Miesch_et_al_00,BMT04}. ASH is a well-established 
pseudospectral code which has been used extensively for 3D solar and stellar convection simulations, 
instabilities, tachocline confinement and many other aspects of solar and 
stellar internal dynamics. The ASH code has capability to solve the velocity equation and 
magnetic induction equation together but for our kinematic model we bypass the velocity 
equation solver and only solve the induction equation by providing observationally motivated velocity fields.  
The version of the code used for 3D kinematic dynamo modeling is named 
STABLE (i.e.\ Surface flux Transport And Babcock LEighton Model).

\begin{figure}[!htbp]
\centerline{\includegraphics[width=0.75\textwidth,clip=]{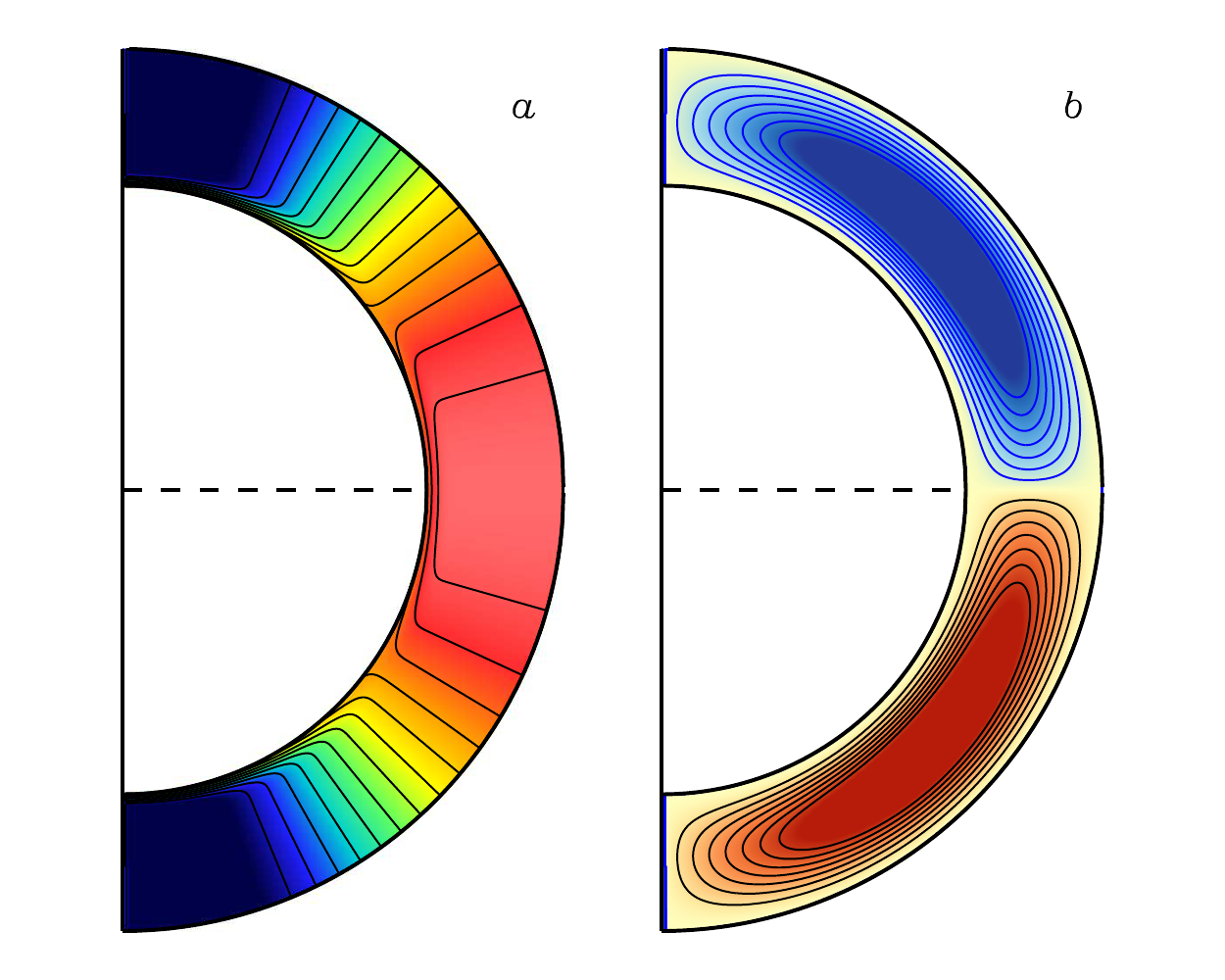}} 
\caption[Differential rotation and meridional circulation used in 3D dynamo model]{(a). Differential rotation profile with color table ranging from 350-480 nHz from blue to red, 
(b) Streamlines for the meridional flow. Blue and red contours show the poleward flow at the surface, and 
an equatorward flow at the bottom of the convection zone in northern and southern hemisphere respectively. 
The amplitude of the meridional circulation is taken as 20.40 m/s on the surface and 1.64 m/s at the lower convection zone.}
\label{fig:MC}
\end{figure}

\begin{figure}[!htbp]
\centerline{\includegraphics[width=0.95\textwidth,clip=]{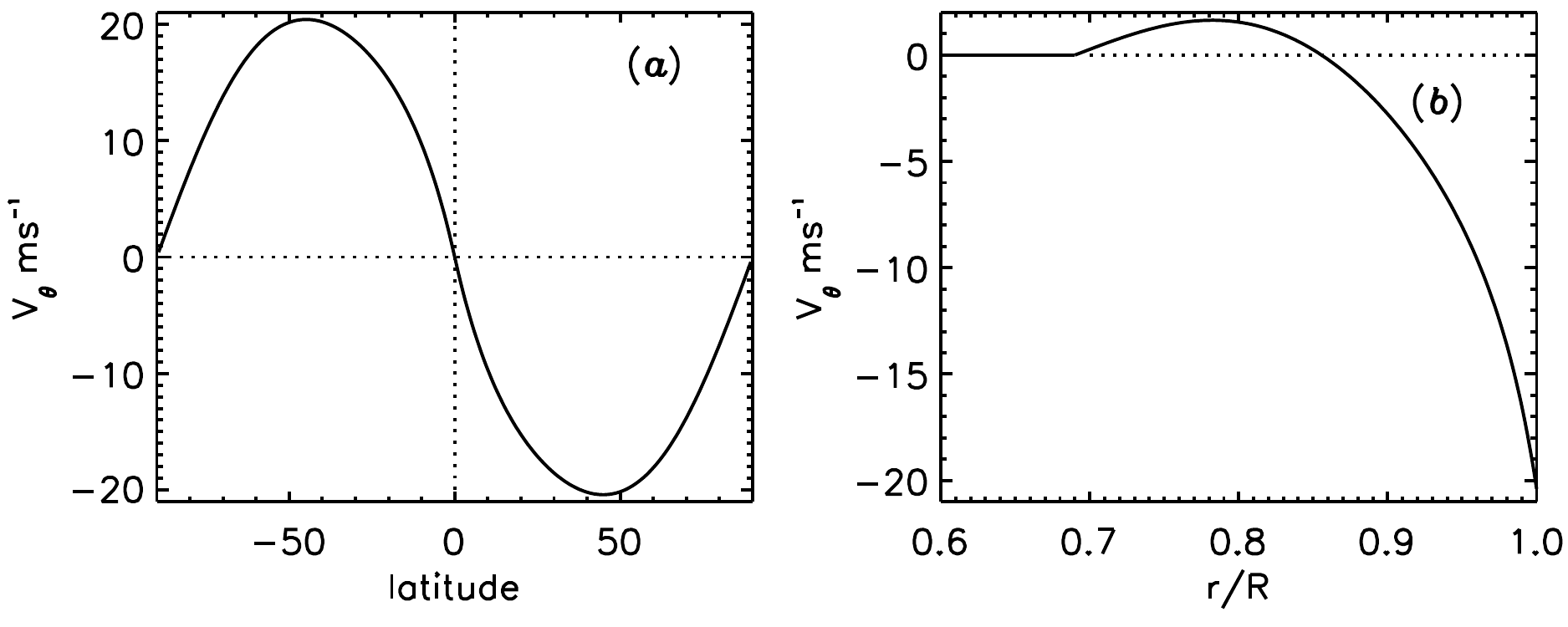}} 
\caption[$V_\theta$ component of meridional circulation]{Variation of latitudinal component of meridional circulation
$V_\theta$ with latitude on the surface (a) and with radius at $45^{\circ}$ latitude (b).}
\label{fig:vtheta}
\end{figure}

The mean velocity field ${\bf v}$ in the Sun can be written as the summation of the part from differential 
rotation $\Omega$ and the meridional circulation $v_p$. Whereas the exact differential 
rotation is mapped from helioseismology quite well \citep{Schou98}, the structure of the
meridional circulation in the solar convection zone is still under study. Recently different 
helioseismology groups have reported substantially different structures of the meridional circulation in 
the solar convection zone \citep{Zhao13,Schad13,RA15}. In response to these observational claims, 
\citet{HKC14} carried on calculations with different types of 
meridional circulation structure and showed that flux transport dynamo works quite well as long as 
there is an equatorward flow at the bottom of the convection zone. Recently \citet{karak16} 
showed that in the presence of appropriate profile of downward pumping in the solar convection zone, 
the flux transport solar dynamo works with even shallow meridional circulation. Since there
is still no compelling reason \citep{RA15} to give up the simple
single-cell profile of meridional circulation used by many previous authors (\citet{CNC04}, \citet{MD14}),
we use a single cell profile of the meridional circulation having 
a poleward flow at the surface and an equatorward return flow at the bottom of the convection 
zone. The stream function corresponding to the meridional circulation which we use here is
\begin{eqnarray}
\label{eq:psi}
\psi r \sin \theta = \psi_0 (r - R_p) \sin \left[ \frac{\pi (r - R_p)}{(R_\odot -R_p)} \right]\{ 1 - e^{- \beta_1 r\theta^{\epsilon}}\}
\{1 - e^{\beta_2 r(\theta - \pi/2)} \} e^{-((r -r_0)/\Gamma)^2} ~~~~
\end{eqnarray}\\
with $\beta_1 = 0.3 \times 10^{-10}~cm^{-1}, \beta_2 = 0.5 \times 10^{-10}~ cm^{-1}$, $\epsilon = 2.0000001$, $r_0 = (R_\odot - R_b)/4.0$, $\Gamma =
3.5 \times 10^{10}$ cm, $R_p = 0.69R_\odot$.
The value of $\psi_0$ determines the amplitude of the meridional circulation. On
taking $\psi_0 = 12.0$, the poleward flow near the surface at mid-latitudes peaks around $v_0=20.40$ m s$^{-1}$.
The contour plot for the meridional circulation is shown in 
figure~\ref{fig:MC}(b) and the variation of $V_\theta$ with latitude on the surface and 
variation $V_\theta$ with radius at mid-latitude $(45^{\circ})$ are shown in figure~\ref{fig:vtheta}(a) and \ref{fig:vtheta}(b) respectively.
For differential rotation we have used the analytical formula given
in \citep{DC99} which is a good fit to the observational data (Figure~\ref{fig:MC}(a)).

Turbulent diffusivity is another important parameter. After the BL process generates
the poloidal field near the solar surface, it has to reach the tachocline where the differential
rotation acts on it to produce the toroidal field.  This can happen in two ways. The poloidal
field may first be advected by the meridional circulation to the pole and then underneath
the surface to the mid-latitude tachocline from where the first sunspots of cycle rise.
The time scale for this is close to 20 yr for a reasonable profile of the meridional circulation.
The second possibility is that the poloidal field diffuses from the surface to the bottom
of the convection zone to be acted upon by the differential rotation of the tachocline. 
The Green's function for the diffusion equation suggests that the diffusion
time across a length $L$ is $L^2/4 \eta_t$ (see, for example, \citet{Parker79}, p.\ 32). If
the turbulent diffusivity within the convection zone is assumed to be $5 \times 10^{11}$
cm$^2$ s$^{-1}$ as we shall do, then this diffusion time
turns out to be about 7 yr if we take $L$ to be the thickness of the convection zone.
The value of the turbulent diffusivity determines whether the poloidal field
is transported across the convection zone primarily by meridional circulation or by
turbulent diffusion, and the behavior of the dynamo is very different in the two
situations \citep{Jiang07,Yeates08}. Over the years, we have got
more and more evidence that the turbulent diffusivity has to be sufficiently high to
make the poloidal field transport primarily by diffusion in order to explain many aspects
of the solar cycle, such as the dipolar parity \citep{CNC04,Hotta10}, the lack of significant hemispheric asymmetry
\citep{CC06,GoelChou09}, the observed correlation between the polar field at the cycle
minimum and the strength of the next cycle \citep{Jiang07}, the Waldmeier effect \citep{KarakChou11}.
Such a value of turbulent diffusivity is also consistent with mixing length arguments
(\citet{Parker79}, p.\ 629) and the theory of mean flows \citep{miesc12b}.

\begin{figure}
\centerline{\includegraphics[width=0.65\textwidth,clip=]{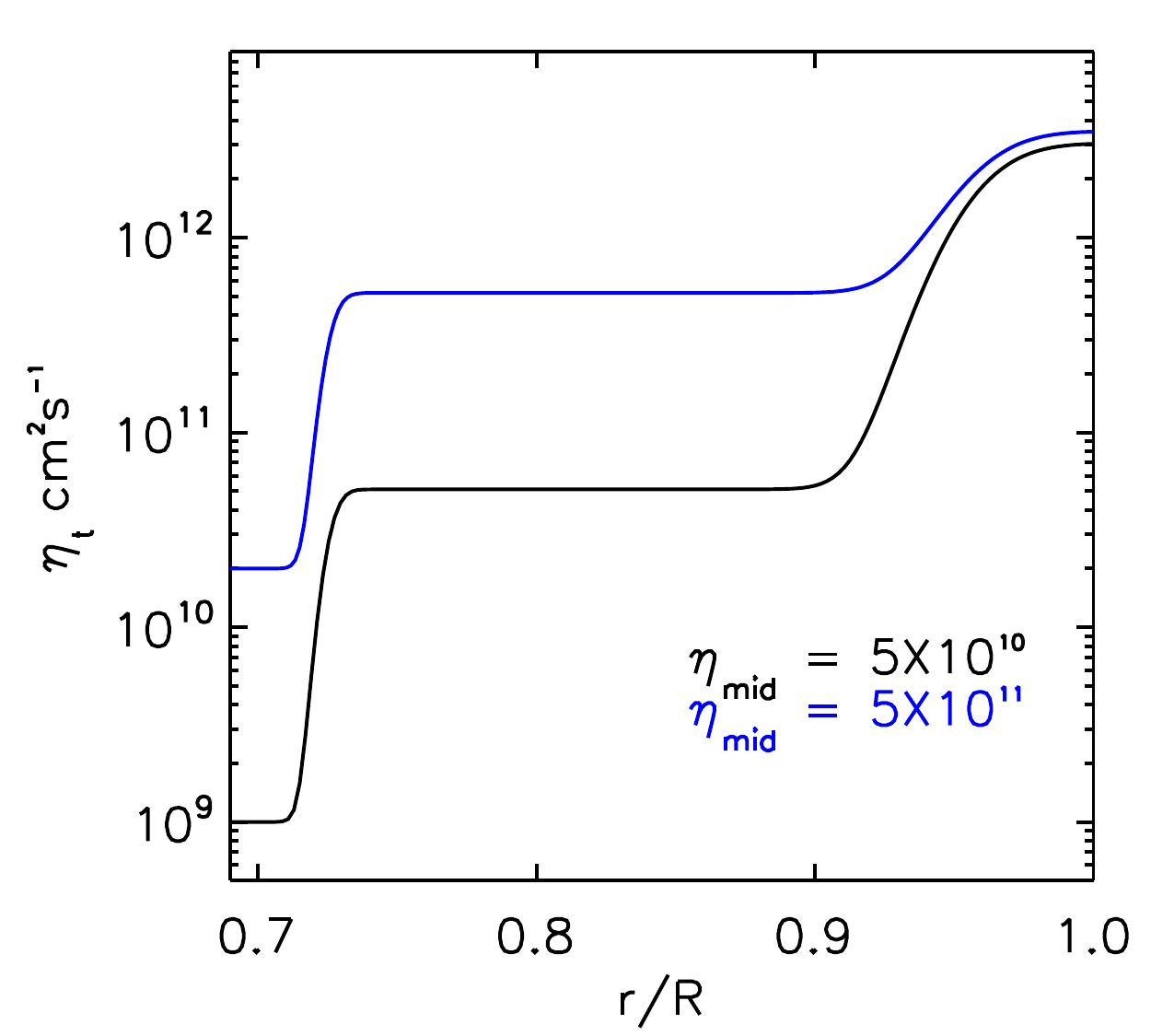}} 
\caption[High diffusivity profile used in our simulation]{High diffusivity profile used for most of our simulation is shown in blue solid line 
and the profile used for advection-dominated regime is shown in black solid line.}
\label{diff}
\end{figure}

In most of the SFT models, a constant diffusivity $(2.5-3.0) \times 10^{12}$ cm$^2$s$^{-1}$ on the 
surface is used \citep{Jiang_review15}. The turbulent diffusivity is expected to be less within the
convection zone and falls drastically at its bottom where convection is less vigorous.
Except when stated explicitly otherwise, the calculations of this chapter assume the diffusivity
to be given by
\begin{eqnarray}
\eta_t = \eta_c + \frac{\eta_{mid}}{2}\left[1 + {\rm erf}\left(2\frac{r-r_{da}}{d_a}\right)\right]
~~~+ \frac{\eta_{top}}{2}\left[1+{\rm erf}\left(2\frac{r-r_{db}}{d_b}\right)\right] 
\end{eqnarray} 
%\begin{equation}
%\label{diff}
%\eta = \eta_c + \frac{\eta_{top}}{2}\left[1 + \rm{erf}\left(\frac{r-r_{MCZ}}{d_t}\right)\right]
%\end{equation}
where $\eta_c = 2\times10^{10}$ cm$^2$ s$^{-1}$, $\eta_{mid}= 5\times10^{11}$ 
cm$^2s^{-1}$,$r_{da} = 0.725 R_\odot$, $r_{db} = 0.956R_{\odot}$, and $d_b = 0.05R$.
In Figure~\ref{diff} we have shown the diffusivity profile by the blue solid line. For 
comparison, the diffusivity profile used by \citet{MD14}
is shown by the black solid line.  It may be noted that some groups \citep{DC99,DG06},
over the years, used a rather low value of diffusivity.
As seen in Figure~\ref{diff}, \citet{MD14} and \citet{MT16} followed these authors in using a diffusivity
which was, within the body of the convection zone, about one order of magnitude smaller
than what we are using. In our case the diffusivity $\eta_{mid}$ in the convection zone
is $5\times 10^{11}$ cm$^2$s$^{-1}$, whereas \citet{MD14} and \citet{MT16} use $5\times10^{10}$ cm$^2$s$^{-1}$.
Such a lower value of diffusivity would make the diffusion time across
the convection zone of the order of 70 yr and the advection by the meridional circulation
would clearly be the dominant process for the transport of the poloidal field. \citet{MD14}
presented a self-excited dynamo solution for this situation.  For the value of diffusivity
we are using, the diffusion across the convection zone is the primary process for bringing
the poloidal field from the solar surface to the tachocline.  We believe that we are the first
to obtain a self-excited 3D kinematic dynamo solution for this case, which we contend is closer to reality.

We now discuss how the source term $S(\theta,\phi,t)$ in Eq.~\ref{eqc6:induction1} is specified with the help
of the SpotMaker algorithm to treat the BL process.
This algorithm is mainly a 3D generalization of the Durney's double ring algorithm 
\citep{Durney95,Durney97}. In this algorithm, two suitable opposite-polarity spots are placed 
on the surface of the Sun in response to the dynamo-generated field near the base of the 
convection zone and then they are allowed to decay in the presence of mean flows (meridional 
circulation and differential rotation) and diffusivity. The first aim of this algorithm 
is to find out the suitable position for these spots to be placed on the surface. To do so, 
we calculate the mean toroidal flux $\bar{B}(\theta,\phi)$ near the bottom of the 
convection zone averaged over the tachocline thickness \citep{MT16} and find out where this 
field is crossing the threshold value $B_t$. It is believed that if magnetic fields near 
the bottom of the convection zone are stronger than the threshold value $B_t$ then they become 
magnetically buoyant and create the bipolar sunspots on the surface \citep{Parker75}. 
So the latitude and longitude of the spot pair is chosen randomly from all grid points
where the mean toroidal flux exceeds $B_t$, subject to a mask that suppresses spots at high latitudes. When 
we are able to find the $\theta_s$ and $\phi_s$ where the dynamo-generated toroidal field 
is more than the threshold value $B_t$, we put two spots on the surface at that position. Once 
the position of the bipolar sunspots is decided, the next step is to specify the magnetic field
there, by putting some tilt angle between the two sunspots according to Joy's law.
For that we use the polynomial profile as given in \citet{MD14} and 
for tilt angle we follow the procedure given in \citet{SK12}. We choose the tilt angle to be 
$\delta = 32.1^{\circ} \cos\theta$. We do not want to put these spots at each time step of our simulation. 
There are always certain time differences between the appearances of different sunspot groups. 
So we have used a time delay probability density function which allows us to put 
successive sunspot pairs having a random time delay between their appearances
\citep{MT16}. Another thing we should mention here is that, as seen in the observed butterfly
diagram, sunspots are found mostly on the lower latitudes and in our model we artificially
suppress the sunspot formation at higher latitude using some masking function as given
in equation (3) of the \citet{MD14}. The flux content in the spots and the strength of the radial field 
are chosen based on the dynamo-generated field $\bar{B}$ and the observed strength of the 
sunspots as given below.
\begin{equation}
\label{eqc6:flux}
\Phi = 2\alpha_{spot}\frac{|{\hat{B}(\theta_s,\phi_s,t)}|}{B_q}\frac{10^{23}}{1 + (\hat{B}(\theta,\phi)/B_q)^2} \rm{Mx}
\end{equation} 
where $\hat{B} = g(\theta)\bar{B}$ is the toroidal field after using the masking function $g(\theta)$ 
to suppress sunspots at high latitude and $B_q$ is the quenching field strength.
Here $\alpha_{spot}$ is the parameter which determines whether the dynamo will be sub-critical or 
super-critical. Our ultimate aim would be to make the dynamo work with
$\alpha_{spot} =1$ so that the flux in a particular BMR will have a value of $10^{23}$ \rm{Mx} as 
observed in case of the subsurface field strength equivalent to the quenching field strength.
But if the subsurface field at the bottom of the convection zone is not close to the quenching field, 
then we have to increase the value of $\alpha_{spot}$ in order to get a working dynamo with bigger 
spots. In case of the diffusion-dominated dynamo, we are able to get a working dynamo with $\alpha_{spot}=100$. 
While creating sunspot pairs by the SpotMaker algorithm, once the total flux is fixed by (\ref{eqc6:flux}), we
have the freedom of selecting either the magnetic field strength or the size. We choose the
magnetic field strength inside the sunspots to be 3000 G, which fixes the size.

Since we are solving magnetic fields in 3D spherical shell, so we must have to specify the subsurface structure of the 
sunspots which are put on the surface using SpotMaker algorithm. As it is argued by \citet{Longcope02} and
\citet{SM05} that the sunspots get quickly disconnected from the parent flux tube, we make a very simple potential 
field approximation for the sunspot fields (see Figure~2(a,b) of \citet{MT16}). We ensure that the 
radial field becomes zero at some penetration depth ($r =0.90R$)
and it is equal to the imposed sunspot field at the surface ($r = R$). For the upper boundary condition, we
take the magnetic field to be radial at the solar surface. Throughout our simulation we use $N_r =200$, $N_\theta = 256$ and $N_\phi = 512.$
All of the cases where we show the field lines above the solar surface ($r = R$) are the extrapolated fields using a free potential approximation.
\begin{figure*}[!htbp]
\centerline{\includegraphics[width=1.0\textwidth,clip=]{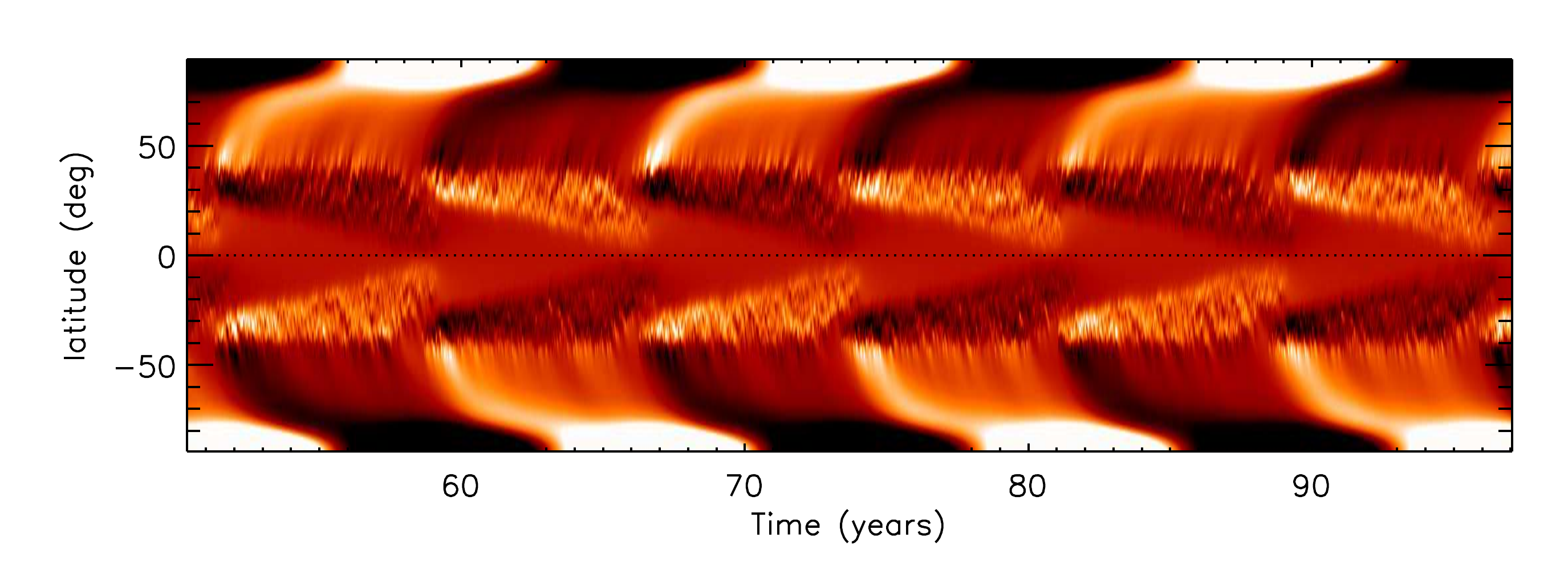}} 
\caption[Butterfly diagram from our 3D model]{Time-latitude plot of longitudinally averaged radial magnetic field on the surface $(r = R)$ of Sun. Color table is set at $\pm10$ kG. }
%Peak amplitude can exceed $\pm 40$ kG but color tables saturates at $\pm 10$ kG}
\label{bfly}
\end{figure*}

\begin{figure}[!t]
\centerline{\includegraphics[width=0.95\textwidth,clip=]{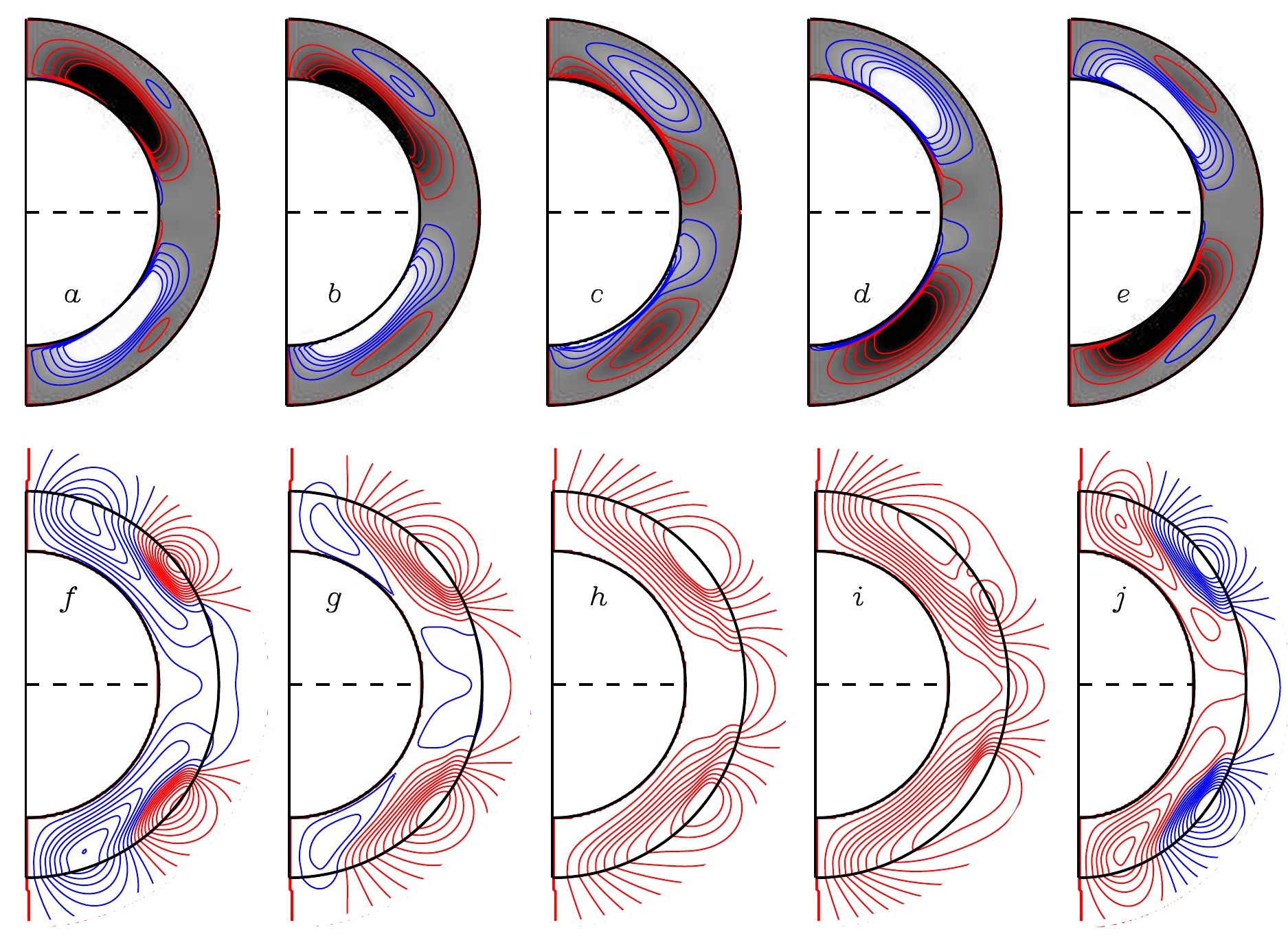}} 
\caption[Mean toroidal and poloidal field lines]{Mean toroidal and poloidal field lines are shown for a particular solar cycle at five different 
time: t= 60.0 (a,f), 62.0 (b,g), 64.0 (c,h), 66.0 (d,i) and 68.0 years (e,j). Frames (a-e) show mean 
toroidal fields with red and blue lines indicating eastward and westward field respectively. 
Filled color also represents the mean toroidal fields. Color table is set in this case at $\pm 500$ kG. 
Frames (f-g) represent poloidal magnetic potential with potential field extrapolation above the surface (upto $r = 1.25R$) where red 
and blue lines represent clockwise and anticlockwise directions. The maximum and minimum contour level is set corresponding to the poloidal field strength of $\pm49$ kG}
\label{fields_ref}
\end{figure}

\section{Our reference model}
\label{C6:S3}
Now we present a self-excited solution from our reference model with parameters as prescribed
in the previous section.  To the best of our knowledge, this is the first self-excited
3D kinematic dynamo solution in which the diffusivity has been assumed sufficiently high to
make sure that the poloidal field is transported from the surface to the tachocline primarily
by diffusion.  The earlier results presented by \citet{MD14} and \citet{MT16} were obtained
with a diffusivity one order of magnitude smaller and the transport of the poloidal field was
due to the meridional circulation.

Figure~\ref{bfly} shows a butterfly diagram obtained by putting the longitude-averaged $B_r$ in a time-latitude
plot.  One clearly sees the butterfly diagram of sunspots at lower latitudes and the poleward
advection of the magnetic field by the meridional circulation at higher latitudes.  Superficially this resembles
Figure~6(a) of \citet{MT16}, although our solution is for the diffusion-dominated case in contrast
to the solution of \citet{MT16} obtained for the case dominated by advection due to the meridional
circulation. 
The differences between the two cases become clear on looking at the distribution
of the magnetic field. Figure~\ref{fields_ref} shows the evolution of the toroidal and the poloidal fields during
a cycle. Comparing with Figure~8 of \citet{MT16}, we see some obvious differences.  In the solution
of \citet{MT16}, the oppositely directed toroidal fields on the two sides of the equator almost
pressed against each other.  Due to the low diffusivity, there would not be much diffusion of the
toroidal field even when two opposite bands are brought so close to each other.  In our model with
higher diffusivity, however, there would be more diffusion of the toroidal field across the equator, making
sure that the bands of concentrated opposite polarity are kept somewhat apart, as seen in Fig.~\ref{fields_ref}.

\begin{figure}[!htbb]
\centerline{\includegraphics[width=1.0\textwidth,clip=]{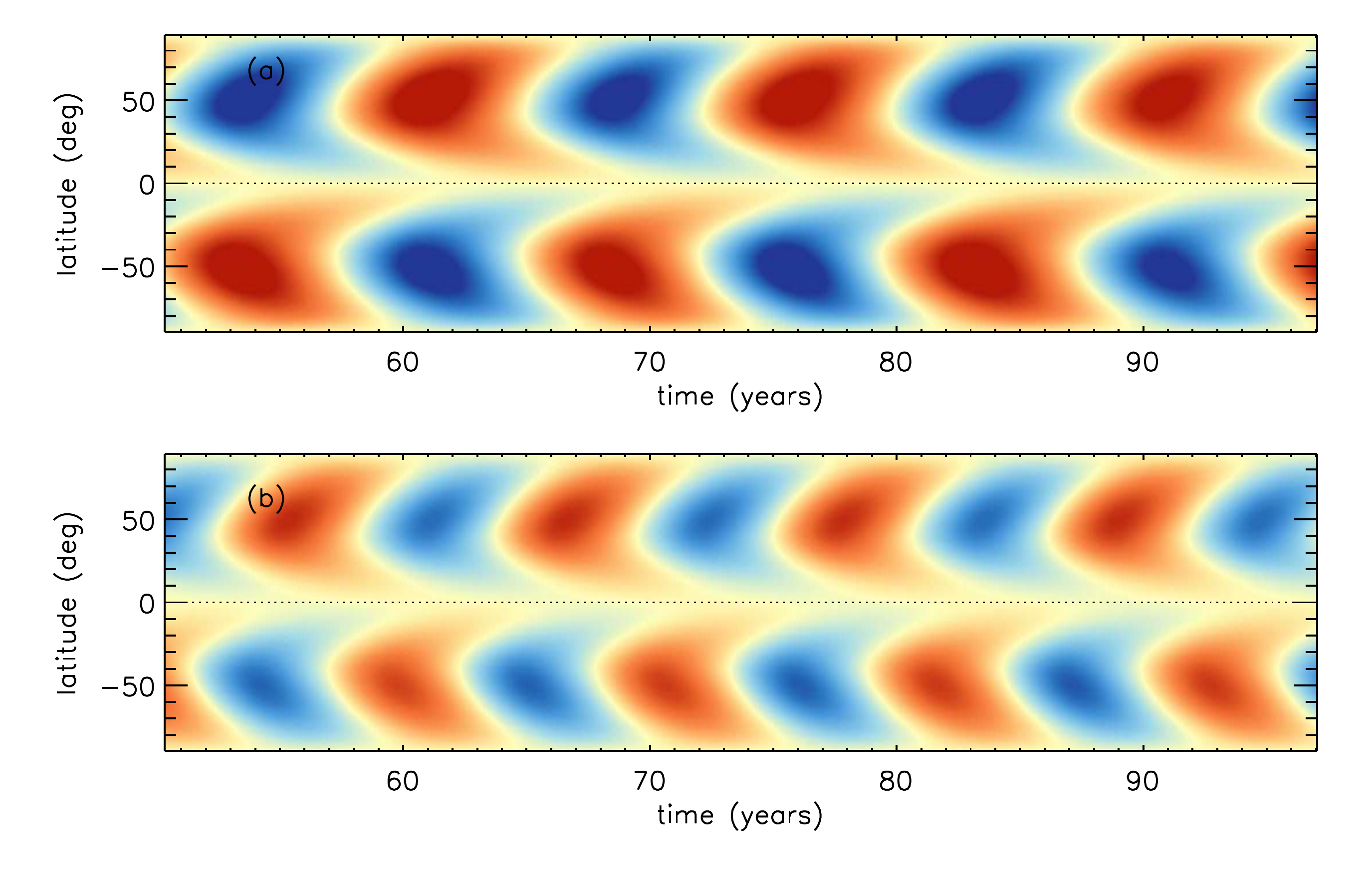}} 
\caption[Azimuthal averaged toroidal fields for different turbulent diffusivity]{Azimuthal averaged toroidal fields $B_{\phi}$ at the bottom of the convection zone $(r = 0.71R)$
for two different values of diffusivity $\eta_{mid}$ at the convection zone
(a) $5\times10^{11}$ cm$^2$ s$^{-1}$ and (b) $7\times10^{11}$ cm$^2$ s$^{-1}$. Color table is set at $\pm  400$ kG.}
\label{bfly2}
\end{figure}

The solar magnetic field is predominantly dipolar.  One requirement of a theoretical solar dynamo model is
that it should have dipolar parity. We have run our reference model for several cycles to ensure
that the dipolar parity persisted. One important question is under what circumstances we
would expect dipolar parity.  This question has been studied thoroughly by
\citet{CNC04} and \citet{Hotta10} for the 2D kinematic dynamo model.  A full study
of this question requires running the code for many different combinations of parameters and
running it for a large number of cycles for each such combination. This would require a huge
amount of computer time for the 3D model.  Because of the limited computer time available to
us, we have not been able to study this question exhaustively.  However, we have made a limited 
number of runs to explore the issue of parity a little bit.  The best way to look at the issue
of parity is to make a butterfly diagram of longitudinally averaged $B_{\phi}$ at the bottom of the convection zone, as
done in Figure~7(a) of \citet{CNC04}. In Figure~\ref{bfly2}(a) we show such a plot for
our reference model, whereas Figure~\ref{bfly2}(b) shows a similar plot for the case in which the value
of diffusivity within the convection zone has been changed from  $5 \times 10^{11}$
cm$^2$ s$^{-1}$ to  $7 \times 10^{11}$ cm$^2$ s$^{-1}$ while keeping all the other
parameters exactly the same as in our reference model. We clearly see in Figure~\ref{bfly2}(b) that the 
nature of the solution is changing from a dipolar parity to a quadrupolar parity.

The surprising fact is that we now seem to get a result which is opposite of what
\citet{CNC04} and \citet{Hotta10} obtained for the 2D kinematic dynamo model. These
authors found that the dipolar parity is preferred on increasing the diffusivity, whereas we
now are finding the opposite of that. Let us look at the physics of the problem.  In a dipolar
mode, the poloidal magnetic field lines connect across the equator, whereas the toroidal field
on the two sides of the equator has to be directed oppositely.  In order for this to happen, we need
diffusivity to have a big effect on the poloidal field, but not much effect on the toroidal field.
In the model of \citet{CNC04}, any strong toroidal field within the 
convection zone was removed by magnetic buoyancy and the toroidal field at the bottom of the
convection zone was also depleted continuously to account for flux loss due to magnetic buoyancy.
As a result, the toroidal magnetic field near the equator was naturally weak and the effect of
diffusion was more important on the poloidal field than on the toroidal field.  This ensured
that higher diffusivity favored the dipolar mode.  In the present calculation, the situation
is rather different.  The strong parts of the toroidal field are now allowed to hover in the 
middle of the convection zone and near the equator.  On increasing diffusivity, the quadrupolar
mode in which the toroidal field on the two sides of the equator has the same sign is favored.
Here we should mention that the magnetic pumping can play a very important role to promote
dipolar parity \citep{Guerrero08}. Another point to note is that \citet{CNC04} used a lower
diffusivity of the toroidal field compared to the poloidal field, to account for the quenching
of turbulent diffusion due to the stronger toroidal magnetic field.  This could be done in a
2D mean field model in which the evolution equations for the toroidal and poloidal fields neatly
separate out, and one could use different values of diffusivity in the two equations. Since it
is not possible to do this in a 3D non-axisymmetric model in which the equations for the toroidal and 
poloidal components do not split in this way, we have used a single diffusivity.  It is possible
that the weaker diffusivity of the toroidal field in \citet{CNC04} helped in producing a dipolar
parity by allowing toroidal fields of opposite sign to exist on the two sides of the equator
more easily.  One way of capturing the physics of this in a 3D non-axisymmetric model may be
to include a quenching of turbulent diffusivity in the regions of strong magnetic field.  We
plan to explore the effect of this in future.

With these two opposite results at hand, one crucial question is: which of the two results is closer
to reality?  Although we cannot assert this with confidence at this stage, we believe
that the 2D kinematic dynamo result that the dipolar parity is preferred on increasing diffusivity
is the more appropriate result.  Although in this study we are taking account of the 3D nature of
magnetic buoyancy and, in that sense, treating magnetic buoyancy more realistically, we still
have not taken account of flux depletion from the convection zone and its bottom in an appropriate
way. This is probably one important reason why our results are not matching with 
the results of previous 2D models \citep{CNC04,Hotta10}.
%Because of this we are getting a result which is probably unphysical.
We are right now exploring possible schemes to take account of the flux depletion due to magnetic buoyancy in a
realistic way.  We believe that this flux depletion is quite important in the solar dynamo.
\citet{CH16} found that the Waldmeier effect cannot be reproduced from a theoretical
dynamo model unless the flux depletion is taken into account. We have a future plan of incorporating
flux loss due to magnetic buoyancy in a realistic way in our 3D kinematic dynamo model and then
studying the parity issue more carefully.

Since we are interested in a dynamo solution which has dipolar parity, we have converged on
the reference solution presented here.  If we decrease diffusivity, then we are led to the case
where the meridional circulation provides the main transport mechanism for the poloidal field.
On the other hand, if we increase diffusivity, then we obtain the quadrupolar mode.  This is what
has led us to choose the value $5 \times 10^{11}$ cm$^2$ s$^{-1}$ for diffusivity inside the
convection zone. 

\section{The build-up of the polar field}
\label{C6:S4}
After constructing the self-excited dynamo model, we now study 
how individual sunspot pairs contribute to the building up of
the polar field and address the question whether our understanding gained from
this study necessitates the revision of some insights we have from surface flux
transport (SFT) models. For this study, we shall put individual sunspot pairs on
the solar surface by hand and look at the evolution of the magnetic field.  In
other words, we shall now not try to construct self-excited periodic solutions,
although we shall keep using the same values of different parameters that we had
used for constructing the self-excited periodic solution.

We start our simulation by putting a single pair of bipolar sunspots in the northern hemisphere at 
different emergence angles $\lambda_{emg}$ and let it evolve under the axisymmetric mean 
flows and diffusion to see the development of the polar field. We have chosen magnetic flux 
of $1\times10^{22}$ \rm{Mx} in each spot and its radius is taken to be 21.71 Mm (somewhat
larger than actual sunspot radii, to make the results of the simulation more clearly visible)
throughout our simulations. In the next set of our simulations, we shall put two pairs of sunspots  
symmetrically in the two hemispheres, which have the same amount of flux and radius 
as in the case of the single pair to see the effects of cross-equator diffusion of magnetic flux. 
%Finally, 
%we shall put an `anti-Hale' sunspot pair in different latitudes and at different stages of the cycle to see 
%how a anti-Hale sunspot pair can effect the polar field.

\subsection{Polar field from one sunspot pair}
\label{C6:S4_1}
\begin{figure}[!htbp]
%\centerline{\includegraphics[width=1.0\textwidth,clip=]{spot_1P_20.pdf}} 
\centerline{\includegraphics[width=1.0\textwidth,clip=]{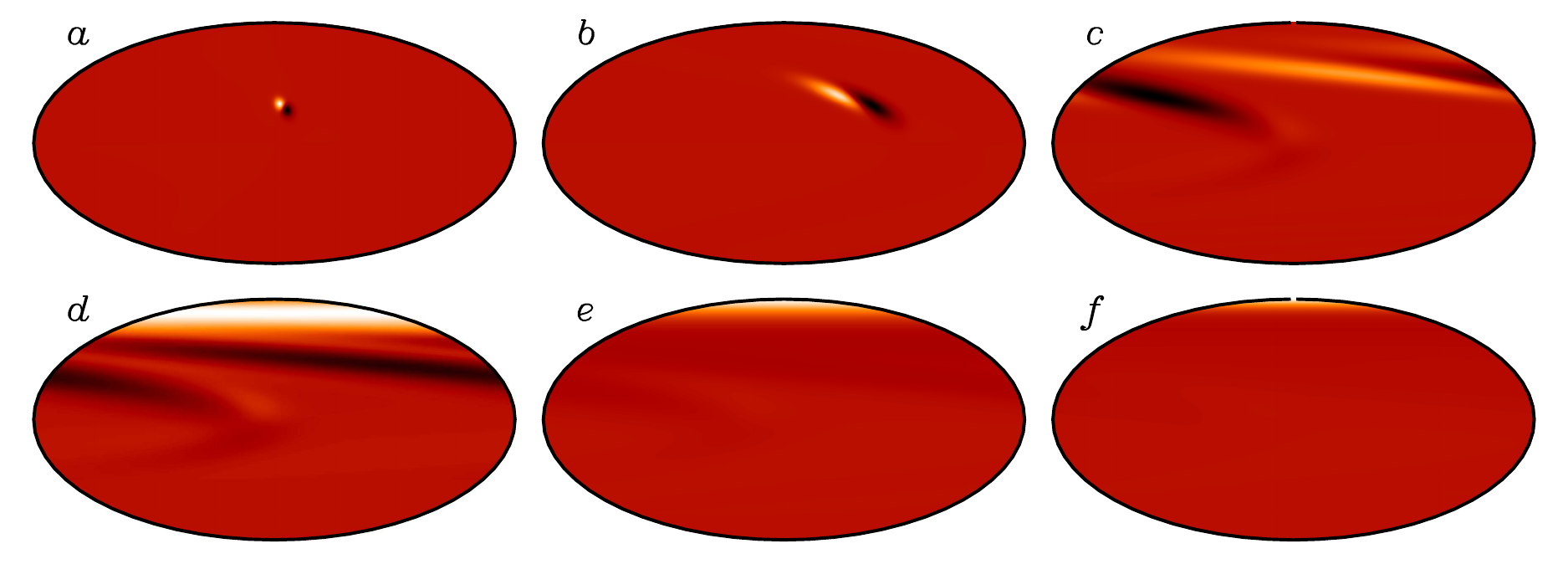}}
\caption[Time evolution of radial fields for single pair sunspots]{Time evolution of radial fields on the surface of the sun with a single pair in the 
northern hemisphere for (a) 0.025 years, (b) 0.25 years, (c) 1.02 years, (d) 2.03 years, (e) 3.05 
years and (f) 4.06 years. Here white color shows the outwards going radial field and black color 
represents inward going radial field. The color scale is set at $\pm$maximum values of the magnetic fields for 
each case. For example $\pm4.66$ G is the color scale for (a) and $\pm0.10$ G is the color scale for (f).}
\label{sfield_1}
\end{figure}

\begin{figure}[!t]
\centerline{\includegraphics[width=0.95\textwidth,clip=]{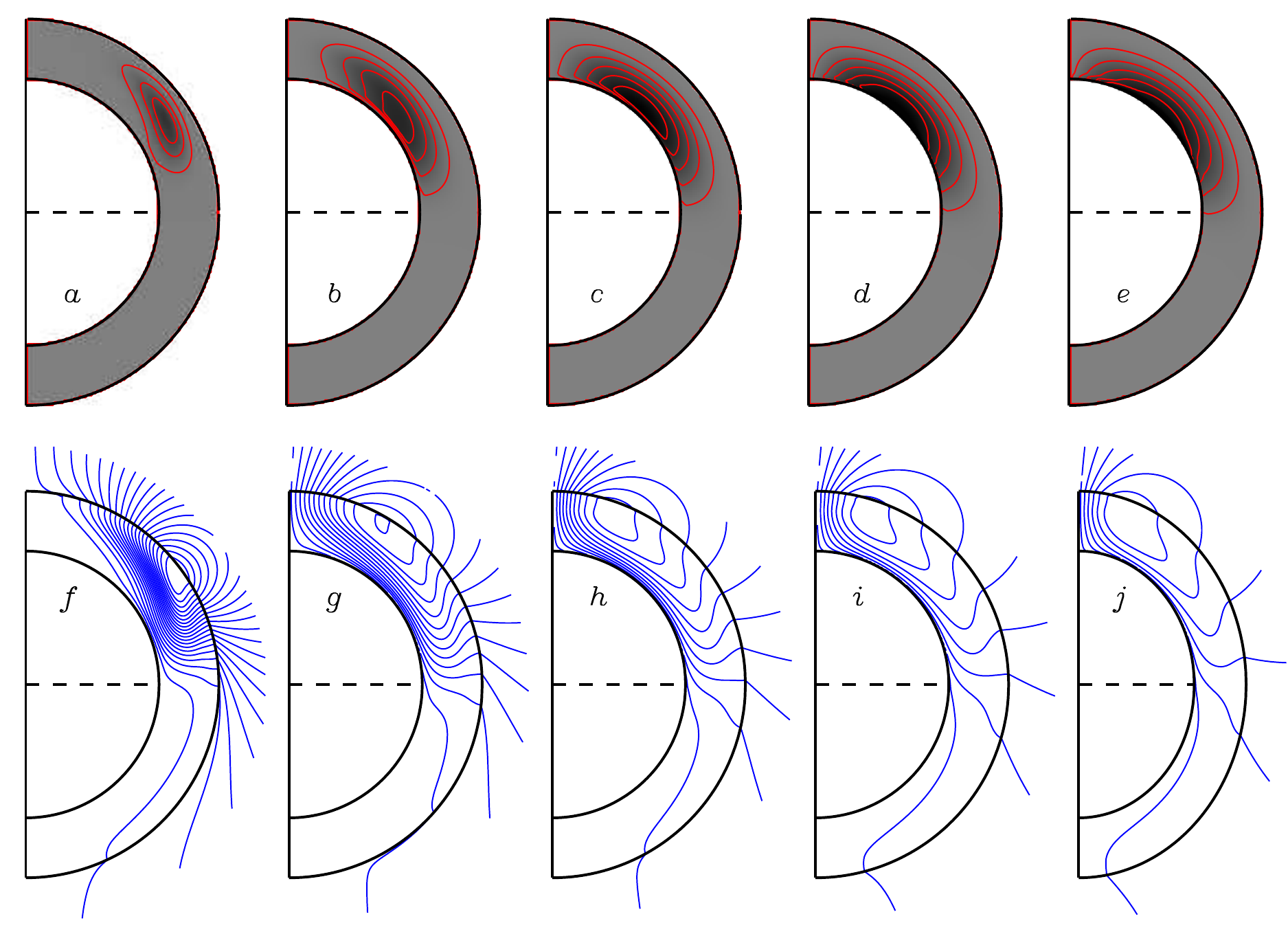}} 
\caption[Toroidal and poloidal field lines for single pair sunspots at different times]{Axisymmetric toroidal field lines (a-e) and axisymmetric poloidal field lines (f-j) are 
shown for 5 different times. Time spans are (a,f) = 1.02 years, (b,g)= 3.05 years, (c,h) = 5.08 years, 
(d,i) = 7.11 years and (e,j) = 9.15 years. Frames (a-e) represent $<B_{\phi}>$ (azimuthal averaged) 
with red and blue indicating eastward and westward fields respectively. Filled contour also
represents the mean toroidal fields. Here color scale is set 
at $\pm1.5$ G. Frames (f-j) represent the square root of poloidal magnetic potential with potential field extrapolation
above the surface (upto $r = 1.25R$) and blue color contours denote the clockwise direction of the field. 
Maximum and minimum contour levels are set corresponding to potential field strength of $\pm0.3$ G respectively.}
\label{mfield_1}
\end{figure}

We use the SpotMaker algorithm to put one sunspot pair at latitude $20^{\circ}$ with tilt
angle $40^{\circ}$. Then we allow our code to evolve the magnetic field from this sunspot
pair leading to the build-up of the polar field. Figure~\ref{sfield_1} shows snapshots of $B_r$ on
the solar surface at different times during the evolution process. This figure can
be compared with Figure~6 of \citet{YM13}. Although \citet{YM13} assumed the sunspot
pair to be initially connected to the toroidal flux system at the bottom of the
convection zone, eventually this connection would be disrupted and the evolution
of the magnetic field on the surface due to the sunspot pair in the northern hemisphere
appears to be very similar to the evolution that
we get by assuming a disconnection from the very beginning. The following
sunspot at the higher latitude has the positive polarity and we clearly see that
this positive polarity is preferentially transported to the higher latitudes. This
positive polarity region gets stretched by the differential rotation into a belt going
around the polar axis. When this belt reaches sufficiently high latitude, we see
that it is followed by a belt of negative polarity coming from the leading sunspot
which was taken at a lower latitude. The meridional circulation takes about 3
yr to bring the flux of $B_r$ to create a positive patch on the pole surrounded a
ring of negative polarity. The formation process of the negative polarity ring is clearly visible in
Figure~\ref{sfield_1}(d), but at later times it becomes weaker due to the action of diffusion
and is not clearly visible. Since the meridional circulation sinks downward at the
polar region, eventually both the positive and negative polarity magnetic fields
are advected simultaneously below the surface. This becomes clear from the field
line plots shown in Figure \ref{mfield_1}. At certain instants of time, we have averaged $B_r$ and
$B_{\theta}$ over the azimuthal direction $\phi$ to obtain the field lines.

It may be noted that the color scale for each plot in Figure~\ref{sfield_1}
is set at $\pm$maximum values of the magnetic field in each case.  This was necessary
because the magnetic field becomes weak with time. Had we used the color scale of 
Figure~\ref{sfield_1} (a) for all cases shown in Figure~\ref{sfield_1}, then the magnetic
field would be completely invisible for the plots at later times.  Though the magnetic
fields in the sunspot pair remain concentrated for shorter time than what one may suspect
from a casual look at Figure~\ref{sfield_1}, the sunspots in our simulations are
nevertheless live longer than real sunspots.  This is expected because we have assumed
the sizes of sunspots in our calculation to be larger than real sunspots.  If sunspots 
decay by the action of turbulent diffusion, then a simple application of 
the diffusion equation suggests that 
the lifetime should go as the square of the size.  A hypothetical sunspot 5 times larger
than a real sunspot should live 25 times longer than a real sunspot.

It should be kept in mind that $\int {\bf B}. d{\bf S}$ integrated over the whole solar surface has
to be zero at any time (since $\nabla. {\bf B} = 0$). This means that, during any time
interval, equal amounts of positive and negative magnetic fluxes have to disappear below 
the surface due to the subduction process. As a result, we see in Figure~\ref{sfield_1} that
the white patch at the pole (representing positive flux) remains there till all fluxes 
disappear and is not replaced by the poleward migrating dark ring (representing negative
flux, which gets subducted along with the positive flux).  In the real Sun 
undergoing successive cycles, the polar field reverses only when fluxes
of the following sunspots from the next cycle reach the pole. 
This subduction process has been seen before in axisymmetric
Babcock-Leighton / Flux-Transport dynamo models but has not been well studied within
the context of SFT models (though see \citet{Cameron12,YM13}).  
It relies on the upward advection of poloidal flux near the equator, which
leads to the emergence of oppositely-signed radial field, as shown in Figure~\ref{mfield_1}(f-j). 
This changes the net radial flux through the surface in each hemisphere
and eats away at the polar field as it is advected poleward.  
Without this low-latitude emergence, the subduction of poloidal flux
at the poles could not change the net flux through the outer surface.


In the SFT model also, a tilted sunspot pair gives rise to a polar field with the
polarity of the following sunspot surrounded by a belt of the opposite polarity.
However, since the low-latitude emergence and subsequent subduction of the mean
poloidal field is not included in the model, the net flux through each hemisphere
can only change by means of cross-equatorial transport and diffusion.
In a model of the solar magnetic field dynamics with
realistic values of various parameters, usually the diffusion time for neutralizing
the opposite magnetic polarities turns out to
be much longer than the time for their disappearance due to low-latitude emergence and subduction by the meridional circulation. 
\begin{figure}[!htbp]
\centerline{\includegraphics[width=1.0\textwidth,clip=]{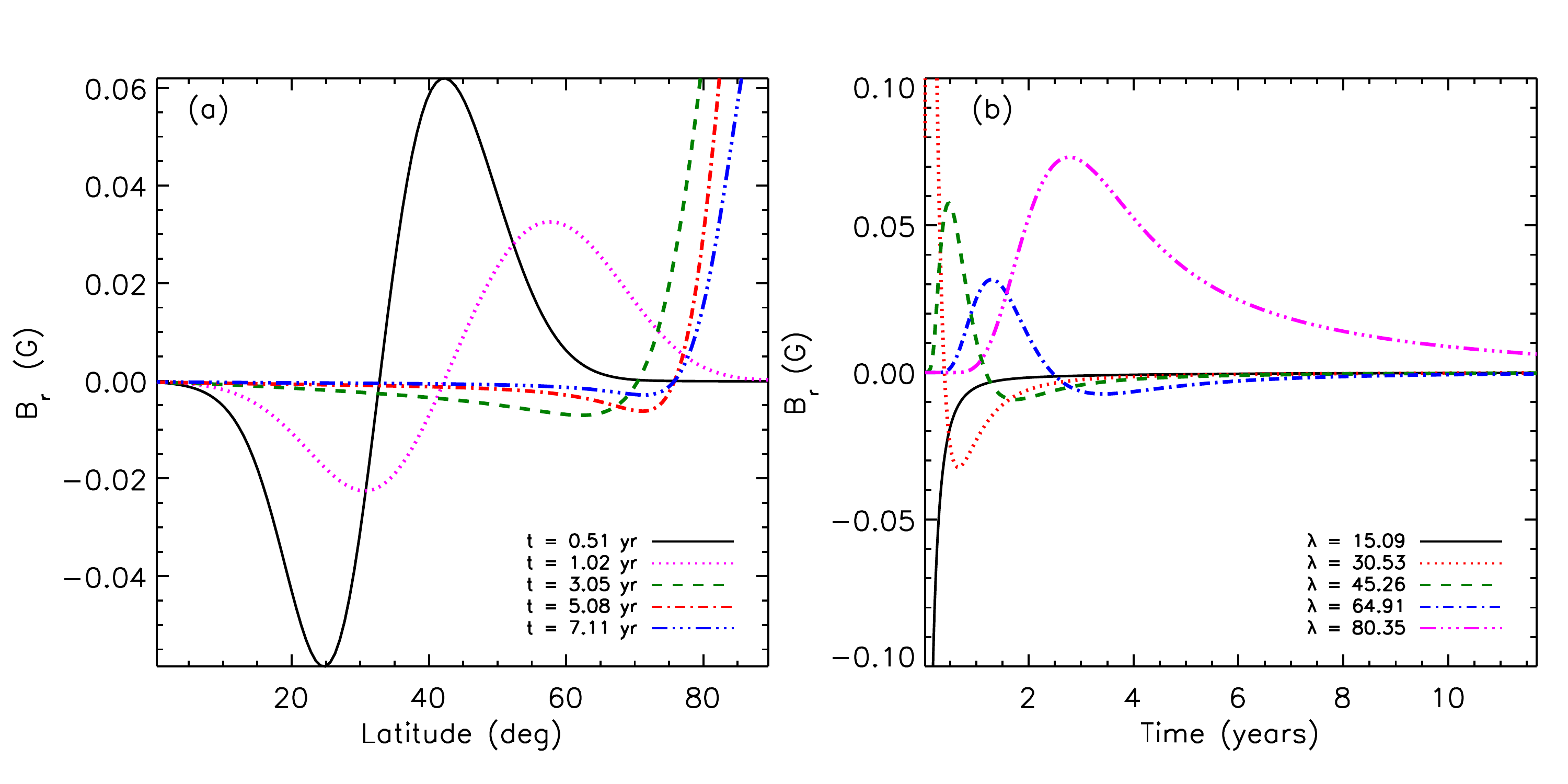}} 
\caption[Behavior of radial field at different times and at different latitudes]{(a) Behavior of radial field with latitude is plotted for different times. Radial field 
just after the emergence of the sunspot (at $20^{\circ}$ latitude) are shown in black line at 
time t = 0.51 years. Magenta dotted, green dashed, red dash dotted and blue long dash dotted lines represent variation of 
radial magnetic field with latitude at time 1.02 years, 3.05 years, 5.08 years and 7.11 years respectively. (b) Time variations 
of radial magnetic field for different latitudes are plotted. Solid black, red dotted, green dash dotted, blue dash dotted 
and magenta long dash dotted lines are for latitude $15.09^{\circ}$, $30.53^{\circ}$, $45.26^{\circ}$, $64.91^{\circ}$ 
and $80.35^{\circ}$ respectively. All units of magnetic fields are given in gauss.}
\label{rfield_1}
\end{figure}
\begin{figure}[!htbp]
%\centerline{\includegraphics[width=1.0\textwidth,clip=]{spot_2P_10.pdf}}
\centerline{\includegraphics[width=1.0\textwidth,clip=]{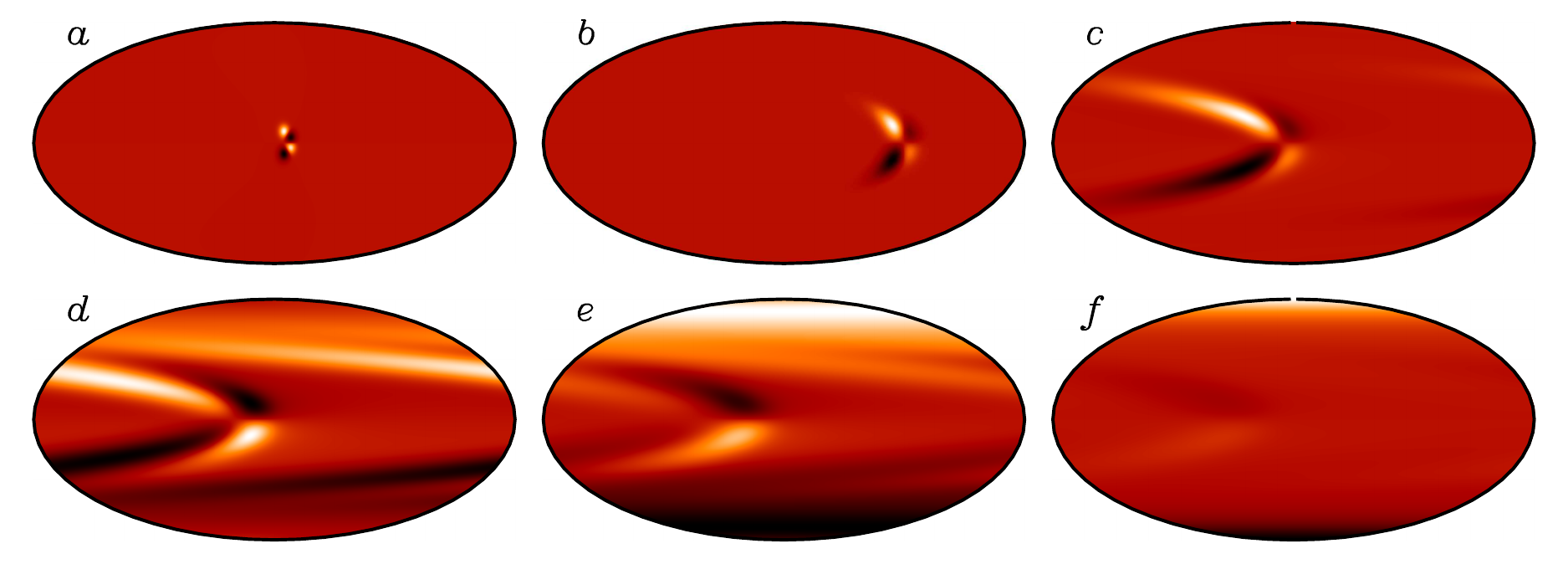}} 
\caption[Radial field evolution for two pair of sunspots on both the hemispheres]{Same as Figure~\ref{sfield_1} but for sunspot emergence in two hemispheres at $\pm5^{\circ}$ latitudes.
In this figure also color scale is set at $\pm$maximum value of the magnetic fields for 
each case.}
\label{sfield_2}
\end{figure}
As we see in Figure \ref{mfield_1}, the magnetic fields tend to sink below the surface 
while they diffuse and the disappearance of the magnetic fields at the surface
takes place in a time scale shorter than the diffusion time scale. We thus see that
the evolution of the polar field in our 3D model is qualitatively different from
what it is in the SFT model.

Figure~\ref{mfield_1} also shows the toroidal field generated in the convection zone.
Since the poloidal field has not yet reached the tachocline
to be acted upon by the radial differential rotation there, it may be worthwhile
to comment how the toroidal field is generated. As soon as we put a sunspot pair
on the surface by the SpotMaker algorithm, some toroidal field arises below the
surface at once because the magnetic loop connecting the two sunspots below the surface
would have a toroidal component.  Additionally, more toroidal field is
produced by the latitudinal differential rotation within the convection
zone.  It has been known that the latitudinal differential rotation can play an
important role in generating the toroidal field \citep{Guerrero07}.
In reality, any strong magnetic field generated within the convection
zone is expected to be quickly removed by magnetic buoyancy which is particularly
effective within the convection zone \citep{Parker75, Moreno83}. Since we
do not allow magnetic buoyancy to remove the toroidal field in the present version
of the code, the toroidal field remains where it is created. However, it may be 
noted that some fully dynamical simulations suggest persistent rings of toroidal
flux within the convection zone \citep{Brown10}.

Finally, Figure~\ref{rfield_1}(a) shows $B_r$ (averaged over $\phi$) as a function of latitude for
different times, whereas Figure~\ref{rfield_1}(b) shows $B_r$ as a function of time at different
latitudes.  A careful scrutiny of Figure~\ref{rfield_1}(a) makes it clear that the poleward meridional
circulation transports the magnetic flux to higher latitudes with time.  After about
3 yr, the polar field starts building up.  It is clear that the polar field becomes
much stronger than the fields at mid-latitudes.  This is purely a geometrical effect.
Since magnetic flux from different longitudes is brought by the meridional circulation
to the pole where it converges, it is natural that the magnetic field becomes stronger
at the pole.  It is also to be noted that we only have the polar field with polarity
corresponding to the polarity of the following sunspot at the higher latitude (positive
in the present case).  Turning to Figure~\ref{rfield_1}(b) now, we first look at the plots corresponding
to the mid-latitudes ($\approx 30^{\circ} - 65^{\circ}$).  At a mid-latitude, first the magnetic field corresponding to
the polarity of the following sunspot (positive in the present case) is brought by
the meridional circulation, followed by the magnetic field with opposite polarity
from the leading sunspot (negative in the present case) a little bit later. This is
seen in all the mid-latitude plots in Figure~\ref{rfield_1}(b).  But we should pay a special attention
to the plots for latitudes $15^{\circ}$ and $80^{\circ}$.  At the latitude of  $15^{\circ}$,
the positive magnetic field from the following sunspot is never seen, because the
following sunspot appeared at a higher latitude and the meridional circulation transported
the flux from its decay towards the pole. On the other hand, at the latitude of  $80^{\circ}$,
we see only the positive magnetic field which has been brought there from following
sunspot.  The negative magnetic field from the leading sunspot forms a negative
polarity belt around the pole, as we have already seen, and then it sinks below
the surface, so the negative magnetic field is never seen at sufficiently high latitudes.
Also, note that, although the peak value of the positive polarity field at $65^{\circ}$ is
less than that at $45^{\circ}$ (due to the action of diffusion while the magnetic field
is transported to higher latitudes), the positive polarity field again becomes strong
at $80^{\circ}$ due to the geometrical effect of converging flow bringing magnetic flux
from different longitudes.

\begin{figure}[!htbp]
%\centerline{\includegraphics[width=1.0\textwidth,clip=]{Bmean_2P_10.pdf}}
\centerline{\includegraphics[width=0.95\textwidth,clip=]{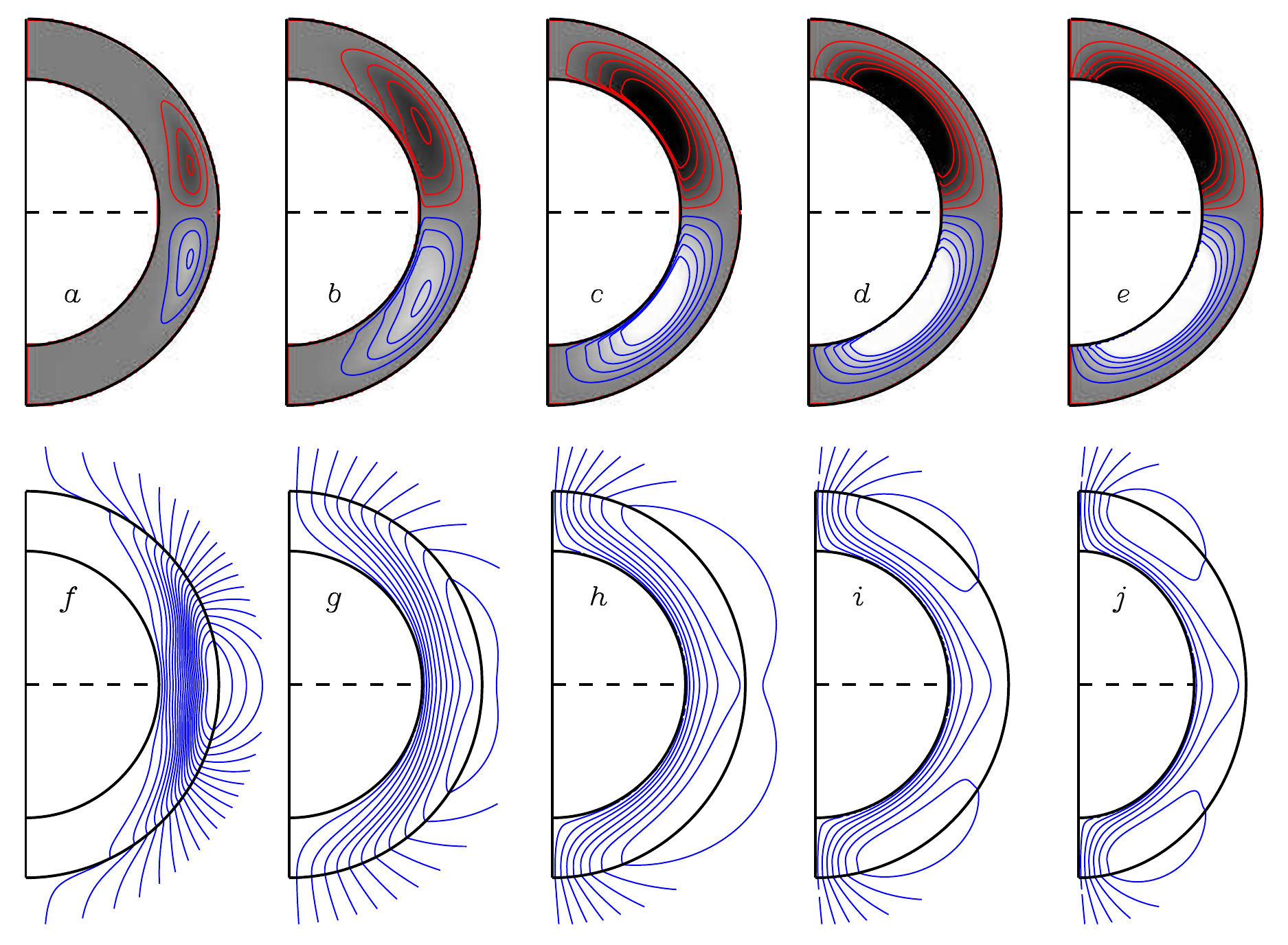}}
\caption[Toroidal and poloidal field lines are shown for two sunspots pairs]{Same as Figure~\ref{mfield_1} but for two pairs at two hemispheres at $\pm5^{\circ}$ latitudes. 
Color scale for toroidal fields is set at $\pm 1.5 G$ and contour levels corresponding to the poloidal fields strengths of $\pm0.02$ G are set as maximum and minimum respectively.}
\label{mfield_2}
\end{figure}
\begin{figure}[!htbp]
\centerline{\includegraphics[width=1.0\textwidth,clip=]{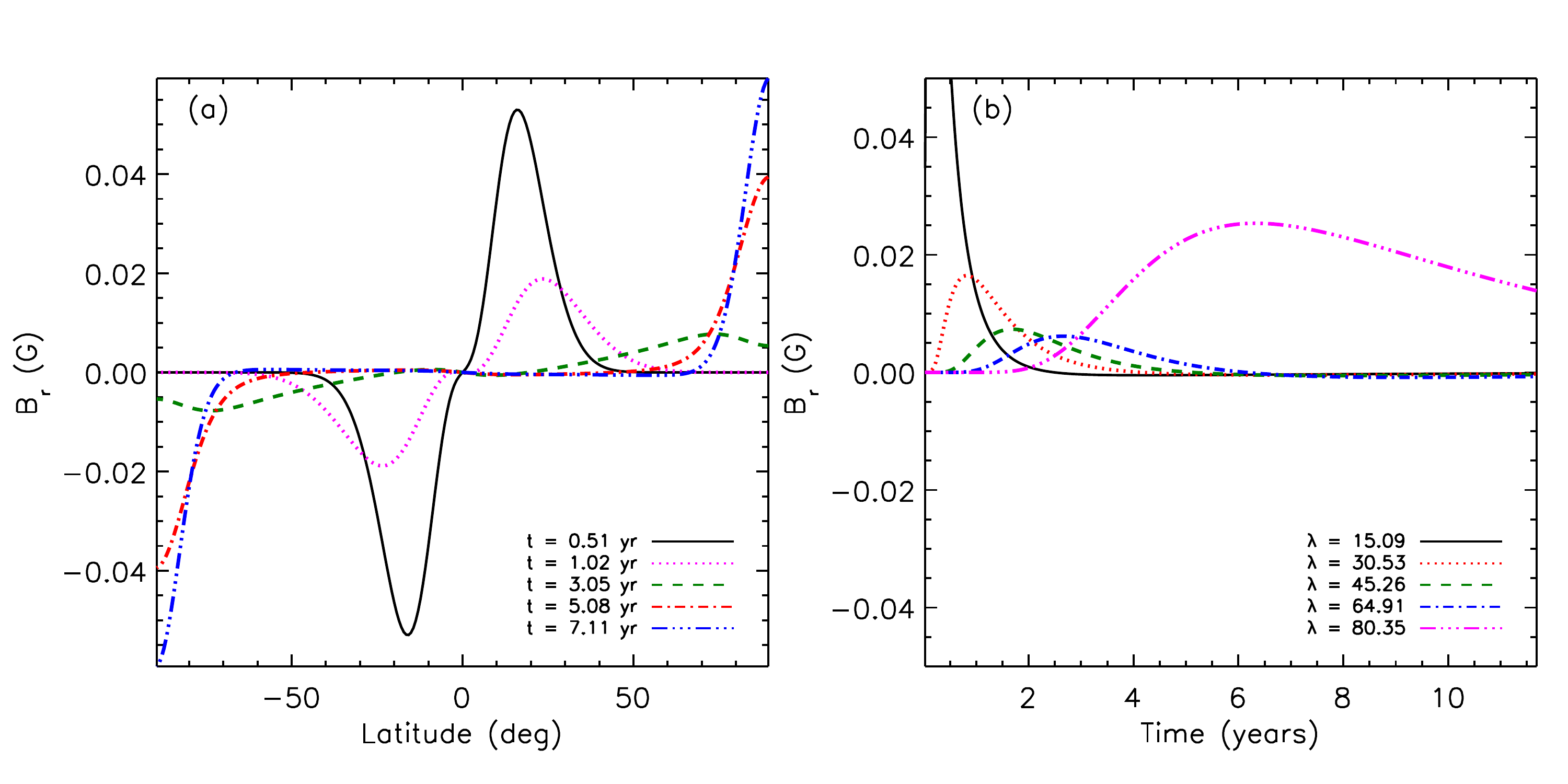}} 
\caption[Same as Figure~\ref{rfield_1} but with two pairs in two hemispheres]{Same as Figure~\ref{rfield_1} but with two pairs in two hemispheres at $\pm5^{\circ}$ latitudes.}
\label{rfield_2}
\end{figure}

\subsection{Polar fields from two sunspot pairs in two hemispheres}
\label{C6:S4_2}
The results of the 3D model differ more dramatically from the results of the SFT model 
when we put two pairs of sunspots located symmetrically in the two hemispheres.  If the 
two pairs are sufficiently close to the equator, then magnetic fluxes of the two leading 
sunspots get canceled by diffusion across the equator. In the SFT model, only the fluxes
from the following polarities are advected to the two poles and we eventually get polar
patches which are not surrounded by rings of opposite polarity as we found in the case
of the single sunspot pair. When the outward spreading of magnetic field from the polar
patches by diffusion is eventually balanced
by the inward advection by the meridional circulation, we reach an asymptotic steady
state in the SFT model, with an asymptotic magnetic dipole which does not change with time.
This is seen in Figure~6 of \citet{Jiang14}.  As we shall discuss now, we get a completely
different result from the 3D model.

We see in Figure~\ref{sfield_2} that polar magnetic patches form with the polarity of
the succeeding sunspots. A careful look at this figure, shows some evidence of opposite 
polarity (i.e.\ opposite of what we see in the poles) at mid-latitudes 
even when we start from two sunspots placed symmetrically at sufficiently low latitudes in
both the hemispheres. The physics of what is happening becomes clear from the plot of
field lines shown in Figure~\ref{mfield_2}. After the fluxes from the leading sunspots near the equator
cancel, we see that initially we get poloidal field lines spanning both the hemispheres.
A look at the field line plots makes it clear that we shall have $B_r$ only at high latitudes
in the early stages of the evolution of the magnetic field. As the meridional circulation
drags the poloidal field towards the poles, we find that eventually the polar fields in the
two hemispheres get detached, as a result of which $B_r$ again appears at lower latitudes
having the opposite polarity of $B_r$ at high latitudes. This is purely a result of the
3D structure of the magnetic field and cannot happen in the SFT model. There would not be
a source for creating $B_r$ at low latitudes in the SFT model and such fields would never
appear in that model.  Because of the breakup of the poloidal field in the two hemispheres
and the appearance of $B_r$ with opposite polarity in the low latitudes, it is possible
for the poloidal magnetic field in the 3D model to be subducted below the surface as the
meridional circulation sinks downward in the polar regions. Thus, in contrast to the SFT
model in which polar fields have nothing to cancel them and therefore persist, the polar
field disappears after some time in the 3D model. 

Though this result is notable, it may be offset to some extent by efficient magnetic 
pumping.  Using a 2D (axisymmetric) model \citet{karak16} have shown that downward 
magnetic pumping due to strongly stratified convection in the solar surface layers can 
suppress the upward diffusion and advection of toroidal and poloidal fields.  This, 
in turn, can produce steady polar fields that might persist indefinitely.  We will 
investigate the role of magnetic pumping in future work.

Figure~\ref{rfield_2}(a) is similar to Figure~\ref{rfield_1}(a) except that latitudes now cover from 
$-90^{\circ}$ to $90^{\circ}$. In this figure, we clearly see that around 1 yr, we had only 
positive $B_r$ in the northern hemisphere and negative $B_r$ in the southern hemisphere, 
but afterwards very weak $B_r$ having sign opposite to the sign at the high latitudes
developed at low latitudes. Figure~\ref{rfield_2}(b), which is similar to Figure~\ref{rfield_1}(b), shows 
that eventually $B_r$ disappears at the surface in this 3D model, exactly similar to what happens when we
put only one sunspot pair on the solar surface.

\begin{figure}[!t]
\centering

\begin{tabular}{cc}
\includegraphics*[width=0.5\linewidth]{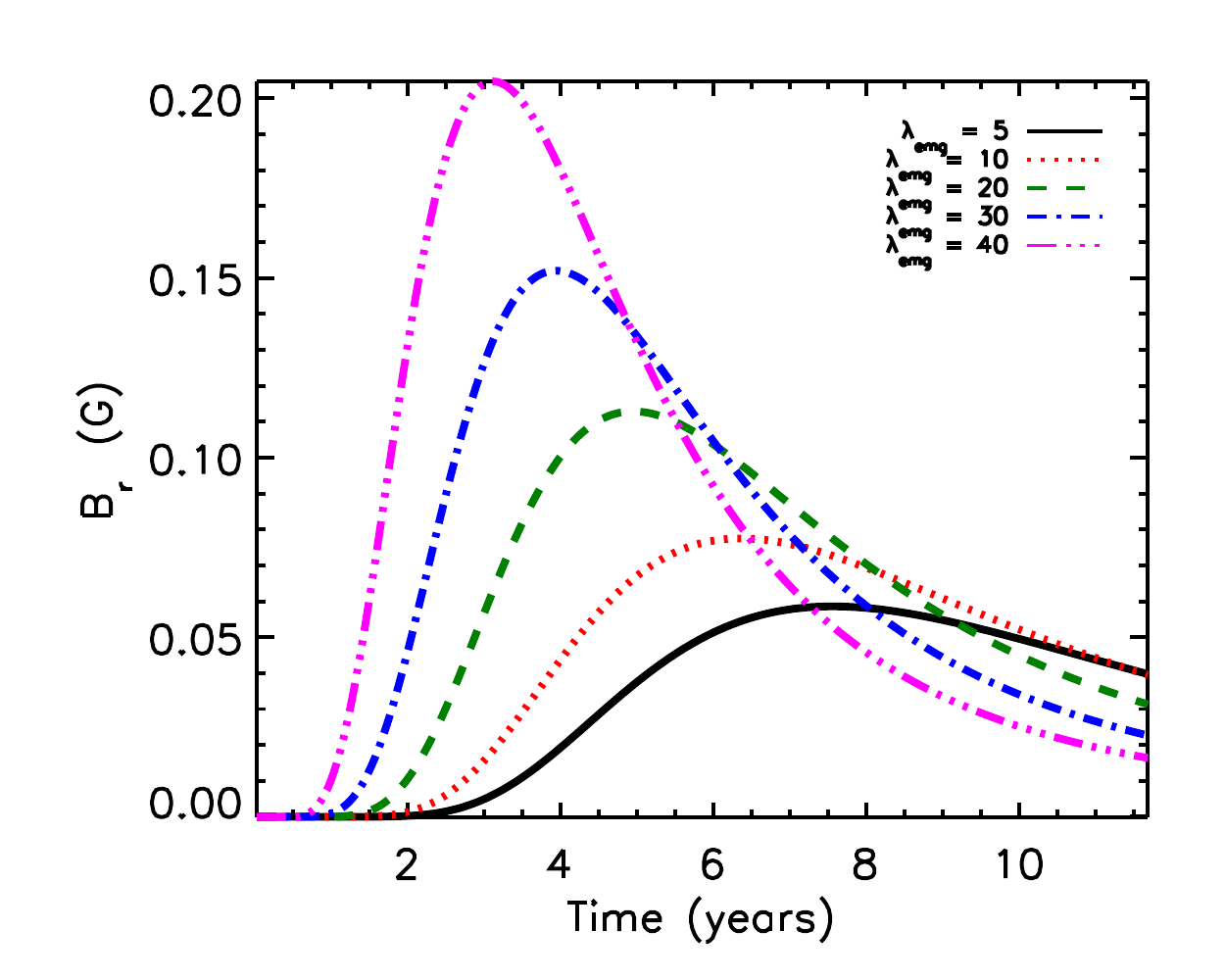} & \includegraphics*[width=0.47\linewidth]{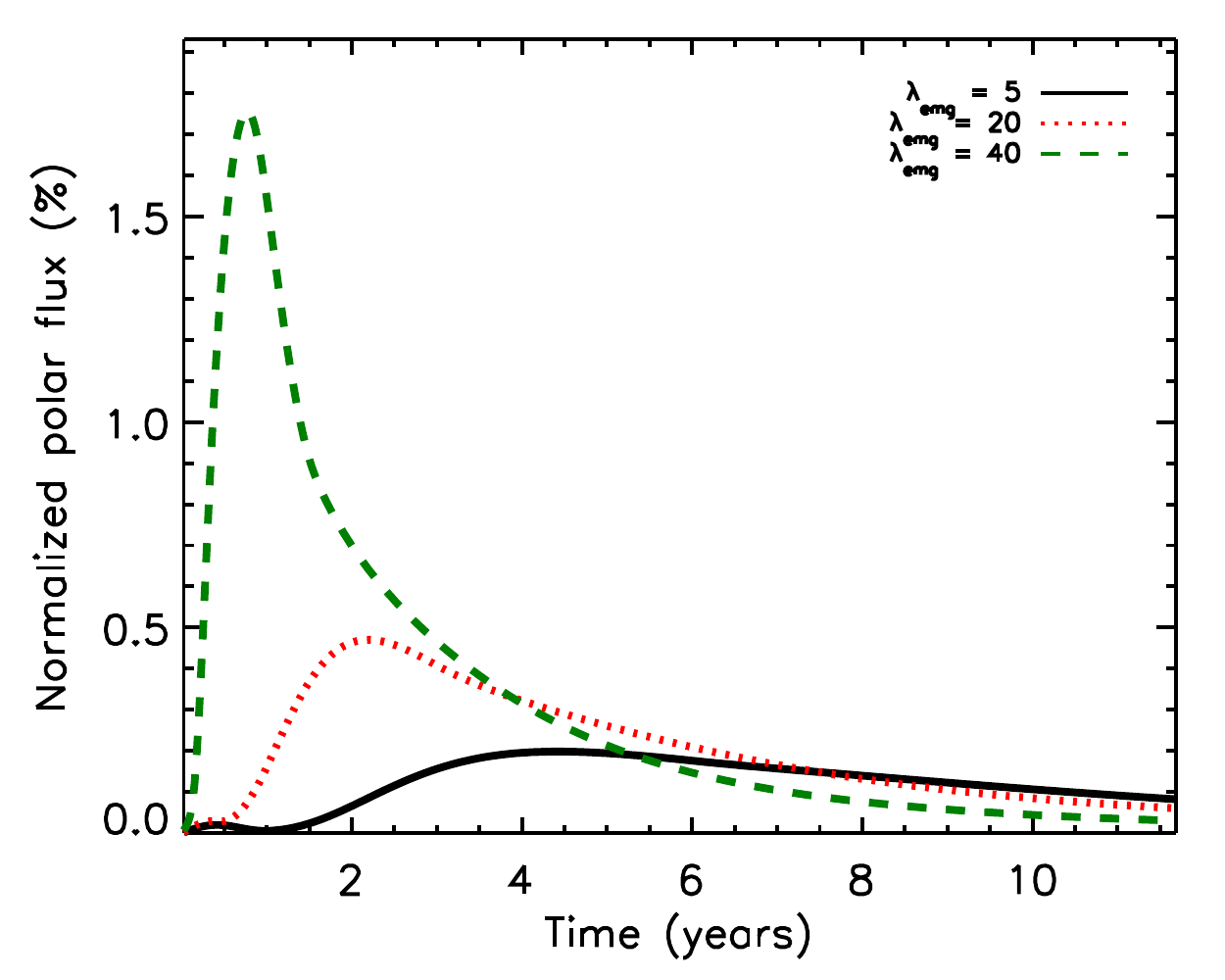}\\
                (a)                                           &                   (b)\\
 \end{tabular}                
\caption[Polar field and Polar flux evolution for two sunspot pairs]{ (a) Polar field evolution with time for different emergence angle $\lambda_{emg}$ of sunspot pairs
in both the hemisphere. Black solid, red dotted, green dashed, blue dash dotted and
magenta long dash dotted lines represent the polar field for the sunspot emergence at $5^{\circ}$, 
$10^{\circ}$, $20^{\circ}$, $30^{\circ}$ and $40^{\circ}$ respectively. Magnetic field is in mG and time is given in years. (b) Polar flux evolution with time for different emergence angle $\lambda_{emg}$ of sunspot pairs
in both the hemisphere. Black solid, red dotted, green dashed lines 
represent the percentage of normalized polar flux able to reach the pole for the sunspot emergence at $5^{\circ}$, 
$20^{\circ}$ and $40^{\circ}$ respectively.}
\label{pfield_2}
\end{figure}

We carry on such calculations by putting two sunspot pairs symmetrically at different
latitudes in the two hemispheres. Figure~\ref{pfield_2}(a) shows how the polar field evolves with time
for sunspot pairs placed at different latitudes.  When the sunspot pairs are placed at
high latitudes, the magnetic flux is brought to the poles without too much diffusion and
the polar field is stronger. Eventually the polar field disappears in all the cases due to
emergence and subduction by the meridional circulation, as we have already discussed.
This figure can be compared with the left panel of Figure~6 of \citet{Jiang14}.  Such
a comparison makes the difference between the 3D model and SFT model completely clear.
In the SFT model, only if the sunspot pairs are put at sufficiently high latitudes so
that cross-equatorial diffusion is negligible, fluxes of both polarity are advected to
the polar regions and eventually the axial dipole moment becomes zero.  If the sunspot
pairs are put at low latitudes in the SFT model, only the fluxes from the following sunspots
reach the poles and give rise to an asymptotic axial dipole.  The situation is completely
different in the 3D model, although we see that the polar field persists for a longer time
when the initial sunspot pairs are put at lower latitudes. So, in that sense, sunspot
pairs appearing in lower latitudes are somewhat more effective in creating the polar
field even in the 3D model. This is in agreement with the claim of \citet{DasiEspuig10}
that we have a better correlation between the average tilt of a cycle and the strength of
the next cycle if more weight is given to sunspot pairs at low latitudes when computing
the average tilt. 
%\begin{figure}[!htbp]
%\centerline{\includegraphics[width=0.5\textwidth,clip=]{Figures/chap6/fig13.pdf}} 
%\caption{Polar field evolution with time for different emergence angle $\lambda_{emg}$ of sunspot pairs
%in both the hemisphere. Black solid, red dotted, green dashed, blue dash dotted and
%magenta long dash dotted lines represent the polar field for the sunspot emergence at $5^{\circ}$, 
%$10^{\circ}$, $20^{\circ}$, $30^{\circ}$ and $40^{\circ}$ respectively. Magnetic field is in mG and time is given in years.}
%\label{pfield_2}
%\end{figure}

%\begin{figure}[!htbp]
%\centerline{\includegraphics[width=0.5\textwidth,clip=]{Figures/chap6/fig14.pdf}} 
%\caption{Polar flux evolution with time for different emergence angle $\lambda_{emg}$ of sunspot pairs
%in both the hemisphere. Black solid, red dotted, green dashed lines 
%represent the percentage of normalized polar flux able to reach the pole for the sunspot emergence at $5^{\circ}$, 
%$20^{\circ}$ and $40^{\circ}$ respectively.}
%\label{pflux_2}
%\end{figure}

\begin{figure}[!htbp]
\centerline{\includegraphics[width=1.0\textwidth,clip=]{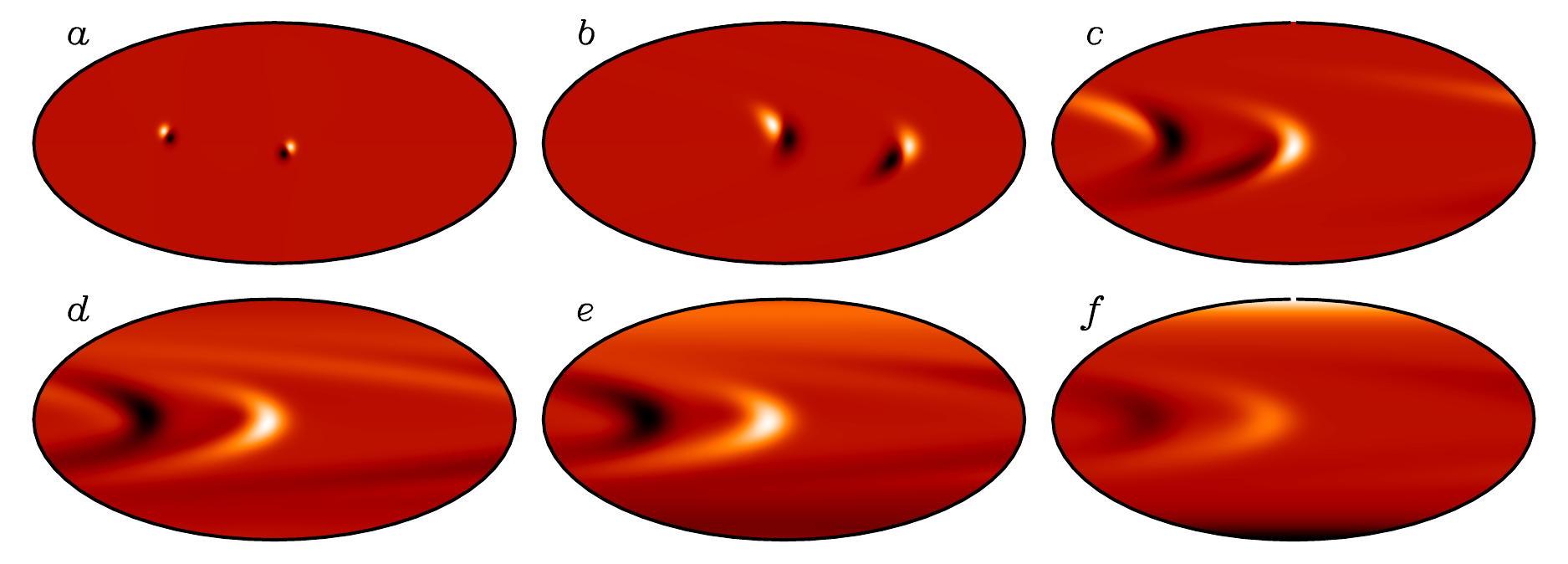}} 
\caption[Same as Figure~\ref{sfield_2} but sunspot pairs are placed at different longitudes]{Same as Figure~\ref{sfield_2} but sunspot pairs are placed at longitude $90^{\circ}$ on
northern hemisphere and at longitude $180^{\circ}$ on southern hemisphere.}
\label{sfield_3}
\end{figure}

We have also calculated the polar magnetic flux for two sunspot pairs emerging on the two hemispheres,
to find out how much flux from the sunspots reaches the poles.
We calculate the polar flux by integrating $B_r$ over only those regions of the surface between
$60^ {\circ}$ latitude and the pole where $B_r$ has one sign (positive in the north pole). While putting the sunspot pairs by hand using
SpotMaker algorithm, we injected $1\times 10^{22}$ \rm{Mx} flux in each spot.
A normalized polar flux is estimated by dividing the signed flux  
by the input flux ($1\times10^{22}$ \rm{Mx}). In Figure~\ref{pfield_2}(b), we have shown the percentage of normalized polar flux with time
for the spot pairs emerging at different latitudes. It is evident from this figure that 
around $1.76 \%$ of the input flux can reach the pole when the spot pair emerges at a high latitude like $40^{\circ}$,
whereas $0.2 \%$ of the input flux can reach the pole when the spot pair is at a low latitude like $5^{\circ}$.
Keeping in mind that we have used an unrealistically high tilt of $40^{\circ}$, we point out that the flux reaching
the poles will be less for more realistic tilts. It is instructive to compare our result with relevant
observational data. \citet{Schrijver94} and \citet{Solanki02a} analyzed the 
NSO Kitt Peak magnetograph data and estimated the maximum active regions flux during solar maxima to be around $5 \times 10^{23}$ \rm{Mx}.
\citet{Munoz12} calibrated century long polar faculae data from Mount Wilson Observatory and estimated the time evolution of the polar flux, 
finding its maximum value to be around $1.5\times10^{22}$ \rm{Mx} for an average cycle. Although these values are not from 
a single dataset and many other observational constraints should be taken into account, a simple division of these values of
flux quoted above suggests that around $3\%$ of the sunspots flux can contribute in the
polar flux. Our theoretical model gives a value having the same order of magnitude, although our theoretical
values are a little bit on the lower side.

All the results presented so far for two sunspot pairs in different hemispheres were obtained by putting
both the pairs in the same longitude. This helped in magnetic fluxes of the two leading sunspots canceling
each other by diffusing across the equator. One important question is whether the final outcome will be
different if the two sunspot pairs in the two hemispheres are widely separated in longitude. Figure~\ref{sfield_3} shows
the surface evolution of magnetic flux in such a case, which can be compared with Figure~\ref{sfield_2}.
We find that the magnetic fluxes from the following sunspots in the two hemispheres are carried towards
the pole exactly as in Figure~\ref{sfield_2}.  However, the evolution of magnetic fluxes from the leading
sunspots is quite intriguing. Because of the gap in longitude, these fluxes cannot cancel with each other
across the equator so easily. However, these fluxes still diffuse across the equator, as seen in Figure~\ref{sfield_3}  ,
and, if we average over longitude, the averaged values are found to be virtually identical with the averaged
values that we get in the case of Figure~\ref{sfield_2}.  When we plotted figures similar to Figure~\ref{mfield_2} 
and Figure~\ref{rfield_2} for this case, they turned out to be indistinguishable from Figure~\ref{mfield_2} 
and Figure~\ref{rfield_2}.

\begin{figure}[!htbp]
\centerline{\includegraphics[width=0.64\textwidth,clip=]{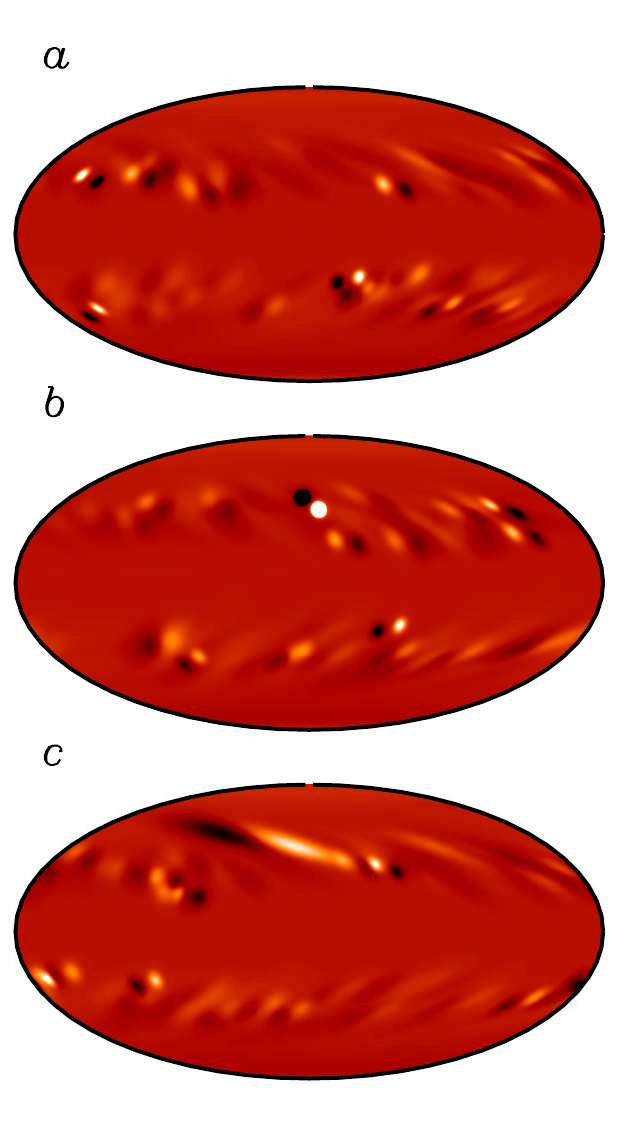}} 
\caption[Radial Magnetic field structures with an ``anti-Hale" sunspot pair]{Radial Magnetic field structures are shown for the case when ``anti-Hale" sunspot pair appears at $40^{\circ}$ latitude and at middle phase of the cycle. (a) Prior to 3 month before the anti-Hale sunspot pair to be appeared, (b) During the emergence of anti-Hale sunspot pair and (c) After 3 month of the anti-Hale sunspot pair has emerged. Here white color shows the outwards going radial fields and black color represents inward going radial fields. The color scale is set at $\pm100$ kG for all three cases.}
\label{ss_ahale}
\end{figure}

\begin{figure}[!t]
\centerline{\includegraphics[width=1.0\textwidth,clip=]{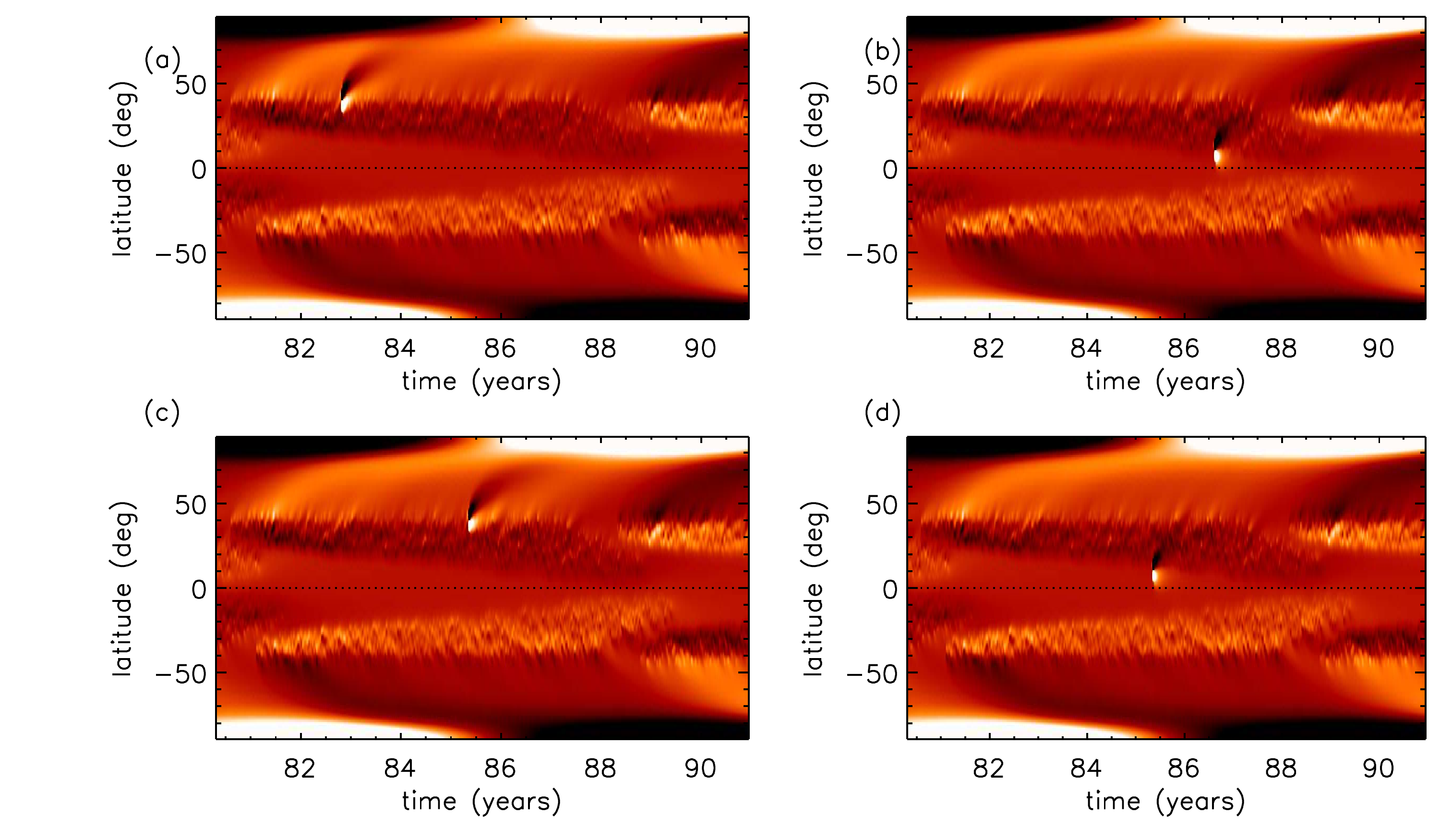}} 
\caption[Butterfly diagram with an ``anti-Hale" sunspot pair at different latitudes and at different phases of the solar cycle]
{Butterfly diagram with an ``anti-Hale" sunspot pair at different latitude and at different phase of the solar cycle. 
(a) At early phase of the cycle and at $40^{\circ}$ latitude. (b) Late phase of the cycle and at $10^{\circ}$ latitude. 
(c) Middle phase and at $40^{\circ}$ and (d) Middle phase and at $10^{\circ}$ latitude. Color scale is set at $\pm15$ kG for all four cases.}
\label{bfly_ahale}
\end{figure}

\section{The contribution of bipolar sunspots not obeying Hale's law}
\label{C6:S5}
Joy's law for tilts of sunspot pairs is only a statistically average law.  We see a spread
of tilt angles around Joy's law. This spread is believed to be caused by the action of
turbulence on rising flux tubes \citep{Longcope02,weber11} and is one of the main sources
of irregularity in the solar cycle \citep{CCJ07, CK09,Chou13}.
It is well known that some bipolar sunspots appear with wrong magnetic polarities not 
obeying Hale's polarity law.  Because of the spread in tilt angles around Joy's law, 
it is certainly expected that a few outliers in this spread would violate Hale's law.
\citet{SK12} estimated that about 4\% of medium and large sunspots violate Hale's 
law---see their Figure 7.  Since the number of such sunspots is small, it is not 
surprising that due to statistical fluctuations more of such sunspots 
violating Hale's law may appear in some particular cycles compared to other cycles. 
This fact assumes significance in the light of the suggestion made by \citet{Jiang15} on the basis of 
their SFT calculations that a few large ``anti-Hale'' sunspot pairs may significantly decrease
the strength of the polar field produced at the end of the cycle.  Especially, \citet{Jiang15}
suggested that the weak polar field at the end of cycle 23 was caused by a few prominent 
anti-Hale sunspot pairs present in that cycle.  In contrast, they argue that not too many such
anti-Hale sunspot pairs appeared in cycles 21 and 22, as a result of which such a decrease
of the polar field did not happen in those cycles.

Since we have seen that some insights gained from SFT calculations have to be modified---especially 
results connected with the build-up of the polar field---on the basis of more
realistic and complete 3D kinematic dynamo calculations, we now address the question whether
anti-Hale sunspot pairs have a large effect on the polar field even in 3D kinematic dynamo 
models. We now use our reference model presented in Section~\ref{C6:S3} and put a large anti-Hale sunspot
pair by hand to study its effect on the build-up of the polar field. To make its effect
visible, we take this anti-Hale sunspot pair to carry 25 times
the magnetic flux carried by the other regular sunspots and to have tilt angle $30^{\circ}$.
We can say that the tilt angle is $-30^{\circ}$, if we define the tilt angle by
following the convention that its value is positive for sunspot pairs obeying Hale's
law.
 
\begin{figure}[!t]
\centerline{\includegraphics[width=0.75\textwidth,clip=]{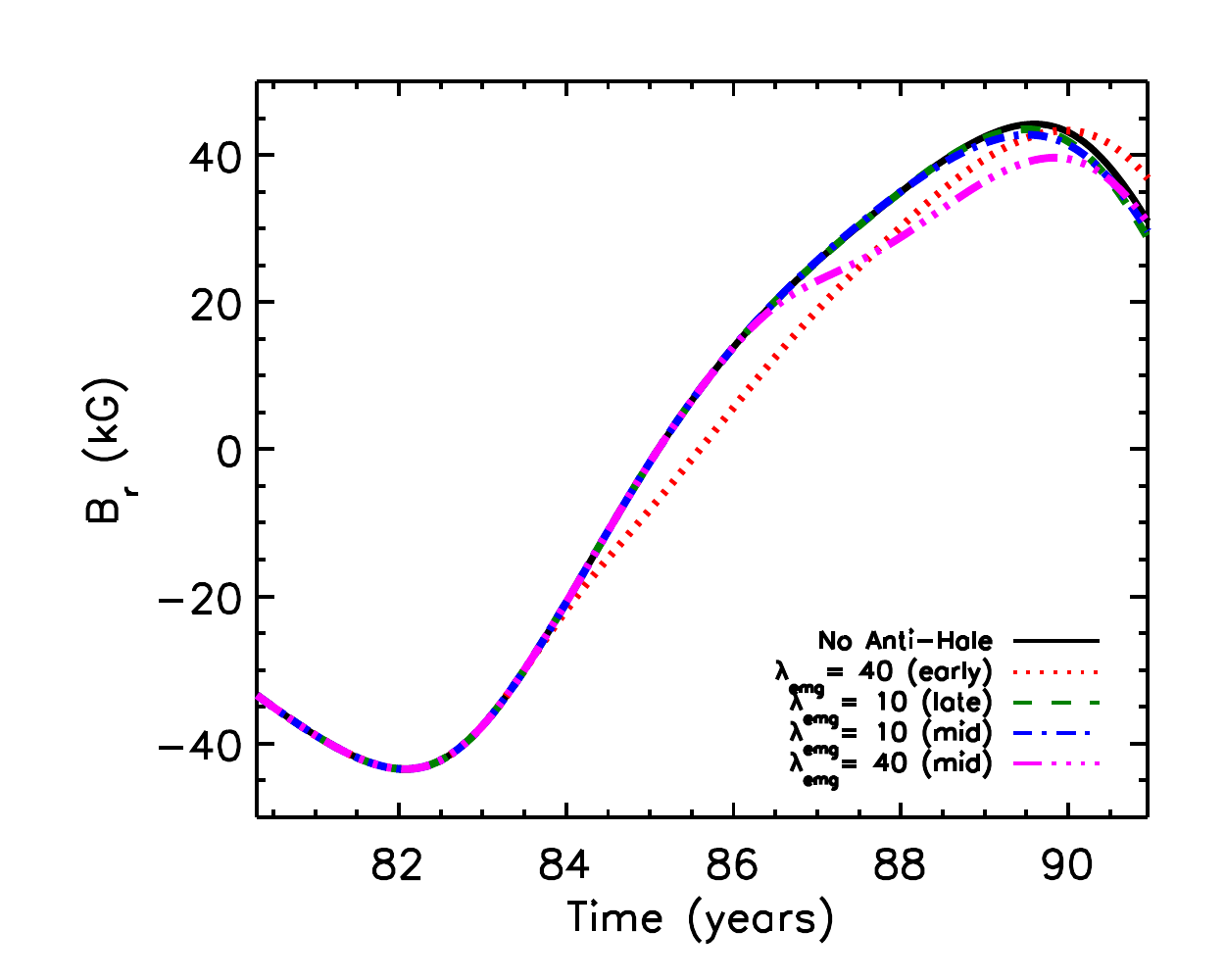}} 
\caption[Polar field evolution with time for one complete solar cycle with ``anti-Hale" sunspot pair at different 
location and different time]{Polar field evolution with time for one complete solar cycle with ``anti-Hale" sunspot pair at different 
location and different time of the cycle. Solid black line represents the regular cycle with no anti-Hale sunspot pair. 
Red dotted line indicates the poloidal field evolution with an anti-Hale pair at $40^{\circ}$ latitude at early phase 
of the cycle. Green dashed line represents poloidal field with an anti-Hale pair at $10^{\circ}$ and late phase. blue 
dashed and magenta long dashed lines indicate the poloidal field with an anti-Hale pair at middle of the cycle but 
at $10^{\circ}$ and at $40^{\circ}$ latitude respectively.}
\label{pfield_ahle}
\end{figure}

We want to understand how the effect of the anti-Hale sunspot pair depends on the emergence
latitude as well as the phase of the cycle when it makes its appearance. So we consider
four different cases. Since sunspots appear at high latitudes in the early phase of the
cycle and at low latitudes in the late phase, we consider one case by putting the anti-Hale
sunspot pair at the high latitude of $40^{\circ}$ in the early phase and another case
by putting the pair at the low latitude of $10^{\circ}$ in the late phase. The two other
cases considered involve putting the large anti-Hale sunspot pair at $40^{\circ}$ and
$10^{\circ}$ (in separate case studies) in the middle phase of the cycle. 
The radial fields on the surface in mollweide projection is shown for
a case where ``anti-Hale" sunspot pair is placed at $40^{\circ}$ latitude during the middle phase of the
cycle in the Figure~\ref{ss_ahale}.
Figure~\ref{bfly_ahale} shows how $B_r$ evolves in a time-latitude plot (a ``butterfly diagram'') for these four cases.
The effect on the polar field can be seen more clearly in Figure~\ref{pfield_ahle} where we plot the time evolution
of the polar field for these four cases, along with the reference case without an anti-Hale
sunspot pair.

It is clear from Figure~\ref{pfield_ahle} even a very large anti-Hale sunspot pair placed at a low latitude
like $10^{\circ}$ does not have much effect on the polar field. Presumably the opposite
fluxes from the two sunspots neutralize each other before they reach the poles. This becomes
quite apparent by looking at Figures~\ref{bfly_ahale}(b) and \ref{bfly_ahale}(d). We see that the sunspot pairs 
at low latitudes produce a kind of ``surge'' behind them, but it does not reach the poles. The effect
of anti-Hale pairs at higher latitudes is certainly much more pronounced. We see in Figures~\ref{bfly_ahale}(a)
and \ref{bfly_ahale}(c) that the surges behind these anti-Hale pairs reach the pole in these situations. If 
an anti-Hale sunspot pair appears at $40^{\circ}$ in the early phase of the cycle, then we
see in Figure~\ref{pfield_ahle} that the build-up of the polar field is weakened and delayed, but eventually
the polar field reaches almost the strength we would expect in the absence of the anti-Hale
sunspot pair.  However, when the anti-Hale sunspot pair is put at $40^{\circ}$ in the middle 
phase of the cycle, it is clear from Figure~\ref{pfield_ahle} that polar field can be reduced by about
17\%. But remember that we get this large reduction by assuming the anti-Hale sunspot pair to be unrealistically
large. Our conclusion is that anti-Hale sunspot pairs do
affect the build-up of the polar field---especially if they appear at high latitudes in
the middle phase of the cycle---but the effect does not appear to be very dramatic.
As for the suggestion of \citet{Jiang15} that the weakness of the polar field at
the end of cycle~23 was due to the appearance of several anti-Hale sunspot pairs, we feel that this is an interesting suggestion
which merits further detailed study in order to arrive at a firm conclusion.
%\begin{figure*}[!htbp]
%\centerline{\includegraphics[width=0.7\textwidth,clip=]{fields_ahale.pdf}} 
%\caption{}
%\label{pfield_ahle}
%\end{figure*}

\section{Conclusion}
\label{C6:S6}
Historically the evolution of the Sun's magnetic field with the solar cycle has been studied
extensively through two classes of 2D theoretical models: the 2D kinematic dynamo model and
the surface flux transport (SFT) model. We argue that the 3D kinematic dynamo model incorporates
the attractive aspects of both, while being free from the limitations of both.  On the one hand,
this model can treat the Babcock--Leighton mechanism more realistically in 3D, which is not
possible in the 2D kinematic dynamo model.  On the other hand, it includes the vectorial nature
of the magnetic field and various subsurface processes which are left out in SFT models.
\citet{Cameron12} have pointed out that the results of SFT model agree with the results of 2D flux
transport dynamo model on the inclusion of a downward pumping.  

In order to study the build-up of the Sun's polar field with a 3D kinematic dynamo model, we
first construct an appropriate self-excited model. The poloidal field generated by the Babcock--Leighton
mechanism near the solar surface has to be transported to the tachocline in order for the solar
dynamo to work.  This transport can be achieved in two ways: (i) due to advection by the
meridional circulation; or (ii) due to diffusion across the convection zone. There are reasons
to believe that (ii) is the appropriate transport mechanism inside the Sun. The earlier papers
by \citet{MD14} and \citet{MT16} presented self-excited dynamo models dominated by advection by
the meridional circulation. We believe that we are the first to construct self-excited 3D kinematic
dynamo model dominated by diffusion. We have briefly looked at the question of parity, although
the limitation of computer time prevented us from an exhaustive study of the subject.

We use this dynamo model to study how the polar field builds up from the decay of one tilted
bipolar sunspot pair and two symmetrically situated bipolar sunspot pairs in the two hemispheres.
We find that the polar field which arises from such sunspot pairs ultimately disappears due to
the emergence of poloidal flux at low latitudes and its subsequent subduction by the meridional flow.
%subduction by the meridional circulation which sinks underneath the surface in the polar region.
This process is not included in the SFT models, in which the polar field can only be neutralized by
diffusion with field of opposite polarity. So we conclude that SFT models do not capture the dynamics
of polar fields realistically and one has to be cautious in interpreting the SFT results pertaining
to polar magnetic fields. Our results differ most dramatically from the SFT results when we put two
symmetric bipolar sunspot pairs in the two hemispheres very near the equator. Then the magnetic
fields of the two leading sunspots on the two sides of the equator cancel each other. At the same
time, the magnetic fields of the following sunspots which formed at higher latitudes are advected
by the meridional circulation to the poles, ultimately causing a magnetic dipole of the Sun. In
the SFT model, this is the whole story and we get an asymptotically steady dipole. In our 3D kinematic
model, on the other hand, magnetic field lines between the two hemispheres can get detached when
they are pulled by the meridional circulation in the opposite directions. As a result, radial magnetic
fields with signs opposite to the polar fields develop in the lower latitudes. This is not possible
in the SFT model in the absence of any source of radial magnetic field in the lower latitudes.
Finally the detached magnetic loops in the two hemispheres are subducted underneath the surface
by the meridional circulation, contradicting the SFT result that the magnetic dipole of
the Sun would be asymptotically steady in this state. While the SFT models played a tremendously
important historical role in our understanding of how the magnetic field on the solar surface
evolves, we should keep in mind that these models cannot capture certain aspects of the dynamics of
the Sun's polar magnetic fields due the intrinsic limitations of these models.

Finally, we look into the provocative question of whether a few large sunspot pairs violating
Hale's law could have a large effect on the strength of the polar field. We find that such
anti-Hale sunspot pairs do produce some effect on the Sun's polar field---especially if they
appear at higher latitudes during the mid-phase of the solar cycle---but the effect is not 
very dramatic.  The question of whether a few large anti-Hale sunspot pairs could be the principal
cause behind the weakness of the polar field at the end of some cycles like cycle~23 needs to
be analyzed carefully.

%footnote
\blfootnote{This chapter is based on \citet{HCM17}.}

%% file: chapter7.tex
\begin{savequote}[100mm]
``There is nothing new under the sun
but there are lots of old things we
don't know"
\qauthor{--Ambrose Bierce, The Devil's Dictionary}
\end{savequote}

\def\pf{poloidal field}
\def\bl{Babcock--Leighton}
\def\ftdm{flux transport dynamo model}
\def\Rs{R_{\odot}}
\def\mps{m~s$^{-1}$}
\def\mc{meridional circulation}
\def\ftd{flux transport dynamo}
\def\pa{\partial}
\def\er{\mbox{erf}}
%\newcommand{\ov}{\overline}

% Mark's additions
\newcommand{\uvr}{\mbox{\boldmath $\hat{r}$}}
\newcommand{\uvt}{\mbox{\boldmath $\hat{\theta}$}}
\newcommand{\uvp}{\mbox{\boldmath $\hat{\phi}$}}
\newcommand{\del}{\mbox{\boldmath $\nabla$}}
\newcommand{\emf}{\mbox{\boldmath ${\cal E}$}}
\newcommand{\demf}{\mbox{\boldmath ${\cal D}$}}
\newcommand{\curl}{\mbox{\boldmath $\nabla \times$}}
\newcommand{\cross}{\mbox{\boldmath $\times$}}
\newcommand{\pump}{\mbox{\boldmath $\gamma$}}
\newcommand{\etens}{\mbox{\boldmath $\eta$}}
\newcommand{\pd}{\partial}

%-----------------------------------
\newcommand{\bdot}{{\bf .}}

\chapter{Incorporating Surface Convection into a 3D Flux Transport Dynamo Model}
\label{C7}
\begin{quote}\small
The observed convective flows on the photosphere (e.g., supergranulation, granulation) play a key role in the Babcock-Leighton (BL) process to generate large scale polar fields from sunspots fields. In most surface flux transport (SFT) and BL dynamo models, the dispersal and migration of surface fields is modeled as an effective turbulent diffusion.  Recent SFT models have incorporated explicit, realistic convective flows in order to improve the fidelity of convective transport but, to our knowledge, this has not yet been implemented in previous BL models.  Since most Flux-Transport (FT)/BL models are axisymmetric, they do not have the capacity to include such flows.  We present the first kinematic 3D FT/BL model to explicitly incorporate realistic convective flows based on solar observations.  Though we describe a means to generalize these flows to 3D, we find that the kinematic small-scale dynamo action they produce disrupts the operation of the cyclic dynamo.  Cyclic solution is found by limiting the convective flow to surface flux transport.  The results obtained are generally in good agreement with the observed surface flux evolution and with non-convective models that have a turbulent diffusivity on the order of $3 \times 10^{12}$ cm$^2$ s$^{-1}$ (300 km$^2$ s$^{-1}$).  However, we find that the use of a turbulent diffusivity underestimates the dynamo efficiency, producing weaker mean fields than in the convective models.  Also, the convective models exhibit mixed polarity bands in the polar regions that have no counterpart in solar observations.  Also, the explicitly computed turbulent electromotive force (emf) bears little resemblance to a diffusive flux.  We also find that the poleward migration speed of poloidal flux is determined mainly by the meridional flow and the vertical diffusion.
\end{quote}
\section{Introduction}\label{sec:intro}
Since it was discovered over a century ago that sunspots are regions of strong magnetic field \citep{Hale1909} and solar cycles are thus manifestations of solar magnetic activity, researchers have been seeking an explanation for this cyclic magnetic activity.  A major breakthrough in this history of research came with the work of \citet{Parker55a} who first demonstrated how helical plasma motions can generate large-scale magnetic fields. \citet{SKR66} later gave Parker's theory a more mathematical foundation with the development of mean field dynamo theory. The first dynamo simulations to produce a solar-like butterfly diagram (cyclic equatorward migration of toroidal field) were made by \citet{SK69} and subsequently by \citet{Roberts72}. There are likely other papers from that time period achieving the same but the most notable in this time period would be \citet{Yoshimura75} and \citet{Stix76}. All of these calculations were based on the kinematic, mean field formulation of the magnetohydrodynamic (MHD) induction equation. 

Prior to the mid-1980s, when there was little observational information about the internal rotation profile of the Sun, theorists were free to adopt any profile that was needed in order to reproduce the observed butterfly diagram using the $\alpha$-$\Omega$ dynamo waves of mean-field theory.  However, by the mid 1990's the internal rotation profile of the Sun was well established \citep{thomp03}.  But, the use of this differential rotation profile in mean-field dynamo models was problematic because it adversely affected the propagation of $\alpha$-$\Omega$ dynamo waves.  \citet{CSD95} showed that the inclusion of a meridional circulation can help to solve this problem by advecting toroidal flux toward the equator, thus promoting more solar-like butterfly diagrams.  About the same time it was also argued that helical turbulence may not be as efficient at generating poloidal field as previously thought.  This was based on non-kinematic effects associated with Lorentz-force back-reactions on both large and small scales.  On large scales, it was argued that the toroidal field concentrations that give rise to sunspots may be too strong (super-equipartition fields of $\geq 10^4$ G) for convection to twist \citep{Dsilva93}.  On small scales it was argued that the buildup of small-scale magnetic helicity may dramatically suppress the turbulent $\alpha$-effect \citep{vains92,gruzi94,brand01}.

These challenges to traditional mean-field dynamo theory led to the resurgence of solar dynamo models based on the so-called Babcock-Leighton (BL) process.  The BL process (or BL mechanism) had been proposed decades earlier \citep{Bab61,Leighton69} as an alternative candidate for the poloidal field generation that does not rely on turbulent convection.  In short, toroidal flux concentrations can destabilize and rise due to magnetic buoyancy, emerging from the photosphere as bipolar magnetic regions (BMRs).  The action of the Coriolis force on the rising flux tube induces a twist that is manifested upon emergence as a preferential tilt of BMRs, known as Joy's law, such that the trailing edge is displaced poleward relative to the leading edge.  The subsequent fragmentation and dispersal of these tilted BMRs after emergence due to turbulent diffusion, differential rotation, and meridional flow generates a dipole moment that, together with the $\Omega$-effect due to differential rotation, sustains the dynamo \citep{dikpa09,Charbonneau10,Karakreview14}. Note that the fragmentation and dispersal of the BMRs are executed by the convective flows on the photosphere (e.g, supergranulation and granulation) and this process is  modeled as a simple random of walk process which is treated as an effective turbulent diffusion \citep{Leighton64}. In the last two decades Babcock-Leighton (BL) solar dynamo models have grown to become the most promising paradigms to explain the origin of solar magnetic cycle and its irregularities \citep{CSD95,DC99,CNC04,dikpa09,Charbonneau10,Karak10,KarakChou11,Karakreview14}.   Yet, most of these models are still kinematic in the sense that they solve only the magnetohydrodynamic (MHD) induction equation with specified mean flows (differential rotation and meridional circulation) derived from observations.

Efforts to solve the fundamental MHD equations in the solar convection zone (CZ) from first principles in a more self-consistent manner date back to the pioneering work of \citet{GM81}, \citet{gilma83} and \citet{glatz84,glatz85a} and have made dramatic progress just in the last seven years.  Convective dynamo simulations now exhibit many solar-like features, including self-organization of large-scale fields, magnetic cycles with periods on the order of a decade, equatorward migration of toroidal flux, torsional oscillations, long-term cycle modulation, and even hints of magnetic flux emergence \citep{ghiza10,Racine11,brown11,kapyl12,kapyl13,nelso13,nelso13b,passo14,fan14,warne14,Karak15,Augustson15,kapyl16,hotta16}.  However, these models operate in parameter regimes far removed from the real solar interior and they still fall short of reproducing the solar cycle with high fidelity.  One reason may be that they do not yet have sufficient resolution to capture the full scope and complexity of the BL process.  This would require faithfully capturing the formation, rise, and emergence of the magnetic flux structures in the deep CZ, as well as the compressible, radiative MHD and small-scale convection in the surface layers responsible for the formation, fragmentation, and dispersal of active regions.

The amount of available flux emerging in BMRs each solar cycle, \citep[$> 2 \times 10^{24}$ Mx][]{schri94,thorn11} is two orders of magnitude larger than the amount of flux needed to reverse the dipole moment of the Sun, at least at the surface \citep[$\sim 5 \times 10^{22}$ Mx][]{Munoz12}.  This alone suggests that the BL process may play an essential role in the solar dynamo.  This conclusion is further supported by the observed evolution of magnetic flux in the solar photosphere as represented by magnetic butterfly diagrams, which suggest that the polar field reversals are triggered by the poleward migration of magnetic flux originating from the trailing edge of BMRs \citep{Hathaway10a}.  Other lines of evidence in favor of the BL process include an observed correlation between the BL source term and cycle strength \citep{dasi10,mccli13}, the flux budget in active regions \citep{camer15}, and the phase relationship between poloidal and toroidal fields \citep{dikpa09,Charbonneau10,Karakreview14}.  

The observed evolution of magnetic flux in the solar photospheric is well captured by Surface Flux Transport (SFT) models. SFT models are not dynamo models. Magnetic flux is injected by means of artificial source terms or 2D magnetograms which provide the radial field $B_r$ as a function of latitude and longitude in some data assimilation window (typically the near side of the Sun).  The model then follows the subsequent evolution of this flux by solving a 2D (latitude-longitude) version of the kinematic MHD induction equation [eq.\ (\ref{induction1}) below] which includes differential rotation, meridional circulation, and turbulent diffusion by photospheric convection \citep{devor84,wang91}. As the evolution proceeds, the leading-polarity flux in low-latitude BMRs cancels across the equator while residual trailing-polarity flux is transported to the poles. The polarity of this trailing flux is opposite to the pre-existing polar flux so its accumulation at the poles from multiple BMRs eventually reverses the polar field and the global dipole moment.  This is the surface manifestation of the BL process.

\citet{upton14a,upton14b} have recently developed an SFT model called AFT (Advective Flux Transport) that used the observed surface flows to improve the fidelity of the model.  The distinguishing feature of AFT is that it uses explicit convective flow fields where other models use turbulent diffusion.  These 2D convective flow fields ($v_\theta$, $v_\phi$) are designed to reproduce the observed photospheric power spectrum and cell lifetimes, with randomized phases, as described by \citet{hatha00,Hathaway12a,Hathaway12}. \citet{upton14a,upton14b} showed that the use of these convective flows greatly improved the realism of the flux transport, reproducing the magnetic network and introducing stochastic variations that are not captured by a turbulent diffusion.

In this study we use explicit three-dimensional (3D) convective flow fields for the first time in a Babcock-Leighton dynamo model of the solar cycle.  On the surface, these flow fields are identical to the empirical flow fields used in the AFT SFT model \citep{Hathaway12,upton14a,upton14b}.  However, here we extrapolate these flows below the surface to create a 3D rendition of surface convection that is responsible for the magnetic flux transport in the upper CZ.  Furthermore, in this initial implementation, we use a time-independent snapshot of the convective flow instead of the evolving flow fields used by Upton \& Hathaway.  We will implement evolving convective and mean flows in future work.

We achieve this through the use of the Surface flux Transport And Babcock-LEighton (STABLE) solar dynamo model \citep{MD14,MT16,HCM17}.  STABLE is a 3D model that explicitly places BMRs on the surface in response to the dynamo-generated toroidal field near the base of the CZ so the BL process can operate more realistically than in previous 2D (axisymmetric) models that employ, for example, a non-local $\alpha$-effect.  STABLE can function both as a BL dynamo model and as an SFT model, although we have not yet assimilated observational data in the latter context.  It is part of a next generation of 3D solar dynamo models that seek to capture the operation of the solar cycle with high fidelity by incorporating observational data and insights wherever possible.  In the future we will include Lorenz-force feedbacks by solving the full MHD equations but here we take the pragmatic approach of previous 2D models and solve the kinematic MHD induction equation with specified, realistic flow fields.  Since STABLE is 3D, these specified flow fields can include surface convection as well as differential rotation and meridional circulation.  Note that no existing global MHD convection simulation has sufficient resolution and scope to realistically capture surface convection on the scales of granulation to supergranulation.  So, our approach of specifying the convective velocity spectrum based on photospheric observations ensures that the surface convective motions are represented as realistically as is currently feasible.

Turbulent transport plays an important role in BL dynamo models, helping to regulate the cycle period and amplitude. 2D models typically represent it by a turbulent diffusivity and (sometimes) magnetic pumping but there are few observational constraints on how these coefficients vary with depth so models vary widely in the profiles they use \citep[see, e.g.][]{Munoz11}. Even SFT models vary in the coefficient they use to describe the observed dispersal of magnetic flux on the solar surface \citep{Jiang_review15}. In this study we investigate how realistic surface convection influences the behavior of a BL dynamo model.  Furthermore, by comparing convective models with models based on turbulent diffusion, we assess the viability of the turbulent diffusion paradigm and estimate the effective diffusion coefficient, $\eta_t$. Our results suggest that the surface convection is approximately equivalent to a turbulent diffusion: $\eta_t \sim 3 \times 10^{12}$ cm$^2$ s$^{-1}$ which is comparable with the values obtained from observations: $\eta_t \sim$ 2--5 $\times 10^{12}$ cm$^2$ s$^{-1}$ \citep{mosher77,topka82,Schrijver96,Chae08,Jiang_review15}

The organization of the chapter is as follows. In Section~\ref{sec:STABLE} and \ref{sec:cflow}, we describe the STABLE model and how we implement convective flow fields.  In Sections~\ref{sec:nocon} and \ref{sec:con} we describe dynamo simulations based on turbulent diffusion and explicit convective motions respectively.  In Section~\ref{sec:discussion} we compare and contrast these simulations in order to quantify the role of convective transport and estimate the effective turbulent diffusivity of the convective motions.  We conclude and summarize our results in Section~\ref{sec:summary}.
       
\section{The Numerical Model}\label{sec:STABLE}
The Surface flux Transport And Babcock-LEighton (STABLE) dynamo model is a 3D generalization of previous 2D BL models.  In particular, it can be classified as a Flux-Transport Dynamo (FTD) model \citep{WSN91,CSD95,DC99,dikpa09}.  FTD models are BL dynamo models in which the imposed meridional circulation plays an important role in transporting toroidal flux toward the equator, thus establishing the butterfly diagram.  Toroidal field is generated by the stretching of poloidal field by the differential rotation (the $\Omega$-effect) and the poloidal field is generated by the BL process (Sec.\ \ref{sec:intro}).  

Many previous 2D models parameterize the BL process by means of an $\alpha$-effect or similar poloidal source term that is nonlocal in the sense that the poloidal field generation is confined to the surface layers but depends on the toroidal flux near the base of the CZ \cite[e.g.][]{DC99,Rempel06}.  We take a different approach here, exploiting the 3D capabilities of STABLE to treat the inherently 3D BL process more realistically.  In particular, we explicitly place tilted BMRs on the surface of the model and follow their subsequent evolution by solving the 3D kinematic MHD induction equation:
\begin{equation}\label{induction1}
\frac{\partial {\bf B}}{\partial t} = \nabla \times ({\bf v}\times{\bf B} -\eta_t \nabla\times {\bf B}) ~~~.
\end{equation} 
The BMRs are sheared out and dispersed by differential rotation and turbulent diffusion and advected poleward by the differential rotation, naturally generating mean poloidal field as originally described by \citet{Bab61} and \citet{Leighton69} and as captured by SFT models (Sec.\ \ref{sec:intro}).  

The numerical algorithm that places BMRs on the surface is called SpotMaker and is described in detail by \citet{MD14} and \citet[][hereafter MT16]{MT16}.  Briefly, we define a spot-producing toroidal field $B^*(\theta,\phi,t)$ at any time $t$ by integrating the longitudinal field component $B_\phi(\theta,\phi,r,t)$ over a radial range near the base of the CZ (here spanning 0.7--0.71$R$, where $R$ is the solar radius) and applying a latitudinal mask that suppresses high latitudes.  We then choose a random latitude and longitude from all points where $B^*(\theta,\phi,t)$ exceeds a threshold value (here equal to 1 kG).  This is where we place a tilted BMR on the surface.

The tilt angle of each BMR follows Joy's law ($\delta = 32^\circ.1 \cos\theta$; Stenflo \& Kosovichev 2012\nocite{SK12})  and the subsurface spot structure is specified by means of a potential field extrapolation down to $r_p = 0.90 R$.  As discussed by MT16 \citep[see also][]{Longcope02,SM05}, this corresponds to the limit in which BMRs decouple quickly from their roots in the tachocline/lower CZ (on a time scale short compared with the solar cycle).  The subsurface structure is a very idealized assumption but it serves to localize the BL process in the near-surface layers as in previous FTD models.  In the future we will consider the opposite limit in which the BMR retains connectivity to the tachocline by implementing a lifting and twisting flow in SpotMaker as described by \citet{YM13}.  

In its current rendition, SpotMaker is essentially a 3D generalization of Durney's Double ring algorithm \citep{Durney95,Durney97,Munoz10}.  It effectively serves as an explicit source term $S(r,\theta,\phi,t)$ added to equation (\ref{induction1}): 
\begin{equation}
\label{induction2}
\frac{\partial {\bf B}}{\partial t} = \nabla \times ({\bf v}\times{\bf B} -\eta_t \nabla\times {\bf B}) + S(r,\theta,\phi, t)  ~~~.
\end{equation}  
Expressed in this way, $S(r,\theta,\phi,t)$ would be composed of a series of $\delta$ functions in time, with several thousand instances in each 11-year cycle.  The time interval between spot appearances in each hemisphere is chosen randomly from a lognormal probability distribution with a mean of 3 days and a mode of 2 days (see Fig.\ 2\textit{c} in MT16).  

The magnetic flux in each BMR is proportional to the toroidal flux at the base of the CZ, $B^*(\theta,\phi,t)$, and at the chosen location, $\theta_s$, $\phi_s$:
\begin{equation}
\label{flux}
\Phi = 2\alpha_{spot}\frac{|{\hat{B}(\theta_s,\phi_s,t)}|}{B_q}\frac{10^{23}}{1 + (\hat{B}(\theta,\phi)/B_q)^2} \rm{Mx}
\end{equation}  
We neglect the rise time of a flux tube, which is short ($\sim$ 1 month) compared to a cycle period ($\sim$ 11 years).  We also currently neglect the deflection of the tube as it rises.  So, the flux in each BMR is proportional to the instantaneous value of $\hat{B}$ at the same time, latitude, and longitude, as described by MT16.  For the simulations reported here we use a quenching field strength $B_q$ of 100 kG.

Defining the flux content of the spots in this way (equation \ref{flux}) has various advantages. The poloidal field generation is proportional to $\hat{B}$ for weak fields, so it correctly mimics the Babcock-Leighton $\alpha$ effect. For $\alpha_{spot}$ = 1, the subsurface field at the quenching strength $B_q$ will generate the biggest spots that are observed which have a flux content $\sim 10^{23} \rm{Mx}$.  However, as discussed by MT16, it is sometimes necessary to artificially increase the BMR flux in order to achieve supercritical (non-decaying) dynamo solutions.  This is done by choosing a  value for $\alpha_{spot} > 1$.  Hence, this parameter determines how much poloidal field will be generated from the subsurface toroidal field and helps to regulate the overall amplitude of the dynamo fields.  We define the radius of a spot $r_r$ based on its flux content, such that $\Phi = B_0 r_r^2$ \citep[neglecting geometric factors of order unity; see][]{MD14}.  For the value of $B_0$ we choose a typical sunspot field strength of 3 kG.  We impose a minimum value of $r_r$ based on the spatial resolution. 

We use the well-tested Anelastic Spherical Harmonic (ASH) code \citep{Miesch_et_al_00,BMT04} to solve only the 3D magnetic induction equation (\ref{induction2}), with fixed velocity fields.  This kinematic formulation follows the pragmatic approach of previous 2D FTD dynamo models that take advantage of solar observations and helioseismic inversions to make the flow fields as realistic as possible.  We will consider Lorentz-force feedbacks in future work.  The ASH code has been verified against both convective dynamo benchmarks \citep{Jones11} and axisymmetric FTD benchmarks (MT16).   

The velocity field ${\bf v}$ in eq.\ (\ref{induction2}) includes differential rotation (DR) and meridional circulation (MC) profiles chosen to match helioseismic inversions and photospheric observations, where available.  Despite a recent proliferation of research fueled by the availability of high-resolution HMI/SDO data and long time series from MDI/SOHO, the subsurface structure and amplitude of the MC is still uncertain.  There is some possible evidence for multiple cells per hemisphere in latitude and radius but different methods do not yet provide a consistent picture \citep{Hathaway12,Zhao13,Schad13,RA15,jacki15}.  In light of this uncertainty and in order to be consistent with the vast majority of papers in the literature on FTD dynamos, we adopt an MC profile here with a single cell per hemisphere, with counter-clockwise circulation in the northern hemisphere (NH) and clockwise circulation in the southern hemisphere (SH).  This yields a peak poleward flow of 25.5 m s$^{-1}$ at mid-latitudes at the surface and an equatorward flow of $\sim $ 2 m s$^{-1}$ near the base of the CZ.  For further details on the MC profile we use see \citet{CNC04} and \citet[][hereafter HCM17]{HCM17}.  

The DR profile is chosen to be independent of radius in the bulk of the CZ, with a latitudinal rotation rate varying from 460.7 nHz at the equator to 330.67 nHz at the poles.  We do not include a near-surface shear layer but we do include a tachocline, expressed by means of an error function centered at $r_c = 0.7 R$, with a thickness of 0.05 $R$.  This produces a sharp transition at the base of the CZ to uniform rotation rate of the radiative zone, which is set equal to 432.8 nHz.  These values are based on the previous 2D FTD models of \citet{DC99} and \citet{CNC04} and our implementation is described further by MT16 (see Fig.\ 1$a$).  \citet{Dikpati02} have argued that the near-surface shear layer does not contribute significantly to the operation of flux-transport solar dynamo, but \citet{karak16} have shown that it can promote equatorward propagation in conjunction with efficient magnetic pumping in the surface layers. 

\begin{figure}[!h]
\centering
\includegraphics[width=8.5 cm,height = 8 cm, angle=0]{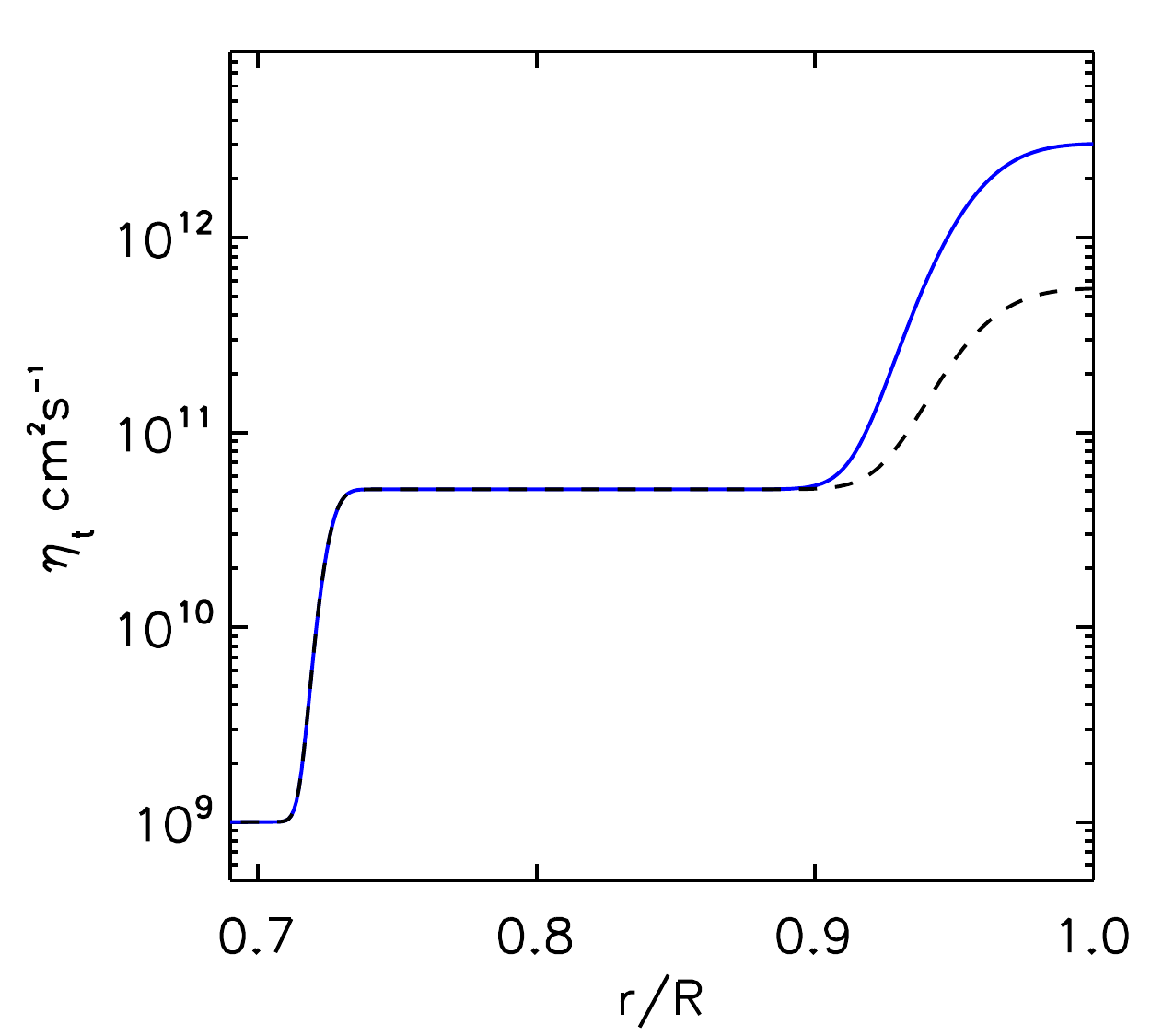}
\caption[Diffusivity profiles used for different cases]{Two step diffusivity profile used in our simulations. Shown are example diffusivity profiles used for simulations with convection (black dotted line) and without explicit convective motions (solid blue line), with $\eta_{top} = 3\times 10^{12}$ cm s$^{-1}$ in the latter.  The convective motions are confined to the upper convection zone and their role in the global dynamo is functionally equivalent to an enhanced turbulent diffusion.}
\label{fig:eta}
\end{figure}

In this study for the first time, we also include a non-axisymmetric convective velocity component in ${\bf v}$.  Our methodology for achieving this is described in Sec.\ \ref{sec:cflow}.  This convective flow is confined to the upper CZ ($r > 0.9R$) and is intended to capture magnetic flux transport by small-scale surface convection on scales from granulation to super-granulation.  Transport in the deeper regions of the CZ (giant cells) is still handled by means of the turbulent diffusivity $\eta_t$, which is a function of radius alone.  For the simulations described in Section \ref{sec:nocon} that do not include 3D convective motions we use a two-step diffusivity profile as in previous 2D FTD models \citep[e.g.][]{Hotta10}.  Example profiles are shown in Fig.\ \ref{fig:eta} (see also Fig.\ 1\textit{c} in MT16 and Fig.\ 3 in HCM17).  The value of $\eta_t(r)$ at $r=R$, which we refer to as $\eta_{top}$, is varied between 1$\times 10^{10}$ cm$^2$ s$^{-1}$ and $3.5 \times 10^{13}$ cm$^2$ s$^{-1}$, as described in Sec.\ \ref{sec:nocon}.  The values of $\eta_t(r)$ in the lower CZ and radiative interior are respectively $5\times 10^{10}$ cm$^2$ s$^{-1}$, and 10$^9$ cm$^2$ s$^{-1}$.  For the simulations described in Sec.\ \ref{sec:results}, we reduce the surface diffusion to $5\times 10^{11}$ cm$^2$ s$^{-1}$, one order of magnitude higher than the value in the mid CZ (black dotted line).  This ensures that the transport in the upper CZ is dominated by the imposed convective motions.

Note that the value of $\eta_t$ that we use in the mid CZ, $5 \times 10^{10}$ cm$^2$ s$^{-1}$ puts us in the so-called advection-dominated regime of FTD models in the sense that the MC dominates the transport of poloidal magnetic flux from the surface layers to the base of the CZ \citep{Jiang07,Yeates08,Karakreview14}.  STABLE can also operate in the diffusion-dominated regime in which turbulent diffusion is the dominant transport mechanism; for an example see HCM17.  It has been argued on several different grounds that the diffusion-dominated regime may be more realistic \citep[e.g.][]{CCJ07,Jiang07,KarakChou11,miesc12} and is less sensitive to the detailed structure of the MC, capable of producing solar-like cycles even for multi-cellular MC profiles \citep{HKC14}.  However, an investigation of these issues lies outside the scope of this present study.  Here we focus on how explicit convective transport affects the operation of FTD dynamo models and our chosen $\eta_t(r)$ profiles are sufficient for this purpose.

We also include a weak horizontal hyperdiffusion on the right-hand-side of eq.\ (\ref{induction2}) to dissipate spurious small-scale fields and thus keep the code numerically stable.  In spectral space this is expressed as $- \eta_h R^{-2} [\ell (\ell+1)]^2 \widetilde{\bf B}$, where $\widetilde{\bf B}$ is the spherical harmonic transform of ${\bf B}$.  We use $\eta_h = 2 \times 10^8$ cm$^2$ s$^{-1}$ for the diffusive Cases A1-A8 (Sec.\ \ref{sec:nocon}) and $\eta_h = 3\times 10^{10}$ cm$^2$ s$^{-1}$ for the convective Cases C1-C3 (Sec.\ \ref{sec:con}).

\begin{figure*}[!t]
\centering{
\includegraphics[width=0.95\textwidth]{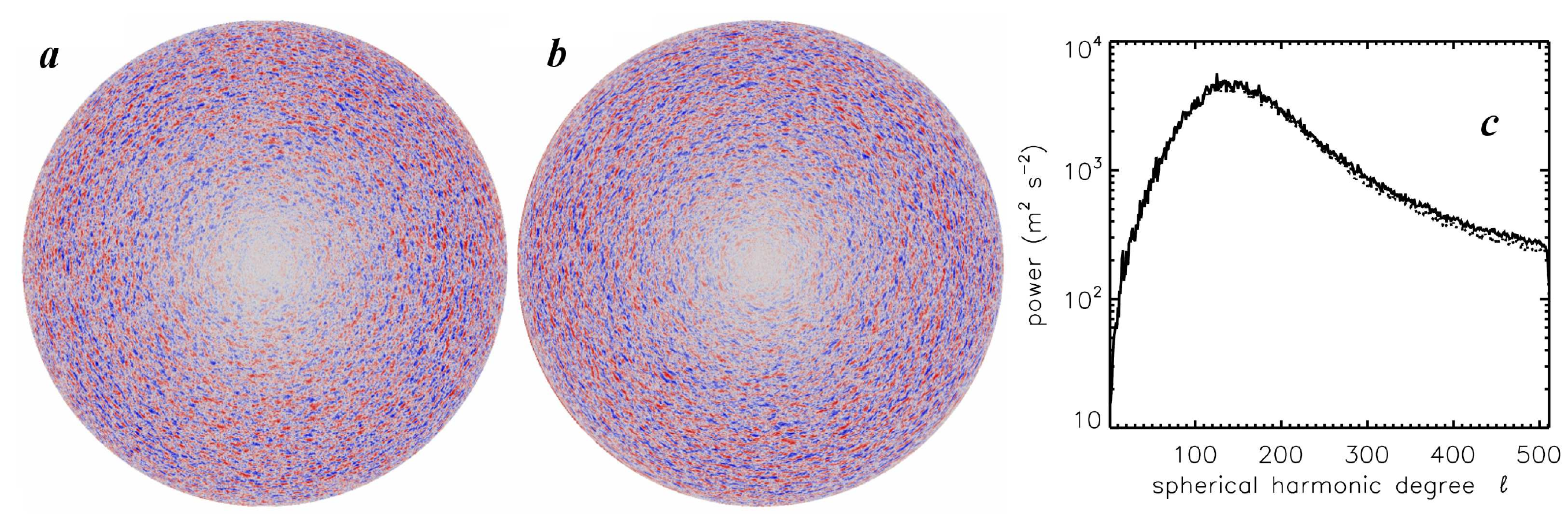}}
\caption[Observed convective flows]{(\textit{a}) Line-of-sight (Doppler) velocity in the solar photosphere measured with the SOHO/MDI instrument on June 4, 1996. (\textit{b}) Simulated line-of sight velocity constructed from the empirical model of \citet{hatha00} and \citet{Hathaway12a,Hathaway12}, which is designed to reproduce the observed horizontal Doppler velocity power spectrum, with randomized phases.  (\textit{c}) Horizontal velocity spectra for Hathaway's simulated flow field (solid line) and for the convective flow field used here (dotted line), at $r=R$.  Small discrepancies at large $\ell$ arise because we only extract the divergent component (see text).  Images and data courtesy of David Hathaway and Lisa Upton.}
\label{fig:conhath}
\end{figure*}

\section{Incorporating Non-Axisymmetric Convective Flows into STABLE}\label{sec:cflow}

\subsection{Empirical Model for Surface Convection}\label{sec:emp}
Though ASH is fully capable of simulating global-scale convective motions, no global solar convection model can accurately capture smaller-scale convective motions near the surface such as granulation and supergranulation.  This would require both extremely high resolution and including physical processes that are often neglected in global models, such as non-LTE radiative transfer, ionization, and the breakdown of the anelastic approximation.  However, it is these small-scale convective motions that contribute most to the breakup and dispersal of BMRs and are thus most important from the perspective of the BL process.

For this reason, we wish to incorporate convection in a manner that is consistent with the pragmatic approach of kinematic dynamo modeling.  In particular, we wish to exploit observational measurements of the convective power spectrum in the solar photosphere in order to ensure that the imposed convective flow fields near the surface are as realistic as possible.  High-quality, full-disk measurements of the photospheric convection spectrum are now available from such instruments as the Helioseismic Magnetic Imager (HMI) onboard NASA's Solar Dynamics Observatory and the Michelson Doppler Imager (MDI) onboard NASA's SOlar and Heliospheric Observatory (SOHO).   These measurements are typically based on Dopplergrams; 2D (latitude-longitude) maps of the line-of-sight velocity component on the solar surface (Fig.\ \ref{fig:conhath}\textit{a}).

Since the velocities in the solar photosphere are predominantly horizontal, these are the velocity components that are mainly sampled by the Dopplergrams; this leads to the dearth of power at disc center in Fig.\ \ref{fig:conhath}\textit{a}.  Furthermore, at any point on the solar disc, only one horizontal direction is sampled.  For example, Doppler velocities near the eastern limb are dominated by $V_\phi$.  In order to reconstruct the complete 2D flow field $V_\theta(\theta,\phi,t)$ and $V_\phi(\theta,\phi,t)$, some modeling is needed. \citet{Hathaway12a,Hathaway12} has devised such a model \citep[see also][]{hatha00}.  Hathaway's model is based on first subtracting off the differential rotation, the meridional circulation, and other unwanted signals such as convective blueshift, spacecraft motion and instrumental artifacts.  Then the convective power spectrum is computed.   A simulated horizontal velocity field is then produced based on that observed spectrum using a series of vector spherical harmonics with randomized complex coefficients.  

Though Hathaway's original implementation is time-evolving, with correlation times chosen to match observations, we consider here a static horizontal velocity field with a resolution of $N_\theta = 512$ and $N_\phi = 1024$.  A sample Dopplergram computed from the simulated horizontal flow field is shown in Fig.\ \ref{fig:conhath}\textit{b} and the horizontal power spectrum is shown in Fig.\ \ref{fig:conhath}\textit{c}.  The peak at $\ell \approx $ 130 represents supergranulation.  This data set does not resolve the broad peak in power beyond $\ell \sim 1000$ due to granulation, as discussed by \citet{hatha00}.  We will refer to this empirical surface velocity field as ${\bf V}_s(\theta,\phi,t)$.  Though we consider only static flows fields in this study, we will retain the explicit time dependence in this section in order to illustrate how our approach can be readily generalized to evolving convective flows.

Our aim is to incorporate the empirical surface velocity field ${\bf V}_s(\theta,\phi,t)$ into STABLE.  Since this velocities are non-axisymmetric, this exploits the 3D capabilities of our dynamo model.  In particular, this implies that the transport and amplification of the non-axisymmetric magnetic field components can influence the time evolution of the mean fields.  This is not the case for axisymmetric, kinematic flows fields in which the $m=0$ component of the magnetic induction equation decouples from the $m>0$ components, where $m$ is the azimuthal wavenumber.

In order to incorporate the 2D empirical surface flow field of Fig.\ \ref{fig:conhath} into our 3D model, we must specify some radial structure.  In specifying this radial structure we have made several assumptions and approximations.  The first assumption is that the small-scale convection in the solar surface layers is much more vigorous than the convection in the deeper convection zone.  This is justified by our theoretical understanding of solar convection, which attributes it to the relatively low density, steep density stratification, and steep superadiabatic entropy gradient in the solar surface layers \citep{miesc05,nordl09}.  It is also justified by recent attempts to estimate the convective velocity amplitudes in the deep solar interior by means of helioseismic inversions, photospheric observations, and numerical modeling \citep{hanas10,miesc12,lord14,greer15,hanas16,omara16,cosse16}.  And, it is consistent with previous FTD models that often use an enhanced turbulent diffusivity in the solar surface layers as illustrated in Fig.\ \ref{fig:eta}.  Thus, we wish to extrapolate the surface velocity field downward but no deeper than $r \sim 0.9$, so that it effectively replaces the enhanced turbulent diffusion near the surface as shown by the dashed line in Fig.\ \ref{fig:eta}.

The second assumption is that the mass flux is divergenceless: $\nabla \cdot (\overline{\rho} {\bf V}) = 0$.  Here $\overline{\rho} = \overline{\rho}(r)$ is the background density stratification, neglecting density fluctuations associated with the convection.  Thus, this is essentially an anelastic approximation, valid for low Mach numbers.  This means that the mass flux can be decomposed into poloidal $(W)$ and toroidal $(Z)$ components defined by:
\begin{equation}\label{eq:poltor}
\overline{\rho} {\bf V} = \nabla \times \nabla \times \left(W \uvr\right) + \nabla \times \left(Z\uvr\right)  ~~~.
\end{equation}

\begin{figure}[!h]
\centering
\includegraphics[width=0.85\textwidth]{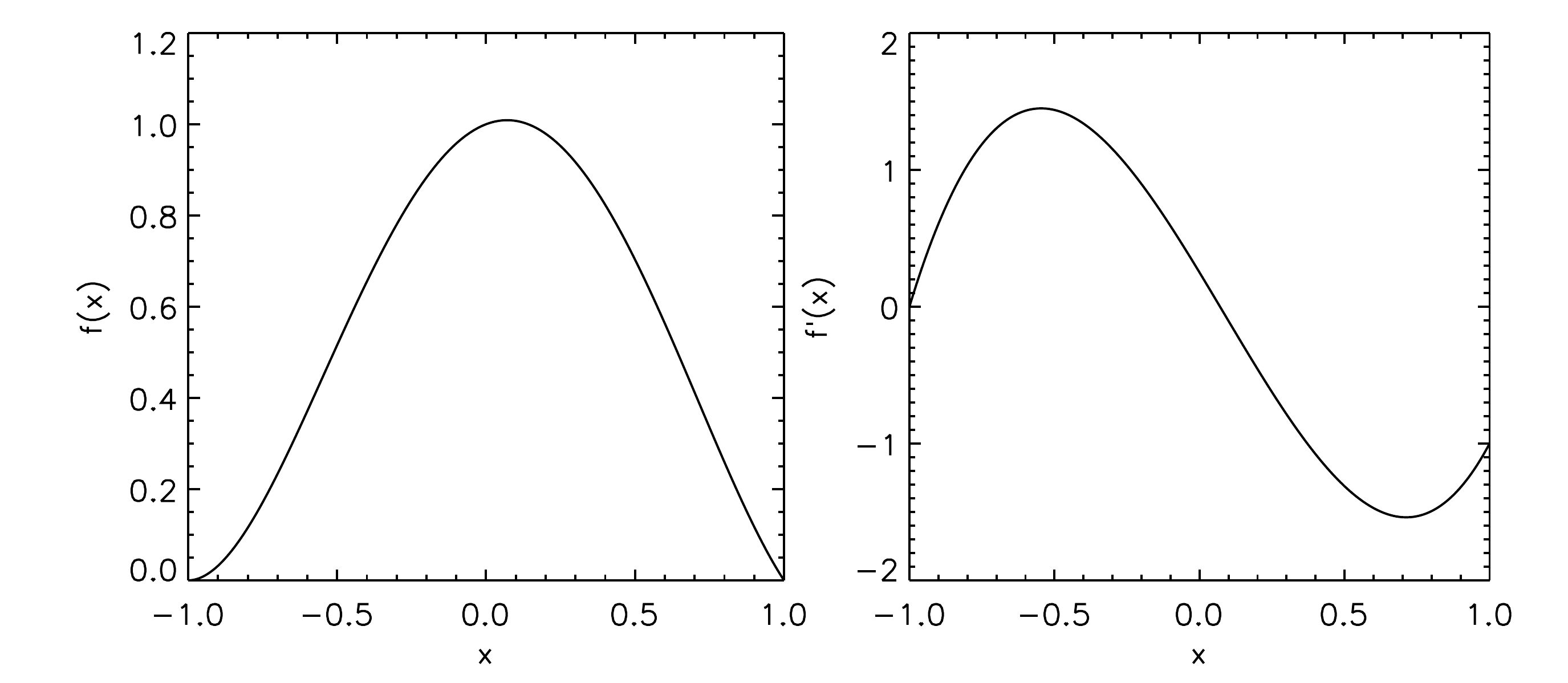}
\caption[Radial extrapolation function]{Radial polynomial used for extrapolation of the surface flow to the deeper convection zone. Shown are (\textit{a}) $f_\ell(x)$ and (\textit{b}) $f_\ell^\prime$(x), where $x = (r-r_w)/(R-r_w)$. The plots extend from $r=r_p$ ($x=-1$) to $r=R$ ($x=1$).}
\label{fig:radfnc}
%\end{center}
\end{figure}

Our third assumption is that the convective velocities are purely poloidal, so $Z = 0$.  This implies that the vertical component of the fluid vorticity is zero.  This is justified mainly by the steep density stratification in the solar surface layers, which imparts a strong horizontal divergence to vertical motions, and the clear signature of supergranulation in the horizontal divergence of surface flows \citep{gizon10}.  There is also clear evidence that the action of the Coriolis force imparts a systematic vertical vorticity to supergranular-scale convective motions \citep{gizon10} but this is a relatively weak effect and it is justified to neglect it as a first approximation.  In future work we will include a vortical (toroidal) component to the flow field and we will investigate its influence both on the turbulent transport and on the potential amplification of magnetic flux.  

This in effect means that we only assimilate the divergent component of the surface flow field into STABLE.  Thus, we define $D(\theta,\phi,t)$ as the horizontal divergence of the imposed surface flow field: $D(\theta,\phi,t) = \del_h \cdot {\bf V}_s$.  Recall that ${\bf V}_s$ has only horizontal components so $D$ is also equal to the full divergence of the surface flow field.  Now we relate the horizontal divergence of the full 3D convective flow field to the poloidal stream function $W$ as follows:
\begin{equation}\label{vh}
\rho\del_h \cdot {\bf V}_h = -\frac{1}{r^2}\frac{\partial}{\partial r}(r^2\rho V_r) 
= - \nabla_h^2 \left(\frac{\partial W}{\partial r}\right) 
\end{equation}
Where $\nabla_h^2$ and ${\bf V}_h = V_\theta \uvt + V_\phi \uvp$ are the horizontal Laplacian and the horizontal velocity respectively.

Now we expand $W$ in a spherical harmonic series as follows
\begin{equation}\label{wspectral}
W(r,\theta,\phi,t) = \sum_{\ell=0}^{\ell_{max}} \sum_{m=-\ell}^\ell \tilde{g}_{\ell m}(t) f_\ell(r) Y_{\ell m}(\theta,\phi)  ~~~.
\end{equation}
Here $\tilde{g}_{\ell m}(t)$ describes the horizontal structure of the convective pattern at the surface and $f_\ell(r)$ describes its downward extrapolation.  We have included an explicit $\ell$ dependence for $f_\ell(r)$ to allow for the possibility that different horizontal scales of convection may have different radial profiles; see Sec.\ \ref{sec:fr}.  Note also that
\begin{equation}\label{eq:hlap}
- \nabla_h^2 W = \sum_{\ell=0}^{\ell_{max}} \sum_{m=-\ell}^\ell \frac{\ell (\ell+1)}{r^2} ~ \tilde{g}_{\ell m} f_\ell Y_{\ell m}  ~~~.
\end{equation}

An expression for $\tilde{g}_{\ell m}$ can be obtained by applying a spherical harmonic transform to eq.\ (\ref{vh}) and evaluating it at the surface, $r=R$, with the help of eq.\ (\ref{eq:hlap}).  This yields:
\begin{equation}\label{gtheta}
\tilde{g}_{\ell m}(t) = \frac{R^2 \overline{\rho}(R)}{\ell(\ell+1)} ~ \tilde{D}_{\ell m}(t) ~ \left[f_\ell^\prime(R)\right]^{-1}  ~~~,
\end{equation}
where $\tilde{D}_{\ell m}(t)$ are the spherical harmonic coefficients for the surface divergence $D(\theta,\phi,t)$ and $f_\ell^\prime(r) = df_\ell(r)/dr$. All that remains is to define the vertical profile $f_\ell(r)$, which we discuss in Section \ref{sec:fr}.  

Once we define $\tilde{g}_{\ell m}(t)$, $f_\ell(r)$, and $\overline{\rho}(r)$, then we can obtain all three components of the convective flow field throughout the entire 3D computational domain by means of equations (\ref{wspectral}) and (\ref{eq:poltor}):
\begin{eqnarray}
V_r(r,\theta,\phi,t) &=& - \frac{1}{\overline{\rho}} \nabla_h^2 W(r,\theta,\phi,t) \\
V_\theta(r,\theta,\phi,t) &=& \frac{1}{\overline{\rho} r}\frac{\partial}{\partial \theta} \left(\frac{\partial W}{\partial r}\right) \\
V_\phi(r,\theta,\phi,t) &=& \frac{1}{\overline{\rho} r \sin\theta}\frac{\partial}{\partial \phi} \left(\frac{\partial W}{\partial r}\right)
\end{eqnarray}
Again, in this study we consider a time-independent convective flow field but this is easily generalizable to evolving flows in which the time evolution is governed by the imposed, empirical surface flow field ${\bf V}_s(\theta,\phi,t)$.  This is in turn reflected in $\tilde{g}_{\ell m}(t)$ through eq.\ (\ref{gtheta}).

Equation (\ref{gtheta}) is the means by which Hathaway's simulated surface flow field, ${\bf V}_s$, is assimilated into STABLE.  As noted above, only the horizontal divergence is used so any non-divergent components of ${\bf V}_s$ will be omitted from our flow field.  Hathaway's horizontal flow field is also constructed on the assumption that the flow is strictly poloidal ($\curl {\bf V}_s = 0$) so this procedure should in principle ensure that $V_\theta$ and $V_\phi$ reproduce ${\bf V}_s$ exactly at $r=R$.  However, in practice there are numerical errors associated with the discretization of ${\bf V}_s$, the interpolation onto a Legendre grid for incorporation into STABLE, and the numerical computation of the derivatives (though the latter computation is spectrally accurate).  So, the resulting horizontal power spectrum lies slightly below Hathaway's spectrum at high wavenumbers, as shown in Fig.\ \ref{fig:conhath}\textit{c}.

For the background density we use a polytropic, hydrostatic, adiabatic stratification \citep{Jones11}:
\begin{equation}\label{eq:rho}
\overline{\rho} = \rho_i ~ \left(\frac{\zeta(r)}{\zeta(r_1)}\right)^n
\end{equation}
where
\begin{equation}
\zeta(r) = c_0 + c_1 \frac{r_2-r_1}{r}
\end{equation}
\begin{equation}
c_0 = \frac{2 \zeta_0 - \beta - 1}{1-\beta}
\end{equation}
\begin{equation}
c_1 = \frac{(1+\beta)(1-\zeta_0)}{(1-\beta)^2}
\end{equation}
and
\begin{equation}
\zeta_0 = \frac{\beta+1}{\beta \exp(N_\rho/n) + 1}  ~~~.
\end{equation}
Here $\rho_i = \rho(r_1) = 0.1788$ g cm$^{-3}$, $\beta = r_1/r_2 = 0.69$ is the aspect ratio,
$n = 1.5$ is the polytropic index, and $N_\rho = 5$ is the number of density scale heights
across the computational domain, which extends from $r_1 = 0.69 R$ to $r_2 = R$.

\begin{figure}[]
\centering
\includegraphics[width=0.95\textwidth]{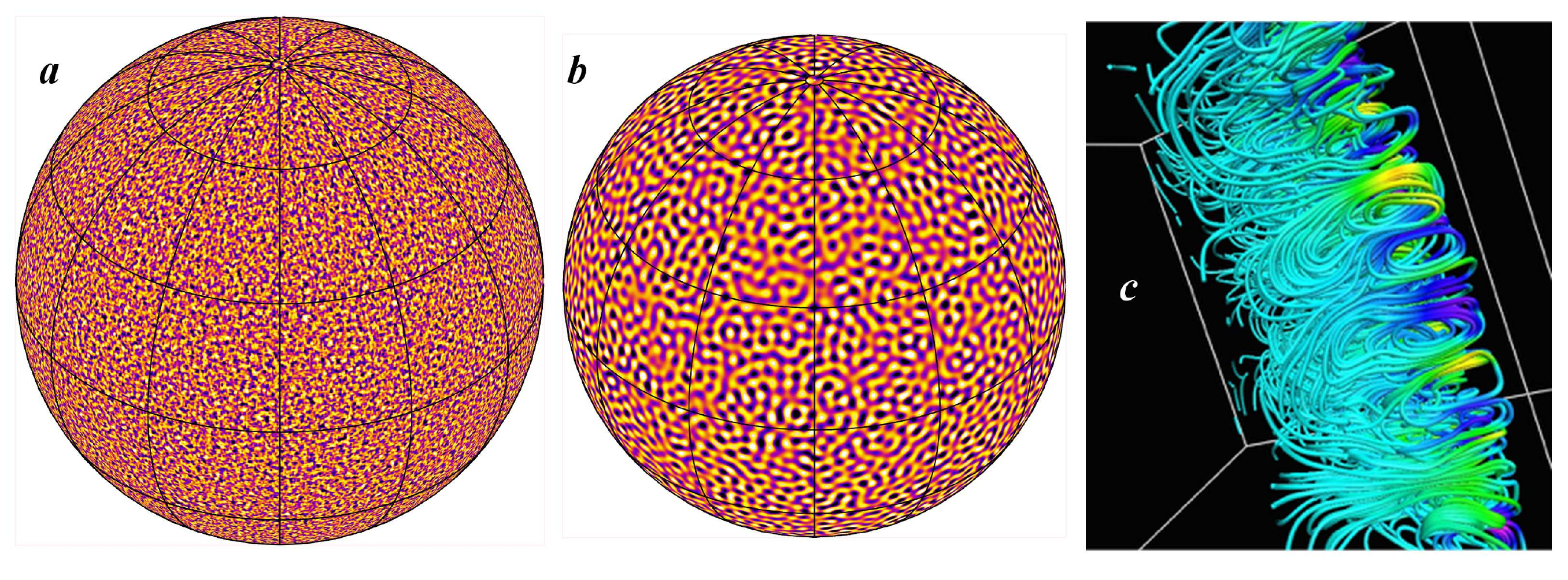}
\caption[Visualization of the imposed convective flow field]{Visualization of the imposed convective flow field.  (\textit{a}) The horizontal divergence at the surface $r=R$ is set equal to the divergence of the empirical flow field ${\bf V}_S(\theta,\phi)$.  Here it is shown in an orthographic projection with yellow and blue representing divergence and convergence respectively. (\textit{b}) As in (\textit{a}) but for $r=0.952$. (\textit{c}) 3D rendering of a zoomed-in portion of the flow field showing streamlines of the mass flux colored by the vertical velocity (yellow upward, blue downward).  Velocity amplitudes decrease with depth due to the increasing density.}
\label{fig:viz1}
\end{figure}
\begin{figure}[]
\centering
\includegraphics[width=\textwidth]{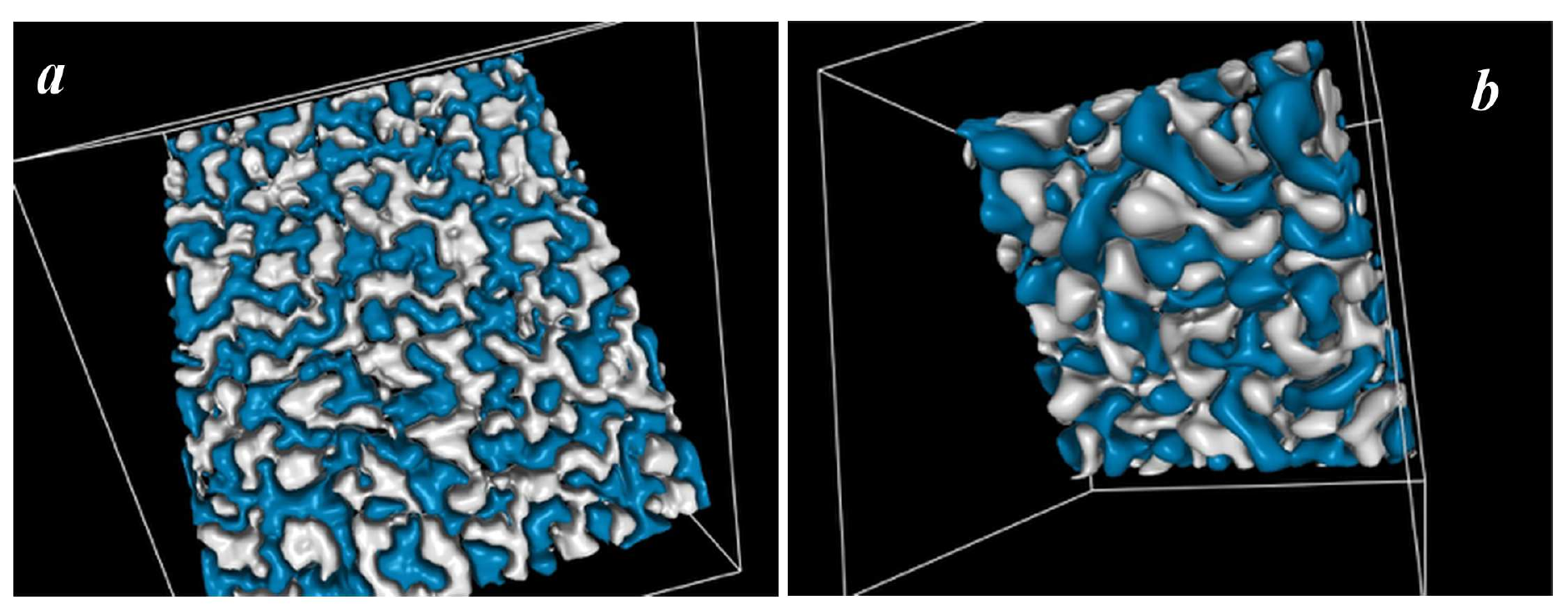}
\caption[Isosurfaces of the convective radial velocity $V_r$ for a zoomed-in patch spanning 20$^\circ$ in latitude and longitude.]{Isosurfaces of the convective radial velocity $V_r$ for a zoomed-in patch spanning 20$^\circ$ in latitude and longitude.  Silver and blue denote upflow and downflow respectively.  Shown are vantage points from (\textit{a}) above and (\textit{b}) below.}
\label{fig:viz2}
\end{figure}
%\clearpage

\subsection{Subsurface Extrapolation of the Convective Flow}\label{sec:fr}

In order to extrapolate the empirical convective flow field at the surface 
downward, a suitable radial function is necessary which confines the convective 
motions to the upper convection zone, as discussed in Sec.\ \ref{sec:emp}.
To achieve this, choose a function which satisfies the following boundary conditions:
\begin{eqnarray}
f_\ell(r)=0 ~ & \& & ~ f_\ell^\prime(r)=-1, ~~ r=R\label{eq:rcon1} \\
f_\ell(r)=1 ~ & \& & ~~  r=r_w\\
f_\ell(r)=0 ~ & \& & ~ f_\ell^\prime(r)=0,~~ r=r_p \\
f_\ell(r) = 0 & & r \leq r_p \label{eq:rcon4}
\end{eqnarray}
Here $r_p$ is the penetration depth and $r_w = (r_p + R)/2$ is the middle of the convective layer, near the point where $f_\ell(r)$ peaks and $f_\ell^\prime(r)$ changes sign. It can thus regarded approximately as the convective turnover depth (see Fig.\ \ref{fig:radfnc}).   The conditions (\ref{eq:rcon1})--(\ref{eq:rcon4}) ensure that there is no convective mass flux through the surface or through the penetration radius $r_p$.  Furthermore, the normalization ensures that the amplitudes of $W$ and $\tilde{g}_{\ell m}$ are on the order of $\rho(R) U$ where $U$ is a characteristic convective velocity amplitude at the surface.  

These conditions can be satisfied with a fourth-order polynomial:
\begin{equation*}
f(x)=a + bx + cx^2 + dx^3 +ex^4
\end{equation*}
where $x=(r-r_w)/(R-r_w)$, $a=1$, $b=0.25$, $c=-1.75$, $d=-0.25$, and $e=0.75$.  The function $f_\ell(x)$ and its first derivative $f^\prime_\ell(x)$ are shown in Fig.\ \ref{fig:radfnc}.  Note that $x=-1$, $x=0$ and $x=1$ correspond to $r_p$, $r_w$ and $R$ respectively.

We have yet to specify the penetration depth $r_p$.  In doing so, it is reasonable to assume that larger-scale motions will penetrate more deeply than smaller-scale motions.  The horizontal length scale $L_h$ of a spectral mode at the surface with total wavenumber $\ell$ is approximately given by
\begin{equation}
\left(\frac{2\pi}{L_h}\right)^2 \sim \frac{\ell (\ell+1)}{R^2} ~~~.
\end{equation}
If we assume that the vertical length scale $L_v$ of a convective mode is comparable to its horizontal length scale, and if we assume that $\ell >> 1$ as it is for most of the convective power (Fig.\ \ref{fig:conhath}), then this gives 
\begin{equation}
L_v \approx L_h \approx \frac{2\pi}{\ell} ~ R  ~~~.
\end{equation}
Thus, we set $r_p = R - L_v = R (1 - 2\pi/\ell)$.  However, we set a minimum value of $r_p = 0.9$ so even the largest motions do not penetrate below this. 

The resulting convective flow fields are illustrated in Figures \ref{fig:viz1} and \ref{fig:viz2}.  The increase of the horizontal length scale of the convection with depth is clear by comparing Fig.\ \ref{fig:viz1}\textit{a} and Fig.\ \ref{fig:viz1}\textit{b} and by comparing the two frames in Fig.\ \ref{fig:viz2}. Fig.\ \ref{fig:viz1}\textit{c} highlights the overturning nature of the motions.  Note that at this resolution, the asymmetry between upflows and downflows is not apparent.  In all calculations reported here we have used radial grid $N_r=340$, latitudinal grid $N_\theta = 512$ and longitudinal grid $N_\phi = 1024$.  

Once this 3D convective flow, ${\bf v}_c$, is defined, we add it to the axisymmetric flows ${\bf v}_a$ such that ${\bf v} = {\bf v}_a + {\bf v}_c$.  Here ${\bf v}_a$ includes the differential rotation and the meridional flow described in Section \ref{sec:STABLE}.  As mentioned in Section \ref{sec:intro}, we restrict our attention in this study to the case in which ${\bf v}_c(r,\theta,\phi)$ is independent of time.  In future work we will implement an evolving flow field as in the AFT model of \citet{upton14a,upton14b}.

\begin{table}[]
%\centering
\centering
{\scriptsize\renewcommand{\arraystretch}{.8}{
\caption{Simulation Summary\label{cases}}
\begin{tabular}{c|c|c|c|c|c|c|c|c|c|c|c}
\hline
\hline
Cases\tablenotemark{1} & $\alpha_{spot}$ & $\eta_{top}$ & Cycle & Migration & \multicolumn{2}{|c|}{$B_{pol}$ (G)} & \multicolumn{2}{|c|}{$B_{tor}$ (kG)} & \multicolumn{2}{|c|}{$B_{nax}$ (kG)}& $\frac{B_{tor}}{B_{pol}}$\\
\cline{6-11}
 & & (cm$^2$ s$^{-1}$) & Period (years) & Speed\tablenotemark{2} (m s$^{-1}$) & Mean & $\sigma$&Mean&$\sigma$&Mean&$\sigma$&\\
\hline
 A1 & 15.0 & $3.0\times 10^{12}$ & 13.3 & 11.6 & 596.6 & 118.3 & 33.54 & 1.91 & 1.76 & 0.51 & 56.45 \\ %A1 = sand1
 A2 & 25.0 & $1.0\times 10^{13}$ & 13.8 & 8.8  & 67.0 & 13.6  & 4.04 & 0.61 & 0.21 & 0.07 & 60.31 \\ %A2 = sand2
 A3 & 25.0 & $3.5\times 10^{13}$ & sub-critical & -- & -- & -- & -- & --  & --   &  --  & --   \\ %A3=sand3
 A4 & 15.0 & $8.0\times 10^{11}$ & 14.0 & 16.6 & 1523.2 & 292.8 & 71.18 & 4.47 & 4.64 & 0.93 & 46.73 \\ %A4 = sand11
 A5 & 15.0 & $3.0\times 10^{11}$ & 14.7 & 18.4 & 2075.5 & 435.2 & 84.72 & 9.88 & 6.97 & 1.10 & 40.94 \\ %A5=sand5
 A6 & 15.0 & $1.0\times 10^{11}$ & 15.1 & 19.4 & 2431.4 & 430.1 & 89.24 & 9.58 & 9.37 & 1.15 & 36.68 \\ %A6 = sand9
 A7 & 15.0 & $5.0\times 10^{10}$ & 15.1 & 20.1 & 2546.7 & 455.2 & 85.73 & 8.37 & 11.13 & 1.60 & 33.64 \\ %A7 = sand6
 A8 & 15.0 & $1.0\times 10^{10}$ & 15.9 & 20.5 & 2531.7 & 388.2 & 89.74 & 12.33 & 12.08 & 1.35 & 35.56 \\ %A8 = sand10
\hline
 C1 & 15.0 & $5.0\times 10^{11}$ & 13.3 & 20.1 & 1541.2 & 226.0 & 77.70 & 7.40 & 2.29 & 0.27 & 50.42\\ %C1=sparrow5
 C2 & 1.0  & $5.0\times 10^{11}$ & sub-critical & -- &--&--&--&--&--&--&--\\ %C2 = sparrow4
% C3 & 15.0 & $3.0\times 10^{11}$ & quadrupolar  & -- &--&--&--&--&--&--&--\\
% C4 & 15.0 & $1.0\times 10^{11}$ & quadrupolar  & -- &--&--&--&--&--&--&--\\
 C3 & 15.0 & $5.0\times 10^{10}$ & 14.0 & 22.7 &1515.72 & 195.3 & 83.5 & 11.0 & 2.48 & 0.40 & 55.08 \\ 
% & C3 & 15.0 & $5.0\times 10^{9}$ & longer period & --&--&&&&&&\\       % C3 = spinifexbird
\hline
\hline
\end{tabular}}}
\tablenotetext{1}{Cases A1-A8 include only axisymmetric flows (DR and MC).  Cases C1-C3 also incorporate explicit convection.}
\tablenotetext{2}{See Section \ref{sec:migration}.}
\end{table}

\section{Axisymmetric Flows}\label{sec:nocon}
\begin{figure*}[!t]
\centering
\includegraphics[width=0.95\textwidth, angle=0]{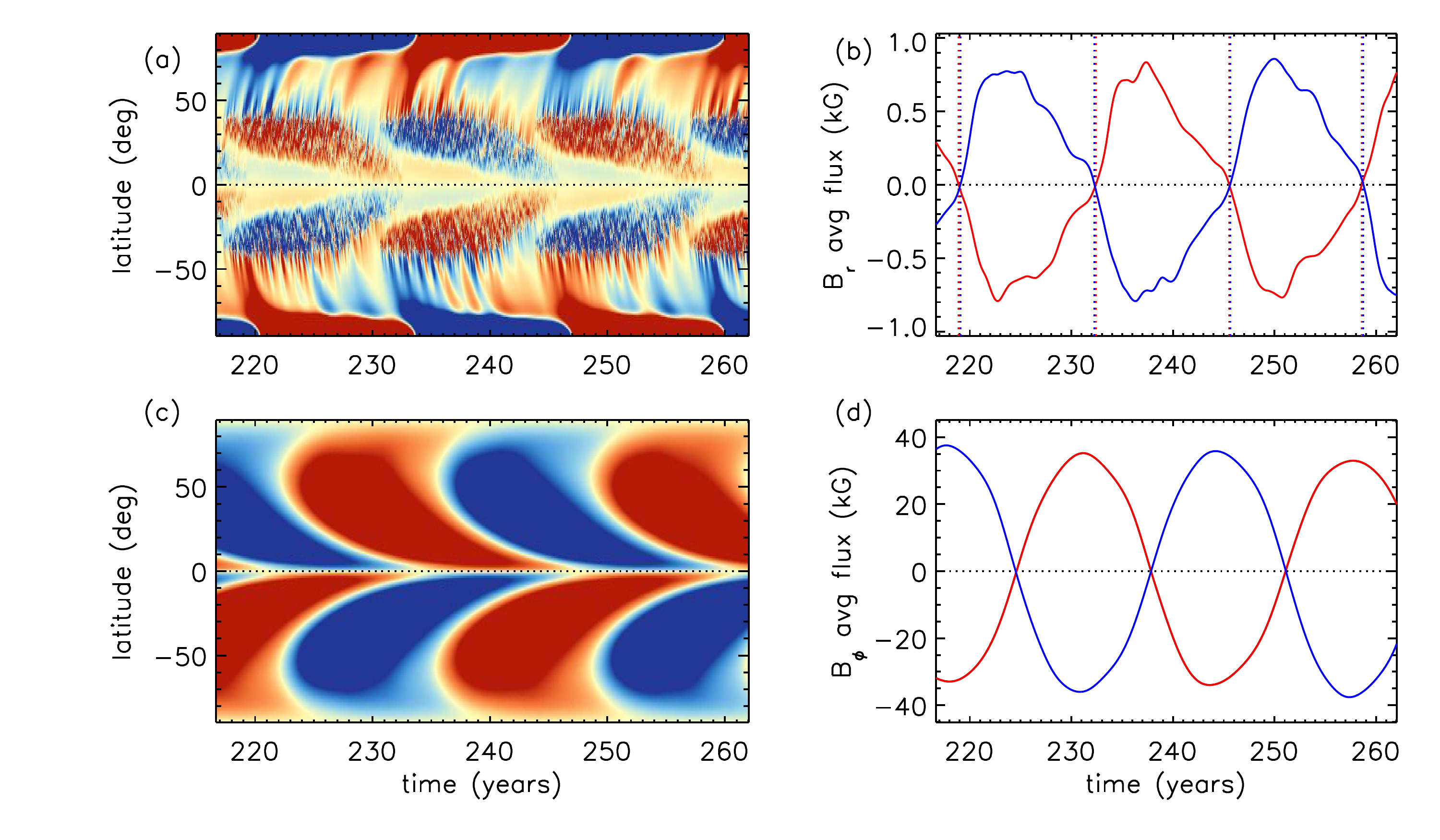}
\caption[Time-latitude plot of the mean radial magnetic field on the solar surface, $\left<B_r\right>$, for Case A1]{(\textit{a}) Time-latitude plot of the mean radial magnetic field on the solar surface, $\left<B_r\right>$, for Case A1. The color scale is set from $-200$G (blue) to +200 G (red). (\textit{b}) Mean polar field calculated by averaging the radial field in (\textit{a}) over latitudes poleward of $\pm 88^\circ$.  Blue and red curves correspond to the northern and southern hemispheres respectively and dotted lines of each color indicate polar field reversals.  (\textit{c}) Time-latitude plot of the mean toroidal field $\left<B_\phi\right>$ at the bottom of the convection zone $r = 0.71 R$.  The color scale saturates at $\pm$ 50 kG, with red and blue denoting eastward and westward field respectively.  (\textit{d}) Mean toroidal flux at low latitudes near the base of the CZ, obtained by averaging the plots in (\textit{c}) over the northern (blue) and southern (red) hemispheres.}
\label{fig:bfly1}
\end{figure*}

In this section, we present some results for the case in which ${\bf v}$ is axisymmetric, consisting only of differential rotation and meridional circulation. These will be compared to the results presented in Section \ref{sec:results} that include 3D convective motions.   All of the simulations presented in this chapter are summarized in Table \ref{cases}.  $B_{pol}$ and $B_{tor}$ refer to the amplitude of the mean poloidal and toroidal fields and $B_{nax}$ refers to the amplitude of the non-axisymmetric field components.  In each case we list the mean and standard deviation $\sigma$ of the time series.  The cycle period is calculated by reversals of the mean toroidal field in the lower CZ (e.g.\ Figs.\ \ref{fig:bfly1}\textit{d} and \ref{fig:bfly_con}\textit{d}).  The primary motivation for the range of diffusive simulations A1-A8 is to assess what value of $\eta_{top}$ most closely corresponds to explicit convective transport, as discussed in Sections \ref{sec:results} and \ref{sec:discussion}.  

The initial conditions for all cases is a weak dipole field.  This gets stretched out by the differential rotation and when the toroidal field strength exceeds the threshold value of 1 kG, BMRs begin to appear.  Soon the field strengths become strong enough to saturation the dynamo by means of eq.\ (\ref{flux}).  Case A1 has been run for over 260 years, which corresponds to about 10-15 cycles after the dynamo saturates and settles down to a steady cycle.  Cases A2-A8 have each been run for about 120 years, around 6-7 cycles after saturation.  Cases C1 and C3 have also been run for nearly 190 years each (7-8 cycles after saturation, see Sec.~\ref{sec:results}).

As discussed in Sec.\ \ref{sec:intro}, SFT models rely on flux injection through artificial or observed BMR databases and solve only the radial component of the induction equation. Meanwhile, previous 2D FTD models were not able to reproduce the evolution of the non-axisymmetric components of the surface field.  STABLE unifies these two models by spontaneously producing BMRs based on the dynamo-generated toroidal field and by capturing their subsequent evolution after emergence.  

One of the main aims of STABLE is to reproduce the observed butterfly diagram (radial field) of the solar photosphere.  Though producing solar-like butterfly diagrams is a valuable test of any solar dynamo model, most rely on the mean toroidal field near the base of the convection zone as a proxy for the surface field.  Since STABLE produces explicit BMRs, there is no need for such a proxy.  Both quantities are shown in Fig.\ \ref{fig:bfly1}.  Notable solar-like features include equatorward propagation of active bands at low latitudes and poleward migration of trailing BMR flux at high latitudes, which eventually reverses the polar fields. The equatorward propagation of active bands at the surface (Fig.\ \ref{fig:bfly1}\textit{a}) is a consequence of the subsurface toroidal field propagation (Fig.\ \ref{fig:bfly1}\textit{c}). Note that polar field reversals occur a few years after the peak toroidal field, comparable to solar observations (Fig.\ \ref{fig:bfly1} \textit{c},\textit{d}). 

\begin{figure*}[]
\centering
\includegraphics[width=0.95\textwidth, angle=0]{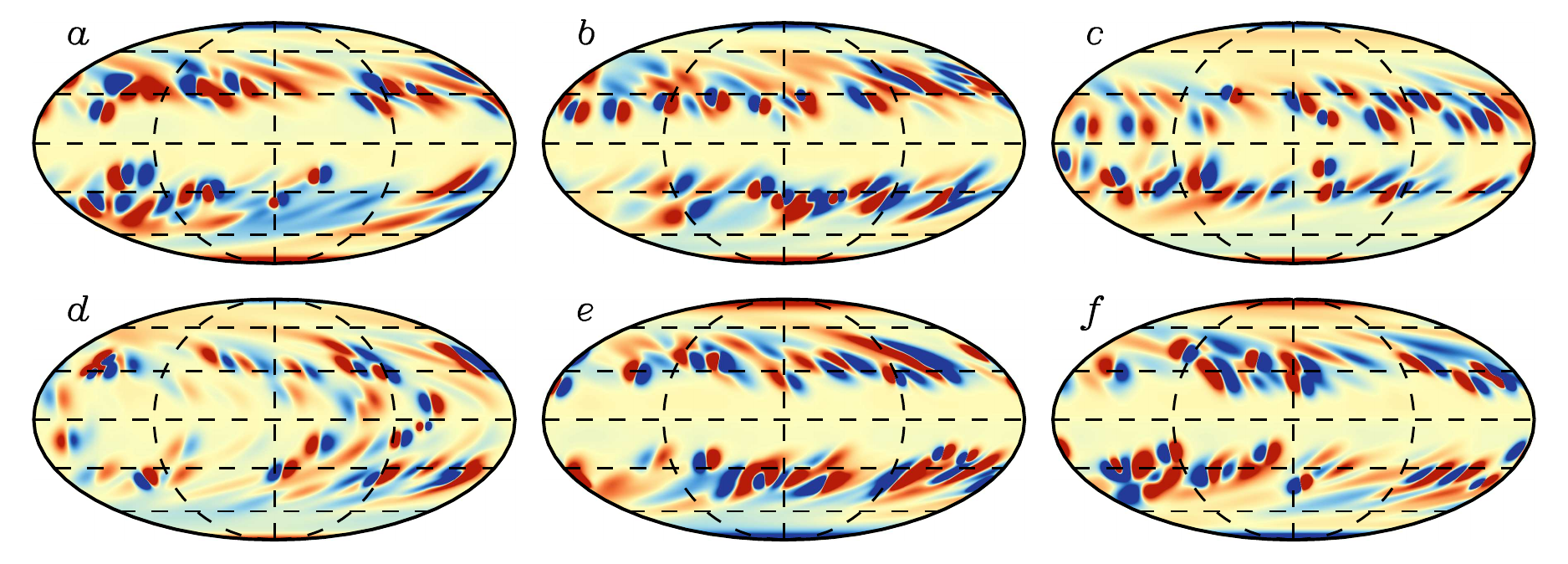}
\caption[Surface transport of sunspots with diffusivity 300 km$^2$s$^{-1}$]{Molleweide projection of $B_r$ at $r=R$ in Case A1 at six different times that span a full magnetic cycle: (\textit{a}) 223.9 yr, (\textit{b}) 226.0 yr, (\textit{c}) 229.0 yr, (\textit{d}) 232.0 yr, (\textit{e}) 234.0 yr and (\textit{f}) 237.0 yr. Red and blue denote outward and inward field respectively and the saturation level on the color table is $\pm $ 1 kG.  Dashed lines denote latitudes of 0$^\circ$, $\pm 30^\circ$, and $\pm 60^\circ$.}
\label{fig:ss_A1}
\end{figure*}

\begin{figure*}[]
\centering
\includegraphics[width=0.95\textwidth, angle=0]{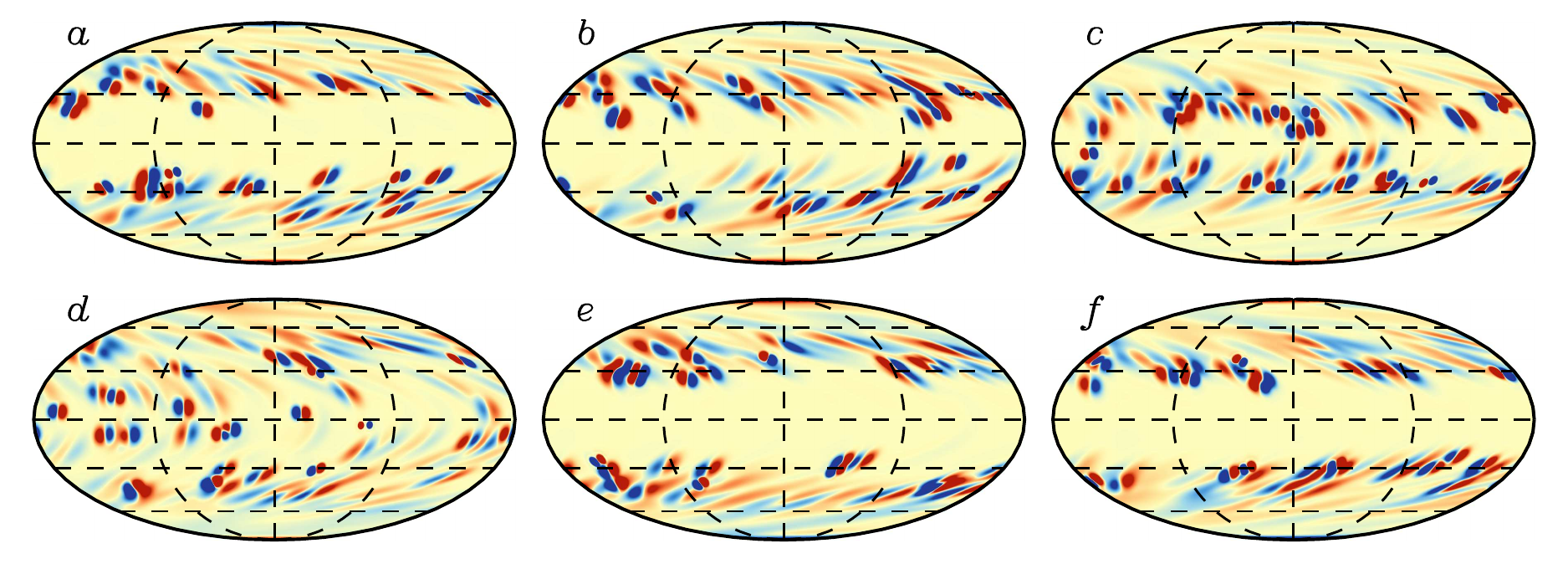}
\caption[Surface transport of sunspots with diffusivity 80 km$^2$s$^{-1}$]{Similar to Fig.\ \ref{fig:ss_A1} but for Case A4, with a color saturation level of $\pm$ 3 kG.  The times again span a full magnetic cycle: (\textit{a}) 124.0yr, (\textit{b}) 127.0 yr, (\textit{c}) 130.0 yr, (\textit{d}) 133.0 yr, (\textit{e}) 136.0 yr and (\textit{f}) 139.0 yr.}
%\caption{Similar to Fig.\ \ref{fig:ss_A1} but for Case A5, with a color saturation level of $\pm$ 5 kG.  The times again span a full magnetic cycle: (\textit{a}) 83.9 yr, (\textit{b}) 85.8 yr, (\textit{c}) 88.9 yr, (\textit{d}) 91.9 yr, (\textit{e}) 93.8 yr and (\textit{e}) 96.1 yr.}
\label{fig:ss_A4}
\end{figure*}

\begin{figure*}[]
\centering
\includegraphics[width=1.0\textwidth, angle=0]{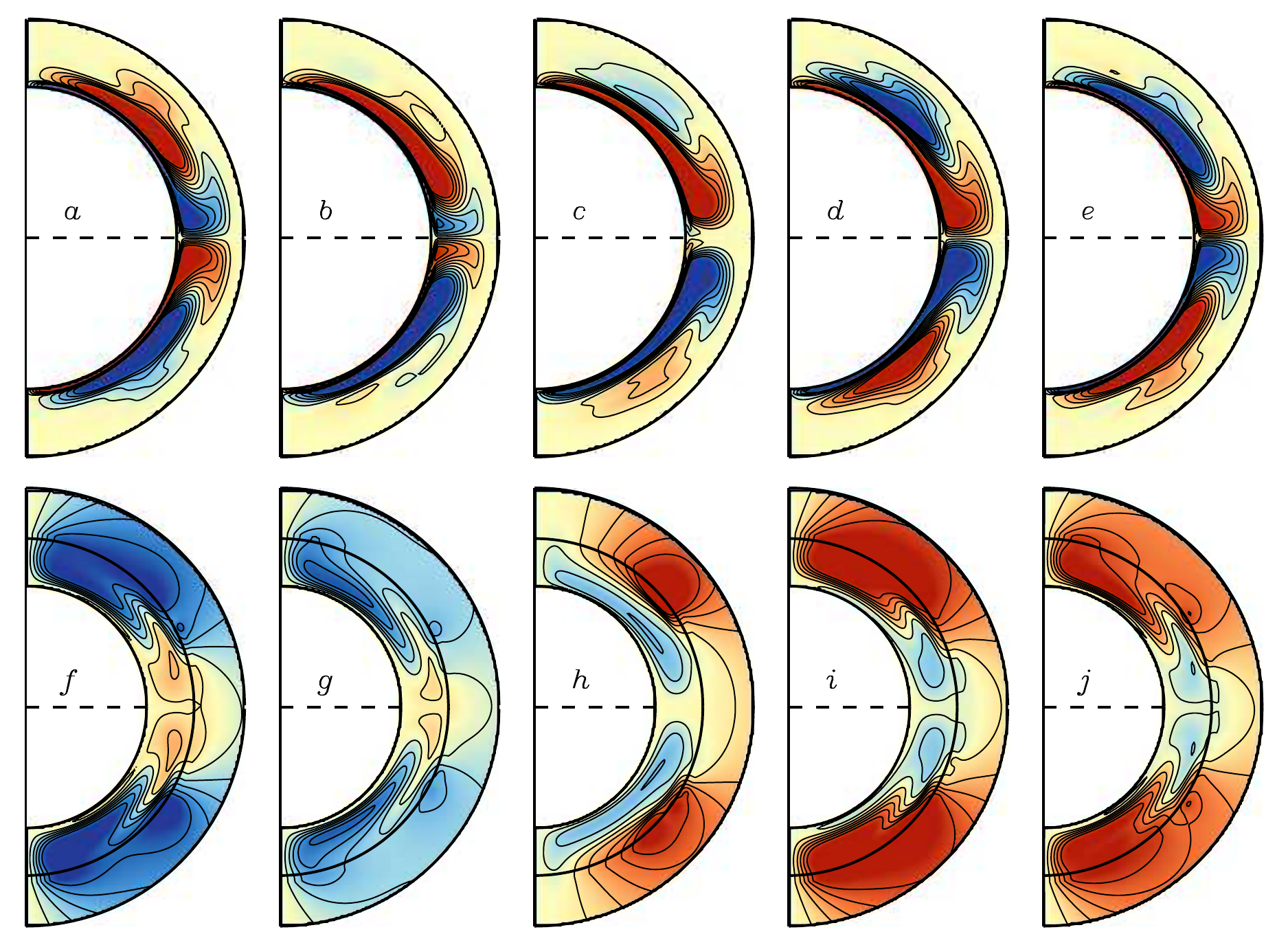}
\caption[Mean magnetic field lines with diffusion]{Evolution of mean (\textit{a}--\textit{e}) toroidal and (\textit{f}--\textit{j}) poloidal fields in Case A1, spanning one magnetic cycle.  Snapshots are shown for the same magnetic cycle as in Fig~\ref{fig:ss_A1}: $t =$ (\textit{a},\textit{f}) 226.0 yr, (\textit{b},\textit{g}) 229.2 yr, (\textit{c},\textit{h}) 233.0 yr, (\textit{d},\textit{i}) 237.1 yr and (\textit{e},\textit{j}) 239.9 yr.  Frames (\textit{a}--\textit{e} show $\left<B_\phi\right>$ with red and blue indicating eastward and westward field respectively.  Frames (\textit{f}--\textit{j}) show the poloidal magnetic potential with a potential field extrapolation above $r=R$ to $r=1.25 R$.  Colors indicate (red) clockwise and (blue) counter-clockwise field orientations.  Average toroidal and poloidal field strengths are about 33 kG and 596 G respectively, as listed in Table \ref{cases}, though peak toroidal fields can exceed 100 kG (the color table for $\left<B_\phi\right>$ is clipped at $\pm 50$ kG).}
\label{fig:bmean_sand1}
\end{figure*}

Figure \ref{fig:ss_A1} highlights the surface flux transport in Case A1.  Compare this with Fig.\ \ref{fig:ss_A4}, which shows the same thing but for a case with a lower surface diffusion (Case A4).  The structures in the former case tend to be wider and spread out, commensurate with the higher diffusion. This is reflected in lower average field strengths for the non-axisymmetric field components: 1.76 kG in Case A1 versus 4.64 kG in Case A4 (Table \ref{cases}). Higher diffusion always make a dynamo less efficient. The mean fields in Case A5-A8 are about 3-4 times stronger than in Case A1 even though the quenching field strength $B_q$ is the same in both Cases (see eq.\ \ref{flux}).  The evolution of the mean fields for Case A1 is shown in Fig.\ \ref{fig:bmean_sand1}, which highlights the general FTD aspects of these simulations. The basic operation of the dynamo is similar to that described in \citet{MD14} and \citet{MT16}. Differences between those previous results and the results shown here are mainly due to modified meridional circulation profile described in \citet{HCM17} and the varying values of $\eta_{top}$.

The cycle period increases somewhat as $\eta_{top}$ is decreased (Table \ref{cases}).  We attribute this to a ``short-circuiting'' of the flux-transport dynamo by diffusive mixing.  Even though all cases have the same value of $\eta$ in the mid and lower CZ, a greater value of $\eta_{top}$ promotes a more efficient downward transport of poloidal flux from the surface to the base of the CZ.  The difference is significant, but not substantial.  A decrease in $\eta_{top}$ by a factor of 300 (from A1 to A8) lengthens the cycle period by about 20\%.  

Note that this result is in contrast to other FTD dynamo parameter studies that report a slight increase in the cycle period as the diffusion is increased \citep{DC99,Yeates08}.  However, these studies mainly focus on increasing the diffusivity in the bulk of the convection zone, not in the surface layers.  For example, in terms of our notation, \citet{Yeates08} vary $\eta_{mid}$ while keeping $\eta_{top}$ fixed, whereas we do the opposite.  This accounts for the difference in our results, as we now explain.

There are at least three ways in which an increase in the turbulent diffusion can influence the cycle period in a Flux-Transport dynamo model:  (A) The ``short-circuit'' effect noted above, by which diffusion enhances the efficiency of poloidal flux transport across the CZ, (B) a decrease in the efficiency of equatorward transport of toroidal flux at the base of the CZ, and (C) a decrease in the efficiency of toroidal flux generation at the base of the CZ.  Effect (A) tends to decrease the cycle period, as described above.  Meanwhile, effects (B) and (C) tend to increase the cycle period.  

Effect (B) arises because of a decrease in the magnetic Reynolds number associated with the MC near the base of the CZ, Rm $ = U_{mc} L_{mc}/\eta_{mid}$, where $U_{mc}$ and $L_{mc}$ are the relevant velocity and length scales.  Lower Rm inhibits transport, as the magnetic field slips through the plasma.  This slows down the equatorward advection of both toroidal and poloidal flux.  It also tends to smooth out the structure of the field, which decreases the field amplitudes and inhibits toroidal field generation by the $\Omega$-effect.  This is effect (C) referred to in the previous paragraph.  In a purely kinematic model, the toroidal field generation time scale would be independent of the field strength.  However, many parameter studies such as \citet{DC99} and \citet{Yeates08} include a quenching field strength for the toroidal field which limits the amplitude of the dynamo.  At high values of $\eta_{mid}$, it takes the dynamo longer to reach this quenching field strength.  

The simulations by \citet{DC99} and \citet{Yeates08} include all three effects, but the latter two dominate.  For example, a comparison of Figs.\ 8\textit{d} and 8\textit{i} in \citet{Yeates08} clearly shows effect (A).  By contrast, our simulations only include effect (A) because we keep the value of $\eta_{mid}$ (and thus the value of Rm in the lower CZ) fixed in all of our simulations.  We are aware of one previous study that investigated the influence of $\eta_{top}$ on the operation of an FTD dynamo with fixed $\eta_{mid}$, namely \citep{Hotta10}.  They find, as do we, that the cycle period decreases with increasing $\eta_{top}$ (see their Fig.\ 5).

\section{Convective Flows}\label{sec:con}

In this section we describe dynamo simulations that include 3D convective flow fields as described in Section \ref{sec:cflow}.  We remind the reader that these are not full MHD simulations.  Rather, the convective velocity field is prescribed in a kinematic sense based on observations of the photospheric power spectrum.  We also remind the reader that the imposed convection does not occupy the entire convection zone.  Rather, it is intended to mimic the relatively vigorous, small-scale convective motions in the solar surface layers ($r > 0.9R$) that are well understood both observationally and theoretically.  But As described in the section \ref{sec:ssd}, we eventually keep the convective flow confined near the surface layer only ($r\sim R$). This imposed convective flow is in lieu of the turbulent diffusion that is used to represent surface convection in most SFT and FTD models \citep{Jiang_review15}.  Thus, we correspondingly reduce the turbulent diffusion coefficient near the surface, $\eta_{top}$, as shown in Fig.\ \ref{fig:eta}.  This follows the approach used by \citet{upton14a,upton14b} in their AFT model. In the deeper CZ we continue to represent convective transport as a turbulent diffusion.  For further details on the implementation, see Section \ref{sec:cflow}.
\label{sec:results}

\subsection{Amplification of fields by convective flows}\label{sec:ssd}
Babcock-Leighton (BL) dynamo models rely on the dispersal and migration of magnetic field by near-surface convection but they typically neglect the fact that these same convective motions can operate as a small-scale dynamo \citep[e.g.][]{Cattaneo99}.   In our model, when we include 3D convective motions as described in Sec.\ \ref{sec:cflow}, we find that this flow does indeed operate as a small-scale dynamo.

A characteristic length scale of the convection can be estimated as $L \sim 2 \pi R / \ell_0 \sim$ 34 Mm, where $\ell_0 \sim 130$ is the spherical harmonic degree at which the spectrum peaks (Fig.\ \ref{fig:conhath}\textit{c}).  For a velocity scale, we can use the rms value of $U \sim 250$ m s$^{-1}$.  Together with $\eta_{top} = 5 \times 10^{10}$ cm$^2$ s$^{-1}$, this implies a magnetic Reynolds number of $R_m = UL/\eta_{top} \sim 1700$.  So, it is not surprising that this flow is supercritical to small-scale dynamo action.  We find that it is even supercritical for $\eta_{top} = 5 \times 10^{11}$ ($R_m \sim 170$).  Though this behavior is reasonable, it is not conducive to establishing magnetic cycles.

We find that small scale dynamo action disrupts the operation of the BL dynamo.  BMRs are quickly overwhelmed by random small-scale fields that grow exponentially. This inhibits poloidal field generation by the BL mechanism and thus suppresses the magnetic cycles.  This problem is not found in the AFT model \citep{upton14a,upton14b}, which uses the same surface flows, because AFT is 2D and therefore cannot exhibit sustained dynamo action.

We have considered several approaches to suppressing this small-scale dynamo action in order to get a functioning BL dynamo.  First, we enhanced the dissipation on small (convective) scales by imposing a horizontal hyperdiffusion $\nabla_h^4$ or $\nabla_h^8$.  But we were unable to find appropriate hyperdiffusion coefficients to yield a cyclic dynamo solution.   Then, we attempted to saturate the small-scale fields by introducing a threshold field strength $B_s$, applied to the non-axisymmetric field components.  However, it was not possible to apply this selectively to suppress dynamo-generated fields but not the remnant fields from emerging BMRs.  We also tried nonlinear quenching of the non-axisymmetric field, with similar results.

We were able to achieve some degree of BL activity by using a drag term of the form
\begin{equation}
\frac{{\pd \bf B}^\prime}{\pd t} = - \frac{{\bf B}^\prime}{\tau} + \ldots
\end{equation}
where ${\bf B}^\prime$ is the non-axisymmetric field and $\tau \sim$ 24 hrs is a suppression time scale that is comparable to the small-scale dynamo growth rate.   This is effectively a scale-independent variation of the hyperviscosity approach.  However, the solutions were not solar-like, exhibiting a large asymmetry about the equator, with BMRs in one hemisphere and not the other.

The only effective way we found to suppress small-scale dynamo action is to essentially make the convective flow 2D, as in AFT.  Thus, for Cases C1, C2, and C3 presented in this chapter, we have only applied the convective flow to the radial field $B_r$ at the radial grid point that is adjacent to the top boundary.  Thus, the upper boundary condition (radial field) is applied at radial grid point 1 and the convective flow is included by adding it to the radial component of the induction equation (\ref{induction2}) at grid point 2.  And, even in this case, we still found it beneficial to use a higher value of the hyperdiffusion ($\eta_h = 3\times 10^{10}$ cm$^2$ s$^{-1}$) relative to the diffusive Cases A1-A8 ($\eta_h = 2 \times 10^8$ cm$^2$ s$^{-1}$) in order to suppress small-scale fields (see Sec.\ \ref{sec:STABLE}).  In this way we were able to achieve a viable BL dynamo model as described in section \ref{sec:transport}.

We appreciate that this solution is not ideal, but it is required by the kinematic nature of our model.  In the Sun, small-scale dynamo action surely occurs but it is saturated by Lorentz-force feedbacks.  The only way to self-consistently capture this in our model is to solve the full set of anelastic MHD equations.  We do intend to implement this, but it would require a substantial increase in the sophistication of the model so we defer it to a future publication.

\subsection{Results with convective flows}\label{sec:transport}
In this section, we present results with explicit convective motions as the effective transport mechanism of large scale magnetic field on the surface of the Sun, in addition to meridional circulation and differential rotation.  This convective transport is limited to the horizontal advection of vertical field near the upper boundary as described in Sec.\ \ref{sec:ssd}, though we still decrease the explicit diffusion $\eta_{top}$ as illustrated in Fig.\ \ref{fig:eta}.  Thus, when we refer to including {\em convective flows} in our models we really mean explicit horizontal flow fields chosen to mimic the observed properties of convection on the solar surface.  Convective transport in the mid convection zone is still modeled with a turbulent diffusivity.

It is worth noting that the introduction of the convective flow field ${\bf v}_c$ increases the computational expense substantially.   This is because of the Courant-Friedrichs-Lewy (CFL) constraint on the time step.  At the horizontal resolution used here ($N_\theta = $ 512, $N_\phi = 1024$), the peak convective velocity is on the order of 900 m s$^{-1}$.  This is a factor of 6.7 larger than the peak axisymmetric flow speed of 135 m s$^{-1}$, requiring a commensurate decrease in the time step.  If vertical convective motions are included this requires a further decrease of the time step since the vertical grid spacing is about a factor of 7 smaller than the horizontal grid spacing.  Turbulent diffusion is not subject to a CFL constraint because it is handled by a semi-implicit Crank-Nicolson scheme.

We have performed three simulations with convection, as summarized in Table \ref{cases}.  Case C1 is a fiducial solar-like dynamo solution that we will discuss further below.  Case C2 is similar to Case C1 but with $\alpha_{spot} = 1$.  This is found to be sub-critical for sustained dynamo action, as found in many diffusive cases (MT16).  Case C3 is similar to Case C1 but with a lower value of $\eta_{top}$; $5 \times 10^{10}$ cm$^2$ s$^{-1}$ instead of $\eta_{top}$; $5 \times 10^{11}$ cm$^2$ s$^{-1}$.  The rationale for choosing a low value is because $\eta_{top}$ is intended to parameterize the convective transport that we are now capturing explicitly, as emphasized in Sec.\ \ref{sec:cflow}.  However, we find that if we make $\eta_{top}$ too low the dynamo flips into a quadrupolar parity ($\left<B_r\right>$ and $\left<B_\phi\right>$ symmetric about the equator).  We believe that Case C3 is in that quadrupolar regime.  Though it's parity is still largely negative (dipolar) after 190 years of evolution, it is tending toward quadrupolar.  Furthermore, a longer simulation with lower resolution (otherwise the same parameters) shifts to quadrupolar after about 300 years.

This is a known feature of BL dynamo models.  Efficient diffusive coupling between the two hemispheres is known to promote dipolar parity ($\left<B_r\right>$ and $\left<B_\phi\right>$ antisymmetric about the equator), as demonstrated by \citet{CNC04} and \citet{Hotta10}.  Evidently, the convective transport alone is not sufficient to establish dipolar parity in our models.  This may be because the explicit convective flows are limited to the surface whereas the 3D structure of the BMRs upon emergence extends down to $0.9R$.  In short, the net turbulent transport (advection plus diffusion) in Case C3 appears to be insufficient to yield dipolar parity but in Case C1 it is.  The value of $\eta_{top}$ in Case C1 is high enough to enhance the subsurface diffusive coupling between hemispheres but low enough that the explicit convective flows dominate the surface flux transport (as will be demonstrated below).  Since the Sun exhibits dipolar parity, we will hereafter focus on Case C1.

\begin{figure*}[]
\centering
\includegraphics[width=0.95\textwidth, angle=0]{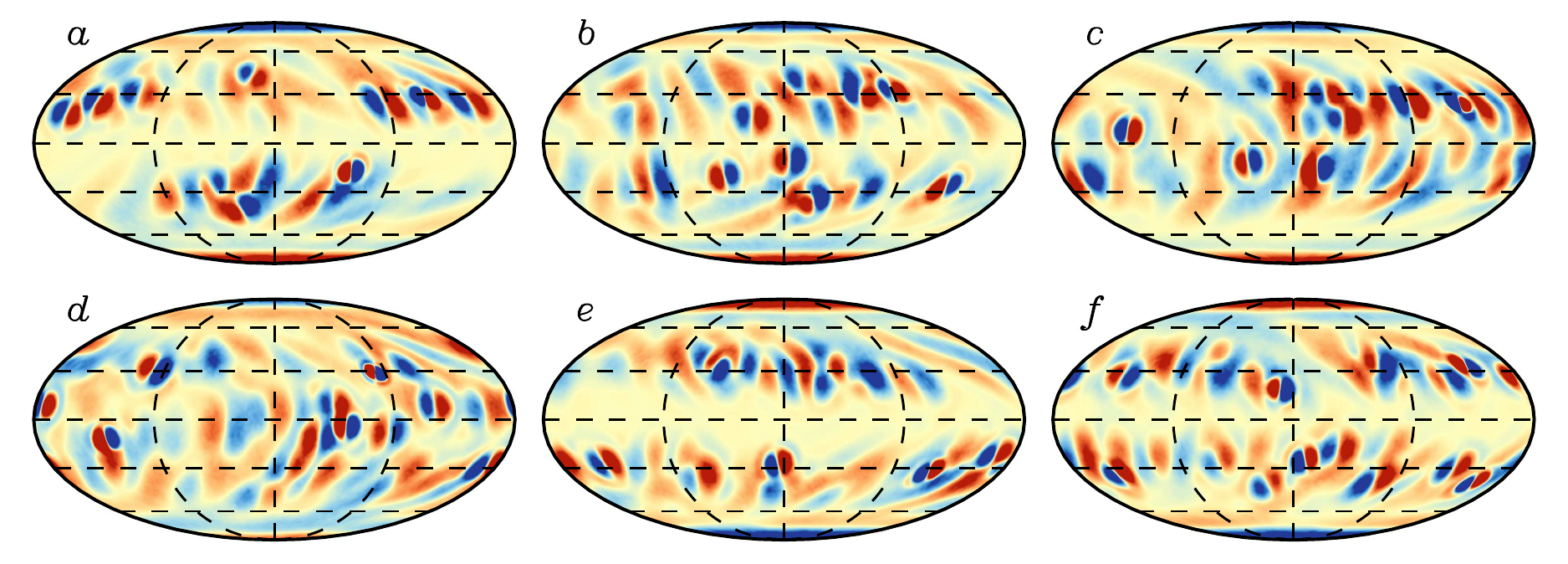}
\caption[Similar to Figures \ref{fig:ss_A1} and \ref{fig:ss_A4} but for the convective Case]{Similar to Figures \ref{fig:ss_A1} and \ref{fig:ss_A4} but for the convective Case C1, with a color saturation level of $\pm$ 3 kG.  Snapshots are shown for $t =$ (\textit{a}) 165.0 yr, (\textit{b}) 168.1 yr, (\textit{c}) 171.0 yr, (\textit{d}) 173.9 yr, (\textit{e}) 176.0 yr and (\textit{f}) 180.0 yr.}
\label{fig:ss_con}
\end{figure*}

\begin{figure*}[t!]
\centering
\includegraphics[width=0.95\textwidth, angle=0]{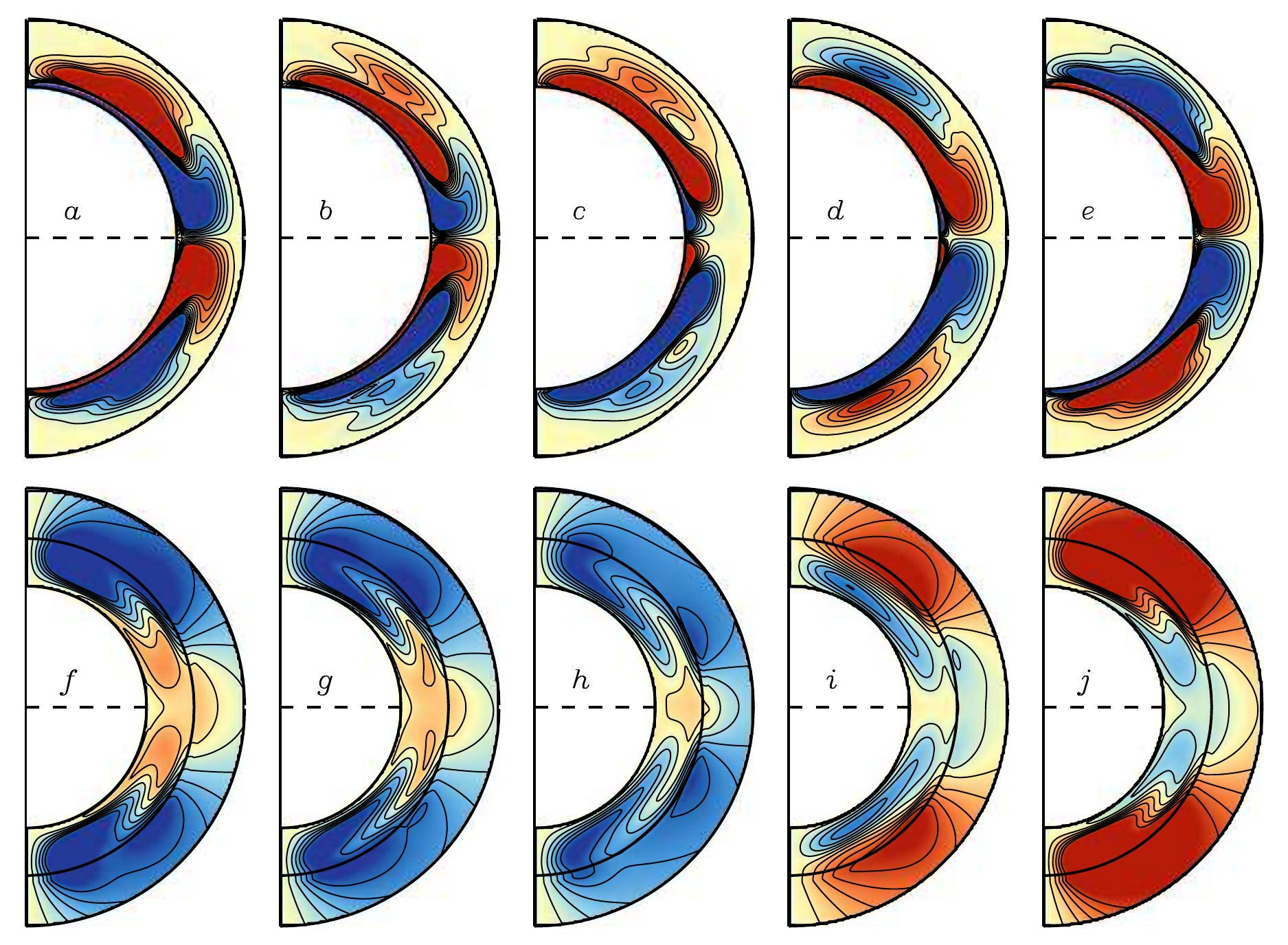}
\caption[Similar to Fig.\ \ref{fig:bmean_sand1} but for convective case]{As in Fig.\ \ref{fig:bmean_sand1} but for Case C1.  This spans the same magnetic cycle as in Fig.\ \ref{fig:ss_con}, at times of $t = $ (\textit{a},\textit{f}) 166.0 yr, (\textit{b},\textit{g}) 169.0 yr, (\textit{c},\textit{h}) 172.0 yr, (\textit{d},\textit{i}) 175.0 yr and (\textit{e},\textit{j}) 179.0 yr.}
\label{fig:bmean_con}
\end{figure*}

\begin{figure*}[t!]
\centering
\includegraphics[width=0.95\textwidth, angle=0]{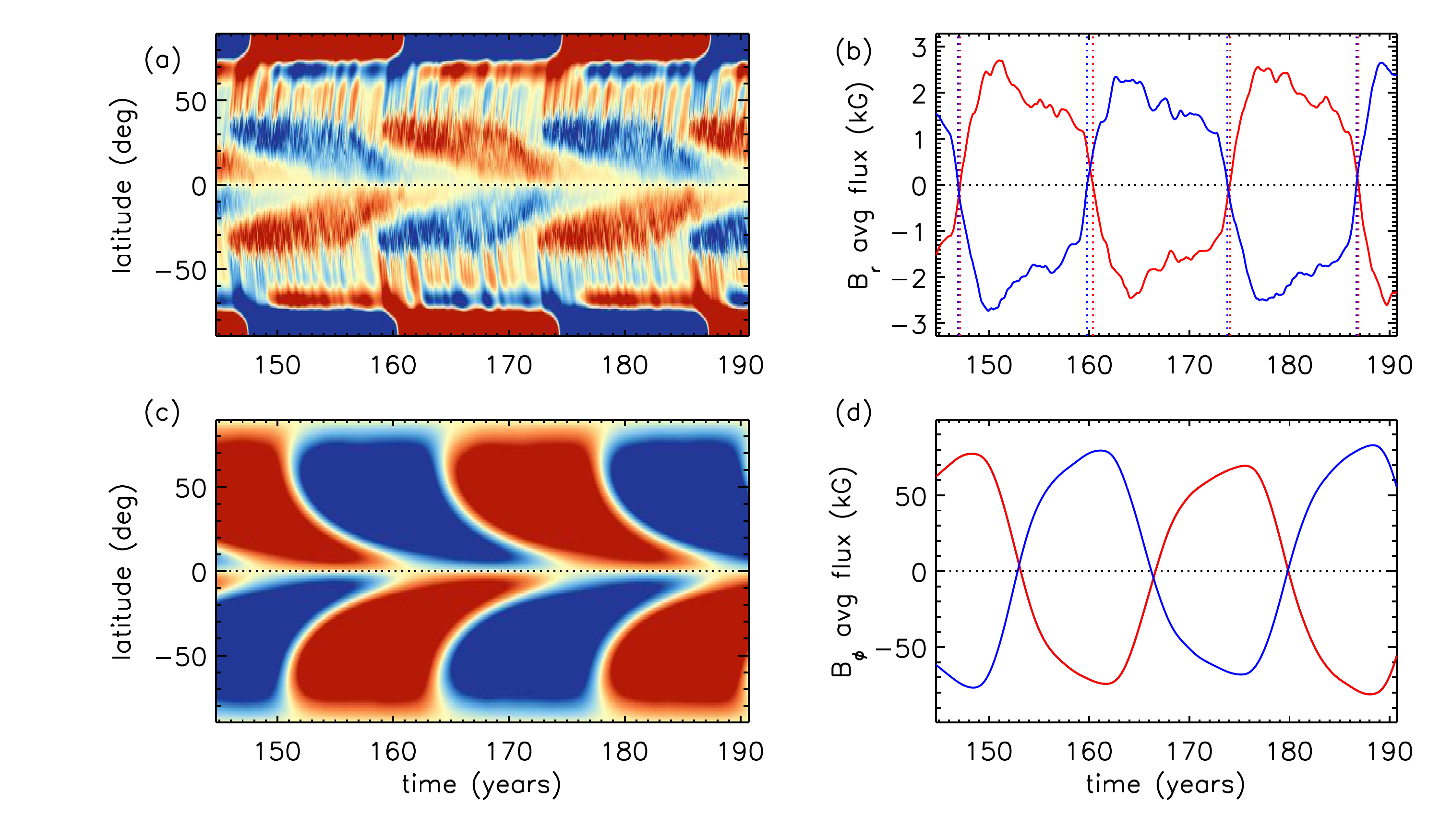}
\caption[As in Fig.\ \ref{fig:bfly1} but for convective case]{As in Fig.\ \ref{fig:bfly1} but for Case C1.  Here the color tables saturate at (\textit{a}) $\pm $ 500 G and (\textit{c}) $\pm$ 100 kG.}
\label{fig:bfly_con}
\end{figure*}

Fig.\ \ref{fig:ss_con} shows the evolution of the surface flux in Case C1.  Qualitatively it looks more like Case A1 (Figs.\ \ref{fig:ss_A1}) than A4 (Figs.\ \ref{fig:ss_A4}), with more diffuse bipolar structures at low latitudes, though Case C1 has stronger fields (Table \ref{cases}).   The evolution of the mean fields in Case C1 is also similar to Case A1, as shown in Fig.\ \ref{fig:bmean_con}.   This suggests that the efficiency of convective flux transport is comparable to that of the turbulent diffusion in Case A1. We will see below that this first impression is borne out by a more thorough analysis. 

The poloidal and toroidal butterfly diagrams for Case C1 are shown in Fig.\ \ref{fig:bfly_con}.  The most apparent differences in the surface flux transport (Fig.\ \ref{fig:bfly_con}\textit{a}) relative to Case A1 (Fig.\ \ref{fig:bfly1}\textit{a}) are less pronounced bipolar structures in the low latitudes and mix polarity polar caps. In Case C1, active regions are shredded by convective flows which results a less concentrated active regions compared to Case A1. A more subtle difference is that the poleward migration rate of the streams is somewhat more rapid (see Section \ref{sec:migration}).  Though the broader polar regions in Case C1 possess mixed polarities, the polar fields are stronger than the unipolar polar regions of Case A1; compare Figs.\ \ref{fig:bfly1}\textit{b} and \ref{fig:bfly_con}\textit{b}.  Such behavior may be attributed to the tendency for the convective motions to disperse and transport BMR fields without dissipating them (see Sec.\ \ref{sec:energetics}).  In the case of A1, BMRs emerge and the opposite polarities partially cancel each other as they disperse.  By contrast, in Case C1, the fields disperse but cancellation is less efficient as a result of the smaller ohmic diffusion.  Thus, both polarities are transported poleward and concentrated into strong, alternating bands.

The shape of the polar flux plot for Case C1 (\ref{fig:bfly_con}\textit{b}) is similar to Case A1 (Fig.\ \ref{fig:bfly1}\textit{b}) but with somewhat more variation and a sharper decay at the end of each cycle.  This variability reflects the mixed polarity fields that cross into the polar regions before they cancel one another. The slower decay phase for Case C1 implies a longer interval of polar flux generation by poleward migrating streams which persists for almost the entire cycle, as seen in Fig.\ \ref{fig:bfly_con}\textit{a}. By contrast, poleward migrating streams are less prominent late in the cycle for Case A1 (Fig.\ \ref{fig:bfly1}\textit{a}).   As BMRs emerge at progressively lower latitudes, most of the emerging flux in this more diffusive case cancels locally before it can migrate to higher latitudes.  This sustained flux emergence at mid latitudes is not supported by solar observations, which show few mid-latitude BMRs in the declining phase of the cycle. This discrepancy may be due in part to our simple flux emergence algorithm.  Resolving it will likely require a better understanding of the flux emergence process.

The sustained supply of poloidal flux to the poles throughout most of the cycle in Case C1 also has consequences for the toroidal field generation.  As this flux is transported to the base of the CZ by meridional circulation and turbulent diffusion, it promotes sustained toroidal flux generation through the $\Omega$-effect, particularly at mid-latitudes where the latitudinal shear is strongest. This is evident in Fig.\ \ref{fig:bfly_con}\textit{c} which shows strong toroidal fields persist near $\pm 50-60^\circ$ throughout most of the cycle.   Compare this to Case A1, (Fig.~\ref{fig:bfly1}\textit(c)), where the mid-latitude toroidal flux diminishes late in the cycle as the bands propagate equatorward.  This excess of mid-latitude flux in Case C1 also accounts for the distortion of the integrated toroidal flux curve (Fig.\ \ref{fig:bfly_con}\textit{d}), which peaks later in the cycle than Case A1 (Fig.\ Fig.\ \ref{fig:bfly1}\textit{d}).  

\begin{figure*}[t!]
\centering
\includegraphics[width=0.95\textwidth, angle=0]{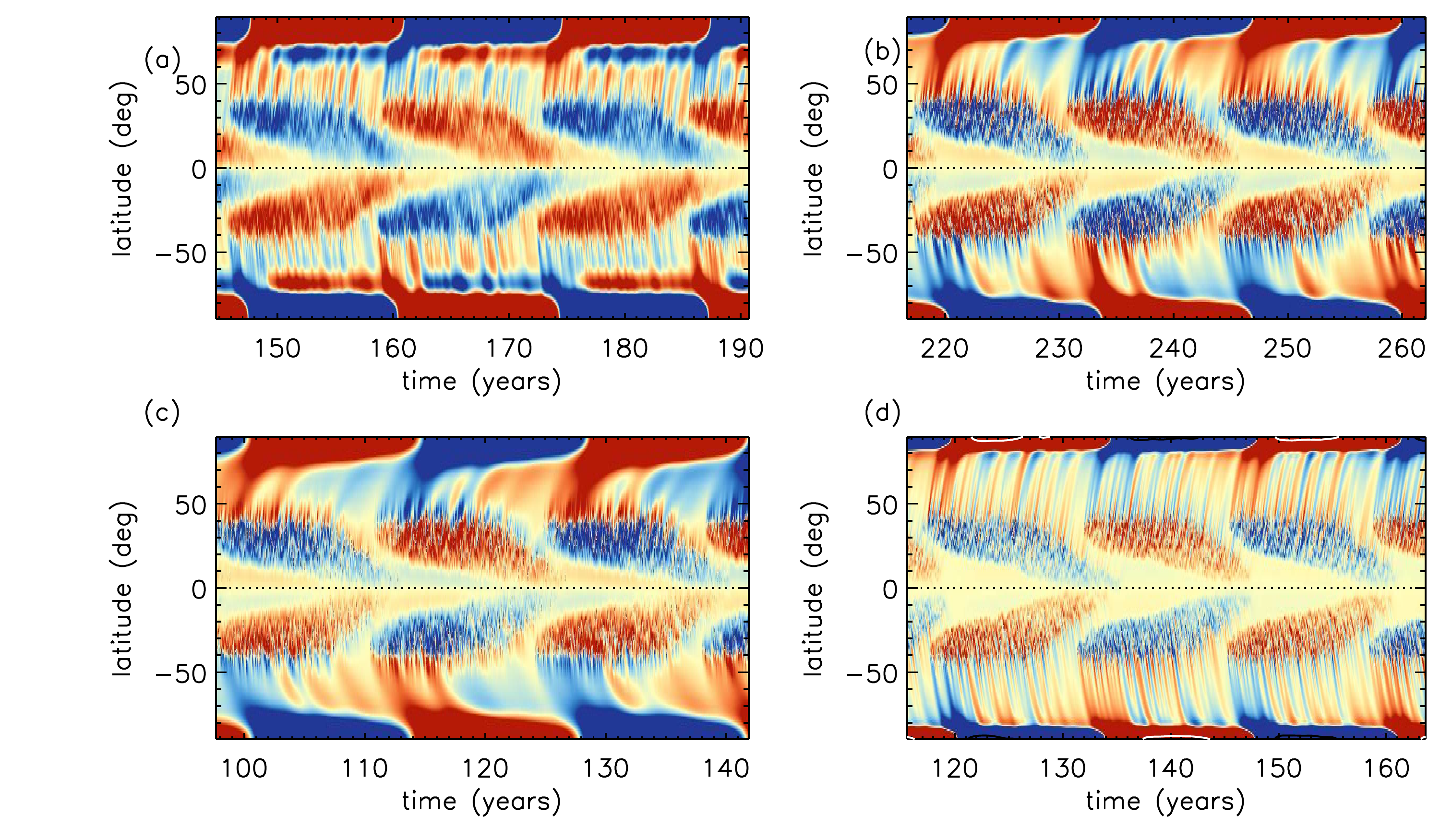}
\caption[butterfly diagram for different important cases]{Time latitude plot of $\left<B_r\right>$ at $r=R$ for: (a) Case C1 ($B_r$ scale $=\pm 500$G). (b) Case A1 with $\eta_{top} = 3\times 10^{12}$ ($Br$ scale $=\pm 200$ G) (c) Case A2 with $\eta_{top} = 10^{13}$ ($Br$ scale $= \pm 30$ G)  and (d) Case A4 with $\eta_{top} = 8\times 10^{11}$ ($B_r$ scale $=\pm 2000$ G).} 
\label{fig:all_bfly}
\end{figure*} 

Figure \ref{fig:all_bfly} compares the surface butterfly diagram in Case C1 to several diffusive cases with relatively high and low values of $\eta_{top}$, ranging from $1 \times 10^{13}$ cm$^2$ s$^{-1}$ to $8 \times 10^{11}$ cm$^2$ s$^{-1}$.  Qualitatively, it bears the greatest resemblance to Case A1. This conclusion is based on the width of the poleward-migrating streams (though note the different ranges for the time axes), the structure and width of the polar flux concentrations, the relative strength of the polar and low-latitude fields, and the location of the active latitudes.  In Section \ref{sec:migration} we take a more quantitative approach to comparing the poleward migration speeds.

It is worth noting that the polar field strength of $\sim$ 1.5 kG in Case C1 is significantly higher than typical measurements of the polar fields in the solar photosphere, which are on the order of 1-3 G \citep[e.g.]{Munoz12}. In our model, we calculate the polar field by averaging the radial magnetic field over the region $88^\circ$ to pole.  However, observational measurements usually quote the line-of-sight field strength averaged over a large region, often down to latitudes of 70$^\circ$ or even 55$^\circ$.  Projection and averaging effects therefore diminish the observed field strengths.  After correcting for these effects, the observed value of the polar field strength is on the order of 10 G, which is consistent with the backwards extrapolation of coronal and heliospheric measurements and models \citep{gibso99}.  Also, it must be remembered that the BMR fluxes that we use here are artificially large.  As discussed in Section \ref{sec:STABLE}, we enhance the photospheric flux budget by setting $\alpha_{\rm spot} > 1$ in order to ensure that the dynamo solution is super-critical.  For Case C1 $\alpha_{\rm spot} = 15$.  We have confirmed that setting $\alpha_{\rm spot} = 1$ leads to decaying solutions (Case C2 in Table \ref{cases}).  Still, the high polar field strengths are a concern and are a known issue with BL/FT models.   A possible remedy might be to take into account the connectivity of emerging BMRs with their deep-seated roots.  \citet{YM13} show that this can shift the region of poloidal field generation more to the interior, producing weaker poloidal fields at the surface and particularly at the poles.

In summary, the convective Case C1 behaves in many ways like the diffusive Case A1, which has $\eta_{top}$ $\sim 3 \times 10^{12}$ cm$^2$ s$^{-1}$. However, the mean fields in Case C1 are much stronger.  In the next section we investigate why this is and we take a closer look at the similarities and differences between Case C1 and the cases in which convective transport is approximated by turbulent diffusion.

\section{Discussion: Does Convection Operate as a Turbulent Diffusion?}\label{sec:discussion}

The main objective of our study is to improve the fidelity of the surface flux transport in our 3D Babcock-Leighton dynamo model by replacing turbulent diffusion with a more realistic depiction of photospheric convection.  However, as mentioned in Section \ref{sec:results}, this introduces challenges, including small-scale dynamo action and reduced computational efficiency.  For this reason, and also for understanding the nature of convective transport, it is important to ask how much we gain from the explicit convective flows.  Is convective transport accurately parameterized by a turbulent diffusion or is is fundamentally non-diffusive?  If the former, what value of $\eta_{top}$ is optimal?  These are the questions we address in this section.

In Section \ref{sec:results} we argued that, at least qualitatively, the convective Case C1 resembles Case A1, which has $\eta_{top} = 3 \times 10^{12}$ cm$^2$ s$^{-1}$.  This is comparable to estimates of the turbulent diffusion from photospheric observations of magnetic flux elements, which suggest $\eta_{top} \sim$ 2--6 $\times 10^{12}$ cm$^2$ s$^{-1}$ \citep{mosher77,topka82,schri96,abram11}.  In mean-field theory, the value of $\eta_t$ is linked to the kinetic energy of the turbulence.  In particular, if $U$ and $L$ are characteristic velocity and length scales of ${\bf v}^\prime$, then $\eta_t \sim U L / 3$ \citep[e.g.][]{ossen03}.  Here we have $U \sim 250$ m$^{-1}$ and $L \sim$ 34 Mm (Sec.\ \ref{sec:ssd}), which suggests a value of $\eta_{MFT} \sim 10^{13}$ cm$^2$ s$^{-1}$ - somewhat higher than the observational estimate.

In the remainder of this section, we take a closer look at these estimates.

\subsection{The turbulent electromotive force (emf)}\label{sec:emf}

In kinematic mean-field dynamo theory, turbulent diffusion is only one component to the turbulent electromotive force (emf), $\emf$.  Though more general averaging procedures are sometimes used, here we define the turbulent emf as
\begin{equation}\label{eq:emf}
\emf = \left<{\bf v}^\prime \cross {\bf B}^\prime\right> 
\end{equation}
where the angular brackets indicate averages over longitude and primes indicate non-axisymmetric components, e.g.\ ${\bf v}^\prime = {\bf v} - \left<{\bf v}\right>$.  

The ansatz of turbulent diffusion as expressed in eq.\ (\ref{induction1}) assumes that
\begin{equation}\label{eq:demf}
\demf \approx -\eta_t \curl \left<{\bf B}\right> ~~~~.
\end{equation}
This is the hypothesis we wish to test.  In particular, we can compute $\emf$ explicitly in Case C1 from the non-axisymetric components of ${\bf v}$ and ${\bf B}$ and then see if it is essentially diffusive in nature as expressed in eq.\ (\ref{eq:demf}).  

If we focus on the longitudinally-averaged radial field $\left<B_r\right>$ at the surface ($r=R$), then the relevant component of the emf in eq.\ (\ref{eq:emf}) is ${\cal E}_\phi = -\left<V_\theta^\prime B_r^\prime\right>$.  When comparing this to the diffusive parameterization in eq.\ \ref{eq:demf}, we wish to focus on the component of $\demf$ that captures the horizontal diffusion of vertical field at the surface, which is
\begin{equation}\label{eq:semf}
{\cal D}_\phi \approx \frac{\eta_t}{R}\left(\frac{\partial <B_r>}{\partial \theta}\right)_{r=R} ~~~.
\end{equation}     

We find that a value of $\eta_t \sim 10^{12}$ cm$^2$ s$^{-1}$ gives similar amplitudes for ${\cal E}_\phi$ and ${\cal D}_\phi$.  This is an order of magnitude smaller than the mean-field estimate of $10^{13}$ cm$^2$ s$^{-1}$ given at the beginning of Section \ref{sec:discussion} but comparable to Case A1 and to observational estimates.  

However, we found little correlation between ${\cal E}_\phi$ and ${\cal D}_\phi$.  In fact, there is a weak anti-correlation.  Averaging each over a one-month interval in the early and late phases of a typical cycle yields correlation coefficients of about -0.06 and -0.08 respectively.  This result was somewhat of a surprise but it arises because this measure highlights the non-diffusive aspects of the convective transport.  

To appreciate this, consider that ${\cal E}_\phi$ is largest where $B_r^\prime$ is largest, namely regions of unipolar field within BMRs and in the polar caps.  Here the convective flow tends to advect vertical magnetic field into regions of horizontal convergence--the equivalent of downflow lanes, though $v_r^\prime$ is effectively zero in Case C1 (see Sec.\ \ref{sec:ssd}).  Thus, $B_r^\prime$ is squeezed before it is dispersed.  This squeezing action is anti-diffusive and opposes the background diffusion, which is weak but non-negligible on the scale of the convergence lanes ($\eta_{top} = 5 \times 10^{11}$ cm$^2$ s$^{-1}$).  

This mimics the evolution of vertical magnetic flux in the solar photosphere, which is non-diffusive at early times, as the flux is advected toward supergranular boundaries to form the magnetic network, and diffusive at spatial and temporal scales much larger than those of the convective motions \citep{cadavid99}.  This long-term diffusive behavior is difficult to capture by means of the pointwise comparison of ${\cal E}_\phi$ and ${\cal D}_\phi$. More sophisticated techniques are needed to quantify the effective diffusion coefficient, such as correlation tracking of magnetic flux elements.  In the next Section (Sec.\ \ref{sec:migration}) we consider an alternative approach, which is to quantify the speed at which residual BMR flux is transported to the poles.

\subsection{Poleward Migration}\label{sec:migration}

One way to quantify the efficiency of surface flux transport is by estimating the rate at which trailing flux migrates from mid-latitudes to the poles.  In particular, we see that the butterfly diagrams in Figures \ref{fig:bfly1}\textit{a}, \ref{fig:bfly_con}\textit{a} and \ref{fig:all_bfly} are dominated by successive streams of mixed polarities that reflect the propagation of radial field from latitudes below about $\pm 40^\circ$ to latitudes above $\pm 70^\circ$.  How do these propagation/migration rates vary among the different cases?  In this section we address that question quantitatively by means of a correlation function.

We proceed by first constructing a 2D map of the mean, radial, surface magnetic field $\left<B_r\right>(r=R,\theta, t)$ for each case as in Fig.\ \ref{fig:all_bfly} but now limited to the latitude range between 40$^\circ$ and 70$^\circ$.  For shorthand, we will refer to this 2D map as $B(\theta_k,t)$ where $k$ is the discrete colatitudinal index.  Then we compute a revised map, $B(\theta_k,t^\prime)$ by shifting each row in time such that $t^\prime = t - (\lambda - 40^\circ)/V_p$ where $\lambda = 90^\circ - \theta$ is the latitude in degrees and $V_p$ is a specified tracking speed.  

We then compute a correlation function between different rows of the shifted map as
\begin{equation}\label{cc1}
c(i,k; V_P) = \frac{\int B(\theta_i,t^\prime) B(\theta_k,t^\prime) dt^\prime}{\sqrt{\int B^2(\theta_i,t^\prime) dt^\prime \int B^2(\theta_k,t^\prime) dt'}}
\end{equation}
Here indices $i$ and $k$ span all latitudes between 40$^\circ$ and 70$^\circ$.  The time integration extends over three cycles, as shown in Fig.\ \ref{fig:all_bfly}.

\begin{figure}[!h]
\centering
\includegraphics[width=0.75\textwidth, angle=0]{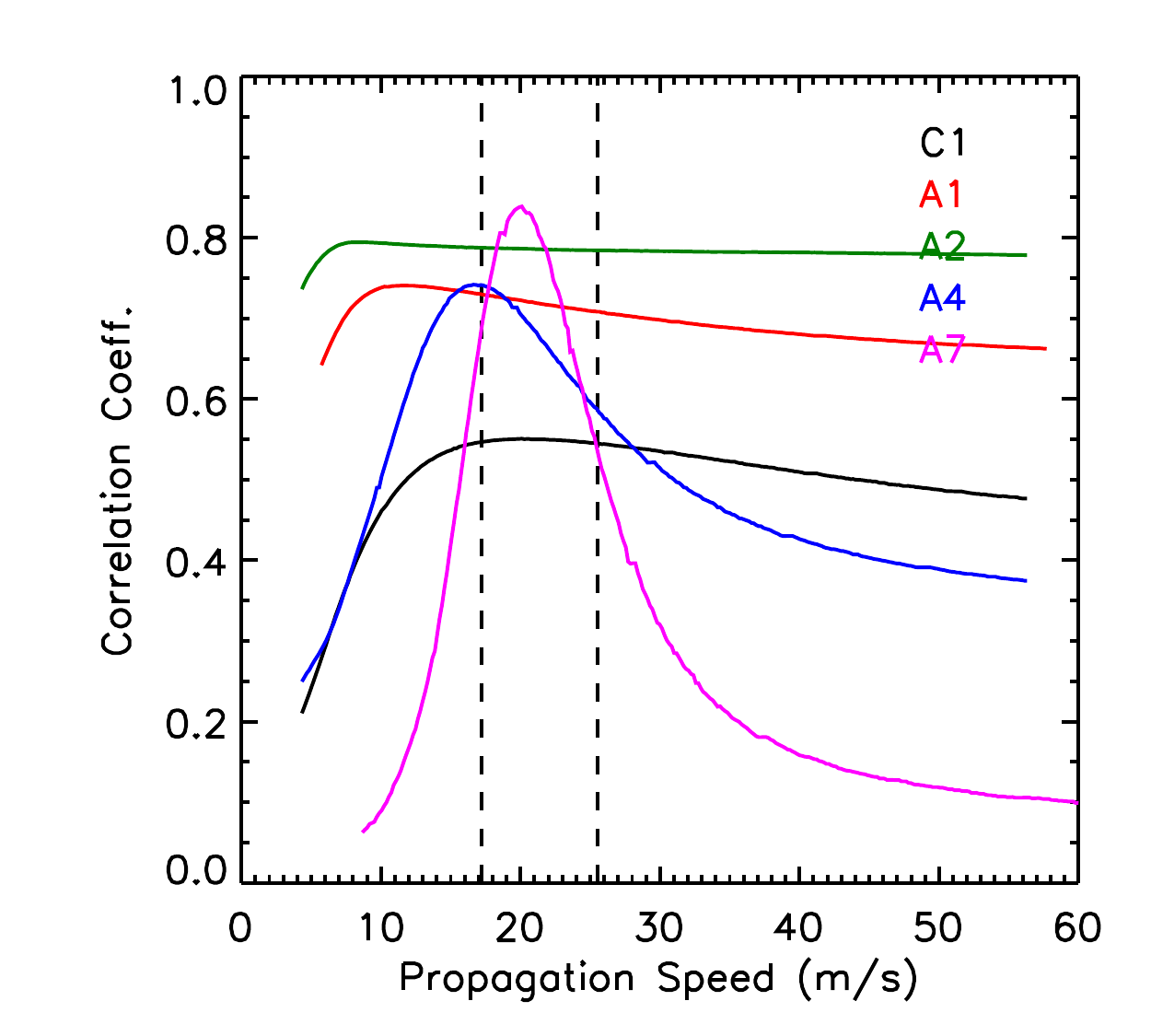}
\caption[Correlation coefficient versus migration speed for different cases]{The correlation coefficient $C$ versus migration speed $V_P$ is plotted for five different cases. The black solid line represents Case C1.  Red, green, blue and magenta lines represent Cases A1, A2, A4, and A7, as indicated.  The two dotted vertical lines show the minimum and maximum speed of the meridional circulation in the chosen latitude range of $40^\circ$--$70^\circ$.}\label{fig:migration}
\end{figure}

Then we sum the result of eq.\ (\ref{cc1}) over all $i$, $k$ pairs to obtain a single correlation coefficient for each value of $V_p$:
\begin{equation}
C(V_p) = \frac{1}{N} \sum_i \sum_{k>i} c(i,k;V_p)
\end{equation}
where N is the total number of pairs in the summation ($k>i$).  The final step is to plot $C(V_P)$ and choose the value of $V_p$ that maximizes the correlation as the characteristic flux migration speed for each case.  The results are listed in Table \ref{cases} and the correlation plots for several cases are compared in Fig.\ \ref{fig:migration}.

The meridional circulation is mainly responsible for the poleward transport of magnetic flux but convective transport can also play an important role. For those cases in which the convective transport is represented by a turbulent diffusion (A1-A8), one might expect that an increase in the diffusion coefficient $\eta_{top}$ might enhance the transport of flux toward the poles because both transport mechanisms (diffusion and MC) will work in concert.  However, we find the opposite; cases with higher diffusion have slower poleward migration speed (Table \ref{cases}).

The broad shape of the curve in Fig.\ \ref{fig:migration} for Case A1 reflects the general tendency for high levels of diffusion to smooth out the field.  The width of the streams is broad, so a range of $V_p$ values will give similar correlation coefficients.  However, there is a clear peak at about 11.6 m s$^{-1}$ that is substantially slower than the corresponding values of 19--20 m s$^{-1}$ for the less diffusive Cases A5, A7 and A8.  Furthermore, this migration speed is substantially slower than the meridional circulation speed (dashed lines), indicating that the presence of strong diffusion inhibits the poleward transport.

This result is akin to that discussed at the end of Section \ref{sec:nocon} with regard to the cycle period.  The higher value of diffusion implies a lower effective magnetic Reynolds number for flux transport near the surface.  For a velocity scale $U_{mc} \sim $ 20 m s$^{-1}$ at the surface and $L \sim \pi R / 2$, a value of $\eta_{top} = 3\times 10^{12}$ yields Rm $\sim 73$.  At this value of Rm, the magnetic field can slip through the plasma, violating Alfv\'en's theorem and rendering the poleward advection by the meridional flow less effective.  

\textit{The suppression of poleward transport by downward turbulent diffusion is an effect that is not seen in 2D SFT models}. In 2D SFT models, once the dipole has started building up, there is a pole-to-equator downgradient in $B_r$ at high latitude with which is associated an equatorward diffusive flux, which clearly opposes poleward transport by the meridional flow. But the suppression of poleward transport by the downward turbulent diffusion depends on the radial structure of the radial field, so the radial dimension is needed to capture it. While the surface meridional flow acts to advect flux poleward, the subsurface field lags behind, resisting this advection. As the diffusion is decreased, the migration rate approaches the rate associated with advection by the meridional flow. 

This result was previously demonstrated for axisymmetric fields by \citet{guerr12} using 1D (latitude) and 2D (latitude, radius) advection-diffusion models for $\left<B_r\right>$.  They showed that the 1D surface transport model produced poleward flux migration at the rate of the meridional flow, regardless of the value of $\eta_{top}$.  However, in the 2D model, the poleward migration rate was systematically slower than the meridional flow of the plasma and the discrepancy increased with increasing $\eta_{top}$.  This is consistent with the parameter study of \citet{Baumann04} who considered the effects of varying diffusion on 2D SFT models.  Close scrutiny of their Figure 6 reveals that the trailing polarity flux from mid-latitude BMRs begins to erode the pre-existing polar fields earlier in the cycle when the diffusion is large.  However, the timing of the polar field reversal is insensitive to $\eta_{top}$.  This is consistent with the idea that flux migrates poleward at a mean rate determined by the meridional flow but it also spreads, so that the leading edge of the flux migration reaches the poles sooner when the diffusion is large.

In some ways, solar observations appear to favor relatively high diffusion, as in Case A1.  The observed poleward migration speed of magnetic flux is about 10 m s$^{-1}$ \citep{howar81,topka82,Hathaway10b}, which is only about 60-70\% of the meridional flow speed \citep{ulric10}.  Furthermore, many 2D SFT models report optimal results when $\eta_{top}$ is on the order of 2--6 $\times 10^{12}$ cm$^2$ s$^{-1}$ \citep{wang89,jiang10b,Jiang_review15,lemer15}.  This latter value is justified by estimates of the diffusion rate from solar observations \citep{mosher77,topka82,schri96,abram11}.

The correlation curve in Figure \ref{fig:migration} for our convective Case C1 is suggestive of a high-diffusion case.  It is widely spread out and the peak value of 20.1 m s$^{-1}$ is slower than low-diffusive cases such as A7.  However, the correlation coefficient is significantly lower than in any of the diffusive cases and the characteristic migration speed is notably faster than comparable cases such as Case A1 (11.6 m s$^{-1}$).  Furthermore, despite the higher migration speed, Case C1 has the same period as Case A1. How can these features be explained?

The higher poleward migration speed in Case C1 relative to Case A1 can be attributed to the lower value of the diffusion coefficient, $\eta_{top}$.  Though the convection enhances the horizontal transport of $B_r$, we have neglected the vertical convective flow component (Sec.\ \ref{sec:ssd}) so the vertical flux transport occurs only through the parameterized turbulent diffusion.  As described earlier in this section, inefficient vertical transport allows the surface flux to be advected poleward at a speed comparable to the meridional flow speed.  This is also consistent with Case C3, which has a lower value of $\eta_{top}$ than Case C1 and a slightly higher migration speed (Table \ref{cases}).

The broad profile of the correlation curve for Case C1 in Fig.\ \ref{fig:migration} reflects the efficient horizontal transport.  As in Cases A1 and A2, the poleward streams are broad and diffuse, giving similar correlation coefficients for a range of tracking speeds.  The low value of the correlation coefficient ($< 0.55$) reflects the presence of residual mixed polarity from both leading and trailing spots, as we will discuss further in Section \ref{sec:energetics}.  We also address the cycle period of Case C1 in Section \ref{sec:energetics}, which is linked to the issue of mixed polarity and dynamo efficiency.

\subsection{Dynamo Efficiency}\label{sec:energetics}
As in any hydromagnetic dynamo, the field strengths achieved in our simulations are determined by a balance between magnetic field generation, nonlinear saturation, and ohmic diffusion.  In our model, the ohmic diffusion operates through the turbulent diffusivity $\eta$ which is varied by altering its value in the upper CZ, $\eta_{top}$.  All simulations have the same nonlinear saturation mechanism, as expressed in eq.\ (\ref{flux}), the same mean flows, and the same parameters for the flux emergence algorithm, SpotMaker, as discussed in Section \ref{sec:STABLE}.  So, one would expect simulations with lower diffusion to achieve higher field strengths.  And, this is indeed the trend seen for Cases A1-A8 in Table \ref{cases}.

Case C1 also falls within this general picture.  The mean field strengths achieved here are similar to Case A4 (Table \ref{cases}), which has a comparable value of $\eta_{top}$.  Thus, convection can transport fields, as demonstrated by the diffuse appearance of Fig.\ \ref{fig:all_bfly}\textit{a} compared to Fig.\ \ref{fig:all_bfly}\textit{d} and the broad profile of the correlation curve in Fig.\ \ref{fig:migration}, but it does not greatly enhance the ohmic dissipation.  Although Case C1 resembles Case A1 in other respects (see Sections \ref{sec:results} and \ref{sec:migration}), it's efficiency (as measured by the strength of the mean fields) is more comparable to Case A4.

The tendency for the convection to disperse fields without dissipating them is also reflected in the lower magnetic energy in the non-axisymmetric field component in Case C1 (2.29 kG) relative to Case A4 (4.47 kG).  This aspect of turbulent diffusion was emphasized in a series of papers by \citet{Piddington75, Piddington76, Piddington81}.   He argued that turbulent diffusion may not be good representation of merging and cancellation of fields, possibly over-estimating the cancellation rate by as much as 0.3 $R_m$ \citep{Piddington81}.  

In the context of our simulations, this implies that both polarities present in a particular BMR will be dispersed by convective motions, with little cancellation.   The mixed polarity in Case C1 is particularly apparent at the poles.  As mentioned in Section \ref{sec:transport}, the polar regions are surrounded by a region of opposite polarity formed from leading flux that has been advected poleward along with the trailing flux.  The low value of $\eta_{top}$ implies a small ohmic dissipation scale; flux elements must come into close proximity in order to cancel.  So, the polar field strength in Case C1 (Fig.\ \ref{fig:bfly_con}\textit{b}) is about a factor of three higher than in Case A1 (Fig.\ \ref{fig:bfly1}\textit{b}; as is the mean poloidal field--see Table \ref{cases}), and this in turn promotes the generation of strong toroidal fields.

However, when the flux in Case C1 is concentrated at the poles, it does eventually cancel.  This may explain why the cycle period in Case C1 (13.3 yrs) is less than in Case A4 (15.1 yrs) even though they have similar values of $\eta_{top}$.  In particular, the enhanced transport of mixed polarity flux to the poles by the convective motions leads to faster polar field reversals.  Thus, the similarity between the cycle periods in Cases C1 and A1 may be fortuitous.   The former may be shorter than in Case A4 because of the enhanced horizontal transport by convective motions while the latter may be shorter than in Case A4 because of higher vertical diffusion (see Section \ref{sec:migration}).

In summary, a turbulent diffusion coefficient of $\sim 3 \times 10^{12}$ cm$^2$ s$^{-1}$ as in Case A1 adequately captures the surface flux transport in Case C1 but it does not adequately capture the dissipation of magnetic energy.  Approximating convective transport with a turbulent diffusion will likely have an adverse effect on the dynamo efficiency, producing artificially weak mean fields.

\section{Summary and Conclusion}\label{sec:summary}

Turbulent transport of vertical magnetic flux by near-surface convective motions is an essential component of BL dynamo models.  In particular, it plays an important role in generating the polar fields that are the seed for the next cycle \citep{CCJ07}.  This process is captured with high fidelity by the AFT Surface Flux Transport (SFT) model, which simulates the advection of vertical magnetic flux by a horizontal flow field that is based on the observed convective power spectrum in the solar photosphere \citep{upton14a,upton14b,ugart15}.  In this study we have taken a substantial step forward in the unification of BL dynamo models and SFT models by presenting the first BL dynamo model that incorporates a realistic photospheric convection spectrum.  

Our 3D BL dynamo model, STABLE, uses the same surface flow field as AFT, though unlike AFT, ours is independent of time.  The generalization to an evolving field as used by AFT is straightforward and will be implemented in future work.  In our initial implementation, we extrapolated this surface flow field downward based on mass conservation, producing a vigorous 3D convective flow that permeated the upper portion of the convection zone  (Sec.\ \ref{sec:fr}).  However, we quickly found that this caused problems for our kinematic framework.  In particular, we found that this 3D convective flow field is an efficient small-scale dynamo, producing a chaotic field component that grew exponentially, overwhelming the cyclic field component (Sec.\ \ref{sec:ssd}).  Note that our implemented convective flow is non-helical, so there is no turbulent $\alpha$ effect, but it is capable of generating small scale fields by stretching and shearing of the vertical fields on the surface. One could in principle suppress this by increasing $\eta_{top}$ until the effective value of the magnetic Reynolds number is subcritical to dynamo action.  However, achieving $R_m \sim 50$ would require a value of $\eta_{top}$ of about $1.7 \times 10^{12}$ cm$^2$ s$^{-1}$, which is comparable to the estimate of convective transport from photospheric observations (see Sec.\ \ref{sec:intro}).  Since our objective is to {\em replace} turbulent diffusion with a more realistic depiction of convective transport, such a large value of $\eta_{top}$ is undesirable.

Instead we chose to suppress small-scale dynamo action by only implementing the horizontal convective flow field at the surface, abandoning our subsurface extrapolation.  So, the link to AFT is even stronger than in our initial implementation.  Since this horizontal flow does not capture vertical transport, we did keep a background value of the turbulent diffusion of $\eta_{top} = 5 \times 10^{11}$ cm$^2$ s$^{-1}$ in the upper layers of the convection zone ($r > 0.9R$), in addition to the explicit convective motions.  Without this, we found that the horizontal transport by the 2D surface convection was insufficient to establish dipolar parity.

Through this approach we were able to achieve viable, cyclic, solar-like dynamo models.  The general appearance and behavior of these models is similar to non-convective cases with a surface diffusion $\eta_{top} \sim 3 \times 10^{12}$ cm$^2$ s$^{-1}$ (Case A1; Secs.\ \ref{sec:transport}, \ref{sec:migration}).  This is demonstrated in particular by the flat profile of the correlation function for Case C1 shown in Figure \ref{fig:migration}.  This value of $\eta_{top}$ is comparable to the range of 2--6 $\times 10^{12}$ cm$^2$ s$^{-1}$ that is estimated from solar observations and that is often used in SFT models \citep{mosher77,topka82,wang89,schri96,jiang10b,abram11,Jiang_review15,lemer15}.

Treating the BL mechanism with explicit convective transport (Case C1) gives us more realistic surface flux transport in comparison to the case with an equivalent surface diffusion (Case A1). The BMRs that emerge at active latitudes are fragmented and dispersed by the convective flows, with less flux cancellation.  However, Case C1 does exhibit a band of opposite polarity surrounding the polar cap that is not seen is solar observations, at least not to this extent (Fig.\ \ref{fig:bfly_con}\textit{a}).  This mixed polarity in the polar region is not seen in the case with turbulent diffusion. We attribute its existence to residual leading flux that is advected poleward along with the trailing flux.  

In our models, the speed at which residual poloidal flux from BMRs migrates to the poles is determined mainly by the vertical diffusion, $\eta_{top}$.  For small values of the $\eta_{top}$ ($\lesssim 8 \times 10^{11}$ cm s$^{-1}$), this poleward migration speed approaches the meridional flow speed.  However, for high vertical diffusion (efficient mixing), the migration speed is slower, as surface flux transport is impeded by the subsurface field that is being ``dragged along''.

Several aspects of Case C1 highlight the limitations of parameterizing convective transport as a turbulent diffusion.  First, the spatial structure of the turbulent emf does not in general take the form of a downgradient diffusive flux (Sec.\ \ref{sec:emf}).  We attribute this to the tendency of convective flows to advect vertical flux into ``downflow lanes'', or more precisely, regions of horizontal convergence, before dispersing it.  The diffusive character of the convective transport only manifests itself on scales larger than that of the convection ($\gtrsim$ 34 Mm), which is almost an order of magnitude larger than the horizontal grid spacing ($\sim $ 4.3 Mm). 

A second limitation of the turbulent diffusion parameterization is an over-estimate of the ohmic dissipation (Sec.\ \ref{sec:energetics}).  The mean fields in the convective Case C1 are about a factor of three stronger than in Case A1, which has $\eta_{top} = 3 \times 10^{12}$ cm$^2$ s$^{-1}$ and the same cycle period.  This is particularly true for the peak polar fields, which are an important factor in determining the strength of the following cycle.  This over-estimate of the ohmic dissipation, or alternatively, an under-estimate of the dynamo efficiency, cannot be addressed with traditional SFT models; since these are not dynamo models, the field strengths are not regulated by the same interplay between field generation, nonlinear saturation, and ohmic dissipation.  Thus, it is a new result that has not been identified in previous studies.

In conclusion, the use of explicit convective motions is a promising way to improve the fidelity of BL dynamo models that have the capability to model 3D flows.  However, the main challenge to producing viable models of the solar cycle with this approach is to properly handle the small-scale dynamo action that will likely ensue.  This may ultimately require a more consistent MHD formulation that takes into account flow suppression by Lorentz force feedbacks.

%footnote
\blfootnote{This chapter is based on \citet{HM17}}

%% file: chapter8.tex
\begin{savequote}[110mm]
``Let me not pray to be sheltered from dangers,
but to be fearless in facing them.

Let me not beg for the stilling of my pain, but
for the heart to conquer it."
\qauthor{--Rabindranath Tagore, Geetabitan}
\end{savequote}

\chapter{Concluding Remarks}
\label{C8}
The results presented in this thesis are broadly classified into two parts. 
In the first part of the thesis, we explain various important issues regarding
the treatment of magnetic buoyancy, irregularities of the solar cycle, the effect of the different spatial
structure of meridional flow on the dynamo and how dynamo generated fields would affect the meridional flow
using 2D axisymmetric Flux Transport Dynamo model. In the second part, 
the build up of polar fields from the decay of sunspots and a proper treatment of the Babcock-Leighton
process by invoking realistic convective flows are presented using 3D Flux Transport Dynamo model.  

In 2D axisymmetric Flux Transport Dynamo models, magnetic buoyancy has been treated mainly in two
ways---a non-local method and a local method. In {\bf Chapter~\ref{C2}}, we have analyzed the advantages 
and disadvantages of both the methods. We find that none of them are satisfactory to depict the correct picture of magnetic buoyancy
because magnetic buoyancy is an inherently 3D process. Unless we go to the 3D framework of Flux Transport Dynamo models, we have
to treat the magnetic buoyancy in such simplistic way. We find that the non-local treatment of magnetic buoyancy is very robust
for a large span of parameter space but it does not take into account the depletion of flux from the bottom of the convection zone
which has a significant importance in irregularity study of the solar cycle. The local treatment of magnetic buoyancy includes the flux
depletion from the bottom of the convection zone and treats the magnetic buoyancy much realistically than the non-local treatment. But
this local treatment of magnetic buoyancy is not so robust. We also pointed out that the long-standing issue about the appearance of sunspots
in the low-latitudes needs to be studied carefully.

In {\bf Chapter~\ref{C3}}, we have studied various irregularities of the solar cycle during its decaying phase. We have reported
that the decay rate of the cycle is strongly correlated with the amplitude of the same cycle as well as the amplitude of the next
cycle from different sunspot proxies like sunspot number, sunspot area, and 10.7 cm radio flux data. We explain these correlations 
from flux transport dynamo models. We find that the correlations can only be reproduced if we introduce stochastic fluctuations in the
meridional circulations. We have also reproduced most of the correlations found in ascending and descending phase of the solar cycle from 
century long sunspot area data \citep{Sudip17} from Kodaikanal Observatory, India which are in great agreement with the correlations found earlier 
from Greenwich sunspots data.     

With the recent development in Helioseismology, plenty of results have come out about the spatial structure of meridional
circulation \citep{Zhao13,Schad13,RA15,Jackiewicz15}. Some helioseismology groups \citep{Zhao13} reported that the meridional circulation have a double cell structure in solar 
convection zone and some groups \citep{Schad13,Jackiewicz15} reported a multi-cellular structure  of meridional circulation in the convection zone. By probing
the supergranular motion \citet{Hathaway12} estimated that the meridional flow has an equatorward return flow in the upper convection zone~70 Mm below the surface.
In view of the above observed results, we have discussed in {\bf Chapter~\ref{C4}} what will happen to Flux Transport Dynamo model if we consider other structure of meridional circulation instead
of a single cell meridional circulation encompassing whole convection zone. We find that our dynamo model works perfectly fine as long as there is an equatorward propagation in the bottom of
the convection zone. Our model also works with shallow meridional circulation as found by \citet{Hathaway12}, if we consider the latitudinal pumping in our model (Section~\ref{C4:S6}).
 
The temporal variation of meridional circulation on the surface is also observed from various measurement techniques. \citet{CD01} first observed a variation with the solar cycle
from their helioseismic measurements. By measuring the magnetic elements on the surface of the Sun \citet{Hathaway10b} found a variation up to 5 m s$^{-1}$ for the solar cycle 23. Recently
\citet{Komm15} have analyzed MDI and HMI dopplergram data and reported a solar cyclic variation with detail latitudinal dependence. In {\bf Chapter~\ref{C5}} we have developed a theoretical
model coupled with the equations of the dynamo model to explain this variation of the meridional circulation with the solar cycle. In this prescribed model, we have considered the Lorentz force feedback on the meridional flow as a perturbation over the steady flows. Considering the magnetic fields have a small effect on the thermodynamics of the Sun, we have decoupled the equation of the perturbed vorticity corresponding to the perturbed meridional flow, from the unperturbed vorticity and solve it simultaneously along with the dynamo equations. We obtained a good match with the observation with a magnetic Prandtl number $P_m =1$. Since it is difficult to drive a flow below the bottom of the convection zone, we take perturbed vorticity $w_1=0$ at $r=0.70R$.

In the second part of the thesis which includes {\bf Chapter~\ref{C6}} and {\bf Chapter~\ref{C7}}, we have studied some of the aspects of solar magnetic field generation process using 3D dynamo model that were not possible to study earlier using axisymmetric 2D Flux Transport dynamo models. We have used the 3D dynamo model developed by Mark Miesch \citep{MD14,MT16} and study how polar fields build up from 
the decay of sunspots more realistically in {\bf Chapter~\ref{C6}}. We first reproduce the observed butterfly diagram and periodic solution considering higher diffusivity value than reported in earlier studies \citep{MD14,MT16} to consider it as a reference model.  
Then we study the build up of the polar fields by putting a single sunspot pair in one hemisphere and two sunspot pairs in both the hemispheres using this reference model. The build up of the polar field from the decay of sunspots is studied earlier using Surface Flux Transport model \citep{Wang89b,van98,Schrijver02,Baumann04,Baumann06,Cameron10} which solve only radial component of the induction equation on the surface of the Sun ($\theta$---$\phi$ plane). But these 2D SFT models have some inherent limitation for not considering the 3D vectorial nature of the magnetic fields and subsurface processes. We have shown that not considering the vectorial nature and subsurface process has an important effect on the development of the polar fields. We have also studied the effect of a few large sunspot pairs violating Hale's law on the strength of the polar field in this Chapter. We find that such anti-Hale sunspot pairs do produce some effect on the polar field, if they appear at higher latitudes during the mid-phase of the solar cycle---but the effect is not dramatic. 

In {\bf Chapter~\ref{C7}}, we have incorporated observed surface convective flows directly in our 3D dynamo model. As we know that the dispersal and migration of the sunspots fields in Babcock-Leighton process are mostly done by the convective flows i.e. by supergranulation and granulation. In most of the SFT and FTD models, dispersal and migration of sunspot fields are modeled as a random walk process with an effective diffusion coefficient \citep{Leighton64,devor84}. We have directly incorporated the observed convective flows on the surface of the Sun in our model to capture the B-L process much realistically than before instead of modeling them as a turbulent diffusion. While incorporating them in our 3D model, we make the convective flows to be non-helical in order to avoid the generation of the fields by helical turbulent $\alpha$ effect and extrapolated them radially inwards from the surface guided by convective simulations. Though our incorporated convective flows are non-helical, we find that they are capable of generating small scale fields by stretching the sunspot fields. Unless we apply a higher hyperdiffusion and confine the convective motion on the surface layers only, the exponential growth of small scale fields can disrupt the operation of large scale dynamo. The results obtained by confining the convective flows on the surface only are generally in good agreement with the observed surface flux evolution and with non-convective models that have a turbulent diffusivity on the order of $3 \times 10^{12}$ cm$^2$ s$^{-1}$ (300 km$^2$ s$^{-1}$). However, we find that the use of a turbulent diffusivity underestimates the dynamo efficiency, producing weaker mean fields than in the convective models. Also, the convective models exhibit mixed polarity bands in the polar regions that have no counterpart in solar observations. Also, the explicitly computed turbulent electromotive force (emf) bears little resemblance to a diffusive flux.  We also find that the poleward migration speed of poloidal flux is determined mainly by the meridional flow and the vertical diffusion.    

\section*{Future work}

The works that we have performed in this thesis, have initiated many interesting possibilities for doing further investigations. 

The various works presented throughout the thesis starting from Chapter~2 are performed at different times and the turbulent diffusivity profiles used in those studies are not identical. Since the nature of turbulent diffusivity is not known from the first principle, we have this freedom to choose various profile of the turbulent diffusivity. But what we have learned from the different studies in this thesis is various results (e.g., parity, irregular properties of the solar cycle) are sensitive to the chosen value of turbulent diffusivity in the convection zone. Therefore, our aim would be to {\bf study the turbulent transport co-efficients in the solar convection zone from 3D magnetoconvection simulations} that help to understand transport coefficient much better than what mixing length theory estimates, and combine them with the mean-field dynamo simulations to better understand the origin of solar magnetic field and its cycle.  

We would like to {\bf develop a good predictive model for solar cycle 25} using flux transport dynamo model. As we know that the main source of irregularities of the solar cycle is the fluctuation in the poloidal field  and fluctuation in the meridional circulation. The measurement of the polar field during the solar minimum gives us an estimate of the value of the polar field which is going to determine the strength of the next cycle. But we don't have any idea about the behavior of the meridional circulation in the next cycle which is also an important parameter to determine the strength of the cycle \citep{Yeates08}. As discussed in Chapter~\ref{C3}, we find an idea to model the meridional circulation of the next cycle using the information available in the present cycle. Based on that we want to develop a model which would take the information of polar field as well as the meridional circulation to predict the amplitude of the next cycle with more accuracy.

As discussed in Chapter~\ref{C5}, we find that the perturbed amplitude of the meridional circulation due to Lorentz force feedback near the bottom of the convection zone is not negligible in comparison to the unperturbed meridional flow there. Hence, the back-reaction due to the Lorentz force on the velocity fields is really important which was not considered in the most of the Flux Transport Dynamo models. Therefore, {\bf a non-kinematic Flux Transport Dynamo models are necessary to develop} which would include the back-reaction due to the Lorentz force on the velocity fields. 

The development of the 3D Flux Transport Dynamo models opens up many opportunities to study the solar magnetic fields with much more detail. As we know that sunspots appear mostly on the higher latitudes during the beginning of the solar cycle and eventually migrate towards equator with time as cycle progresses. Similar to these active latitudes, sunspots mostly appear in two longitudes of the Sun separated by 180$^{\circ}$ which is known as the active longitudes of the Sun \citep{Usoskin05,Mandal17}. It is believed that the non-axisymmetric component of the solar magnetic field is responsible for the persistence of the active longitudes. Studying active longitudes of the Sun was not possible by the previous 2D axisymmetric models. But now, we can study their existence by non-axisymmetric 3D flux transport dynamo models. So our next aim would be to {\bf study the persistence of the active longitudes of the Sun using 3D dynamo models}.

As we have explained in the Chapter~\ref{C7}, the Babcock-Leighton Mechanism of the Sun can be modeled much more realistically by invoking direct observed convective flows. But we have used a time independent flow for the results shown in the Chapter~\ref{C7}. Next {\bf a time dependent convective flows can be invoked} in our 3D model to reproduce the solar photospheric magnetic fields that would be much closer to observations. Reproducing the solar photosphere magnetic field is important because it sets the boundary condition for the coronal and heliospheric magnetic fields. So if we can reproduce the solar photospheric magnetic fields, then we can extrapolate it to get some estimate about the coronal magnetic fields structures. Hence, theoretically, we can couple the coronal magnetic fields with dynamo generated fields in the interior of the Sun. In view of current Indian mission {\it Aditya-L1} for the Sun, which is planning to measure the coronal magnetic fields, the 3D model of dynamo would be very relevant to couple the interior of the Sun with the corona.